\documentclass[prd,12pt,tightenlines,superscriptaddress,nofootinbib,notitlepage,preprintnumbers]{revtex4-2}

        \usepackage[normalem]{ulem}       
      \usepackage{xcolor}

\usepackage[colorlinks=true, pdfstartview=FitV,linkcolor=blue, citecolor=violet, urlcolor=magenta]{hyperref}

\usepackage{graphicx}
\usepackage[figuresleft]{rotating}

\usepackage[greek,british]{babel}

\usepackage{graphicx}
\usepackage{dcolumn}
\usepackage{bm}

\usepackage{latexsym,array,theorem,mathrsfs,subfigure,bm,float}
\usepackage{stackengine}
\usepackage{amsmath}
\usepackage{amssymb}
\usepackage{xcolor}
\usepackage{mathtools}
\usepackage{natbib}
\usepackage{tensor}
\usepackage[T1]{fontenc}
\usepackage{graphicx}
\usepackage[utf8]{inputenc}

\newcommand{\be}{\begin{eqnarray}}
\newcommand{\ee}{\end{eqnarray}}

\newcommand{\F}{{B}}
\newcommand{\A}{{B}}

\newcommand{\WW}{{\rm W}}
\newcommand{\nn}{\nonumber}
\newcommand{\thetaw}{\theta_{\mbox{\tiny W}}}
\newcommand{\T}{{\rm T}}
\newcommand{\rr}{{\rm r}}
\newcommand{\trr}{\tilde{\rm r}}

\newcommand{\U}{\mho}
\newcommand{\Uo}{{\rm U}}
\newcommand{\So}{{\rm S}}

\newcommand{\Ko}{{\rm K}}

\newcommand{\N}{{\rm N}}
\newcommand{\rrh}{{\rm r}_{\rm H}}

\newcommand{\xx}{\rr}

\newcommand{\X}{{x}}

\newcommand{\FF}{{\mathcal F}}
\newcommand{\J}{{\mathcal J}}

\newcommand{\mz}{m_{\mbox{\tiny Z}}}
\newcommand{\mh}{m_{\mbox{\tiny H}}}
\newcommand{\mw}{m_{\mbox{\tiny W}}}

\newcommand{\tc}{\textcolor{blue}}

\DeclareMathAlphabet\mathbfcal{OMS}{cmsy}{b}{n}
\renewcommand{\theequation}{\arabic{section}.\arabic{equation}}

\begin{document}


\title{Black holes with electroweak hair --  the detailed derivation}


\author{Romain Gervalle}
\email{romain.gervalle@univ-tours.fr}
\affiliation{
Institut Denis Poisson, UMR - CNRS 7013, Universit\'{e} de Tours, Parc de Grandmont, 37200 Tours, France}

\author{Mikhail~S.~Volkov}
\email{mvolkov@univ-tours.fr}
\affiliation{
Institut Denis Poisson, UMR - CNRS 7013, Universit\'{e} de Tours, Parc de Grandmont, 37200 Tours, France}

\affiliation{CERN, Theoretical Physics Department,  CH-1211 Geneva 23, Switzerland
}

\begin{abstract}

\vspace{1 cm}

We present a very  detailed derivation of solutions describing hairy black holes within 
the gravity-coupled Weinberg-Salam theory, 
which were previously reported in 
\href{https://doi.org/10.1103/PhysRevLett.133.171402}{Phys.Rev.Lett. 133 (2024) 171402}. 
These black holes support a strong magnetic field that polarizes the electroweak vacuum 
and creates a condensate of massive fields 
carrying superconducting currents along the black hole horizon. 
The currents, in turn,  generate a ``corona'' of magnetic vortex segments 
attached to the horizon at both ends. The condensate and corona together constitute the black hole hair. 
The extremal solutions approach, in the far field, 
the magnetic Reissner-Nordström configuration,  with a total mass that is {\it lower} than the total charge, 
$M<|Q|$, due to the negative Zeeman 
energy of the condensate. 
This makes the removal of the hair energetically unfavorable.
The maximally hairy black holes exhibit masses comparable to terrestrial values, with approximately 
11\% of their total mass stored in the hair. 
Given that these solutions arise within a well-tested theoretical framework, they are likely to have physical relevance.

\end{abstract}

\maketitle

\newpage

\tableofcontents

\section{INTRODUCTION}

In 1971, Ruffini and Wheeler formulated the no-hair conjecture according to which the 
only parameters of an isolated black hole should be its mass, angular momentum, and 
electric/magnetic charge \cite{Ruffini}. This conjecture was supported by the uniqueness 
theorems showing that all {\it electrovacuum} black holes are described by the Kerr-Newman 
metrics \cite{Israel:1967wq, Robinson:1975bv, Mazur:1982db} (see \cite{Heusler1996} for a review). 
Besides, the conjecture was confirmed by the no-hair theorems proven for a number of field 
theory models containing other (scalar, vector, etc.) physical fields 
\cite{Chase:1970omy, Bekenstein:1971hc, Bekenstein:1972ny, Bekenstein:1995un, Mayo:1996mv, Bekenstein:1996pn, Saa:1996aw, Herdeiro:2015waa}.

The first broadly recognized counterexample to the conjecture was found in 1989 \cite{Volkov:1989fii}: black holes supporting, 
outside the horizon, a non-trivial Yang-Mills field not characterized by any charge, hence ``a hair'' (see \cite{Volkov:2016ehx} for details). 
This finding was confirmed \cite{Kuenzle:1990is, Bizon:1990sr} and then generalized in many ways, 
leading to a plethora of black holes supporting a Skyrme hair \cite{DROZ1991371, Bizon:1992gb}, a 
Higgs hair \cite{BREITENLOHNER1992357, Greene:1992fw}, various types of stringy hair
 \cite{LAVRELASHVILI1993407, Mignemi:1992nt, Kanti:1995vq, Torii:1996yi}, hairy black holes 
 without spherical symmetry \cite{PhysRevLett.79.1595, Kleihaus:2000kg}, and many others (see \cite{Volkov:1998cc} for a review).

More recent findings include black holes with scalar hair obtained by adjusting the scalar potential 
\cite{Nucamendi:1995ex, Zloshchastiev:2004ny, Anabalon:2013qua} or by considering non-minimally 
coupled scalars \cite{Rinaldi:2012vy, PhysRevD.90.124063, Babichev:2013cya, Kobayashi:2014eva}. 
Besides, there exist interesting solutions describing black holes supporting ``spinning clouds'' made of 
ultra-light bosons \cite{Herdeiro:2014goa, Herdeiro:2016tmi, East:2017ovw}, ``spontaneously 
scalarized'' black holes \cite{Doneva:2022ewd}, black holes with a massive graviton hair \cite{Brito:2013xaa, Gervalle:2020mfr}, etc.

Nowadays, solutions describing hairy black holes have become so numerous that even a brief review of this subject would be difficult. 
This naturally raises the question of the relevance of these solutions for describing something that may really exist in Nature. 
Unfortunately, it seems one cannot be too optimistic in this respect. All known solutions describing hairy black holes 
were obtained either within overly simplified models, or within exotic models relying on the existence of yet 
undiscovered particles and fields, or within alternative theories of gravity whose applicability remains to be confirmed.

It is not completely excluded that gravitons may have a non-zero mass or that there exists a dilaton field. 
It is equally possible that the Dark Matter contains ultralight bosons or that the Dark Energy can be described by 
a scalar-tensor theory. 
However, none of these possibilities have been confirmed yet. Therefore, the physical relevance of solutions describing 
hairy black holes with massive gravitons, or with a dilaton field, etc., is not immediately obvious. 
The same can be said about most of the other known solutions.

Some physical relevance may perhaps be attributed to the black holes supporting a pure Yang-Mills field, 
because the latter arises in QCD. In that case, however, quantum corrections are large and should totally 
destroy the classical contribution. Therefore, it does not make much sense to describe hairy black holes 
by classical solutions in this case, although such solutions may probably be used for some effective description \cite{Alonso-Monsalve:2023brx}.

What could be physically relevant solutions describing hairy black holes? To address this question, 
we explore in this work the coupling between two well-tested physical theories: the Standard Model 
of fundamental interactions and General Relativity. The former contains the QCD sector, where 
quantum effects are dominant, but in the electroweak  sector, the quantum corrections 
are not too large. Therefore, it makes sense to study extended classical configurations in the 
Einstein-Weinberg-Salam (EWS) theory. This theory contains the electrovacuum sector with 
the Kerr-Newman black holes, but does it allow for some other black holes as well~?

In fact, allowing for a net magnetic charge, more general black hole solutions do exist in the EWS theory. 
However, as we shall see, magnetically charged hairy electroweak black holes are in general not spherically 
symmetric, hence they are difficult to construct. Spherical symmetry is possible in the unphysical 
limit of the theory when the weak mixing angle vanishes and the U(1) hypercharge field decouples. 
The corresponding hairy solutions were constructed long ago \cite{Greene:1992fw}. 
For the physical choice of the weak mixing angle, $\sin^2 \thetaw \approx 0.22$, spherical symmetry
 is possible only if the black hole magnetic charge is  $P = \pm1/e$, where $e$ is the electron charge. 
 The corresponding solutions were obtained only recently in Ref. \cite{Bai:2020ezy}
  (they will be reviewed below), but these black holes are too close to the Planck scale,
   which again raises the issue of applicability of the classical description. 
   The latter applies if $|P| \gg 1$, but spherical symmetry is then lost.

An important contribution to the analysis of non-spherical black holes was made in Refs.\,\cite{Lee:1994sk, Ridgway:1995ke}
 by considering perturbations around a magnetically charged Reissner-Nordström (RN) black hole. 
 For simplicity, the analysis was performed not within the full electroweak theory but within 
 its truncated version, not containing the Z and Higgs bosons. However, the remaining field content of the 
 model was sufficient to show that the perturbative black hole hair has a shape similar to a Platonic solid 
 with only discrete symmetries \cite{Ridgway:1995ke}. We were able to carry out a similar analysis 
 within the full electroweak theory and obtained similar results that will be presented below.

A qualitative description of magnetically charged black holes with electroweak hair was given by
 Maldacena \cite{Maldacena:2020skw}. These black holes become hairy and lose spherical 
 symmetry because of electroweak condensation — a change in the vacuum structure 
 induced by very strong magnetic fields \cite{Ambjorn:1988tm, Ambjorn:1989sz, Chernodub:2012fi, Chernodub:2022ywg}. 
 The EWS theory admits the magnetically charged RN solution supporting a
  Coulombian magnetic field $\mathcal{B} = P/r^2$, with the massive Z, W, and Higgs fields being in the vacuum. 
  If the horizon radius $r_h$ is large, then $|\mathcal{B}(r_h)| = |P|/r_h^2$ is small, and the configuration is stable.
   If $r_h$ is small and $|\mathcal{B}(r_h)| > \mw^2/e$, where $\mw$ is the W-boson mass, then the 
   configuration no longer minimizes the energy. The latter can be lowered by developing a condensate of the massive fields around the horizon.

The condensate carries superconducting currents, which form loops along the horizon and produce 
segments of magnetic vortices  orthogonal to the horizon. Due to their mutual repulsion, these vortices are 
expected to be maximally separated from each other, forming a ``corona'' around the black hole horizon. 
This breaks the rotational symmetry.

 If $|\mathcal{B}(r_h)| > \mh^2/e$, where $\mh$ is the Higgs-boson mass, then, in addition to the 
 massive condensate, the horizon should be surrounded by a region where the Higgs field 
 approaches zero, and the full electroweak symmetry is restored. In the far field, all massive fields
  approach their vacuum values, and the magnetic field becomes spherically symmetric.

 Following this scenario, various aspects of magnetic black holes have been discussed, 
 but always at a phenomenological  level \cite{Bai:2020spd,Bai:2021ewf,Ghosh:2020tdu,Estes:2022buj,Diamond:2021scl,Cho:2024vyq}. 
 Explicit solutions describing non-spherically symmetric hairy black holes have not been studied, as this would require 3D 
 numerical simulations -- a challenging task. However, our analysis of perturbations  of the RN background 
 suggests a simpler option.

Although the absolute minimum of the condensate energy corresponds to a configuration without continuous symmetries, the 
energy has another critical point describing an axially symmetric state. In this state, the corona degenerates into two oppositely 
directed multi-vortices, which is possible for any value of the magnetic charge. The individual vortices within 
the multi-vortex configuration repel each other, making the axially symmetric state unstable with respect to 
spreading into a fully fledged corona. However, this state is a stationary point of the energy and corresponds 
to an axially symmetric solution of the equations, whose construction requires only 2D simulations.

As we shall see below, the mass of the extremal and near-extremal axially symmetric hairy black holes is {\it smaller}
 than their charge. Therefore, they cannot shed their hair and transition into RN black holes, as the mass 
 of the latter exceeds the charge. However, they can further lower their energy when the two multi-vortices 
 split into elementary vortices, forming a fully spread  corona. In this state, they appear to approach 
 an absolute energy minimum and become stable.

Thus, axially symmetric hairy black holes are close to the stable hairy configuration and can evolve toward it when perturbed.

Following the above scenario, we will explicitly construct axially symmetric hairy black holes, 
and the main steps of our analysis are as follows. After introducing the EWS theory in 
Section \ref{SecII}, we describe its simplest solutions, including the RN and RN-de Sitter black holes, 
in Section \ref{SecIII}. In Section \ref{SecIV}, we study linear perturbations around the RN background 
and identify the threshold values of the event horizon radius at which electroweak condensation begins. 
The condensate remains spherically symmetric only when the magnetic charge is $P = \pm 1/e$, 
and in Section \ref{Sspher}, we review the corresponding non-perturbative hairy black hole solutions, 
originally obtained in Ref. \cite{Bai:2020ezy}. The structure of the non-spherically symmetric 
condensate for $|P| > 1/e$ is analyzed in Section \ref{Secpert} by minimizing the energy of the {\it non-linear} 
perturbations around the RN background.

Restricting to the axially symmetric sector, we present in Section \ref{Secaxial} the ansatz for the 
gravitational and electroweak fields, along with the essential details: the integral formula for the mass, 
the removal of the Dirac string singularity, gauge fixing, boundary conditions, the numerical procedure,
 and the virial relations used to monitor numerical accuracy.

The profiles of the axially symmetric solutions for $P = \pm 2/e$ are illustrated in Section \ref{Sechairy}, 
while the general properties of non-extremal axially symmetric hairy black holes are discussed in Section \ref{Sprop}. 
Section \ref{Sextr} is pivotal to our discussion: there, we present the extremal hairy solutions, explain the
 vortex structure forming the corona, and describe the phase transition when the horizon symmetry 
changes  from spherical to oblate. We also discuss the ``weakness of gravity'' leading to the relation $M < |Q|$, 
provide estimates for the mass, radius, and magnetic field of the  ``maximally hairy'' black hole, and include many other important details.

Finally, Section \ref{Ssum} contains a brief summary of results, a discussion of possible formation 
mechanisms for hairy electroweak  black holes in the early universe, and potential methods for their detection. 
Many technical subtleties are addressed in the six Appendices. Among them, \ref{AppC},
which presents models of electroweak anti-screening, and \ref{AppF},
which discusses electroweak vortices, contain new results that are of interest in
their own right, independently of their relevance to the main discussion.

We try  to be as explicit as possible and present all the essential aspects of our analysis,
aiming at a reader who may not necessarily be expert  both in General Relativity and electroweak theory. 
We explain how to introduce dimensionless variables, define the black hole mass, charge, 
currents, and quadrupole moments, and handle the constraints. We also describe how to
 switch to the regular gauge, fix the residual gauge freedom, establish the virial relations, 
 and compute the condensate energy, among other key steps. Although our text is long, 
 we believe it is worth keeping everything in one place rather than splitting it into separate parts, 
 as this would not necessarily improve readability.

A short and highly condensed version of this text can be found in \cite{Gervalle:2024yxj}. 
The results presented below generalize the  previous analysis of electroweak monopoles in 
flat space \cite{GVI,GVII}, where the electroweak condensate is generated by the magnetic field of 
a pointlike hypercharge. Electroweak monopoles closely resemble extremal hairy black holes,
 but their mass is infinite because the energy of the pointlike magnetic charge diverges. 
 Gravity provides a natural cutoff by creating an event horizon, which regularizes the 
 charge’s energy and renders the total mass finite.

\section{EINSTEIN-WEINBERG-SALAM THEORY \label{SecII}}
 \setcounter{equation}{0}

Denoting dimensionful quantities by boldfaced letters and dimensionless ones by ordinary letters, 
the action of the bosonic part of the gravity-coupled electroweak theory is
\be                                     \label{00}
{\bm {\mathcal S}}=\frac{1}{\bm c}\int\left(\frac{{\bm c}^4}{16\pi{\bm G}}\,{\bm R}+  {\bm  L}_{\rm WS} \right)\,\sqrt{-{\rm g}}\, d^4 {\bm x}\,.
\ee
Here, 
${\bm c}$ is the speed of light, ${\bm G}$ is Newton's constant, and ${\bm R}$ is the Ricci scalar. 
The electroweak Lagrangian is 
\be                             \label{L}
{\bm L}_{\rm WS}=
-\frac{1}{4}\,{\bm \WW}^a_{\bm\mu\bm\nu}{\bm \WW}^{a\bm\mu\bm\nu}
-\frac{1}{4}\,{\bm\F}_{\bm\mu\bm\nu}{\bm\F}^{\bm\mu\bm\nu} 
-({\bm D}_{\bm\mu}{\bm \Phi})^\dagger {\bm D}^{\bm \mu}{\bm \Phi}
-{\bm \lambda}\left({\bm \Phi}^\dagger{\bm \Phi}-{\bm \Phi}_0^2\right)^2.~~~~~
\ee
Here, the SU(2) and U(1) field strengths are, respectively, 
\be
{\bm \WW}^a_{\bm\mu\bm\nu}=\partial_{\bm\mu}{\bm \WW}^a_{\bm \nu}-\partial_{\bm\nu}{\bm \WW}^a_{\bm \mu}+
{\bm g}\epsilon_{abc}{\bm \WW}^b_{\bm \mu}{\bm \WW}^c_{\bm \nu},~~~~~
{\bm\A}_{\bm\mu\bm\nu}=\partial_{\bm\mu}{\bm\A}_{\bm \nu}-\partial_{\bm\nu}{\bm\A}_{\bm \mu}, 
\ee
with $\partial_{\bm\mu}=\partial/\partial {\bm x^\mu}$. 
The complex dublet scalar Higgs field  and its covariant derivative are 
\be
 {\bm \Phi}=\begin{pmatrix}
{\bm \phi}_1  \\
{\bm \phi}_2
\end{pmatrix},~~~~
{\bm D}_{\bm\mu}{\bm \Phi}=\left(\partial_{\bm \mu}- \frac{i\bm g^\prime }{2}{\bm \A}_{\bm\mu}
- \frac{i\bm g}{2}\,\tau_a {\bm \WW}^a_{\bm\mu}
\right){\bm \Phi}, 
\ee
where $\tau_a$ are the Pauli matrices. Here, ${\bm g}$ and ${\bm g}^\prime$ are the gauge couplings, 
${\bm \lambda}$ is the Higgs self-coupling,  
and 
${\bm \Phi}_0$ is the Higgs field vacuum expectation value. Their values are known from experiments
\cite{Tiesinga:2021myr}. 

Denoting ${\bm g}_0=\sqrt{{\bm g}^2+{\bm g}^{\prime 2}}$, one can introduce  dimensionless quantities:
\be
\WW^a_\mu=\frac{g}{{\bm\Phi}_0}
{\bm \WW}^a_{\bm \mu}\,
~~\A_\mu=\frac{g^\prime}{{\bm\Phi}_0}{{\bm \A}}_{\bm \mu},~~\Phi=\frac{\bm\Phi}{{\bm\Phi}_0},~~
\beta=\frac{8{\bm\lambda}}{{\bm g}_0^2} \,,~~g=\frac{\bm g}{{\bm g}_0},~~g^\prime=\frac{{\bm g}^\prime}{{\bm g}_0}\,,~~
x^\mu=\frac{{\bm x^\mu}}{{\bm l}_0}.
\ee
Here, the length scale  and the corresponding mass scale are 
\be                 \label{2.6}
{\bm l}_0=\frac{1}{{\bm g}_0{\bm\Phi}_0}=1.53\times 10^{-16}~\text{cm},~~~~~
{\bm m}_0{\bm c}^2=\frac{\pmb{\hbar}{\bm c}}{{\bm l}_0}={\pmb\hbar}{\bm c}\,{\bm g}_0{\bm\Phi}_0=128.9~{\rm GeV}.
\ee
The rescaled gauge couplings are expressed in terms of the Weinberg weak mixing angle: 
\be
g^\prime \equiv \sin\thetaw\approx \sqrt{0.223},~~~~~~g\equiv \cos\thetaw. 
\ee
The rescaled Higgs self-coupling is given by 
\be
\beta\approx 1.88.
\ee 
The absolute value of the electron charge and fine structure constant are 
\be 
{\bm e}={\pmb{\hbar} \bm{c g}}_0\times e,~~
e\equiv gg^\prime=\sin\thetaw\cos\thetaw=0.416,~~~~
\alpha=\frac{{\bm e}^2}{4\pi{\pmb{\hbar}\bm { c}}}=\frac{e^2}{4\pi}\,{\pmb{\hbar}\bm {cg}_0^2}\approx \frac{1}{137}. 
\ee
Therefore, 
\be         \label{2.9}
\frac{1}{\bm {cg}_0^2}=\frac{e^2}{4\pi\alpha}\,\pmb{\hbar}=1.89\,{\pmb\hbar},~~~~~~~
\frac{1}{\bm{g}_0}=\frac{\pmb{\hbar}\bm{c}}{\bm e}\,e=\frac{e}{4\pi\alpha}\,\bm{e}.
\ee
The relation to the Planck mass ${\bf M}_{\rm Pl}$ and Planck length ${\bf L}_{\rm Pl}$ is 
\be             \label{Planck}
{\bm m}_{0}=\sqrt{\frac{\alpha\kappa}{2e^2}}\, {\bf M}_{\rm Pl}=1.05\times 10^{-17}\times{\bf M}_{\rm Pl}\,~~~~~
{\bm l}_{0}=\sqrt{\frac{2e^2}{\alpha\kappa}}\, {\bf L}_{\rm Pl}=9.46\times 10^{16}\times{\bf L}_{\rm Pl}\,.~~
\ee
One defines  the dimensionless gravitational coupling,
\be            \label{kappa}
\kappa=\frac{8\pi {\bf G} {\bm \Phi}_0^2}{{\bm c}^4}=\frac{4\,e^2}{\alpha}
\left(\frac{{\bm m}_{\mbox{\tiny Z}}}{{\bf M}_{\rm Pl}}\right)^2=5.30\times 10^{-33}, 
\ee
where ${\bm m}_{\mbox{\tiny Z}}$ is the Z-boson mass.

Substituting  everything into \eqref{00}, 
the action becomes 
$
{\bm {\mathcal S}}={\pmb\hbar}\times {\mathcal S}
$
with 
\be                                     \label{II}
{\mathcal S}&=&\frac{e^2}{4\pi\alpha}\int \left(\frac{1}{2\kappa}\,R+{\cal L}_{\rm WS}
\right)\sqrt{-\rm g}\, d^4x,~~~~\\
{\cal L}_{\rm WS}&=&-\frac{1}{4g^2}\,\WW^a_{\mu\nu}\WW^{a\mu\nu}
-\frac{1}{4g^{\prime 2}}\,{\F}_{\mu\nu}{\F}^{\mu\nu} 
-(D_\mu\Phi)^\dagger D^\mu\Phi
-\frac{\beta}{8}\left(\Phi^\dagger\Phi-1\right)^2, \nn 
\ee
where  ${\F}_{\mu\nu}=\partial_\mu{\A}_\nu
-\partial_\nu{\A}_\mu$
and 
\be
\WW^a_{\mu\nu}=\partial_\mu\WW^a_\nu
-\partial_\nu \WW^a_\mu
+\epsilon_{abc}\WW^b_\mu\WW^c_\nu,~~~D_\mu\Phi
=\left(\partial_\mu-\frac{i}{2}\,{\A}_\mu
-\frac{i}{2}\,\tau_a \WW^a_\mu\right)\Phi.
\ee
This is the theory of the interacting  gravitational field ${\rm g}_{\mu\nu}$, the complex-valued  Higgs field  
$\Phi=(\phi_1,\phi_2)^{\rm T}$, 
the U(1) hypercharge field $\A= \A_\mu dx^\mu$,
and the SU(2) field $\WW=\T_a\WW^a_\mu dx^\mu$, where 
$\T_a=\tau_a/2$  are  the SU(2) generators. 

 The action \eqref{II} is invariant under spacetime diffeomorphisms and  also under 
SU(2)$\times$U(1) gauge transformations, 
\be                               \label{gauge}
\Phi\to {\rm \U}\,\Phi,~~~
{\cal V}\to {\rm \U}\,{\cal V}\,{\rm \U}^{-1}
+2i\,{\rm \U}\,\partial_\mu {\rm \U}^{-1}dx^\mu\,,
\ee
where 
\be                            \label{U}
{\cal V}=(B_\mu+\tau_a\WW^a_\mu)\, dx^\mu\,,~~~~
{\rm \U}=\exp\left(\frac{i}{2}\,\upsilon+\frac{i}{2}\,\theta^a\tau_a\right), 
\ee
with  $\upsilon$ and $\theta^a$ being  functions of $x^\mu$.

\subsection{Field equations} 
 
Varying the action \eqref{II} with respect to $\A_\mu,\WW^a_\mu$, and $\Phi$ 
yields the EW equations:
\begin{align}           
\nabla^\mu {B}_{\mu\nu}&=g^{\prime 2}\,\frac{i}{2}\,
(\Phi^\dagger D_\nu\Phi -(D_\nu\Phi)^\dagger\Phi
),\nn 
\\
{\cal D}^\mu \WW^a_{\mu\nu}
&=g^{2}\,\frac{i}{2}\,
(
\Phi^\dagger\tau^a D_\nu\Phi
-(D_\nu\Phi)^\dagger\tau^a\Phi
)
, \nn 
\\
D_\mu D^\mu\Phi&-\frac{\beta}{4}\,(\Phi^\dagger\Phi-1)\Phi=0,      \label{P2}
\end{align}
where  
\be 
{\cal D}_\mu\WW^a_{\alpha\beta}=\nabla_\mu \WW^a_{\alpha\beta}
+\epsilon_{abc}\WW^b_\mu\WW^c_{\alpha\beta},
\ee 
with  $\nabla_\mu$
 being  the geometrical covariant derivative with respect 
to the spacetime metric ${\rm g}_{\mu\nu}$. 
Varying the action with respect to the latter yields the Einstein equations: 
\be             \label{Einst}
G_{\mu\nu}=\kappa\, T_{\mu\nu} \,,
\ee
where $G_{\mu\nu}=R_{\mu\nu}-(R/2)\,{\rm g}_{\mu\nu}$ is the Einstein tensor, and 
the energy-momentum tensor is given by 
\be                      \label{TT}
T_{\mu\nu}&=&
\frac{1}{g^2}\,\WW^a_{~\mu\sigma}\WW^{a~\sigma}_{~\nu}
+\frac{1}{g^{\prime\,2}}B_{\mu\sigma}B_\nu^{~\sigma}  
+(D_\mu\Phi)^\dagger D_\nu\Phi 
+(D_\nu\Phi)^\dagger D_\mu\Phi
+{\rm g}_{\mu\nu}\mathcal{L}_{\rm WS}\,. 
\ee
The vacuum is defined as the configuration with $T_{\mu\nu}=0$. Modulo gauge transformations, it can be chosen as 
\be
\WW^a_\mu=\A_\mu=0,~~~~~~ 
 \Phi=\begin{pmatrix}
0  \\
1
\end{pmatrix},~~~~~~~{\rm g}_{\mu\nu}=\eta_{\mu\nu},~~~~~
\ee 
where $\eta_{\mu\nu}={\rm diag}[-1,1,1,1]$ is the Minkowski metric. 

Allowing for small fluctuations around the vacuum and 
linearizing the field equations 
with respect to these fluctuations yields the perturbative mass spectrum of the theory.
The spectrum includes the massless photon and graviton, as well as 
the massive Z, W and Higgs bosons with masses
\be                                   \label{masses}
\mz=\frac{1}{\sqrt{2}},~~~
\mw=g\,\mz,~~~
\mh=\sqrt{\beta}\,\mz,
\ee
in units of the mass scale ${\bm m}_0$. 

\subsection{ADM mass}

We shall solve  the field equations to obtain 
asymptotically flat solutions with a finite ADM mass. 
The metric ${\rm g}_{\mu\nu}$ then approaches the Minkowski metric $\eta_{\mu\nu}$ at spatial infinity.
Introducing the Schwarzschild radial coordinate $r$, the mass 
is determined from the asymptotic form of the ${\rm g}_{00}$ metric 
coefficient as $r\to\infty$, 
\be               \label{Ninf}
 -{\rm g_{00}}=1-\frac{2M}{r}+\ldots = 1-\frac{2{\bf GM}}{\bm{c}^2\bm{r}}+\ldots\,,
 \ee
where ${\bm r}={\bm l}_0\, r$, and  the dots denote subleading terms. 
Therefore, 
 \be                 \label{MM1}
\frac{\bf M}{\bm{m}_0}&=&\frac{\bm{c}^2\bm{l}_0 }{{\bf{G}}\bm{m}_0}\, M=
 \frac{8\pi}{\kappa}\times \frac{M}{\pmb{\hbar}\,\bm{c g}_0^2} 
=
 \frac{e^2}{4\pi\alpha}\times
 {\cal M}=1.89\, {\cal M},
 \ee
 where we have used \eqref{2.9}, and 
 where the rescaled dimensionless mass is 
 \be       \label{M1}
 {\cal M}=\frac{8\pi}{\kappa}\,M. 
 \ee

\subsection{Electromagnetic and Z fields, currents, and magnetic charge}

There is no unique way to define electromagnetic and Z fields 
away from the Higgs vacuum, but two special  definitions are 
ofter considered in the literature. 

\subsubsection{Nambu definition}

According to Nambu,
the electromagnetic and Z fields can be defined as \cite{Nambu:1977ag},
\be                                  \label{Nambu}
\FF_{\mu\nu}&=&\frac{g}{g^\prime}\,  
\A_{\mu\nu}-\frac{g^{\prime}}{g}\,n^a\WW^a_{\mu\nu}\,,~~~
{\mathcal Z}_{\mu\nu}=\A_{\mu\nu}+n^a\WW^a_{\mu\nu}\,,~~~
\ee
where 
\be                               \label{Nambu1}
n^a=\frac{(\Phi^\dagger\tau^a\Phi)}{(\Phi^\dagger\Phi)},~~~~~~n^an^a=1.
\ee
The Higgs vacuum can be  defined as a state where, modulo gauge transformations, one has  
\be
\WW^1_\mu=\WW^2_\mu=0,~~~~ \Phi=\begin{pmatrix}
0  \\
1
\end{pmatrix}.
\ee
This choice is compatible with the equations for $\WW^1_\mu$, $\WW^2_\mu$, and $\Phi$ contained in Eqs.\eqref{P2}. 
The field strengths in this state  can be represented as 
\be
 {\cal F}_{\mu\nu}=\partial_\mu {\cal A}_\nu-\partial_\nu {\cal A}_\mu,~~~~
  {\cal Z}_{\mu\nu}=\partial_\mu {\cal Z}_\nu-\partial_\nu {\cal Z}_\mu,~~~~
 \ee
 where the potentials are given by 
 \be
 {\cal A}_\mu =\frac{g}{g^\prime}\,B_\mu+\frac{g^\prime}{g}\,\WW^3_\mu,~~~~~~~~
 {\cal Z}_\mu =B_\mu-\WW^3_\mu\,. 
 \ee
 The remaining non-trivial field equations in \eqref{P2} then reduce to: 
 \be
 \nabla^\mu {\cal F}_{\nu\mu}=0,~~~~~\nabla^\mu {\cal Z}_{\nu\mu}+\mz^2 {\cal Z}_\nu=0. 
 \ee
 Therefore, ${\cal F}_{\mu\nu}$ satisfies the source-free Maxwell equations. 
The 2-forms 
  \be
 \mathcal{F}=\frac12 {\cal F}_{\mu\nu} dx^\mu\wedge dx^\nu\,,~~~~
 \mathcal{Z}=\frac12 {\cal Z}_{\mu\nu} dx^\mu\wedge dx^\nu\,
 \ee
 are closed in the Higgs vacuum, satisfying 
 \be
 d{\cal F}=0,~~~~ d{\cal Z}=0.
 \ee
 Away from the Higgs vacuum, however, the form ${\cal F}$ is not necessarily closed, i.e., $d\FF\neq 0$
 (similarly for ${\cal Z}$), 
 since  there is no 
 fundamental reason for the Maxwell equations to hold in this regime. 
 This  is actually  useful for our purposes, as it allows us to introduce a smooth magnetic charge density.

 \subsubsection{'t\,Hooft definition}

 The electromagnetic and Z fields can be defined as closed forms   even away from the Higgs vacuum
 by using the prescription of 't\,Hooft   \cite{tHooft:1974kcl}. 
 The key observation is  the identity 
\be                             \label{id}
n^a \WW^a_{\mu\nu}+g^2\psi_{\mu\nu}=\partial_\mu{\cal W}_\nu-\partial_\nu{\cal W}_\mu,
\ee
where 
  \be            \label{Hooft}
 g^2\,\psi_{\mu\nu}=-\epsilon_{abc}\, n^a\mathcal{D}_\mu n^b\mathcal{D}_\nu n^c\,,~~~~~
 {\cal W}_\mu=n^a W^a_\mu+i\,\frac{(\Phi^\dagger\partial_\mu\Phi-\partial_\mu\Phi^\dagger\Phi )}{(\Phi^\dagger\Phi)},
 \ee
 with $\mathcal{D}_\mu n^a\equiv \partial_\mu n^a+\epsilon_{abc} \WW^b_\mu n^c$. 
 It is worth noting that the $\Phi$-contribution to $ {\cal W}_\mu$ vanishes if
 $\Phi$  is real-valued. 
 The tensor $\psi_{\mu\nu}$ is gauge-invariant, while $ {\cal W}_\mu$  changes under generic gauge transformations \eqref{gauge},
 \eqref{U} as
\be
 {\cal W}_\mu\to{\cal W}_\mu -\partial_\mu \upsilon, 
 \ee
 so that it is affected only by the U(1) part of the transformation and remains invariant under the SU(2) transformations.

 This allows one to introduce  the closed 2-forms, 
  \be               \label{FF} 
F_{\mu\nu}=\mathcal{F}_{\mu\nu}-e\psi_{\mu\nu}= \partial_\mu A_\nu-\partial_\nu A_\mu\,,~~~~~
Z_{\mu\nu}=\mathcal{Z}_{\mu\nu}+g^2 \psi_{\mu\nu}= \partial_\mu Z_\nu-\partial_\nu Z_\mu\,,~~~~~
 \ee
where the corresponding potentials are given by
 \be                     \label{2_29}
 A_\mu =\frac{g}{g^\prime}\, B_\mu-\frac{g^{\prime}}{g}\,{\cal W}_\mu,~~~~~~
 Z_\mu =B_\mu+{\cal W}_\mu. 
 \ee
 The effect of the gauge transformations \eqref{gauge},\eqref{U} is 
 \be                \label{gU1}
 A_\mu\to A_\mu+\frac{1}{e}\,\partial_\mu\upsilon,~~~~Z_\mu\to Z_\mu.
 \ee

  In the unitary gauge, where $\Phi=(0,\phi)^{\rm T}$ with real-valued $\phi$, one has $n^a=-\delta^a_3$
  and ${\cal W}_\mu=n^aW^a_\mu$, 
  therefore, 
 \be                     \label{psi}
 g^2\,\psi_{\mu\nu}=\WW^1_\mu \WW^2_\nu-\WW^1_\nu\WW^2_\mu,~~~~~~
 A_\mu=\frac{g}{g^\prime}\,\A_\mu+\frac{g^\prime}{g}\,\WW^3_\mu\,,~~~~~~
 Z_\mu=\A_\mu-\WW^3_\mu\,. 
 \ee
  This equation is often used in the literature as the definition of the field potentials.
 However, the unitary gauge is not always suitable when the Higgs field approaches zero.
 For this reason,  it is preferable to use the more general definition \eqref{2_29}, which does not require 
 imposing the unitary gauge. 
 
 Since $\psi_{\mu\nu}$ in \eqref{psi} is constructed from the massive W-field, it vanishes exponentially fast 
  in the far field zone. Consequently,  both  $\FF_{\mu\nu}$ and $F_{\mu\nu}$ 
reduce to the same Maxwell tensor at large distances. 
This implies that the global quantities expressed as surface integrals at spatial infinity 
remain identical  in both the Nambu and 't\,Hooft descriptions. 
However,  at short distances, the two frameworks provide different interpretations,  which, 
 in a certain sense, are dual to each other.

 \subsubsection{Current densities and magnetic fields}

The approach of 't\,Hooft is more commonly used in the literature since $A_\mu$ and $Z_\mu$ 
can serve as fundamental variables, allowing the entire set of electroweak equations to be 
formulated in terms of these fields (see Eqs.\eqref{eqWS} below). 
The same is not possible with the Nambu fields as they do not admit potentials, making 't\,Hooft’s formulation more fundamental in this regard.

However, a key advantage of the Nambu approach is that the 2-form 
${\cal F}_{\mu\nu}$ is not closed and allows for the definition of 
a smooth magnetic charge density, 
which can sometimes offer valuable insights. In 't\,Hooft’s approach, the magnetic charge is always pointlike; 
nevertheless, we shall see that it can effectively ``mimic'' a smooth charge distribution.

Therefore, in what follows, we will consider  both ${\cal F}_{\mu\nu}$  and ${F}_{\mu\nu}$, 
along with their associated currents, 
 \be             \label{cur}
4\pi \J^\mu &=&\frac{1}{\sqrt{-\rm g}}\,\partial_\nu \left(\sqrt{-\rm g}\,\FF^{\mu\nu}\right),~~~~~
4\pi  J^\mu =\frac{1}{\sqrt{-\rm g}}\,\partial_\nu \left(\sqrt{-\rm g}\,F^{\mu\nu}\right).
 \ee
 The rescaled $(-g/g^\prime)J^\mu$ is sometimes referred to as neutral current because it sources the Z field
 (see Eq.\eqref{eqWSb}). 
 Introducing the dual tensors,
  \be                   \label{dual0}
 \tilde{\FF}^{\mu\nu}=\frac{1}{2\sqrt{-\rm g}}\,\epsilon^{\mu\nu\alpha\beta}\FF_{\alpha\beta},~~~~
 \tilde{F}^{\mu\nu}=\frac{1}{2\sqrt{-\rm g}}\,\epsilon^{\mu\nu\alpha\beta}F_{\alpha\beta},~~~~
\ee
the Nambu definition allows us to define the magnetic current,
\be             \label{curm}
4\pi\tilde{ \J}^\mu &=&\frac{1}{\sqrt{-\rm g}}\,\partial_\nu \left(\sqrt{-\rm g}\,\tilde{\FF}^{\mu\nu}\right),~~~~~
 \ee
 which can be non-trivial. 
 In contrast, within the 't\,Hooft   definition, the corresponding  magnetic current  is locally zero. 
 
  We will focus on static and purely magnetic systems. 
 Denoting $\mathcal{N}=\sqrt{|\rm g_{00}|}$ and ${\rm h}=\sqrt{\det({\rm g}_{ik})}$, the magnetic field is defined as: 
 \be            \label{magn}
 {\cal B}^i=\tilde{\FF}^{0i}=\frac{1}{\cal N}\frac{1}{2\rm h}\,\epsilon^{ijk}{\cal F}_{jk}~~~\Rightarrow~~~
 {\cal B}_i=\frac{1}{\cal N}\frac{\rm h}{2}\,\epsilon_{ijk}{\cal F}^{jk}~~~\Rightarrow~~~
 {\cal F}^{ik}=\frac{\cal N}{\rm h}\,\epsilon^{ikm}{\cal B}_m\,. 
 \ee
The magnetic field in the 't\,Hooft   definition will be also denoted by ${\cal B}^i=\tilde{F}^{0i}$,
 with no  ambiguity,  as it will always be clear from the context which definition is being used. 
 Similarly, we define  the massive magnetic fields,
 \be
 {\cal B}^i_Z=\tilde{Z}^{0i},~~~~{\cal B}^i_{\cal Z}=\tilde{\cal Z}^{0i},
 \ee 
 but we introduce 
 a different symbol for the hypermagnetic field:
 \be
 {\rm B}^i=\tilde{B}^{0i}.
 \ee 
 
 \subsubsection{Integrated currents}

 Injecting \eqref{magn} into \eqref{cur} yields the following relations: 
 \be                   \label{xer1}
 4\pi\sqrt{\rm -g}\,{\cal J}^i=\epsilon^{ikm}\partial_k ({{\cal N}^2\cal B}_m)~~~\Rightarrow~~~
 4\pi\sqrt{\rm -g}\,\epsilon_{kmi}\,{\cal J}^i=\partial_k({{\cal N}^2\cal B}_m)-\partial_m({{\cal N}^2\cal B}_k)\,.
 \ee
 Hence, introducing 1-forms ${\cal B}={\cal B}_k dx^k$ and ${\cal J}={\cal J}_k dx^k$, we obtain: 
 \be                      \label{xer2}
 4\pi \mathcal{N}\ast \mathcal{J}=d({{\cal N}^2\cal B})=d({\cal N}\ast \FF), 
 \ee
 where the star $\ast$ denotes the Hodge duality operator on the spacelike 3-surface. 
 Both sides of these relations are 2-forms. Integrating over a 2-surface $\Sigma$ and applying Green's theorem
  yields the 
 total current through $\Sigma$,
 \be             \label{cur1}
 4\pi\, {\cal I}\equiv 4\pi \int_\Sigma \mathcal{N}\ast \mathcal{J} 
 =\oint_{\partial \Sigma} {\cal N}^2 {\cal B}_m dx^m\
 =\frac12\oint_{\partial \Sigma} \sqrt{\rm -g}\, \FF^{ik}\epsilon_{ikm}\, dx^m\,.
  \ee
 This  is the Ampère's law relating the electric current to the the circulation of the magnetic field.
 Replacing everywhere ${\cal F}_{ik}\to F_{ik}$ yields similarly the 't\,Hooft   current $I$. 
The dimensionful currents expressed in amperes are:
 \be
 \mathbfcal{I}={\bm c\bm \Phi}_0\, \mathcal{I}=\frac{e}{4\pi\alpha}\, \frac{\bm e}{\bm{t}_0}\,\mathcal{ I}=
 \mathcal{ I} \times 1.42\times10^8\text{A},
 ~~~~~\bm{I}={\bm c\bm \Phi}_0\, {I}, 
 \ee
 where ${\bm t}_0={\bm l}_0/{\bm c}=5.1\times 10^{-27}$ sec.
 
 \subsubsection{Magnetic charge}
The magnetic charge density in the Nambu definition is given  by:
 \be             \label{cur1a}
 4\pi \tilde{\J}^0 =\frac{1}{\sqrt{-\rm g}}\,\partial_\nu \left(\sqrt{-\rm g}\,\tilde{\FF}^{0\nu}\right)
 =\nabla_k{\cal B}^k=\frac{e}{2}\,\frac{1}{\sqrt{\rm -g}}\, \epsilon^{ijk}\partial_i\psi_{jk}\,,
 \ee
 whereas in the 't\,Hooft   definition,  it  is locally zero. However, 
 the total magnetic charge, defined via the 
 flux  through a closed two-surface at spatial infinity, is the same in both descriptions: 
 \be             \label{PPP}
 P=\frac{1}{4\pi}\oint_{\partial \mathbb{R}^3}  {\cal F}=\frac{1}{4\pi}\oint_{\partial \mathbb{R}^3}  {F},
  \ee
  because  at infinity one has $\FF=F$. 
  
  Since  $F$ is closed, 
 the second integral in \eqref{PPP}  is determined solely by the topology of $F$ and 
 remains unchanged as the integration surface shrinks. Consequently, within the 't\,Hooft   definition, 
 the magnetic charge  is always pointlike, akin to the Dirac monopole. 
 Its  density can only have a distributional support, in which case 
 one can introduce a potential such that, locally, $F=dA$. 
 
 At the same time, since we are working within the non-Abelian theory, 
 it is meaningful to consider a scenario where the magnetic charge is smoothly distributed in space,
 as described by the density in Eq.\eqref{cur1a}. 
 To achieve this, we use the Nambu field ${\cal F}$, which lacks potential but allows for a well-defined 
 magnetic charge within a finite volume $V$:
 \be             \label{PPP1}
 P(V)=\frac{1}{4\pi}\oint_{\partial V}  {\cal F}=\frac{1}{4\pi}\oint_{\partial V}  (F+e\psi)=
 P+\frac{e}{4\pi}\oint_{\partial V} \psi.
  \ee
  Since $\psi$ is exponentially suppressed at infinity, its flux vanishes in the asymptotic limit, ensuring 
  that $P(V)$  approaches $P$ as $V\to\infty$.

\section{ABELIAN SOLUTIONS \label{SecIII}}
\setcounter{equation}{0}

We begin our analysis with the simplest case: spherically symmetric solutions. The assumption 
of spherical symmetry is quite restrictive, allowing  either solutions with known metrics or,
for a specific value of the magnetic charge $|P|=1/e$,  simple hairy solutions. 
However, it is natural to start with the simplest cases. 
Moreover, studying  perturbations 
around the Reissner-Nordstr\"om background   provides valuable insight into more general solutions.
For these reasons, 
 we  shall focus on the spherically symmetric sector for now. 

The static and spherically symmetric metric can be  parameterized   as: 
\be                   \label{geom}
ds^2=-\sigma^2(r)N(r)\,dt^2+\frac{dr^2}{N(r)}+r^2(d\vartheta^2+\sin^2\vartheta d\varphi^2).
\ee
The static, spherically symmetric, and purely magnetic electroweak fields can be expressed as 
(see  \cite{GVI} for details):
\be                \label{RRa}
\WW=f(r)\left[ \T_2\, d\vartheta
-\nu\, T_1\,\sin\vartheta \,d\varphi\right]+\T_3\,\nu \cos\vartheta d\varphi,~~
B=\nu\cos\vartheta\, d\varphi\,,~~
 \Phi=\begin{pmatrix}
0 \\
\phi(r)
\end{pmatrix}.~~~~~~
\ee
The gauge fields exhibit a Dirac string singularity at $\vartheta=0,\pi$ due to their non-vanishing 
azimuthal components. However, as will be explicitly shown below 
in the more general axially symmetric case,   this singularity can be gauged away if the constant $\nu$ is 
given by:
\be                   \label{nun}
\nu=-\frac{n}{2},~~~~~n\in\mathbb{Z}.
\ee
Substituting \eqref{geom},\eqref{RRa}  into the Einstein-Weinberg-Salam equations \eqref{P2},\eqref{Einst},
the angular variables  separate, yielding the following system of 
ordinary differential equations:
\begin{subequations}           \label{e1}
\begin{align}
(N\sigma f^\prime)^\prime&=\sigma\left(\frac{f^2-1}{r^2}+\frac{g^2}{2}\,\phi^2 \right)f\,,  \label{e1a} \\
(r^2N\sigma \phi^\prime)^\prime&=\sigma \left(\frac{\beta r^2}{4}(\phi^2-1)
+ \frac12\,f^2\right)\phi \,,  \label{e1b} \\
m^\prime &=\frac{\kappa}{2}\left( N {\cal U}_1+ {\cal U}_0+\frac{\nu^2}{2g^{\prime 2} r^2} \right)\equiv 
\frac{\kappa}{2}\, r^2 T_{\hat{0}\hat{0}}
 \,,  \label{e1c}  \\
\frac{\sigma^\prime}{\sigma}&=\frac{\kappa}{r}\, {\cal U}_1\,,  \label{e1d} \\
(\nu^2-1)f^\prime&=(\nu^2-1)f\phi=0,\label{e1e} 
\end{align}
\end{subequations}
where 
\be           \label{e2}
{\cal U}_1=\frac{1}{g^2} f^{\prime 2} +r^2\phi^{\prime 2},~~~~~
 {\cal U}_0=\nu^2\frac{(f^2-1)^2}{2g^2\, r^2}+ \frac12\,f^2\phi^2 +\frac{\beta r^2}{8}\,(\phi^2-1)^2.
\ee
We have introduced the mass function $m(r)$, with $N(r)=1-2m(r)/r$, and denote 
$T_{\hat{0}\hat{0}}=-T^0_{~0}$ the projection of $T_{\mu\nu}$ onto  the timelike  vector of the orthonormal tetrad. 

These equations admit simple analytical solutions describing {\it Abelian} 
configurations, for which the non-linear part of the SU(2) field strength vanishes.

\subsection{Reissner-Nordstr\"om solution}
The simplest solution to Eqs.\eqref{e1} is given by 
\be           \label{QQ}
f=0,~~\phi=1,~~\sigma=1,~~N(r)=1-\frac{2M}{r}+\frac{Q^2}{r^2}~~\text{with}~~ Q^2=\frac{\kappa}{2}\,P^2,~~P=\frac{n}{2e},
\ee
where $M$ is an integration constant. 
This describes the magnetic RN black hole, 
 \be                 \label{RN} 
 ds^2&=&-N(r)\, dt^2+\frac{dr^2}{N(r)}+r^2(d\vartheta^2+\sin^2\vartheta\,d\varphi^2),~~~~~ \nn \\
 \A&=&-(n/2)\,\cos\vartheta\, d\varphi,
 ~~\WW=\T_3 \A,~~~\Phi=\begin{pmatrix}
0  \\
1
\end{pmatrix}.
 \ee
 The event horizon is located at $r_h=M+\sqrt{M^2-Q^2}$. 
  The gauge fields exhibit a Dirac string 
 singularity at $\vartheta=0,\pi$, which can be removed by 
 gauge transformations \eqref{gauge} with 
 \be
{\rm \U}_\pm=\exp\left(\pm i\nu\varphi(1+\tau_3)/2\right),
\ee
whose effect is   
\be
B\to B_\pm=B\pm \nu d\varphi,~~~~~~\WW\to \WW_\pm=\T_3 B_\pm.
\ee
The fields $W_{-}, B_{-}$ and $W_{+}, B_{+}$
should be used, respectively, in the northern and southern hemispheres, where they remain regular at the axis. 
These two locally regular gauges are related to each other in the equatorial region 
 by the  gauge transformation 
 \be
 {\rm \U}=\exp\left(i\nu\varphi\,(1+\tau_3)\right),
 \ee 
 which is regular, because  $2\nu=-n\in \mathbb{Z}$ \cite{GVII}. 
However, for notational simplicity, we shall continue using  the singular gauge as in \eqref{RN}. 

We have 
\be
dB=\frac{n}{2}\,\sin\vartheta\, d\vartheta\wedge d\varphi,
\ee 
while the electromagnetic Nambu  tensor 
$\FF$ from \eqref{Nambu} and the 't\,Hooft   tensor $F$ from \eqref{FF} are given by 
\be
\FF=F=\frac{1}{e}\,dB=dA~~~\text{with}~~~~A=\frac{1}{e}\,B,
\ee
while ${\cal Z}=Z=0$.  
This describes the radial hypermagnetic field and radial magnetic field,
\be               \label{BB}
{\rm B}\equiv {\rm B}^r=\frac{n}{2r^2},~~~~~~\mathcal{B}\equiv{\mathcal B}^r=\frac{\rm B}{e}=\frac{P}{r^2}. 
\ee
 The magnetic field is the identical to  that of a 
 Dirac monopole with magnetic charge 
 \be
 P=\frac{n}{2e},~~~n\in\mathbb{Z},
 \ee
 in agreement with Dirac's quantization condition
 \cite{Dirac:1931kp}. 
 
 The RN solution \eqref{RN} can thus be interpreted as the gravitational extension
of the flat space Dirac monopole, to which it
 reduces in the $\kappa\to 0$ limit, where one can set $N(r)=1$. 
 
 In what follows, we shall use the term ``magnetic charge'' 
 to refer not only  to $P$ but also to $Q$ defined in \eqref{QQ}, as well as to the integer $n$.

 The mass of the RN black hole  is bounded below,
 \be \label{RNeq}
 M\geq |Q|,
 \ee
 where the  bound is saturated in the extremal limit, where $r_h=M=|Q|$. In this limit,  the function $N(r)=(1-r_h/r)^2$ 
 develops a double zero at the horizon, making it degenerate and causing the surface gravity to vanish. 
                                                          
The hairy black holes studied below approach the RN solution \eqref{RN} in the far-field region. 
Remarkably,  the inequality \eqref{RNeq} 
 is violated for the extremal and near-extremal hairy black holes:  we shall see later that 
 their mass $M$ is smaller than their charge $|Q|$.        

\subsubsection*{Bertotti-Robinson solution}

The metric of the extremal RN solution, where $N=(1-r_h/r)^2$, can be rewritten in terms of a new radial coordinate 
$\eta\in(-\infty,\infty)$ defined by 
\be           \label{3_10}
d\eta =\frac{dr}{r\sqrt{N}}~~\Rightarrow~~~
r=(1+e^\eta)\,r_h\,.
\ee
Rescaling the time coordinate by a constant factor, 
the line element takes the form
\be                \label{extrBR}
ds^2&=&r_h^2\left\{-\frac{e^{2\eta}}{(1+e^\eta)^2 }\, dt^2 +(1+e^\eta)^2\left( d\eta^2+d\vartheta^2+\sin^2\vartheta d\varphi^2 \right)\right\}. 
\ee
The horizon is at $\eta\to-\infty$, where the line element reduces to 
\be                \label{BR}
ds^2&=&r_h^2\left\{-e^{2\eta}\, dt^2 +d\eta^2+d\vartheta^2+\sin^2\vartheta d\varphi^2 \right\}. 
\ee
This is the Bertotti-Robinson metric \cite{Bertotti:1959pf,Robinson:1959ev},
a  solution valid for all $\eta\in(-\infty,\infty)$, describing 
the  direct product geometry $AdS_2\times S^2$. 
The $\A,\WW,\Phi$ fields remain as in  \eqref{RN}, while the radial magnetic and hypermagnetic fields are given by \eqref{BB}
with $r=r_h=|Q|$. 


\subsection{Reissner-Nordstr\"om-de Sitter solution}
Another simple solution of \eqref{e1},\eqref{e2} arises when the Higgs field vanishes:
\be               \label{RNdSo}
f=1,~~~~\phi=0,~~~~\sigma=1,~~~~N=1-\frac{2M}{r}+\frac{g^2 Q^2}{r^2}-\frac{\Lambda}{3}\, r^2,
\ee
with the same $Q$ as in \eqref{QQ} and
\be
\Lambda=\frac{\kappa\beta}{8}. 
\ee
Here,  the Higgs field remains in the false vacuum, where $\Phi=0$, and its potential 
acts as an effective cosmological constant, making the spacetime geometry 
RN-de Sitter (RNdS). The $B$-field remains identical to that of the RN solution, while the $\WW$-field 
is pure gauge.  By performing a gauge transformation  (Eq.\eqref{gauge2} below), one can set $\WW=0$,
making the full electroweak symmetry manifest.

Since  $\WW^a_{\mu\nu}=0$, Eq.\eqref{Nambu} gives
\be
\FF_{\mu\nu}=\frac{g}{g^\prime}\,B_{\mu\nu}=\frac{g^2}{e}\, B_{\mu\nu}.
\ee  
This  corresponds to the radial hypermagnetic and magnetic fields: 
\be               \label{BRNdS}
{\rm B}=\frac{n}{2r^2},~~~~~~\mathcal{B}=\frac{g^2{\rm B}}{e}=\frac{g^2P}{r^2}. 
\ee
Thus, the magnetic charge is now $g^2P=g^2n/(2e)=gn/(2g^\prime)$, which 
 does not satisfy the standard Dirac quantization condition. This is expected, as 
the system is not in the Higgs vacuum and 
standard electrodynamics does not apply. 

However, the hypercharge  field $B$ can still be made regular via removing the string singularity 
by using  the same gauge transformations as in  the RN case.    
Notably, while the magnetic charge transforms as               
$P\to g^2 P$ compared to the RN case, in \eqref{RNdSo}  we have 
$Q^2\to g^2Q^2$, not  $Q^2\to g^4 Q^2$.

 \subsubsection*{Extremal limit}

The function $N(r)$ in \eqref{RNdSo} has three positive roots, corresponding to the inner black hole horizon, 
the outer black hole horizon $r_h$, and the 
cosmological horizon $r_{c}$. The value of $r_h$ is bounded below as 
\be             \label{horRNdS}
r_{h}\geq r_{\rm ex}= \sqrt{\frac{1-\sqrt{1-4g^2\Lambda {Q}^2}}{2\Lambda}}
= g|Q|\left(1+\frac{g^2 Q^2\Lambda}{2}+\ldots\right)= g|Q| + {\cal O}(n^3\kappa^{5/2}).~~~~
\ee
Furthermore, we have 
\be           \label{horRNdS1}
1-2\Lambda r_{\rm ex}^2=\sqrt{1-\frac{Q^2}{Q_{\rm m}^2}}\equiv \lambda^2~~~~\text{with}~~~~
Q_{\rm m}=\frac{1}{2g\sqrt{\Lambda}}\approx \frac{1.17}{\sqrt{\kappa}}. 
\ee
The lower bound $r_{h}=r_{\rm ex}$ corresponds to the extremal  limit, 
where  the two black hole horizons merge,  forming a degenerate horizon.
In this case, the function  $N(r)$  factorizes as 
\be                 \label{Nex}
N(r)=k^2(r)\left(1-\frac{r_{\rm ex}}{r}\right)^2,~~~~~k^2(r)=1-\frac{\Lambda}{3}\, \left[r^2+2\,r_{\rm ex}\,r
+3\,(r_{\rm ex})^2\right].
\ee
After a coordinate transformation similar to \eqref{3_10}, 
the horizon limit $r\to r_{\rm ex}$ leads to 
an independent solution of the Einstein equations with an 
$AdS_2\times S^2$ geometry, similar to \eqref{BR}:
\be                \label{BRa}
ds^2&=&r_{\rm ex}^2\left\{-e^{2\lambda \eta}\, dt^2 +d\eta^2+d\vartheta^2+\sin^2\vartheta d\varphi^2 \right\},~~
B=-\frac{n}{2}\,\cos\vartheta\,d\varphi,~~\WW=\Phi=0,~~~
\ee
with $\eta\in (-\infty,\infty)$ and $\lambda$ defined in \eqref{horRNdS1}. 
A more detailed discussion of this will be provided  in \ref{AppE}.

The positive root of $k^2(r)$  determines the position 
of the cosmological horizon $r_c$, which is typically much larger than the black hole horizon 
$r_{\rm ex}$ due to the smallness of $\kappa$: 
\be
r_{c}=\sqrt{\frac{3}{\Lambda}-2\,r_{\rm ex}^2}-r_{\rm ex}
=\sqrt{\frac{24}{\beta\kappa}}+{\cal O}(n\sqrt{\kappa} )\gg r_{\rm ex}= \frac{\sqrt{\kappa}|n|}{2\sqrt{2} g^\prime} + {\cal O}(n^3\kappa^{5/2}). 
\ee
 However, as the magnetic charge increases and  $Q=\sqrt{\kappa/2}\,P$  approaches $Q_{\rm m}$,
 the horizons $r_c$ and $r_{\rm ex}$ approach each other. 
In the $Q\to Q_{\rm m}$ limit,  
$N(r)$ develops a triple zero, and one obtains 
\be                   \label{numax}
P=\frac{n}{2e}=\sqrt{\frac{2}{\kappa}}\,Q_{\rm m}~~\Rightarrow~~
|n|= \frac{4 g^\prime }{\sqrt{\beta}\,\kappa}=2.60\times 10^{32},~~r_{\rm c}=r_{\rm ex}=\frac{1}{\sqrt{2\Lambda}}=\frac{2}{\sqrt{\beta\kappa}}, 
\ee
which corresponds to  the  dimensionful radius 
${\bm r}_c=r_c\,{\bm l}_0=3$ cm. 

 Further increasing the magnetic charge would lead to the 
appearance of a naked singularity. Therefore, the extremal RNdS black holes exist only for $|Q|\leq Q_{\rm m}$. 

Although the extremal (and non-extremal) RNdS solution is not asymptotically flat, we will see in 
the following that its horizon 
limit  is the same as that of the asymptotically flat extremal hairy black holes.

\section{PERTURBATIONS AROUND THE RN BLACK HOLE\label{SecIV}}
\setcounter{equation}{0}

The  RN solution \eqref{RN} encapsulates an important feature: when its
magnetic field becomes sufficiently large at the horizon, this should trigger the electroweak condensation 
\cite{Ambjorn:1988tm,Ambjorn:1989sz}, 
 leading to an instability. 
As the condensate just begins to form,  it can be described perturbatively
and viewed as the ``hair''  that appears around  the black hole. 

 Therefore,  we now consider  small perturbations around the  RN solution \eqref{RN}:
\begin{equation}
{\rm g}_{\mu\nu}\to {\rm g}_{\mu\nu}+\delta {\rm g}_{\mu\nu},~~~~
    \WW^a_\mu\rightarrow \WW^a_\mu+\delta \WW^a_\mu,\quad\quad B_\mu\rightarrow B_\mu+\delta B_\mu,\quad\quad\Phi\rightarrow\Phi+\delta\Phi.
\end{equation}
Assuming the unitary gauge, where $\Phi=(0,\phi_2)^{\rm T}$ with real-valued $\phi_2$, and 
linearizing the field equations with respect to the  perturbations, the equation for 
\be
w_\mu=\frac{1}{g}\,(\delta\WW^1_\mu+i \delta\WW^2_\mu)
\ee
 decouples from the rest 
and  reduce to the complex Proca equation: 
\be              \label{Proca}
{\cal D}^\mu w_{\mu\nu}+ie F_{\nu\sigma}w^\sigma =\mw^2\,w_\nu\,.
\ee
Here $w_{\mu\nu}={\cal D}_\mu w_\nu-{\cal D}_\nu w_\mu$, with 
${\cal D}_\mu=\nabla_\mu +ieA_\mu$, and   $A_\mu=B_\mu/e$ is the background field. 
One has 
$F_{\mu\nu}=\partial_\mu A_\nu-\partial_\nu A_\mu$.
We recall  that $B_\mu dx^\mu=-(n/2)\cos\vartheta\, d\varphi$, where $n\in\mathbb{Z}$. 

To separate the angular variables in \eqref{Proca}, we set 
\be
w_\mu dx^\mu=w_{\rm m}~~~\text{if}~~~ n>0,~~~ \text{and}~~~ w_\mu dx^\mu=\bar{w}_{\rm m}~~~\text{if}~~~n<0,
\ee
where the bar denotes complex conjugation, and 
\be               \label{Wpert}
w_{\rm m}(t,r,\vartheta,\varphi)= e^{i\omega t-i{\rm m}\varphi} \psi(r)\, 
(\sin\vartheta)^{j}\left(\tan\frac{\vartheta}{2}\right)^{\rm m} 
(d\vartheta-i\sin\vartheta d\varphi). 
\ee
Here, the orbital angular momentum is 
\be               \label{jnu}
j=\left|\frac{n}{2}\right|-1\equiv j(n),~~~~~|n|\geq 2, 
\ee
and ${\rm m}\in [-j,j]$ is the azimuthal number. The values $n=\pm 1$ are excluded 
because $w_{\rm m}$ is then unbounded at $\vartheta=0,\pi$,  and the perturbation theory does not apply. 

Injecting \eqref{Wpert} into \eqref{Proca}, the problem reduces to a 
Schr\"odinger-type equation
\be             \label{RNss}
\left(-\frac{d^2}{dr_\star^2}+N(r)\left[\mw^2-\frac{|n|}{2r^2}\right]\right)\psi(r)=\omega^2 \psi(r). 
\ee
Here, $N(r)$ 
is the metric coefficient in the RN solution \eqref{QQ}, and the 
tortoise radial coordinate $r_\star$ is defined by 
$
dr_\star={dr}/{N(r)}. 
$
If this equation  admits negative modes, that is, bounded solutions with $\omega^2<0$, 
then perturbations grow in time, and  the background is unstable.

In the flat space limit, where $N(r)=1$ and the RN solution reduces to the Dirac monopole, 
the equation \eqref{RNss} admits infinitely many bound states with $\omega^2<0$. Indeed, 
if $\omega=0$ and $r\ll 1$, one can neglect $\mw^2$ in the potential, and the solution is
$$
\psi(r)=\sqrt{r}\,\cos\left(\sqrt{2|n |-1}\,\ln(\sqrt{r})\right).
$$
This function oscillates infinitely many times as $r\to 0$, hence it does not correspond to 
the ground state. Therefore, there should be 
eigenstates  with $\omega^2<0$. It follows that  the Dirac monopoles are unstable with respect to developing 
the electroweak condensate \cite{GVI}, except  for 
$n=\pm 1$, when  the perturbation theory does not apply.

In curved space, the range of the radial coordinate $r$ in 
\eqref{RNss} is restricted to the outer black hole region, $r\geq r_{h}$, which bounds the 
attractive term  $-|n|/(2r^2)$  in the potential. As a result, large black holes are stable, while the 
small ones exhibit instabilities characterized by negative modes -- normalizable solutions 
of \eqref{RNss} with $\omega^2<0$. 

By varying the black hole size, one can identify the threshold radius $r_h=r^0_h(n)$, 
for which the first instability just settles in 
as a zero mode -- normalizable solution of \eqref{RNss} with $\omega=0$. 
This should occur when the hypermagnetic field at the horizon becomes sufficiently strong to trigger 
the electroweak condensation, leading to the condition
\be                 \label{Bww}
|{\rm B}(r^0_h)|=\frac{|n|}{2(r^0_h)^2} \simeq \mw^2=\frac{g^2}{2}~~~~~\Rightarrow~~~~r^0_h\simeq \frac{\sqrt|n|}{g}=1.13\times{\sqrt{|n|}}\,. 
\ee
Numerical solutions to Eq.\eqref{RNss} for small  $|n|$  yield  actually slightly smaller values of $r^0_h(n)$
 (see Table \ref{TabII} below). This discrepancy is explained by the fact that  the estimate $|{\rm B}|=\mw^2$ 
is obtained  for a field orthogonal to a plane, not to a spherical horizon. 
However,  when $|n|\gg 1$, the horizon becomes large and locally approximates a plane. In this regime, 
the numerical results recover the expected scaling    $r^0_h(n)=\sqrt{|n|}/g+\ldots$, where the dots denote terms 
that vanish as $|n|\to\infty$. 
\begin{table}
\renewcommand\thetable{I}
  \centering
\caption{Values of $r_h^k$ for which Eq.\eqref{RNss} with $|n|=2$ admits a $k$-node zero mode.}
 \label{TabI}   
\begin{tabular}{ |c|c| }    
 \hline
  $~k$ & $r_h^k$  \\ 
\hline 
 $~0~$ & $0.8942$  \\ 
 $~1~$ & $4.306 \times 10^{-2}$  \\ 
 $~2~$ & $1.1723 \times 10^{-3}$  \\ 
 $~3~$ & $3.1180 \times 10^{-5}$  \\ 
 $~4~$ & $8.2879 \times 10^{-7}$   \\ 
 \hline
\end{tabular}
~~
\begin{tabular}{ |c|c| } 
 \hline
  $~k$ & $r_h^k$  \\ 
\hline 
 $~5~$ & $2.2029 \times 10^{-8}$  \\ 
 $~6~$ & $5.8553  \times 10^{-10}$  \\ 
 $~7~$ & $1.5563 \times 10^{-11}$  \\ 
 $~8~$ & $4.1367 \times 10^{-13}$   \\ 
 $~9~$ & $1.0997 \times 10^{-14}$  \\ 
 \hline
\end{tabular}
~~
\begin{tabular}{ |c|c| } 
 \hline
  $~k$ & $r_h^k$  \\ 
\hline 
 $~10~$ & $3.3758  \times 10^{-16}$  \\
 $~11~$ &   $1.2763  \times 10^{-16}$  \\
  $~12~$ & $1.2379 \times 10^{-16}$   \\ 
   $~13~$ & $1.2369 \times 10^{-16}$   \\ 
     $~14~$ & $1.23688 \times 10^{-16}$   \\ 
 \hline
\end{tabular}

\end{table}

For $|n|=2$, where  $j(n)=0$, the perturbations in \eqref{Wpert}  are spherically symmetric. 
Starting   from a large value of $r_h$ and gradually decreasing  it, 
the first zero mode appears 
for $r_h^0(2)\approx 0.89$, with the corresponding wavefunction $\psi_0(r)$ being 
{\it nodeless}. As $r_h$ decreases further, 
this solution becomes a negative mode, with $\omega_0^2<0$. However, 
when $r_h=r_h^1(2)\approx 0.04$, the equation 
develops a second zero mode $\psi_1(r)$, which has  {\it one node}. 

As $r_h$ continues to decreases, $\psi_1(r)$ also becomes a negative mode, 
alongside the  nodeless $\psi_0(r)$. Further modes with an increasing number of nodes
will continue to develop as the black hole size shrinks. 
In practice, we obtain these 
bound state solutions 
by applying  the numerical multiple shooting method 
(see \cite{Gervalle:2020mfr} for details). 

\begin{figure}
    \centering
    \includegraphics[scale=0.8]{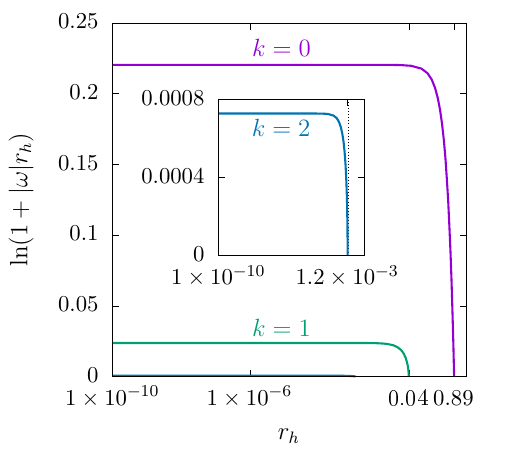}
    \includegraphics[scale=0.8]{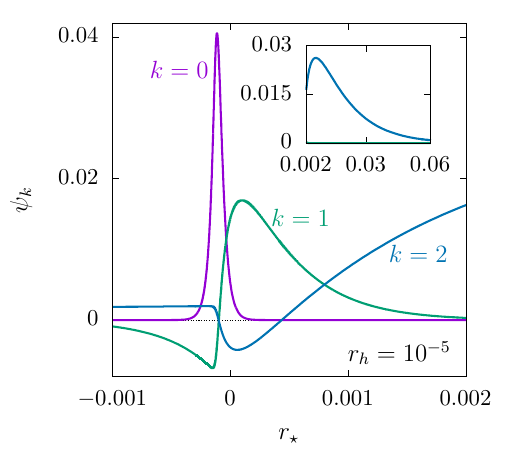}
    \caption{Left: the rescaled eigenvalue  $|\omega|\times r_h$ against $r_h$ for the bound state solutions of \eqref{RNss} 
 with  $k=0,1,2$ nodes. 
 Right: the profiles 
    of the  bound state eigenfunctions $\psi_k(r_\ast)$   with  $k=0,1,2$ nodes against the radial tortoise coordinate $r_\star$ 
    for $r_h=10^{-5}$.}
    \label{Fig1}
\end{figure}

Reducing  $r_h$ all the way 
down to the minimal value $r_h=|Q|$, Eq.\eqref{RNss} develops 
a total of 15 bound states. The values $r_h=r^k_h$ for  which the $k$-node zero mode appears 
are shown in Table \ref{TabI}. For $r_h<r_h^k$, each zero mode becomes negative, and 
the number of instabilities increases. 
For example, at $r_h=10^{-5}$, there are four negative modes; 
their profiles $\psi_k(r_\star)$ and the values of $|\omega_k(r_h)|\times  r_h$  for $k=0,1,2$
are shown  in Fig.\ref{Fig1}.  

The eigenvalues $\omega^2_k(r_h)\leq 0$  start from zero at $r_h=r_h^k$ and then 
decrease as $r_h$ gets smaller. However, 
 the product $|\omega| r_h$ 
rapidly approaches a constant value.

In summary, small RN black holes with $|n|=2$ are unstable with respect to the formation of an electroweak condensate. 
Each new instability manifests as a zero mode  $\psi_k(r)$ at $r_h=r_h^k(2)$, which represents a 
static deformation of the RN black hole by the condensate that begins to form. This 
can be viewed as a perturbative approximation  of a new static hairy solution 
 that merges  with the RN  family for  $r_h=r_h^k$ but deviates from it 
 for $r_h< r_h^k$. The new solutions are spherically symmetric since $j(n)=0$ for $|n|=2$.

\begin{figure}[b]

    \centering
        \includegraphics[scale=0.7]{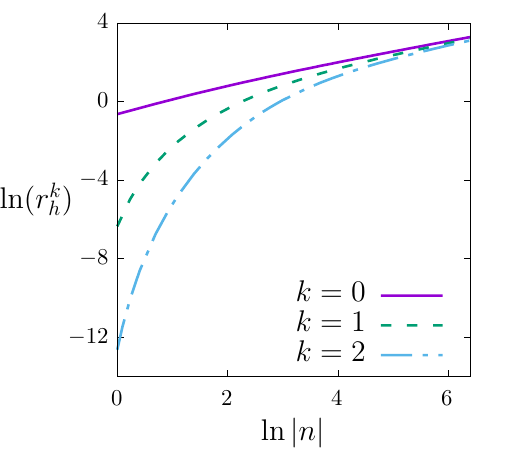}
                \includegraphics[scale=0.7]{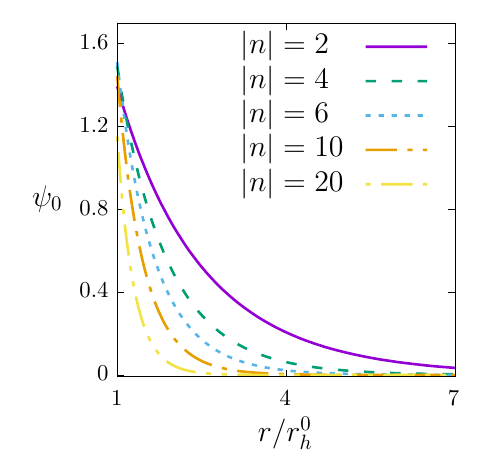}
    
         \caption{\small The RN horizon radius  $r_h^k(n)$, for which there exists a $k$-node zero mode $\psi_{k}(r)$ (left), 
         and the profile of the fundamental zero mode $\psi_0(r)$ (right). 
         }
    \label{Figw}
\end{figure}

The same arguments apply for $|n|>2$, where  $j(n)>0$ and the solutions \eqref{Wpert} are no longer spherically symmetric. 
A zero mode $\psi_k(r)$ that appears for $r_h=r^k_h(n)$ then describes  a static deformation of the RN black hole by a 
non-spherically symmetric condensate. This deformation generalizes to a static, non-perturbative hairy solution for $r_h<r^k_h(n)$. 
The values of $r^k_h(n)$ for $k=0,1,2$  and $n>2$ are given in Table \ref{TabII} and shown in Fig.\ref{Figw}. 
The profiles of the nodeless zero mode $\psi_0$  in Fig.\ref{Figw} show that the solution 
becomes increasingly localized at the horizon when $n$ increases.

\begin{table}
\renewcommand\thetable{II}
  \centering
\caption{Values $r^k_h(n)$ for which Eq.\eqref{RNss} admits a $k$-node zero mode. }
 \label{TabII}   
\begin{tabular}{ |c|c|c|c|c|c|c|c|c| }        
 \hline
  $~n$ & $2$ & $4$ & $6$ & $10$ & $20$ & $40$ & $100$ & $200$\\
  \hline
   $~r^0_h(n)$ & $~0.894~$ & $~1.465~$ & $~1.927~$ & $~2.682~$ & $~4.109~$ & $~6.164~$ & $~10.290~$ & $~14.966~$\\
   \hline
    $~r^1_h(n)$ & $~0.043~$ & $~0.245~$ & $~0.508 ~$ & $~1.054 ~$ & $~ 2.274 ~$ & $~ 4.191~$ & $~ 8.202~$ & $~12.823 ~$\\
   \hline
 $~r^2_h(n)$ & $~0.001~$ & $~0.024 ~$ & $~ 0.089  ~$ & $~0.306 ~$ & $~1.066   ~$ & $~ 2.627 ~$ & $~ 6.342 ~$ & $~10.829 ~$\\
   \hline
\end{tabular}
\end{table}

It is worth mentioning the analytical formula  obtained by solving 
Eq.\eqref{RNss} within the WKB approximation \cite{Hod:2024swn}:
\be               \label{WKB} 
r^k_h(n)=\frac{\sqrt{|n|}}{g}\left(1-\frac{2}{\sqrt{|n|}}\,(k+\frac12) \right),~~~~k=0,1,2\ldots,~~~|n|\gg 1.
\ee
For $n=200$, this gives values of $r^k_h(n)$ that agree with 
those  in the last column of Table \ref{TabII}, with relative discrepancies of
$0.4\%$, $1\%$, and $4\%$ for $k=0,1,2$,  respectively. The  agreement is reasonable.

To recapitulate, Eqs.\eqref{RNss} admits bound state solutions with $\omega=0$ for 
discrete values $r_h^k(n)$ of the black hole size. These zero modes can be used for a perturbative approximation 
of new hairy solutions that exist for $r_h<r_h^k(n)$. The hairy solutions 
are spherically symmetric only for $|n|=2$, while  for $|n|>2$, the spherical symmetry is lost.

\section{SPHERICALLY SYMMETRIC HAIRY SOLUTIONS FOR $n=\pm 2$  \label{Sspher}}
\setcounter{equation}{0}

As explained  above, the RN solution \eqref{RN} with $|n|=2$ becomes  unstable with respect to the formation 
of a spherically symmetric condensate when the  horizon radius is small. 
This instability  is expected to lead to the emergence  of more general static and  spherically 
symmetric non-Abelian solutions describing hairy black holes. Such solutions must satisfy  Eqs.\eqref{e1}. 

Notably, unless for $|n|=2$, Eq.\eqref{e1e} 
requires that $f^\prime=f\phi=0$, which only permits the simplest Abelian solutions described above. 
However, for $|n|=2$, we have $\nu^2-1=0$, so Eq.\eqref{e1e} 
 is automatically satisfied, allowing for solutions with non-constant $f(r)$ and $\phi(r)$.
 
Such solutions were previously reported in Ref.\cite{Bai:2020ezy}; 
below, we reproduce them and highlight their most important features. Although the two signs of $n=\pm 2$
correspond to  two values of the  magnetic charge $P=n/(2e)=\pm 1/e$, 
the radial amplitudes 
$N(r),\sigma(r),f(r),\phi(r)$ are the same in both cases.

Eq.\eqref{e1d} can be  solved as
\be            \label{sig}
\sigma(r)=\exp\left(-\kappa\int_r^\infty \frac{\mathcal{U}_1}{r}\,dr \right),
\ee
using which  the remaining Eqs.\eqref{e1a}--\eqref{e1c} reduce to 
\begin{subequations}           \label{ee1}
\begin{align}
(Nf^\prime)^\prime+\frac{\kappa}{r}\,{\cal U}_1 Nf^\prime&=\left(\frac{f^2-1}{r^2}+\frac{g^2}{2}\,\phi^2 \right)f\,, \label{ee1a} \\
(r^2N\phi^\prime)^\prime+\kappa r\,{\cal U}_1N\phi^\prime &=\left(\frac{\beta r^2}{4}(\phi^2-1)+\frac12\, f^2\right)\phi \,, \label{ee1b} \\
1-(rN)^\prime&=\kappa\left( N {\cal U}_1+{\cal U}_0+\frac{1}{2g^{\prime 2} r^2}\right) \,. \label{ee1c}
\end{align}
\end{subequations} 
The simplest solution is  the RN for which  $f=0$, $\phi=1$, and $N$ is given by \eqref{QQ}. 
Perturbing  this solution as $f\to \delta f$, $\phi\to 1+\delta\phi$, and $N\to N+\delta N$, then linearizing 
the equations 
with respect to the static perturbations, 
yields 
\be                   \label{eqslin}
(N\delta f^\prime)^\prime=\left(-\frac{1}{r^2}+\mw^2 \right)\delta f\,, ~~~~~~
\frac{1}{r^2}\,(r^2N\delta\phi')' = \mh^2\, \delta\phi,~~~~~~
(r\delta N)'=0.
\ee
The first of these equations   is precisely Eq.\eqref{RNss} with $|n|=2$ and $\omega=0$. 
As we know,  this equation admits a non-trivial 
solution  for $r_h=r_h^0(2)=0.894$, and we expect this perturbative zero mode  
to generalize to a non-perturbative solution
of Eqs.\eqref{ee1} when  $r_h<0.894$. 

Assuming that   at the horizon $r=r_h$, where $N(r_h)=0$ and  $N^\prime(r_h)\equiv N^\prime_h> 0$,
the equations imply 
 \be           \label{hor}
  N_h^\prime&=&\frac{1}{r_h}\left(1-\kappa\,{\cal U}_{0h} -\frac{\kappa}{2g^{\prime 2} r_h^2}\right), ~~~~~
 f^\prime_h=\frac{1}{{N_h^\prime}}\left(\frac{f_h^2-1}{r_h^2}+\frac{g^2}{2}\,\phi_h^2 \right)f_h,\nn \\
 \phi^\prime_h&=&\frac{1}{r_h^2 N^\prime_h}\left(\frac{\beta r_h^2}{4}(\phi_h^2-1)+\frac12\, f_h^2\right)\phi_h,
  \ee
 where the subscript $h$ denotes the value at $r=r_h$. 
 This determines the boundary conditions at the horizon
 in terms of  $f_h$ and $\phi_h$.  
 
 For $r\to\infty$, the perturbations should vanish, and the 
 solutions  should approach the RN form. Thus, we have in this limit 
 \be           \label{binf}
 N=1-\frac{2M}{r}+\delta N_{\cal N}, ~~~
f=C_f \, e^{-\mw r}+\delta f_{\cal N}, ~~~
 \phi=1+\frac{C_\phi}{r}\, e^{-\mh r}+\delta\phi_{\cal N}.
 \ee
 Here, $M,C_f,C_\phi$ are integration constants corresponding to 
 the massless Newtonian mode and the two massive linear modes described by \eqref{eqslin}, 
 while
 $\delta N_{\cal N}$, 
  $\delta f_{\cal N}$,  $\delta \phi_{\cal N}$ contain  the subleading terms and nonlinear corrections.
 These expressions  determine the boundary 
 conditions at large $r$. The values of $\delta N_{\cal N}$, 
  $\delta f_{\cal N}$,  $\delta \phi_{\cal N}$ can be obtained by converting \eqref{ee1}
  into integral equations and iterating 
(see \cite{Gervalle:2020mfr} for details). 

Integrating the equations with the above  boundary conditions and 
adjusting the values of the 5 free parameters  $f_h,\phi_h,M,C_f,C_\phi$ 
 within the numerical multiple shooting method \cite{Press:2007:NRE:1403886},
yields global solutions in the region $r>r_h$. 

There always exists the 
 trivial RN solution with $f(r)=0,\phi(r)=1$,  but for $r_h<0.894$,  we also find  
 solutions with non-constant $f(r)$ and $\phi(r)$ that increasingly 
deviate from the RN solution as $r_h$ decreases. 

 The horizon radius  $r_h$ of these   hairy black holes ranges  within the limits  $r_h\in [r_{\rm ex},r_h^0=0.894]$.
 The maximal value $r^0_h$ corresponds to the bifurcation with the RN solution, while 
  the minimal value $r_{\rm ex}$ determines the extremal solution  for which $N^\prime_h=0$. 
  
Notably,  the formulas \eqref{hor} remain valid even as  $N^\prime_h\to 0$, provided that  
 $f_h\to 1$, $\phi_h\to 0$, $f^\prime_h\to 0$, $\phi^\prime_h\to 0$, and $r_h\to r_{\rm ex}$. 
 The value  $r_{\rm ex}$  is determined as the root of the equation $N^\prime_h=0$  in \eqref{hor}. 
 Remarkably,   it coincides with the radius $r_{\rm ex}\approx g|Q|$  of the 
 extremal RNdS solution in  \eqref{horRNdS}. 
 
 For $|n|=2$ one finds 
 $r_{\rm ex}\approx 1.09\times  10^{-16}$, 
 which is close to the Planck length.

\begin{figure}
    \centering
    \includegraphics[scale=0.7]{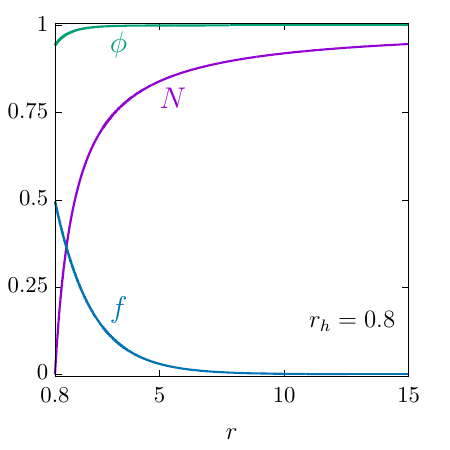}
    \includegraphics[scale=0.7]{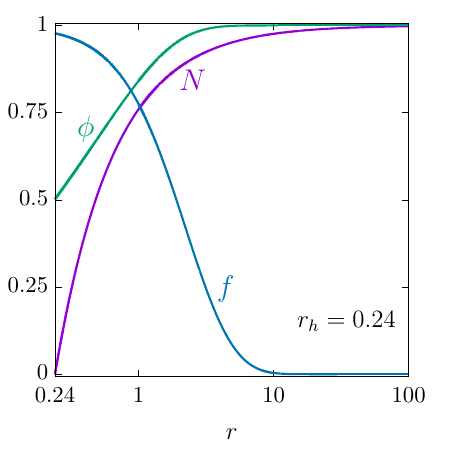}
     \includegraphics[scale=0.7]{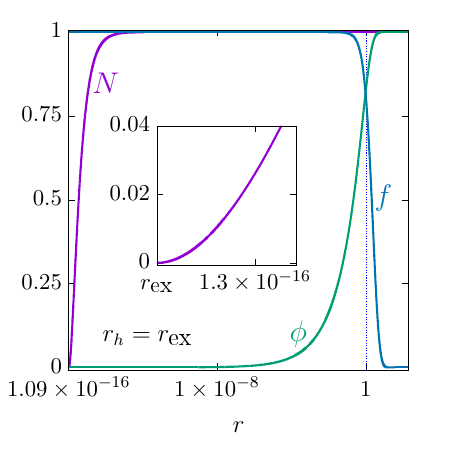}
           \caption{Profiles of the hairy solutions  with $r_h=0.8$ (left), $r_h=0.24$ (center),
           and the extremal solution with $r_h=r_{\rm ex}\approx 1.09\times 10^{-16}$ (right). In the extremal case, $N(r_h)=N^\prime(r_h)=0$. 
 }
    \label{Fig3}
\end{figure}

 Examples of hairy solutions for $r_h=0.8$, $r_h=0.24$, and 
 the extremal solution with $r_h=r_{\rm ex}$ are shown in Fig.\ref{Fig3}.
 The function 
 $\sigma(r)$ (not shown in the figures) is determined by \eqref{sig}
and  remains very close to unity since $\kappa$ is very small. 
As $r_h$ shrinks, the horizon value of the Higgs field $\phi_h$ decreases and approaches zero, 
 while $f_h$ increases and approaches unity, as illustrated in Fig.\ref{Fig3} and Fig.\ref{Fig3aa}. 
  This is a manifestation of 
 symmetry restoration in the presence of a strong magnetic field \cite{Ambjorn:1989bd}.
 When $r_h$ is small,  
the U(1) hypermagnetic field, $|{\rm B}|=n/(2r^2)=\pm 1/r^2$ (as in \eqref{BRNdS}),
becomes very large  at the horizon. As a result, the value of $\phi_h$ is driven to zero, 
along with  the value of the SU(2) field strength, which is  proportional to $(1-f_h^2)$. 
As $f_h$ increases, the 
SU(2) component of the magnetic charge shifts {\it outside} the horizon. This  phenomenon can be 
understood  as follows. 

\subsection{The charge}

Both the hairy black holes and the RN black hole share the same total magnetic charge 
measured  at infinity, given by \eqref{PPP}:
\be             \label{PPPP}
 P=\frac{1}{4\pi}\oint_{S^2}  {\cal F}=\frac{n}{2e}=\pm \frac1e\,.
  \ee
In the RN case,  this charge is entirely confined inside the horizon.
  For the hairy black hole, the charge remains inside the horizon according to the  't\,Hooft description.
 However, in the Nambu description, the charge 
splits  as $P=P_{\rm H}+P_{\rm h}$, where $P_{\rm H}$ is contained inside the horizon, 
while  $P_{\rm h}$ is distributed outside. 

The exterior charge $P_{\rm h}$   is determined by integrating the 
magnetic charge density \eqref{cur1a}:
\be                \label{Pext}
P_{\rm h}&=&\int_{r>r_h} \tilde{\mathcal{J}}^0\,\sqrt{\rm -g}\, d^3x\equiv \int_{r_h}^\infty \rho_{\rm SU(2)}(r)\, dr
=\frac{e}{8\pi}\int_{r>r_h}\epsilon^{ijk}\partial_i\psi_{jk}\, d^3x \nn \\
&=&-\frac{e}{8\pi}\oint_{r=r_h}\psi_{jk}\, dx^j\wedge dx^k
=-\frac{e}{4\pi}\oint_{r=r_h}\psi_{\vartheta\varphi}\, d\vartheta\, d\varphi=f_h^2\, g^{\prime 2} P.
\ee
Here, we used the fact that the fields \eqref{RRa} are already in the unitary gauge. Thus, Eq.\eqref{psi} implies that 
\be
g^2 \psi_{\vartheta\varphi}=\WW^1_\vartheta \WW^2_\varphi-\WW^2_\vartheta \WW^1_\varphi=\nu f^2(r)\sin\vartheta,
\ee 
which 
 decays exponentially fast at infinity. The same result can be derived  
 using the relation  $P_{\rm h}=P-P(V)$, where $P(V)$ is given by \eqref{PPP1} 
and $V$ is the volume for $r<r_h$. 

Since $f_h$ increases when $r_h$ decreases, the hair charge $P_{\rm h}$ grows  from zero to its maximal 
value of $g^{\prime 2}P$ in the extremal limit. On the other hand, 
 the horizon charge $P_{\rm H}$ decreases from $P$ to $g^2 P$ in the 
extremal limit. 

\begin{figure}
    \centering
   \includegraphics[scale=0.65]{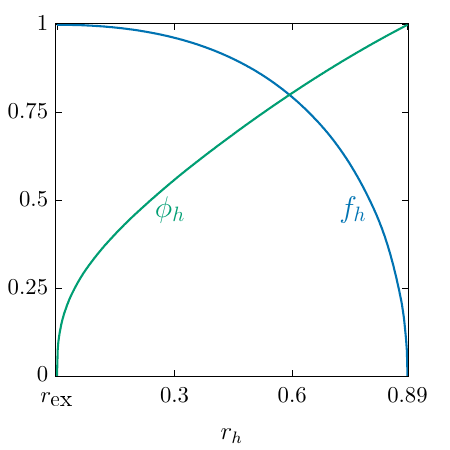}
    \includegraphics[scale=0.65]{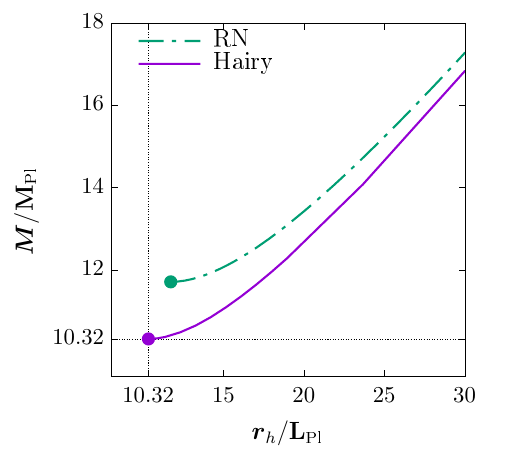}
    \includegraphics[scale=0.64]{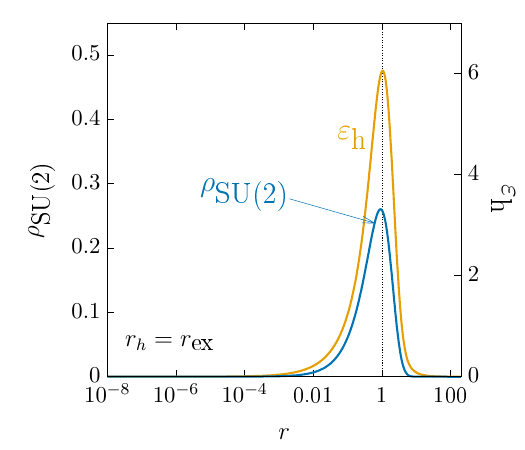}
   
           \caption{
Left: the horizon values $\phi_h=\phi(r_h)$ and $f_h=f(r_h)$ for the fundamental hairy solutions. Center: 
the mass of the hairy and RN solutions, with the former being always less energetic than the latter. 
Right:  the radial densities of the magnetic charge and of the energy 
stored in the hair. 
 }
    \label{Fig3aa}
\end{figure}

To summarize, within the Nambu description, 
the charged condensate outside the black hole horizon carries $22\%$ 
of the total charge in the extremal limit, 
$P_{\rm h}=g^{\prime 2} P\approx 0.22\,P$.  The radial profile of the condensate charge density 
$\rho_{\rm SU(2)}$, as defined in Eq.\eqref{Pext}, exhibits 
 a peak at $r\approx 1$, as shown in Fig.\ref{Fig3aa}.

 Although the descriptions by Nambu and by 't\,Hooft appear to yield different results for $n = \pm 2$, 
 for large $|n|$ 
they become mutually consistent  and, in a certain sense, ``dual'' to each other, as we shall see below.

\subsection{The ADM mass \label{SspherB}} 

The ADM mass is defined  by Eq.\eqref{Ninf}, where $M=m(\infty)$ is the asymptotic value of the mass function 
 appearing in the relation $N(r)=1-2\,m(r)/r$. 
 The mass function  $m(r)$ satisfies Eq.\eqref{e1c}, given by
 \be
m^\prime(r) =\frac{\kappa}{2}\, r^2 T_{\hat{0}\hat{0}}=
\frac{\kappa}{2}\,\left(
\frac{1}{2 g^{\prime 2} r^2}+N{\cal U}_1+{\cal U}_0
\right), 
\ee
and integrating this with the boundary condition   $N(r_h)=0$ yields 
  \be           \label{MMMa}
  M=m(\infty)=\frac{r_h}{2}+
  \frac{\kappa}{4g^{\prime 2} r_h}+
 \frac{\kappa}{2}\,  \int_{r_h}^\infty \left(
 N{\cal U}_1+{\cal U}_0
 \right) dr.
\ee
For the RN solution \eqref{QQ}, we have $f=0,\phi=1$, and $Q^2=\kappa/(2e^2)$ for $n=\pm 2$.
Using this in \eqref{MMMa} yields the mass of the RN solution:
\be          \label{MRN}
M_{\rm RN}=\frac12\left(r_h+\frac{Q^2}{r_h}\right).
\ee
As shown in the central panel of Fig.\ref{Fig3aa}, the mass  $M$ of hairy solutions is always below 
$M_{\rm RN}$ for a given $r_h$, 
so that  hairy black holes are  less energetic than the RN black holes of same size. 

As explained above, hairy black holes contain the 
charge $P_{\rm h}=f_h^2 g^{\prime 2} P$ in the hair and $P_{\rm H}=P-P_{\rm h}$ 
inside the horizon. This allows one to split the black hole mass into the ``horizon mass'' $M_{\rm H}$ and the  ``hair mass''  $M_{\rm h}$, 
\be               \label{MHhor}
M=\frac12\left(r_h+(g^2+(1-f_h^2)^2 g^{\prime 2})\,\frac{Q^2}{r_h}\right)+M_{\rm h}\equiv M_{\rm H}+M_{\rm h}\,,
\ee
where $M_{\rm H}$ is the mass of the RN black hole with charge $P_{\rm H}$ and size $r_h$. 
It turns out that $M_{\rm H}\ggg M_{\rm h}$, 
so $M\approx M_{\rm H}$ with very high precision. 

The formula \eqref{MHhor} will be justified below within a more 
general context. At  present, we simply note that in the extremal limit, where $r_h=r_{\rm ex}\approx g|Q|$ and 
$f_h=1$,  the horizon mass $M_{\rm H}$ in \eqref{MHhor} 
reduces to the sum of the first two terms on the right in \eqref{MMMa}.
Hence, $M=M_{\rm H}+M_{\rm h}$, with 
\be               \label{MHhor1}
M_{\rm H}=\frac12\left(r_{\rm ex}+\frac{g^2Q^2}{r_{\rm ex}}\right)\approx g|Q|,~~~~~
M_{\rm h}=\frac{\kappa}{2}\,  \int_{r_h}^\infty \left(
 N{\cal U}_1+{\cal U}_0
 \right) dr.
\ee
Here, the  hair mass $M_{\rm h}$ is determined by the non-linear $f,\phi$ fields forming the condensate, 
so this represents 
the mass of the condensate. 
It is instructive to consider their  values rescaled according to \eqref{M1}:
\be              \label{MHh}
{\cal M}_{\rm H}=\frac{8\pi}{\kappa}\, M_{\rm H}
=\frac{8\pi g}{\sqrt{2\kappa} e}+\mathcal{O}(\kappa^{7/2})=5\times 10^{17}\,,~~~~\nn \\
{\cal M}_{\rm h}=\frac{8\pi}{\kappa}\, M_{\rm h}=4\pi \int_{r_{\rm ex}}^\infty \left(
 N{\cal U}_1+{\cal U}_0
 \right)dr\equiv \int_{r_{\rm ex}}^\infty \varepsilon_{\rm h}(r)\, dr =15.759.
\ee
The value of ${\mathcal M}_{\rm h}$ is obtained from our numerical solution.  
As a result,  the hair mass is negligible as compared to the horizon mass, 
with ${\cal M}_{\rm h}/{\cal M}_{\rm H}\approx 3\times 10^{-17}$, meaning  $M\approx M_{\rm H}$. 

The latter  relation  has an  important consequence: although the mass of 
the extremal 
 RN black hole  is equal to the charge,
\be
M_{\rm RN}=|Q|,
\ee
the mass of the extremal hairy solution  is {\it smaller}  than the charge:
\be                 \label{MH}
M=M_{\rm H}+M_{\rm h}\approx M_{\rm H}\approx g|Q|\approx 0.88\times |Q|<|Q|.
\ee
Therefore, the hairy black hole is energetically favoured compared to the RN black hole. 
This remarkable feature can be explained by recalling  that the Zeeman energy is negative. 
The W bosons contained in the 
condensate  interact with the magnetic field of the black hole, which shifts their  mass as $\mw^2\to \mw^2-|{\rm B}|$.
This  renders  the condensate almost massless.   Although  
the condensate  carries $22\%$ of the total charge, i.e.,  $Q_{\rm h}\equiv \sqrt{\kappa/2}\,P_{\rm h}=0.22\times Q$, 
its relative contribution to the total mass is only $3\times 10^{-17}$. Therefore, the
mass-to-charge  ratio for the condensate is $M_{\rm h}/|Q_{\rm h}|\sim 10^{-16}$.

The radial energy density of the condensate, $\varepsilon_{\rm h}(r)$, defined in \eqref{MHh}, 
has a sharp peak outside the horizon 
at $r\approx 1$, where the charge density $\rho_{\rm SU(2)}(r)$ is also maximal, as shown in Fig.\ref{Fig3aa}. 
The condensate is gravitationally attracted by the black hole but repelled magnetically, 
as the horizon charge $P_{\rm H}$ 
and the condensate charge $P_{\rm h}$  have the same sign. 
Since $M_{\rm h}/|Q_{\rm h}|\sim 10^{-16}$,  the magnetic repulsion is 
16 orders of magnitude
stronger than gravitational attraction. 

This can be viewed as a manifestation 
of the weak gravity conjecture \cite{Arkani-Hamed:2006emk}. 
However, 
the condensate cannot be pushed away  to infinity because it is composed of massive fields and 
encounters  the Yukawa potential barrier at large $r$.

Since ${\cal M}\approx {\cal M}_{\rm H}$, 
the dimensionful mass and horizon size of the extremal hairy black hole are given by 
\be
\text{hairy:}~~
{\bf M}=\frac{e^2}{4\pi\alpha}\, {\cal M}\times {\bm m}_0\approx \frac{g}{\sqrt{\alpha}}\,{\bf M}_{\rm Pl}
=10.32\times {\bf M}_{\rm Pl},~~
{\bm r}_{\rm ex}=r_{\rm ex}\, {\bm l}_0\approx \frac{g}{\sqrt{\alpha}}\,{\bf L}_{\rm Pl}. ~~
\ee
For 
the extremal  RN black hole with the same magnetic charge, the mass and radius are obtained by dividing by $g$:
\be
\text{RN:}~~~~~~
{\bf M}=\frac{1}{\sqrt{\alpha}}\,{\bf M}_{\rm Pl}=11.71\times {\bf M}_{\rm Pl},~~~~~~
{\bm r}_{\rm ex}=\frac{1}{\sqrt{\alpha}}\,{\bf L}_{\rm Pl}.
\ee

\subsection{Extremal hairy black hole versus  Cho-Maison monopole \label{SspherC}}

For the extremal hairy black hole, one has $f_h=1$ and $\phi_h=0$,
so that the horizon values  of the fields correspond to  the extremal RNdS solution \eqref{RNdSo},\eqref{Nex}. 
Therefore, the function $N(r)$ for the extremal  hairy black hole has the same form as in \eqref{Nex}, 
 \be                \label{Nextr}
 N(r)=k^2(r) \left(1-\frac{r_{\rm ex}}{r}\right)^2\approx \left(1-\frac{r_{\rm ex}}{r}\right)^2.
 \ee
 Here,   $r_{\rm ex}=g|Q|+\mathcal{O}(\kappa^{5/2})\sim 10^{-16}$  is determined by 
 \eqref{horRNdS}, and 
 $k(r)$ interpolates between the same horizon limit as in \eqref{Nex}, $k(r_{\rm ex})=\sqrt{1-2\Lambda r_{\rm ex}^2}=1+\mathcal{O}(\kappa^2)$,  and $k(\infty)=1$. Hence, $k(r)\approx 1$.  
  The function   $N(r)$ 
quickly approaches  unity, and already for $r\simeq 10^2\times  r_{\rm ex}\sim 10^{-14}$,
the geometry is practically flat. As a result, the massive fields described by $f(r)$ and  $\phi(r)$  
actually live in a flat geometry, and their 
 profiles practically coincide with those for the flat space Cho-Maison monopole. 
 
 The Cho-Maison (CM) monopole \cite{Cho:1996qd} 
  is the solution in the flat space limit, when 
 $\kappa=0$ and $N=\sigma=1$, and  Eqs.\eqref{ee1}  reduce to 
  \be
  f^{\prime\prime}=\left(\frac{f^2-1}{r^2}+\frac{g^2}{2}\,\phi^2\right)f,~~~~~~~~
  (r^2\phi^\prime)^\prime=\left(\frac{\beta r^2}{4}(\phi^2-1)+\frac12 f^2\right)\phi.
  \ee
  These equations admit a smooth solution with the  boundary conditions for $0\leftarrow r\to\infty$:
  \be                 \label{CM}
  1+{\cal O}(r^2)~\leftarrow~f(r)~\to~{\cal O}\left(e^{-\mw r} \right),~~~~~~
  {\cal O}\left(r^{\frac{\sqrt{3}-1}{2}}\right)~\leftarrow~\phi(r)~\to~{\cal O}\left(e^{-\mh r} \right). 
  \ee  
  This describes a magnetic monopole with the same magnetic charge $P=\pm 1/e$ 
  as for the hairy black holes.
  The SU(2) part of the charge $g^{\prime 2}P=0.22 \times P$ 
   is smoothly distributed in space in the condensate, 
   while the U(1) part $g^2P=0.78 \times P$   is pointlike and located at the origin.
  The latter carries the energy density $1/(2g^{\prime 2} r^4)$, which 
 renders the total
  monopole energy $E_{\rm mon}$ divergent. However, 
  subtracting the divergent part  leaves a finite value:
  $$
  E_{\rm CM}=E_{\rm mon}-4\pi\int_0^\infty \frac{dr}{2g^{\prime 2} r^2}=15.759,
  $$ 
  (see \cite{GVII} for details). This is the same value as that in Eq.\eqref{MHh}, where it corresponds 
  to the  energy of the condensate
  around the black hole.

 The profiles of $f(r)$ and $\phi(r)$ for the the extremal black hole in the region $r>r_{\rm ex}$ 
 are very close to those for the  CM monopole and converge to the latter pointwise in the  $\kappa\to 0$ limit. 
 Therefore,  the extremal hairy black hole can be viewed  as the 
  CM monopole harbouring in its center a tiny black hole whose radius 
  is microscopically small compared  to the size of the monopole itself. 
  
  As seen in Fig.\ref{Fig3} (right panel), the near-horizon region where the geometry essentially 
  deviates  from Minkowski space, $10^{-16}\leq r\leq 10^{-14}$,  is {\it parametrically} small 
  compared to the region containing the condensate, where $r\simeq 1$. 
  The amplitudes $f(r)$ and $\phi(r)$ stay close to their horizon values $f_h=1,\phi_h=0$ 
  in the interval that exceeds the horizon size by approximately $12$ orders of magnitude, 
  and only for $r\geq  10^{-4}$ they start to change. Therefore,  the massive fields forming the condensate 
  actually live in flat geometry and almost do not feel the presence 
 of the black hole. However, the black hole is  important because it regularizes 
  the divergent part of the monopole energy by replacing it 
with the finite term ${\cal M}_{\rm H}$ in \eqref{MHh}, so that the total energy becomes finite. 

The hair mass ${\cal M}_{\rm h}$  of the extremal  black hole is given by the integral in \eqref{MHh}, where 
the integration starts at $r=r_{\rm ex}\sim \sqrt{\kappa}\sim 10^{-16}$. 
The regularized energy of the Cho-Maison monopole $E_{\rm CM}$ is given by the same integral, but the integration 
starts at $r=0$, hence ${\cal M}_{\rm h}=E_{\rm CM}+\mathcal{O}(\sqrt{\kappa})$. 

To recapitulate, the extremal hairy solution  can be viewed as a superposition of the flat space CM 
monopole of size 
$\sim 1$ and a black hole of radius  $\sim 10^{-16}$ (in electroweak units). 
The tiny black hole almost does not affect the monopole 
configuration but renders its energy $E_{\rm mon}$ 
finite by providing the cutoff at the horizon. This  can be 
illustrated as follows:
\be
E_{\rm mon}=\lim_{\kappa\to 0}\frac{8\pi g}{\sqrt{2\kappa} e}+15.759=\infty~~\Rightarrow~~
\mathcal{M}=\frac{8\pi g}{\sqrt{2\kappa} e}+15.759=5\times 10^{17}+15.759. ~~~~~~
\ee
When the gravity 
is taken into account,
the divergent part of the monopole energy becomes finite, albeit large, and reduces to the black hole horizon mass 
${\mathcal M}_{\rm H}=8\pi g/(\sqrt{2\kappa} e)$. 
The finite part of the monopole energy $E_{\rm CM}=15.759$ remains the same up to corrections of order $10^{-16}$ 
and becomes  the mass of the black hole hair 
$\mathcal{M}_{\rm h}=15.759$.

\begin{figure}[b]
    \centering
    \includegraphics[scale=0.7]{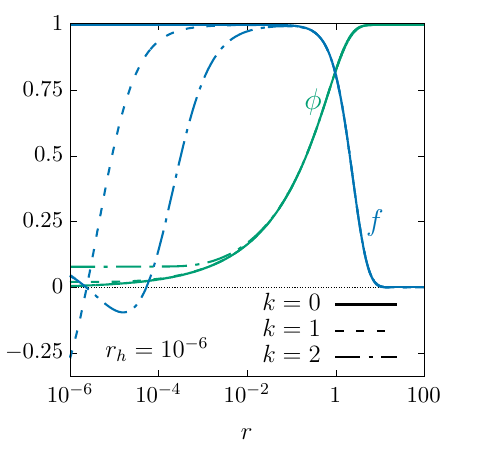}
    \caption{The amplitudes $f(r)$ and $\phi(r)$ for the  
    fundamental hairy solution  with $r_h=10^{-6}$ and 
    for its first two radial excitations.
   }
    \label{Fig4}
\end{figure}

\subsection{Stability of the extremal hairy solution \label{SspherS}}

The extremal hairy black hole with $n = \pm 2$ is expected to be stable. This expectation is motivated 
by its similarity to the flat-space Cho–Maison monopole, which has been proven to be stable \cite{GVI}. 
Although a complete stability analysis that includes gravitational effects is not yet available,
 the close resemblance between the electroweak field configurations with and without gravity 
 suggests that the stability properties should be consistent.

Next, since the condensate significantly reduces the black hole mass, the hairy black hole with $M\approx 0.88\,|Q|$ 
cannot  transition to a RN black hole, for which  $M\geq |Q|$. This would require either to increase the energy
or to reduce the magnetic charge, but the latter  is conserved due to the field topology. It is conceivable that 
the hairy black hole could decay into a lower-energy state. However, the perturbative analysis 
in the next Section suggests that the spherical condensate with $n=\pm 2$ already represents 
the absolute energy minimum. 

Therefore,  the extremal hairy 
black holes  with $n=\pm 2$ are likely to be stable.

\subsection{The excited solutions} 
For the sake of completeness, 
 it is important to mention also the excited zero mode solutions of 
Eq.\eqref{RNss}, 
$\psi(r)=\psi_k(r)$  with $k>0$, which exist for specific horizon values $r_h=r_h^k(n)$ 
shown in Tables \ref{TabI} and \ref{TabII}. 
These excited solutions  can also be promoted to static hairy black holes at the non-perturbative level. 

As an example, consider  the one-node zero mode solution $\psi_1(r)$, which exists for 
 $r_h^1(2)=4.3\times 10^{-2}$. When this solution is used  as the input for the numerical  solver 
 and the horizon radius is set to $r_h=4\times 10^{-2}<r^1_h(2)$, we find 
 a static hairy solution for which $f(r)$ is not monotonic, but instead exhibits a node. 
 Additionally, for the same value of $r_h$, there is also the black hole solution with a monotonic  $f(r)$.
 
 Thus, for $r_h=4\times 10^{-2}$,  there are two distinct hairy solutions: 
 one fundamental  and one excited. Similarly,  
 for $r_h<r^2_h(2)=1.1\times 10^{-3}$, a second radial excitation is found, and so on. 
 
 The excited solutions have not been described  in the literature. Their profiles  for $r_h=10^{-6}$ 
 are shown in Fig.\ref{Fig4}. However, such  solutions are presumably unstable, and 
 we will not discuss them anymore.

\section{PERTURBATIVE DESCRIPTION  OF HAIRY BLACK HOLES WITH  $|n|>2$  \label{Secpert}}

\setcounter{equation}{0}

We considered above spherically symmetric hairy solutions with magnetic charge $n=\pm 2$, 
beginning with their perturbative description in terms of Eqs.\eqref{Proca}--\eqref{RNss}, 
and later passing to the fully non-perturbative description.
 Now, we aim to extend the analysis to cases where $n>|2|$, when spherical symmetry is no longer preserved. 
 As the first step, we will apply the perturbative approach.

Let us get back to Eqs.\eqref{Proca}--\eqref{RNss}, which describe perturbations of the RN black hole. 
We choose the horizon radius $r_h=r_h^0(n)$ such that Eq.\eqref{RNss} admits 
a zero mode solution $\psi(r)=\psi_0(r)$. Substituting this into \eqref{Wpert} yields a static solution of the Proca equation
\eqref{Proca}, which describes  the electroweak condensate beginning to form on the RN background: 
\be               \label{Wpert1}
w_\mu dx^\mu=w_{\rm m}(r,\vartheta,\varphi)= e^{-i{\rm m}\varphi} \psi(r)\, 
(\sin\vartheta)^{j}\left(\tan\frac{\vartheta}{2}\right)^{\rm m} 
(d\vartheta-i\sin\vartheta d\varphi).
\ee
It is assumed here that 
$n>0$, while for $n<0$, one should replace this expression by its complex conjugate.  
We have  $j=|n/2|-1$. This solution is not unique since the azimuthal number 
${\rm m}\in[-j,j]$ can take several values. Thus, the general solution is given by:
\be                \label{sum}
w_\mu dx^\mu=\sum_{\rm m}c_{\rm m}w_{\rm m}(r,\vartheta,\varphi)\,,
\ee
where $c_{\rm m}$ are constant coefficients. 
This solution 
is defined up to an overall sign, which 
can be flipped by the gauge transformation \eqref{gauge} generated by ${\rm \U}=-\tau_3$.
The effect of this transformation is $\T_1\to-\T_1$ and $\T_2\to-\T_2$, resulting in $w\to -w$, while 
 the RN background remains unchanged. 
 
 Since $w$ must be a single-valued function of $\varphi$ modulo 
 a gauge transformation, it should either be invariant under $\varphi\to\varphi+2\pi$ or change sign. 
Therefore, since 
$w_{\rm m}\sim\exp(-i{\rm m}\varphi)$, the azimuthal number ${\rm m}$ must take either integer or 
half-integer values, regardless  of the value of $j=|n|/2-1$, which can itself be either integer  or half-integer. 

Additionally, 
one must have $|{\rm m}|\leq j$, 
since otherwise the mode \eqref{Wpert1} would diverge at $\vartheta=0$ or $\vartheta=\pi$. 
Consequently, the allowed values for ${\rm m}$ are as follows:

$\bullet$ if $j=0$, then ${\rm m}=0$; 

$\bullet$ if $j=\frac12$, then ${\rm m}=\pm \frac12$  or ${\rm m=0}$; 

$\bullet$  if $j=1$, then ${\rm m}=\pm \frac12$ or ${\rm m}=0,\pm 1$; 

$\bullet$  if $j=\frac32$, then ${\rm m}=\pm \frac32,\pm \frac12$ or ${\rm m}=0,\pm 1$, etc. \\
Therefore, ${\rm m}$ assumes either $2j$ or $2j+1$ distinct values, among which 
${\rm m}=0$ is always present. 

If $|n|=2$, then $j={\rm m}=0$, and there is only one coefficient $c_0$ in \eqref{sum}, which can be fixed by 
normalization,  hence there is no ambiguity. For $|n|>2$  one has $j>0$, 
spherical symmetry is lost, and the azimuthal number ${\rm m}$ can take several values.
As a result,  there are several coefficients $c_{\rm m}$ in \eqref{sum}. 
Within the framework of linear perturbation theory, these coefficients are
arbitrary, but higher-order  perturbation theory imposes conditions on them. 
Specifically, the choice of these coefficients should minimize the energy of the condensate. 

In what follows, we adopt  the approach of \cite{Ridgway:1995ke}, which involves 
expanding around the RN solution up to the second order terms and then minimizing the energy. 
We  neglect  the backreaction of perturbations on the RN geometry, as it is proportional to $\kappa$.

\subsection{Electroweak equations in the unitary gauge}

Switching to the unitary gauge, where $\Phi=(0,\phi)^{\rm T}$ with $\phi$ being real-valued, 
one defines  the complex-valued W field and its strength: 
\be               \label{wdef}
w_\mu=\frac{1}{g}\,(\WW^1_\mu+i\WW^2_\mu),~~~~
w_{\mu\nu}={\cal D}_\mu w_\nu-{\cal D}_\nu w_\mu,
\ee
where ${\cal D}_\mu=\nabla_\mu +i\WW^3_\mu$. 
The electroweak Lagrangian in  \eqref{II} then takes the following form: 
\be           \label{LEW1}
-{\cal L}_{\rm EW}=\frac{1}{4g^{\prime 2}}(B_{\mu\nu})^2+\frac{1}{4g^{2}}(\WW^3_{\mu\nu})^2
+ \frac{1}{4}\,|w_{\mu\nu}|^2 ~~~~~~~~~ \\
+(\partial_\mu\phi)^2+\frac14\left( g^2|w_\mu|^2+(B_\mu-\WW^3_\mu)^2 \right)\phi^2+\frac{\beta}{8}(\phi^2-1)^2,~~\nn
\ee
with $B_{\mu\nu}=\partial_\mu B_\nu-\partial_\nu B_\mu$, where 
\be                \label{psi1}
\WW^3_{\mu\nu}=\partial_\mu \WW^3_\nu-\partial_\nu \WW^3_\mu+ g^2\psi_{\mu\nu},~~~~
\psi_{\mu\nu}=\frac{i}{2}(w_\mu\bar{w}_\nu-w_\nu\bar{w}_\mu). 
\ee
The Nambu electromagnetic field and Z-field, as given in \eqref{Nambu}, become 
\be               \label{psi1a}
e\FF_{\mu\nu}&=&g^2 B_{\mu\nu}+g^{\prime 2}\WW^3_{\mu\nu}= e F_{\mu\nu}+ e^2\psi_{\mu\nu}\,,~~~\nn \\
{\mathcal Z}_{\mu\nu}&=&B_{\mu\nu}-\WW^3_{\mu\nu}=  Z_{\mu\nu}- g^2\psi_{\mu\nu}\,.
\ee
Here, 
$F_{\mu\nu}=\partial_\mu A_\nu-\partial_\nu A_\mu$ 
and $Z_{\mu\nu}=\partial_\mu Z_\nu-\partial_\nu Z_\mu$ are the 't\,Hooft fields, 
with the potentials 
\be
eA_\mu=g^2\,B_\mu+g^{\prime 2}\, \WW^3_\mu,~~~~~~
Z_\mu=B_\mu-\WW^3_\mu\,.
\ee
Notice that $\psi_{\mu\nu}$ in \eqref{psi1} is the same as in \eqref{psi}, while $F_{\mu\nu}$ 
is the same as in  \eqref{FF}. 

The electroweak equations \eqref{P2} then take the form: 
\begin{subequations}           \label{eqWS}
\begin{align}
&\nabla^\mu F_{\nu\mu}=4\pi J_\nu\,, ~~~~~\label{eqWSa} \\
&\nabla^\mu Z_{\nu\mu}+\frac{\phi^2}{2}\,Z_\nu=-4\pi\,\frac{g}{g^\prime}\, J_\nu\,, \label{eqWSb} \\
&{\cal D}^\mu w_{\mu\nu}+i\left( g^2\psi_{\nu\sigma}+gg^\prime F_{\nu\sigma}-g^2 Z_{\nu\sigma}  \right)w^\sigma
=\frac{g^2\phi^2}{2}\, w_\nu\, ,\label{eqWSc} \\
&\nabla^\mu\nabla_\mu \phi=\frac14\left( g^2\,w_\mu \bar{w}^\mu+Z_\mu Z^\mu\right)\phi+\frac{\beta}{4}(\phi^2-1)\phi\,. ~~~~\label{eqWSd}
\end{align}
\end{subequations}
The gauge covariant derivative can be written as
\be 
{\cal D}_\mu=\nabla_\mu+i( e A_\mu- g^2 Z_\mu),
\ee
 and 
the current is 
\be          \label{cur33}
4\pi J_\nu= e\,\nabla^\sigma \psi_{\sigma\nu} + e\,\Im (\bar{w}^\sigma w_{\sigma\nu}).
\ee
One has 
$
\FF_{\mu\nu}=F_{\mu\nu}+ e\,\psi_{\mu\nu}, 
$
whose divergence yields 
\be          \label{cu1r}
4\pi {\mathcal J}_\nu=\nabla^\mu \FF_{\nu\mu}=\nabla^\mu F_{\nu\mu}+e\nabla^\mu\psi_{\nu\mu}
=4\pi J_\nu-e\nabla^\mu\psi_{\mu\nu}\
= e\,\Im (\bar{w}^\sigma w_{\sigma\nu}),
\ee
hence the 't\,Hooft   current and Nambu current  are related to each other as follows, 
\be          \label{cur22}
J_\nu=\J_\nu+\frac{e}{4\pi}\,\nabla^\sigma \psi_{\sigma\nu}.
\ee

\subsection{Condensate energy}

Eqs.\eqref{eqWS} admit the solution 
\be        \label{back}
A_\mu dx^\mu=-\frac{n}{2e}\,\cos\vartheta \, d\varphi,~~Z_\mu=w_\mu=0,~~\phi=1,
\ee
assuming the background geometry to be the RN  given by \eqref{QQ},\eqref{RN}. 

Linearizing the equations around this solution, the only bounded solution for the first order corrections of 
$A_\mu, Z_\mu, \phi$ is the trivial one; hence, these fields remain unchanged in the first perturbative order. 
However, the first order correction for $w_\mu$ can be non-trivial since it fulfils the same Proca equation  
as in \eqref{Proca}, whose bounded  solution is given by Eq.\eqref{Wpert1} and \eqref{sum} above. 
This solution has only two non-vanishing components, $w_\vartheta$ and  $w_\varphi=-iw_\vartheta\sin\vartheta$
(assuming that $n>0$), 
implying   the following relations:
\be        \label{PSI}
|w_\mu|^2&\equiv& w_\mu \bar{w}^\mu=\frac{2}{r^2}\,|\psi(r)|^2\times  \Theta(\vartheta,\varphi), \\
\Theta(\vartheta,\varphi)&=&\left(\sin\vartheta\right)^{2j}\sum_{\rm k,m}c_{\rm k}c_{\rm m}\left(\tan\frac{\vartheta}{2}\right)^{\rm k+m}
e^{i\rm(k-m)\varphi}\,.~~~~\nn
\ee
The tensor $\psi_{\mu\nu}=-\psi_{\nu\mu}$, defined in \eqref{psi1}, also has only two components, 
\be       \label{PSI22}
\psi_{\vartheta\varphi}=-\psi_{\varphi\vartheta}=-\frac{r^2}{2}|w_\mu|^2\sin\vartheta=-|\psi(r)|^2\,\Theta(\vartheta,\varphi)\sin\vartheta. 
\ee
The nonzero components of the strength tensor $w_{\mu\nu}$ are
\be
w_{r\vartheta}=-w_{\vartheta r}=\partial_r w_\vartheta,~~~~w_{r\varphi}=-w_{\varphi r}=\partial_r w_\varphi,
\ee
while 
\be
w_{\vartheta\varphi}=\partial_\vartheta w_\varphi-(\partial_\varphi+ieA_\varphi)w_\vartheta=w_{\varphi\vartheta}=0.
\ee
As a result, 
\be            \label{Omega} 
(\psi_{\mu\nu})^2=\frac{1}{2}\,|w_\mu|^4,~~
|w_{\mu\nu}|^2=2N(r)\left|\frac{\psi^\prime(r)}{\psi(r)}\right|^2\,|w_\mu|^2\equiv 2\,\Omega^2(r)|w_\mu|^2.
\ee
It follows  that  
$\Im (\bar{w}^\sigma w_{\sigma\nu})=0$, hence the Nambu current vanishes, $\mathcal{J}_\mu=0$,
 while 
the 't\,Hooft   current \eqref{cur33} becomes 
\be
4\pi J_\mu= e\nabla^\sigma \psi_{\sigma\mu}\, . 
\ee
At this point, it is important to emphasize  the difference between the Nambu tensor $\FF_{\mu\nu}$ 
and the 't\,Hooft   tensor $F_{\mu\nu}$. The electric current ${\cal J}^\mu$ associated with $\FF_{\mu\nu}$ vanishes in the second 
perturbative order, but the magnetic  current $\tilde{\cal J}^\mu$  does not vanish and determines the magnetic charge density. 
Conversely, for the tensor $F_{\mu\nu}$ the situation is reversed: 
\be            \label{JJJ}
{\mathcal J^\mu}&=&\frac{1}{4\pi}\,\nabla_\sigma \FF^{\mu\sigma}=0,~~~~~~~~~~~~~~~
\tilde{{\mathcal J}}^0=\frac{1}{4\pi}\,\nabla_\sigma \tilde{\FF}^{0\sigma}
=\frac{e}{8\pi\sqrt{\rm -g}}\,\epsilon^{ijk}\partial_i\psi_{jk}\,, \nn \\
J^\mu&=&\frac{1}{4\pi}\,\nabla_\sigma F^{\mu\sigma}=\frac{e}{4\pi}\,\nabla_\sigma\psi^{\sigma\mu},~~~~
\tilde{J}^0=\frac{1}{4\pi}\,\nabla_\sigma \tilde{F}^{0\sigma}=0.  
\ee
Therefore, $F_{\mu\nu}$ defines the electric current, whereas $\FF_{\mu\nu}$ gives rise to the magnetic current. 
However, it should be noted  that  the electric current ${\cal J}^\mu$  starts deviating from zero at higher perturbative orders,
while  the magnetic current $\tilde{J}^\mu$ vanishes at all orders.

The non-zero components of $J_\mu$ are 
\be             \label{J}
J_\vartheta= \frac{e }{8\pi\sin\vartheta}\,\partial_\varphi |w_\mu|^2,~~
J_\varphi= -\frac{e \sin\vartheta }{8\pi}\,\partial_\vartheta |w_\mu|^2. 
\ee
This current is tangent to the horizon  and 
 orthogonal to the gradient of  $|w_\mu|^2$. 
Restricting to $r=r_h$, the condition  $|w_\mu|^2=const.$ determines lines on the horizon that  are 
orthogonal to the gradient of  $|w_\mu|^2$. 
Therefore,  the current $J^\mu$ 
is tangent to the level lines of  $|w_\mu|^2$.

The current, being quadratic in $w_\mu$,  acts as a source for the second order correction 
$f_{\mu\nu}=\partial_\mu f_\nu-\partial_\nu f_\mu$ to the 
background electromagnetic field, leading to
\be              \label{Ftot}
F_{\mu\nu}=\partial_\mu A_\nu-\partial_\nu A_\mu+f_{\mu\nu}\equiv \overset{(0)}{F}_{\mu\nu}+f_{\mu\nu},
\ee
with the same $A_\mu$ as in \eqref{back}. 
The current also induces nonzero 
$Z_\mu$ and a deviation  $\delta\phi=\phi-1$. Expanding Eqs.\eqref{eqWS} to second order gives 
\begin{subequations}               \label{eqWS2}
\begin{align}
\nabla^\mu f_{\nu\mu}&= e\, \nabla^\sigma \psi_{\sigma\nu}\equiv {4\pi} J_\nu^{(f)}\,, \label{e1o} \\ ~~~~~
\nabla^\mu Z_{\nu\mu}+\frac{1}{2}\,Z_\nu&= -g^2\nabla^\sigma \psi_{\sigma\nu}\equiv {4\pi}  J_\nu^{(Z)}\,, \label{e2o}  \\
-\nabla^\mu\nabla_\mu \delta\phi+\frac{\beta}{2}\,\delta\phi&= -\frac{g^2}{4}\,|w_\mu|^2\equiv {4\pi}  J^{(\phi)}\,, \label{e3o} 
\end{align}
\end{subequations}
whose solution is  quadratic in $c_{\rm m}$. Substituting  this  solution  into \eqref{LEW1} yields 
the condensate energy density, containing terms up to fourth order in $c_{\rm m}$:
\be           \label{LEW2}
\mathcal{E}_{\rm cond}\equiv -{\cal L}_{\rm EW}&=&\frac{1}{4}(g F_{\mu\nu}+g^\prime Z_{\mu\nu})^2 
+\frac{1}{4}(g^\prime F_{\mu\nu}-g Z_{\mu\nu}+ g\,\psi_{\mu\nu})^2 \nn \\
&&+ \frac{1}{4}\,|w_{\mu\nu}|^2 
+ \frac{g^2}{4}\,|w_\mu|^2(1+2\delta\phi)+\frac14\,(Z_\mu)^2 
+(\partial_\mu\delta \phi)^2+\frac{\beta}{2}\,\delta\phi^2. 
\ee
This determines the condensate energy 
\be
E_{\rm cond}=\int {\cal E}_{\rm cond}\sqrt{\rm -g}\, d^3x.
\ee
Integrating  by parts, using Eqs.\eqref{eqWS2}, and omitting the zeroth-order contribution yields 
\be
E_{\rm cond}=E_w+E_{fZ\phi},
\ee
where 
\be            \label{ENo}
E_w&=& \int\left(\frac{e}{2}\,\overset{(0)}{F}_{\mu\nu}\psi^{\mu\nu}+
\frac{g^2}{4} (\psi_{\mu\nu})^2+ \frac{1}{4}\, |w_{\mu\nu}|^2
+\frac{g^2}{4}|w_\mu|^2  \right)\sqrt{\rm -g} \,d^3x \nn \\
&&=
\int\left(\frac{g^2}{8}|w_\mu|^4
+\frac{\Omega^2(r)+\mw^2(r)}{2}\,|w_\mu|^2
\right)\sqrt{\rm -g} \,d^3x, \nn \\
E_{fZ\phi}&=&{4\pi} \int\left(
-\frac12 f^\nu J^{(f)}_\nu 
-\frac12 Z^\nu J^{(Z)}_\nu 
-\delta\phi \,J^{(\phi)} 
\right) \sqrt{-\rm g}\, d^3x. 
\ee
Here $\Omega^2(r)$ is defined in \eqref{Omega}, while 
\be          \label{mwr}  
\mw^2(r)=\mw^2-|{\rm B}(r)|=\frac{g^2}{2}-\frac{|n|}{2r^2},
\ee
where the last term arises from the $\overset{(0)}{F}_{\mu\nu}\psi^{\mu\nu}$ contribution to \eqref{ENo}. 
The modulus  $|n|$ appears because 
flipping  the sign of $n$ changes  the signs of both   
$ {F}_{\mu\nu}$ and $\psi_{\mu\nu}$.  This explicitly demonstrates that   the Zeeman interaction 
with the background 
field effectively reduces  the W-boson mass.

The $E_w$ term can be computed  using the already known first order solution for $w_\mu$, 
while $E_{fZ\phi}$ requires solving 
Eqs.\eqref{eqWS2} to determine the second order corrections $f_\mu,Z_\mu,\delta\phi$. 
These can be obtained  through  the following steps. 

First, 
expand the function $\Theta(\vartheta,\varphi)$ defined in \eqref{PSI}
over spherical harmonics:
\be           \label{Theta}
\Theta(\vartheta,\varphi)=\sum_{J=0}^{2j}\sum_{M=-J}^J\,C_{JM}Y_{JM}(\vartheta,\varphi),
\ee
with $j=|n/2|-1$. As explained above after Eq.\eqref{sum}, the azimuthal numbers  ${\rm k},{\rm m}$ 
in  Eq.\eqref{PSI} can either be all integers or all half-integers. Therefore, their difference,  ${\rm k-m}$, is always an integer, which 
implies that the 
azimuthal number $M$ in \eqref{Theta} must  be an integer. Since $2j$ is also an integer, 
it follows that $J$ must be an integer as well. Thus, Eq.\eqref{Theta} involves sums over integer values 
of $J$ and $M$.

The next step is to introduce 
\be
X^{(I)}(r,\vartheta,\varphi)=\sum_{J=0}^{2j}\sum_{M=-J}^J\,R^{(I)}_{JM}(r)Y_{JM}(\vartheta,\varphi),
\ee
where the index $I$ runs over $f,Z,\phi$, and the radial functions satisfy the equation 
\be                \label{rad}
\left(
\frac{1}{q}\frac{d}{dr}\, qN\,\frac{d}{dr}-\frac{J(J+1)}{r^2}-m^2
\right)R^{(I)}_{JM}=a\,\frac{\psi^2(r)}{r^2}\, C_{JM}\,,
\ee
with the following parameters: 
\be
&&I=f:~~~~q=1,~~a=-e,~~m^2=0; \nn \\
&&I=Z:~~~~q=1,~~a=g^2,~~m^2=\mz^2=1/2; \nn \\
&&I=\phi:~~~~q=r^2,~~a=g^2/2,~~m^2=\mh^2=\beta/2.
\ee
Since $f_{\mu\nu}=\partial_\mu f_\nu-\partial_\nu f_\mu$, Eqs.\eqref{eqWS2} are fulfilled by  
\be
f_\vartheta&=&\frac{1}{\sin\vartheta}\,\partial_\varphi X^{(f)},~~~~~f_\varphi=-\sin\vartheta\,\partial_\vartheta X^{(f)}, \\
Z_\vartheta&=&\frac{1}{\sin\vartheta}\,\partial_\varphi X^{(Z)},~~~~~Z_\varphi=-\sin\vartheta\,\partial_\vartheta X^{(Z)}, \nn
~~~~\delta\phi=X^{(\phi)} \, .
\ee
Injecting this solution to $E_{fZ\phi}$ in \eqref{ENo} gives, after some algebra, 
\be
E_{fZ\phi}=\frac12 \sum_{J,M} \bar{C}_{JM}\int_{r_h}^\infty |\psi(r)|^2
\left\{
\frac{J(J+1)}{r^2}\left(g^2 R^{(Z)}_{JM}-e R^{(f)}_{JM}\right) +g^2  R^{(\phi)}_{JM}
\right\} dr\,,~~~
\ee
and one should stress  that the formula contains the complex conjugated $\bar{C}_{JM}$ and not $C_{JM}$. 

As a result, 
the problem reduces to solving the radial equations \eqref{rad}. 
It follows  that 
\be                 \label{fhor}
f_{\vartheta\varphi}=\sin\vartheta \,\sum_{J,M}J(J+1) R^{(f)}_{JM}(r)Y_{JM}(\vartheta,\varphi), 
\ee
hence
\be         \label{fluxf}
\oint \frac12\, f_{ik}\, dx^i\wedge dx^k=\oint f_{\vartheta\varphi} \,d\vartheta\, d\varphi=0. 
\ee
Therefore, the total magnetic flux of $F_{\mu\nu}$, given by \eqref{Ftot}, is the same as for the background:
\be         \label{fluxfa}
\oint \frac12\, F_{ik}\, dx^i\wedge dx^k=\frac{2\pi n}{e}, 
\ee
but the magnetic field is no longer purely radial and homogeneous on the sphere,
since components of $f_{ik}$ show a general  dependance on $\vartheta,\varphi$, while 
the components  $f_{r\vartheta}$, $f_{r\varphi}$ do not vanish. 

As we shall see below, the fields $f_{\mu\nu}$ and $Z_{\mu\nu}$ create $|n|-2$ vortices orthogonal to the horizon,
forming the black hole corona.

\subsection{Energy minimization}

The condensate energy $E_w+E_{fZ\phi}$ should   be minimized 
with respect to the coefficients $c_{\rm m}$. 
Since  computing the $E_{fZ\phi}$ term requires additional effort, we leave this
for future work and focus only on minimizing 
$E_w$. Already, this gives the expected result: a lattice of vortices 
on the horizon, forming the electroweak corona.
 
This suggests  that the minima of $E_w$  coincide with those  of $E_w+E_{fZ\phi}$ or close to them.  Although we 
cannot immediately prove this, it seems natural  that the $E_w$ term should be dominant because it 
contains  only $w_\mu$,  which  is 
the leading correction  to the background configuration, while $E_{fZ\phi}$ 
contains  contributions from the subleading 
second-order corrections for the $f,Z,\phi$ fields. 

Therefore, we minimize 
\be
\langle |w_\mu|^4\rangle\equiv \int |w_\mu|^4 \,\sqrt{-\rm g}\, d^3x\,,
\ee
while keeping the norm fixed: 
\be
\langle |w_\mu|^2\rangle\equiv \int |w_\mu|^2 \,\sqrt{-\rm g}\, d^3x=const. 
\ee
This is equivalent to minimizing 
\be
E_w=const.\times \langle |w_\mu|^4\rangle+const.\times \langle |w_\mu|^2\rangle.
\ee 
Using \eqref{PSI}, 
one finds that  $\langle |w_\mu|^2\rangle=const.\times E_2$ with 
\be              \label{E2}
E_2=\sum_{{\rm m}\in[-j,j]} A_{j,\rm m}\, c_{\rm m}^2\,,
\ee
where 
\be
A_{j,\rm m}&=& \int_0^\pi  (\sin\vartheta)^{2j+1} \left(\tan\frac{\vartheta}{2}\right)^{\rm 2m} d\vartheta 
=2^{2j+1}\frac{\Gamma(j+1+{\rm m})\Gamma(j+1-{\rm m})}{\Gamma(2j+2)}\,.
\ee
Similarly, 
 after some algebra, one finds that 
$\langle |w_\mu|^4\rangle=const.\times E_4$ with 
\be            \label{E4}
E_4= \sum_{\rm k,m,l}A_{2j,\rm k+l}\, c_{\rm m}\, c_{\rm k}\, c_{\rm l}\, c_{\rm k+l-m}\,,
\ee
where it is assumed  that $c_{\rm m}=0$ if $|m|>j$. 

Therefore, omitting constant factors, the problem reduces to the minimization  of 
\be         \label{Emin} 
E_c=E_4+\mu\,(E_2-1),
\ee
with respect to $c_{\rm m}$, 
 where $\mu$ is  the Lagrange multiplier, and with $E_2$ and $E_4$ expressed by Eqs.\eqref{E2},\eqref{E4}. 

\begin{figure}
    \centering

        \includegraphics[scale=0.5]{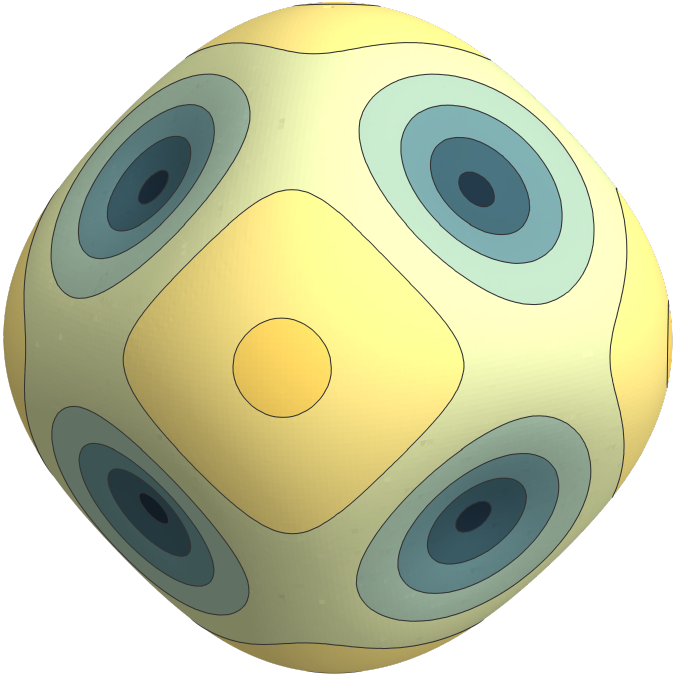}
      \hspace{5 mm}
      \includegraphics[scale=0.5]{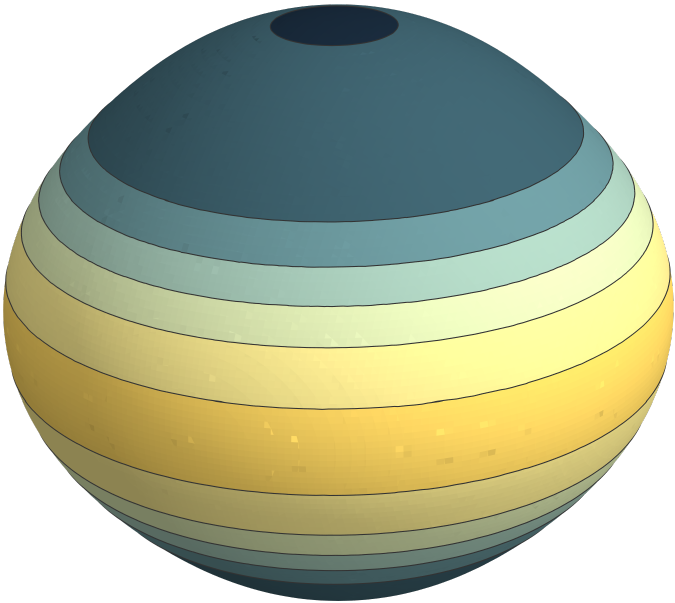}

       \caption{\small        Left:  the horizon distribution of the W-condensate 
$\bar{w}^\mu w_\mu$  minimizing the energy for $n=10$. The level lines coincide with the current 
flow, forming loops around vortices that start at the horizon in the radial direction  (dark spots). 
The current flows in the opposite  direction around the yellow regions, where 
the condensate is maximal and where the vortices return to the horizon. 
Right: the same distribution when 
all vortices merge into two oppositely directed multi-vortices. 
        }
    \label{Fig6}
\end{figure}

Assuming, for example, that  $n=10$, hence $j=4$ and $m=-4,-3,\ldots 4$,  
there are 9 coefficients $c_{\rm m}$ to find. 
Their values corresponding to the absolute minimum of $E_c$ determine the function $|w_\mu|^2$ in \eqref{PSI}, 
whose horizon profile  is shown in Fig.\ref{Fig6} (left panel). 
The figure shows the level  lines,  which 
coincide with the flow lines of the current $J^\mu$.

As seen in Fig.\ref{Fig6}, the flow lines form loops. A loop of current generates a magnetic field in the orthogonal  
direction, so the loops in Fig.\ref{Fig6} encercle radially directed vortices  
located at the points where the condensate $|w_\mu|^2$  vanishes (dark spots). 
There are 4 vortices on the visible side of the surface in Fig.\ref{Fig6} (left panel),
and the same number on the other side, so altogether there are 8 vortices ($8=|n|-2$) 
homogeneously distributed over the horizon. The condensate exhibits, in this case, the same discrete symmetries as an 
octahedron. 

The difference $|n|-2$ arises because, for $|n|=2$, 
the black hole is spherically symmetric, and  the condensate produces no currents, 
meaning there are no vortices. 
We checked that minimization of \eqref{Emin} gives one vortex for $n=3$, two vortices for $n=4$, and so on. 
These vortices repel each other and form a lattice on the horizon such that their mutual separations are maximal. 
These solutions correspond to the global energy minimum. 

The vortices start at the horizon where the condensate $|w_\mu|^2$  vanishes 
(dark  spots in Fig.\ref{Fig6}),
but, as we shall now see, 
they eventually return to the horizon in the regions  where the condensate $|w_\mu|^2$  
is maximal (yellow  spots in Fig.\ref{Fig6}).

 \subsection{Structure of vortices}
 
 The vortices   carry non-zero 
magnetic and Z fluxes, but their net flux through the horizon 
is zero. Therefore,  they start at  the horizon and then return,
being attached to the horizon at both ends, similarly to the 
coronal loops in the Sun's atmosphere.  

The vortices leave the horizon in the dark regions, which are encircled by an electric  current 
flowing in the anticlockwise direction (for $n>0$).
They return back in the yellow regions, where the  current flows in a clockwise direction.
The fluxes through the dark and yellow regions 
mutually compensate, as can be seen as follows.

One notices that the term containing the derivatives in the radial equation 
 \eqref{rad} vanishes at the horizon because the first derivative of $R_{JM}^{(f)}(r)$ vanishes there due to the boundary conditions. 
 This means that  $f_{r\vartheta}=f_{r\varphi}=0$ at the horizon. 
 
Eq.\eqref{rad} then implies that for $J>0$, one has $J(J+1)R_{JM}^{(f)}(r_h)=e\psi^2(r_h) C_{JM}$. Injecting this to 
\eqref{fhor} (where $J>0$) and using \eqref{Theta}, yields the horizon value of the magnetic field, 
\be                 \label{fhor1}
f_{\vartheta\varphi}=e\psi^2(r_h) \sin\vartheta \,\sum_{J>0,M}C_{JM}Y_{JM}(\vartheta,\varphi)= 
e\psi^2(r_h) \sin\vartheta \left\{\Theta(\vartheta,\varphi)-C_{00}Y_{00} \right\},
\ee
where $C_{00}=\oint  \Theta(\vartheta,\varphi)Y_{00}\, d^2\Omega$ with $d^2\Omega=\sin\vartheta\, d\vartheta\, d\varphi$. 

Therefore, the condensate creates a magnetic field that is purely radial at the horizon:
\be                \label{6.40} 
\delta{\cal B}^r=\frac{\epsilon^{r\vartheta\varphi}}{\sqrt{-\rm g}}\,f_{\vartheta\varphi}
=\frac{n}{|n|}\times\frac{e}{2}\left(|w_\mu|^2-\frac{1}{4\pi}\oint |w_\mu|^2\,d^2\Omega  \right), 
\ee
where we have used Eq.\eqref{PSI} and restored the dependence on the sign of $n$. 

The total flux of $\delta{\cal B}^r$ through the horizon is zero, but locally this field is non-zero. 
Specifically, $\delta{\cal B}^r>0$ where 
the condensate density $|w_\mu|^2$ is large, which results in the enhancement of 
the background magnetic field ${\cal B}^r=n/(2e\, r_h^2)$. 
On the other hand,  $\delta{\cal B}^r<0$ near the 
vortex centers where $|w_\mu|^2=0$, leading to the weakening of the background field. 
This phenomenon corresponds to the {\it anti-screening} property of the condensate 
\cite{Ambjorn:1988tm,Ambjorn:1989sz}; see \ref{AppC}.

Therefore, the sign of $\delta{\cal B}^r$ is determined by the value of the condensate density $|w_\mu|^2$,
whose profile is  shown in Fig.\ref{Fig6}. 
The vortices leave   the horizon in regions  marked by dark colours in Fig.\ref{Fig6}, 
where $\delta{\cal B}^r<0$ (for $n>0$) and carry
away a negative magnetic flux. They return to the horizon in the yellow regions, 
bringing back the same flux. 

The total magnetic flux carried by a vortex can be obtained by integrating $\delta{\cal B}^r$ over a dark region where it is 
sign definite. The flux of the massive Z field can be computed similarly and will have an opposite  
sign, because of the opposite sign of  the source term for the Z field 
in Eq.\eqref{rad}.

The magnetic charge density  $\sim \epsilon^{ijk}\partial_i\psi_{jk}$ is also maximal in the yellow regions in Fig.\ref{Fig6}, 
together with the condensate density $|w_\mu|^2$. Therefore, 
these regions are magnetically charged when considered within the Nambu approach. 
There are 8 vortices and 6 charged yellow regions in Fig.\ref{Fig6} (left panel). 

Each vortex center (dark region)  is surrounded by three  yellow regions. 
Since none of the latter is favored,  each  vortex, carrying  a negative (for $n>0$) outgoing magnetic flux,
must split into three flux jets flowing towards the three neighbouring yellow regions. 
Each yellow region is surrounded by four vortices and, hence, absorbs  
incoming  jets  from  four neighbouring vortices. 
Therefore, there are 8 vortices splitting into 24 jets, which together form the black hole corona -- a rather 
complicated pattern, whose complexity increases with growing $n$.

\subsection{Axial symmetry}

The axially symmetric configuration with
 $c_{\rm m}\propto \delta_{\rm m 0}$ is also a stationary point of the energy 
\eqref{Emin}.  It can be viewed  as a system in which  all elementary  vortices       
merge into two oppositely directed multi-vortices,   
 generated by two oppositely directed azimuthal currents. 
 The system  is invariant under $\vartheta\to\pi-\vartheta$, except for the current density, which changes sign. 
 
The condensate $|w_\mu|^2$, the  radial magnetic field $F_{\vartheta\varphi}$, and the magnetic charge density 
 $\sim \epsilon^{ijk}\partial_i\psi_{jk}$ 
 attain maximal values at the equator, where the azimuthal current changes sign, as shown in the right panel of 
 Fig.\ref{Fig6} (see also \eqref{J}). 
 
However, the Hessian matrix has negative eigenvalues in this case, indicating that 
such  solutions are unstable and should decay  into non-axially symmetric  configurations. 
This is because the elementary vortices forming  a multi-vortex repel each other. 

Setting  $c_{\rm m}=c_0\delta_{\rm m 0}$ and minimizing the energy yields  $c_0=1/\sqrt{2}$. 
Eq.\eqref{PSI} then implies that 
\be
\Theta(\vartheta,\varphi)=\frac12\,(\sin\vartheta)^{2j}.
\ee 
Inserting this   Eqs.\eqref{fhor1},\eqref{6.40}, we obtain the horizon value of the magnetic field
(compare with Eq.\eqref{C15} in \ref{AppC}): 
\be               \label{Bhor} 
\delta{\cal  B}^r=e\,\frac{n}{|n|} \,\frac{\psi^2(r_h)}{2r_h^2}\times \left((\sin\vartheta)^{2j}-\frac{\sqrt{\pi}\,\Gamma(j+1)}{2\,\Gamma(j+3/2)} \right),
\ee
where $j=|n|/2-1\geq 0$. This describes two multi-vortices made of $|n|-2$ elementary vortices: one along the $z$-axis 
in the northern hemisphere (the dark spot in Fig.\ref{Fig6}, right panel),  
and the other one along the negative direction of the  $z$-axis in the southern hemisphere. Each vortex 
starts at the horizon in the polar region, extends outside, spreads out like a fountain, and then returns back to 
the horizon in the equatorial region.

\subsection{Special cases}

Let us consider the lowest values of $n$. 

\subsubsection{$|n|=2$}
 In the case where the condensate is spherically symmetric, the currents vanish, and consequently, 
 there are no vortices. This leads to $\delta{\cal B}^r=0$ (see \eqref{Bhor} with $j=0$), meaning 
 the magnetic field is purely radial and homogeneous on the sphere.  
 
 The energy $E_c$ in \eqref{Emin} 
 depends only on one coefficient $c_0$, and the minimization yields a unique solution: $c_0=1/\sqrt{2}$, 
 corresponding to the global minimum. 
 
 Thus, it is plausible that 
 non-perturbative generalizations of this solution -- the hairy black holes with $|n|=2$, 
are stable, as  discussed in Section \ref{SspherS}.

 \subsubsection{$|n|=4$}
 
 If we do not assume the axial symmetry,  the condensate corresponding to the global energy minimum 
 would be determined by three coefficients $c_{0},c_{\pm 1}$. However, having obtained these coefficients by 
 minimizing the energy, they can be transformed to 
 $c_{\rm m}\propto  \delta_{\rm m 0}$ by a global rotation. This means that the global minimum 
remains  axially symmetric.  

In this case, there are $|n|-2=2$ elementary vortices, which position  themselves at 
 opposite sides of the sphere due to their mutual repulsion. Since this perturbative solution is stable, 
 it is plausible  that its  non-perturbative generalizations -- the hairy black holes with $|n|=4$, would also be stable. 
 These solutions will be explored in Section \ref{Sechairy}.

\subsubsection{$|n|=3$}

In this case  there are two axially symmetric solutions. 
The first solution contains a single  vortex along 
 the positive $z$-axis, say. This solution it is not invariant under $\vartheta\to\pi-\vartheta$, 
 since it has a preferred direction for the vortex. 
 Consequently, it must support a nonzero dipole moment. 
 Despite this, it represents the global minimum of the energy. 
 
The second solution  is invariant under $\vartheta\to\pi-\vartheta$, 
and it contains two oppositely directed half-vortices. 
While this solution is symmetric, it is energetically less favorable than the first one and 
does not correspond to the global energy minimum.

Therefore, 
the first solution is less symmetric but stable, while the second one is more symmetric but less stable.

  \subsection{Conclusion of the perturbative analysis}

 Summarizing the above discussion, minimum-energy hairy solutions with higher magnetic charge $|n|$ 
 can have at most only discrete symmetries. As a result, 
 constructing them at the non-perturbative level would require full 3D numerical simulations. However, 
 in the special case of axial symmetry,  the magnetic charge can take arbitrary values, but the complexity 
 of the problem is reduced. In this case,  only 2D simulations are needed, although they do not
 necessarily yield the global energy minimum. 
 
 Therefore,
 in what follows, we shall restrict to the axially symmetric sector.

 \section{AXIAL SYMMETRY \label{Secaxial}}
 \setcounter{equation}{0}
 
 Choosing the spacetime coordinates as $x^\mu=(t,\rr,\vartheta,\varphi)$,
 we assume the invariance under the action of two symmetry generators:
\be             \label{Killings}
\iota=\frac{\partial}{\partial t},~~~~~
\varsigma=\frac{\partial}{\partial \varphi}, 
\ee
where the timelike Killing vector $\iota$ is assumed to be hypersurface-orthogonal. Additionally, we impose 
invariance under reflections across  the equatorial plane, $\vartheta\to \pi-\vartheta$. 

\subsection{Static and axially symmetric gravitational fields}

The line element is parameterized as 
\be                   \label{metr11}
ds^2&=&-e^{2\Uo} {\rm N}(\rr)\, dt^2+e^{-2\Uo} dl^2, ~~~~~~
dl^2=
e^{2\Ko}\left[\frac{d\rr^2}{{\rm N}(\rr)}+\rr^2d\vartheta^2\right]+e^{2 \So}\,\rr^2\sin^2\vartheta d\varphi^2,
\ee
where $\Uo,\Ko,\So$ depend on  $\rr,\vartheta$ and $\N(\rr)$ is an auxiliary function. 
It is important to note that the radial coordinate $\rr$ is not the same as the Schwarzschild coordinate $r$ used 
in previous sections. The radial coordinate is defined 
up to diffeomorphisms $\rr\to\tilde{\rr}=\tilde{\rr}(\rr)$, which lead to the transformations
$\Uo\to \tilde{\Uo}$, $\N\to\tilde{\N}$, $\Ko\to\tilde{\Ko}$, $\So\to\tilde{\So}$, where 
\be           \label{change}
\frac{d\rr}{\rr \sqrt{\N}}=\frac{d\tilde{\rr}}{\tilde{\rr} \sqrt{\tilde{\N}}},~~~~~~~~~e^{\Uo}\sqrt{\N}=e^{\tilde{\Uo}}\sqrt{\tilde{\N}}, \\
\rr\sqrt{\N} e^{\Ko}=\tilde{\rr}\sqrt{\tilde{\N}} e^{\tilde{\Ko}},~~~~~~~~~
\rr\sqrt{\N} e^{\So}=\tilde{\rr}\sqrt{\tilde{\N}} e^{\tilde{\So}}.
\ee
This gauge freedom can be fixed by choosing a specific form for the auxiliary function $\N(\rr)$. 
For example, one can set $\N(\rr)=1$.

However, even after fixing the radial coordinate $\rr$,  the line element still admits a large gauge symmetry. 
Indeed, if $\N(\rr)=1$,  then  the $\rr,\vartheta$ part of the metric  can be represented as 
\be
e^{2\Ko-2\Uo}(d\rr^2+\rr^2 d\vartheta^2)=e^{2\Ko-2\Uo} |d\varpi|^2, 
\ee
where the complex variable is  $\varpi=\rr\, e^{i\vartheta}$. Rewriting  this in terms of a  new 
complex variable $\tilde{\varpi}$ such that  $\varpi=\varpi(\tilde{\varpi})$, the line element retains the same structure, 
up to the replacement 
\be
\Ko\to \tilde{\Ko}=\Ko+\ln\left|\frac{d\varpi}{d\tilde{\varpi}}\right|.
\ee
This reveals that the $\rr,\vartheta$ sector of the geometry  enjoys a conformal symmetry generated by holomorphic functions 
$\varpi(\tilde{\varpi})$. 
However, it turns out that this symmetry is ultimately fixed by the boundary conditions given in Eq.\eqref{bc} below, 
which permit only the identity transformation   $\varpi=\tilde{\varpi}$.

For non-extremal black hole solutions we choose 
$\N(\rr)=1-\rrh/\rr$, where $\rr=\rrh$ 
corresponds to the position of the event horizon.   
 To ensure that the geometry is asymptotically flat and regular outside the event horizon, we impose the following boundary conditions:
\be            \label{bc}
\text{infinity},~~\underline{\rr=\infty}~&:&~\Uo=\Ko=\So=0; \nn \\
\text{horizon},~~\underline{\rr=\rr_{\rm H}}~&:&~|\partial_\rr \Uo|<\infty,~|\partial_\rr \Ko|<\infty,~|\partial_\rr \So|<\infty; \nn \\
\text{symmetry axis},~~\underline{\vartheta=0,\pi}~&:&~\partial_\vartheta \Uo=\partial_\vartheta \Ko=\partial_\vartheta \So=0;~~
\nn \\
\text{equator},~~\underline{\vartheta=\pi/2}~&:&~\partial_\vartheta \Uo=\partial_\vartheta \Ko=\partial_\vartheta \So=0.
\ee
It turns out that imposing these conditions automatically ensures  that 
at the symmetry axis one has 
\be
\underline{\vartheta=0,\pi}~&:&~\Ko=\So, 
\ee
which is needed for the absence of the conical singularity.

 The functions $\Uo,\Ko,\So$  satisfy second-order differential equations, which are 
 obtained by taking suitable linear combinations of the Einstein equations \eqref{Einst}:
  \begin{subequations}  \label{Eeq}
 \begin{align}        
 &\left( \frac{1}{2}\, e^{\So-2\Uo}\rr^2\left(\N e^{2\Uo} \right)^\prime_\rr \sin\vartheta \right)^\prime_\rr
 +\left( e^{\So}\Uo^\prime_\vartheta \sin\vartheta \right)^\prime_\vartheta=-\kappa\,(T^0_{~0}-\frac{1}{2}T)\sqrt{-\rm g},~~~~~~~~~   \label{Eeq1}\\
& \left( \rr\sqrt{\N}\left(\rr\sqrt{\N} e^\So\right)^\prime _\rr\sin\vartheta \right)^\prime_\rr
 +\left(e^\So\sin\vartheta\right)^{\prime\prime}_{\vartheta\vartheta}=\kappa\,(T^r_{~r}+T^\vartheta_{~\vartheta})\sqrt{-\rm g},   \label{Eeq3} \\
& \rr\sqrt{\N}\left(\rr\sqrt{\N} \Ko^\prime_\rr\right)^\prime_\rr+\Ko^{\prime\prime}_{\vartheta\vartheta}
 +\N\rr^2\left(\Uo^\prime_\rr\right)^2+\left(\Uo^\prime_\vartheta\right)^2 
 +\rr^2\N^\prime  \Uo^\prime_\rr+\frac 12\left(\rr^2\N^\prime_\rr\right)^\prime_\rr=\rr^2 e^{2\Ko-2\Uo}\kappa\,
 T^\varphi_{~\varphi} \,, \label{Eeq2} 
   \end{align}
 \end{subequations}
 with $T=T^\mu_{~\mu}$. 
 At large distances, where the theory reduces to  Einstein-Maxwell, one has  $T^r_r+T^\vartheta_\vartheta=0$,
 which allows  for an exact integration of the $\So$-equation  \eqref{Eeq3}. The solution which remains 
 bounded for any $\vartheta$   is given by 
 \be             \label{SSSS}
 \rr\sqrt{\N}e^\So=A_1 e^\xi+A_2 e^{-\xi} ~~~\text{with}~~~~~\xi=\int \frac{d\rr}{\rr\sqrt{\N}}. 
 \ee
The integration constants $A_1,A_2$ can be adjusted to ensure  that $\So\to 0$ as $\rr\to\infty$. 
 At shorter distances, where $T^r_r+T^\vartheta_\vartheta\neq 0$, the solution can no longer be obtained analytically.
 Instead, the $\So$-equation must be solved numerically together with the other  field equations in \eqref{Eeq}. 
  
 Among the remaining Einstein equations \eqref{Einst}, there are also constraints 
\be           \label{eqs1}
{\cal C}_1\equiv \rr\sqrt{\rm-g}\left(G^\rr_{~\rr}-\kappa T^\rr_{~\rr} \right)=0,~~~~~
{\cal C}_2\equiv \frac{\sqrt{\rm-g}}{\rr}\left(G^\rr_{~\vartheta}-\kappa T^\rr_{~\vartheta} \right)=0,
\ee
which do not involve second derivatives of $\Uo,\Ko$, but only those of $\So$. 
These equations can be solved for  $\partial_\rr \Ko$ and $\partial_\vartheta \Ko$, 
giving
\be           \label{K}
\partial_\rr \Ko={\mathcal K}_1,~~~~
\partial_\vartheta \Ko={\mathcal K}_2,
\ee
where ${\mathcal  K}_1,{\mathcal K}_2$ are functions that do not depend on $\Ko$. 
The function $\Ko$ can be obtained by integrating these two equations, as in the vacuum case \cite{Synge}, 
provided that the compatibility condition,
\be
\partial_\vartheta {\mathcal K}_1=\partial_\rr {\mathcal K}_2,
\ee
is satisfied. This condition follows from equations \eqref{Eeq1} and \eqref{Eeq3}, while equation \eqref{Eeq2} is automatically fulfilled as a consequence of \eqref{K}.

 In practice, however, it is preferable to treat the two equations \eqref{eqs1} as constraints rather than solving them explicitly.
 Specifically,  the 
two non-trivial Bianchi identities $\nabla_\mu (G^\mu_{~\nu}-\kappa T^\mu_{~\nu})=0$
for $\nu=\rr,\vartheta$ can be rewritten as 
\be                        \label{CN}
\partial_\vartheta {\cal C}_2=-{\rm N}\,\partial_\rr{\cal C}_1 -\frac12 {\rm N}^\prime\, {\cal C}_1+\ldots,~~~~~
\partial_\vartheta {\cal C}_1=\rr^2\,\partial_\rr {\cal C}_2+\rr\,{\cal C}_2+\ldots ,
\ee
where the dots denote terms that vanish if the second-order equations \eqref{Eeq} are satisfied.
It follows that  if the constraints ${\cal C}_1=0$, ${\cal C}_2=0$ hold 
at the symmetry axis, then they will be satisfied  everywhere. 
At the symmetry axis, we have 
\be                  \label{CNax}
\vartheta=0,\pi :~~ {\cal C}_1=-\rr\, e^{\So}\,\partial_\vartheta(\Ko-2\So),~~~~~
{\cal C}_2=\rr\,{\rm N}e^{\So}\,\partial_\rr (\Ko-\So),
\ee
which vanish owing to the boundary conditions \eqref{bc}. Thus, by 
solving the three second order equations \eqref{Eeq}
with the boundary conditions \eqref{bc}, the two constraint equations  \eqref{eqs1} are automatically satisfied.

\subsection{The integral formula for the ADM mass}

 The ADM mass $M$ in \eqref{Ninf} can be expressed as a surface integral over a sphere at spatial infinity, 
 following Komar  \cite{Komar:1958wp}:
 \be
 M=-\frac{1}{8\pi} \oint_{S^2_\infty} \star\, d(\iota_\mu dx^\mu)\,. 
 \ee
 Here $\iota_\mu$ are the covariant components of the timelike Killing vector, 
  and the star denotes the 4-dimensional Hodge dual. 
  Using Gauss's theorem, 
 the surface integral at infinity can be transformed into a volume integral  over the 3-space, plus a surface integral over
 the event horizon. This gives 
 \be             \label{M0}
 M=\frac{{\rm k}_{\rm H} {\rm A}_{\rm H}}{4\pi}+\frac{1}{4\pi}\int_{\rr>\rr_{\rm H}} R_{\hat{0}\hat{0}}\,\sqrt{\rm-g}\, d^3x\,. 
 \ee
 The first term represents the contribution from the event horizon, where ${\rm k}_{\rm H}$ is the surface gravity
 and ${\rm A}_{\rm H}$ is the horizon area. In the  second term one has 
 $R_{\hat{0}\hat{0}}=R_{\mu\nu}\iota^\mu\iota^\nu/|\iota^\alpha\iota_\alpha|=-R^0_{~0}$. 
 The surface gravity is expressed by 
 \be                \label{sGR}
 \left. {\rm k}^2_{\rm H}=-\frac12 \nabla^\mu {\iota}_\nu\nabla_\mu {\iota}^\nu\right|_{\rr=\rr_{\rm H}}. 
 \ee
 Using the line element \eqref{metr11}, one obtains:
 \be                   \label{kA}
 {\rm k}_{\rm H}=\left.\frac{{\rm N}^\prime}{2}\,e^{2\Uo-\Ko}\right|_{\rr=\rr_{\rm H}},~~~~~~~~~~
 {\rm A}_{\rm H}=\left.2\pi \rr_{\rm H}^2 \int_0^\pi  e^{\Ko-2\Uo+\So}\sin\vartheta\, d\vartheta\,\right|_{\rr={\rm r}_{\rm H}}\,.
 \ee
 The surface gravity should be constant at the horizon, which  follows from the second constraint equation in \eqref{K}. At $\rr=\rr_{\rm H}$ , where $\N=0$, this equation simplifies to
 \be                   \label{kAA}
\left. \partial_{\vartheta} \Ko\right|_{\rr=\rr_{\rm H}}=\left. 2\partial_{\vartheta} \Uo\right|_{\rr=\rrh}~~~~
\Rightarrow~~~~~\left. \partial_{\vartheta} (2\Uo-\Ko)\right|_{\rr=\rr_{\rm H}}=0.
 \ee
Thus, $2\Uo-\Ko$ is independent of $\vartheta$ at the horizon, ensuring that ${\rm k}_{\rm H}$ remains constant.

 The mass formula \eqref{M0} can be easily verified since
 $R_{\hat{0}\hat{0}}\sqrt{\rm -g}$ corresponds to the right-hand side  of Eq.\eqref{Eeq1}.
 Therefore, replacing it by the left-hand side of the the same equation, integrating, and using the 
 boundary condition \eqref{bc} yields
\be              \label{Massrel}
&&\frac{1}{4\pi}
\int_{\rr>\rrh} R_{\hat{0}\hat{0}}\sqrt{\rm -g} \,d^3x=
\left. \frac{1}{4\pi}\int_0^{2\pi} d\varphi \int_0^\pi \frac{1}{2}\, e^{\So-2\Uo}\rr^2 \left(\N e^{2\Uo} \right)^\prime_\rr \sin\vartheta\, d\vartheta\, \right|^{\rr=\infty}_{\rr=\rrh} \nn \\
&=&M-\left. \frac{1}{4\pi} \frac{\N^\prime(\rrh)}{2}\,\rrh^2 \int_0^{2\pi}  d\varphi \int_0^\pi    e^{\So} \sin\vartheta\, d\vartheta\,\right|_{\rr=\rrh}
=M-\frac{{\rm k}_{\rm H}{\rm A}_{\rm H}}{4\pi}\,.
\ee
Here we used the fact that at large $\rr$, one has   $\So,\Uo\to 0$ and $\N e^{2\Uo}=1-2M/\rr+{\cal O}(1/\rr^2)$, 
along with the condition that $2\Uo-\Ko$ remains  constant at the horizon. 

\subsection{Axially symmetric electroweak fields}

Since the two Killing vectors  in Eq.\eqref{Killings} commute with each other, 
there is a gauge where the fields do not depend on $t,\varphi$. Additionally, we require 
the electroweak fields to be purely magnetic and 
their energy density to be invariant under reflections across the equatorial plane,  $\vartheta\to \pi-\vartheta$.
We choose the fields as follows: 
\be               \label{RR}
\WW&=&\T_2\left( F_1\,d\rr+F_2\,d\vartheta \right)+\nu \left( \T_3\,F_3-\T_1 F_4\,\right)d\varphi\,, ~~~
B=\nu\, Yd\varphi\,,~~~
\Phi=\begin{pmatrix}
\phi_1 \\
\phi_2
\end{pmatrix}\,,
\ee
where $F_1,F_2,F_3,F_4,Y,\phi_1,\phi_2$ are 7 real-valued functions of $\rr,\vartheta$,  and 
the constant is given by 
\be      \label{nun1}
\nu=-n/2,~~~~~~~n\in\mathbb{Z}.
\ee 
This is the same condition as in \eqref{nun}, and it will  be justified shortly. 

The SU(2) gauge field in \eqref{RR} corresponds to the static and 
purely magnetic ansatz of Rebbi and Rossi 
\cite{Rebbi:1980yi}. 
The fields \eqref{RR} retain their form under gauge 
transformations \eqref{gauge} generated by ${\rm \U}=\exp\left\{i\chi(\rr,\vartheta)\T_2\right\}$, whose effect  is 
\be               \label{res}
&&F_1\to F_1+\partial_\rr \chi,~~
F_2\to F_2+\partial_\vartheta \chi,~~Y\to Y,\nn \\
&&F_3\to F_3\,\cos\chi-F_4\,\sin\chi,~~~
F_4\to F_4\,\cos\chi+F_3\,\sin\chi\,,~~~~~~\nn \\
&&\phi_1\to \phi_1\,\cos(\chi/2)+ \phi_2\,\sin(\chi/2),~~~
\phi_2\to \phi_2\,\cos(\chi/2)- \phi_1\,\sin(\chi/2).
\ee
Injecting \eqref{metr11},\eqref{RR} into \eqref{TT} yields 
components of the energy-momentum tensor $T_{\mu\nu}$, shown in \ref{AppA}. 
Injecting into \eqref{Nambu},\eqref{cur} yields $\mathcal{J}^\varphi$ and $J^\varphi$  
as the only non-zero components of the 
electric  currents. 
Hence,  the currents are purely azimuthal.  Injecting into \eqref{curm} yields only one non-trivial component of the magnetic current, 
corresponding to the magnetic charge density $\tilde{{\cal J}}^0$. 
The energy-momentum tensor and the currents are  gauge-invariant. 

Modulo gauge transformations \eqref{res}, 
the zero energy configuration is 
\be
F_1=F_2=F_4=\phi_1=0,~~~\phi_2=1,~~~~~F_3=Y=const.\equiv Y_\infty, 
\ee
which retains  its form under gauge transformations generated by 
${\rm \U}=\exp\left\{i\,{\cal C}\,\nu\varphi (1+\tau_3)/2 \right\}$
with a constant ${\cal C}$, whose effect is $Y_\infty\to Y_\infty+{\cal C}$.   

The electroweak Lagrangian is 
\be
L_{\rm WS}=\int \mathcal{L}_{\rm WS} \sqrt{\rm -g}\, d^3x=\int T^0_{~0}\sqrt{\rm -g}\, d^3x
=-2\pi \int  (a_1+a_2+a_3+a_4+a_5)\, d\rr\, d\vartheta\,,~~~
\ee
where $a_1,\ldots ,a_5$ are given  by \eqref{a5} in \ref{AppA}. 
Varying the Lagrangian  with respect to the seven amplitudes $F_1,F_2,F_3,F_4,Y,\phi_1,\phi_2$ yields the 
electroweak  equations. Together with the Einstein equations \eqref{Eeq}, this gives a closed system of $10$ 
second-order partial differential equations, 
whose solutions describe static, axially symmetric, and purely magnetic  EWS fields. 
Before solving these equations, some preliminary work remains  to be done. 

\subsection{Removing the string singularity}

The gauge  fields in \eqref{RR} contain the Dirac string singularity  at the symmetry axis because their 
$\varphi$-components do not vanish there. 
This singularity can be removed by setting 
\be                \label{var}
&&F_1=-\frac{H_1(\rr,\vartheta)}{\rr\sqrt{\rm N}},~~F_2=H_2(\rr,\vartheta),~~
F_3=\cos\vartheta+H_3(\rr,\vartheta)\sin\vartheta\,,~~\nn \\
&&F_4=H_4(\rr,\vartheta)\sin\vartheta\,,~~
Y=\cos\vartheta+y(\rr,\vartheta)\sin\vartheta\,, 
\ee
and then performing 
the gauge transformation 
generated by
\be                   \label{Ureg} 
 {\rm \U}_\pm =
 e^{- i\nu\varphi \T_3 }
 e^{- i\vartheta \T_2}
 e^{\pm i\nu\varphi/2}\,,
  \ee
  which 
brings the SU(2) field to the form 
 \be           \label{gauge2}
\WW&=&\T_\varphi\left(-\frac{H_1}{\rr\sqrt{\N}}\,d\rr+(H_2-1)\,d\vartheta \right) 
+\nu \,\left( \T_r\,H_3+\T_\vartheta\,(1-H_4) \right)\sin\vartheta\, d\varphi\,,
\ee
where the angle-dependent SU(2) generators are 
$\T_r=e^a_r\T_a $, $\T_\vartheta=e^a_\vartheta\T_a $, $\T_\varphi=e^a_\varphi \T_a $  with 
\be                \label{3vec}
\left.\left.e^a_r=\right( \sin\vartheta \cos(\nu\varphi),\sin\vartheta\sin(\nu\varphi),\cos\vartheta \right),~~
~e^a_\vartheta=\partial_\vartheta e^a_r,
~~~~e^a_\varphi=\frac{1}{\nu\sin\vartheta}\,\partial_\varphi e^a_r.~~~~
\ee
This field is $\varphi$-dependent but  regular at the symmetry axis, owing to the boundary conditions 
defined in Eq.\eqref{bcc} below
(see \cite{GVII} for details). 
The U(1) and Higgs fields become 
\be          \label{gauge2a}
B_\pm &=&\nu\,(\cos\vartheta\pm 1+\,y\,\sin\vartheta)\, d\varphi\,, \\
\Phi_\pm&=&
e^{\pm i\nu\varphi/2}\,
\begin{pmatrix}
\left.\left. \,\right(\phi_1\,\cos(\vartheta/2)-\phi_2\,\sin(\vartheta/2)\right) e^{-i\nu\varphi/2} \\
\left.\left. \,\right(\phi_1\,\sin(\vartheta/2)+\phi_2\,\cos(\vartheta/2)\right) e^{+i\nu\varphi/2}
\end{pmatrix}, \nn 
\ee
where the upper $``+"$ and lower $``-"$ signs correspond to two local gauges used, respectively, 
in the southern and northern hemispheres. These two gauges  are related to each other in the equatorial region 
by the U(1) transformation with ${\rm \U}=\exp(i\nu\varphi)$. 
When expressed in these two local gauges, the U(1) and Higgs fields are everywhere regular. 

The above formulas suggest that $\nu=-n/2$ should be integer. However, it can be half-integer as well. In that case, 
the $\varphi$-dependent part of $\T_r$ changes sign under $\varphi\to\varphi+2\pi$, but the correct sign can be restored 
by the global gauge transformation generated by ${\rm \U}=\tau_3$ or ${\rm \U}=-\tau_3$, whose effect is 
\be            \label{TTT}
\T_1\to -\T_1,~~~~~\T_2\to -\T_2\,, 
\ee
so that the field \eqref{gauge2} does not change. Similarly, if $\nu$ is half-integer, then the lower component of $\Phi_{+}$ and the 
upper component of $\Phi_{-}$ change sign under  $\varphi\to\varphi+2\pi$, but the correct sign can be restored
by the gauge transformations generated by ${\rm \U}=\tau_3$ and ${\rm \U}=-\tau_3$, respectively. 
The fields \eqref{gauge2a} then remain invariant. As a result, all integer values of $n$ in \eqref{nun1} 
are allowed. 

To summarize, by 
passing to the parameterization  in terms of the  seven amplitudes $H_1$, $H_2$, $H_3$, $H_4$, $y$, $\phi_1$, 
and $\phi_2$, 
subject to the boundary conditions to be specified below, the electroweak fields become regular at the symmetry axis. 

\subsection{Fixing the gauge}
\vspace{0.2cm}
The  equations for $H_1,H_2,H_3,H_4,y,\phi_1$ and $\phi_2$
admit pure gauge solutions due to the residual gauge invariance \eqref{res}. Such zero modes should be removed 
by fixing the gauge, as otherwise the differential operators in the equations are not invertible. The gauge can be fixed by setting to zero 
the covariant divergence of the two-vector  $F_1\, d\rr+F_2\,d\vartheta$ in \eqref{RR}, 
computed with respect to the $\rr,\vartheta$ part of the metric. 
This requires that 
\be                \label{fix}
\rr\sqrt{\N}\,\partial_\rr H_1=\partial_\vartheta H_2~~~\Rightarrow~~~\partial_\xi H_1=\partial_\vartheta H_2,
~~~~~~~\text{where}~~~~~\xi=\int_{r_H}^\rr \frac{d\rr}{\rr\sqrt{\N}}.
\ee
This gauge condition insures  good numerical convergence, although 
 it gives rise to a spurious long-range mode  in the solutions at large $\rr$ \cite{GVII}.  
The condition \eqref{fix} still allows for residual gauge transformations \eqref{res} with  the gauge 
parameter $\chi$ subject to 
\be                   \label{resid}
\left(\partial^2_{\xi\xi}+\partial^2_{\vartheta\vartheta}\right)\chi=0.
\ee
Since $\xi\in [0,\infty)$, $\vartheta\in[0,\pi]$, this equation possesses bounded solutions
which generate gauge transformations $H_1\to H_1-\partial_\xi\chi$ and  $H_2\to H_2+\partial_\vartheta \chi$.
At the same time, as seen in \eqref{var}, $H_1$ must vanish at the horizon where $\xi=0$ 
since otherwise the gauge field amplitude $F_1$ diverges. As a result, the only allowed solution of \eqref{resid} 
is $\chi=const.$, but this is also ruled out since otherwise the value of the Higgs field components  would change 
according to \eqref{res}, whereas we impose at infinity  $\phi_1=0$ and $\phi_2=1$. 

Therefore, the gauge is completely fixed by the condition \eqref{fix}. 
Using this condition, all equations assume a manifestly elliptic form.

\subsection{Boundary conditions and numerical procedure} 

 The axially symmetric EWS fields are the line element in \eqref{metr11} 
 and the electroweak fields in \eqref{gauge2},\eqref{gauge2a}
 expressed in terms of 10 functions $\Uo,\Ko,\So,H_1,H_2,H_3,H_4,y,\phi_1,\phi_2$
 which satisfy 
  10 elliptic equations. Solutions of these equations 
also satisfy the gravitational constraints \eqref{eqs1} owing to the 
boundary conditions described by Eq.\eqref{bcc} below.

 We use 
 the compact radial variable $\rm x\in[0,1]$ such that 
$$
\rr=\sqrt{\rrh^2+{\rm x}^2/(1-{\rm x})^2},
$$
and since we assume reflection invariance of the energy density under $\vartheta\to \pi-\vartheta$, we can restrict 
the range of the angular variable to $\vartheta\in[0,\pi/2]$. 
This  assumption implies that $\Uo$, $\Ko$, $\So$, $H_2$, $H_4$, and $\phi_2$ remain invariant under $ \vartheta\to  \pi - \vartheta$, 
while $H_1$, $H_3$, $y$, and $\phi_1$ change sign (see Ref.\cite{GVII} for details). Regularity at the horizon and along the symmetry axis, 
together with asymptotic flatness, then imposes the following boundary conditions: 
\be          \label{bcc}
\underline{\text{axis}, \vartheta=0}: &&
~\partial_\vartheta=0~ \text{for}~  \Uo,\So,H_2,H_4,\phi_2;\nn  \\
&&~\Ko-\So=H_1=H_3=y=\phi_1=0;  \nn \\
\underline{\text{equator}, \vartheta=\pi/2}: &&
~\partial_\vartheta=0 \text{~for~}  \Uo,\Ko,\So,H_2,H_4,\phi_2; \nn \\ 
&&~H_1=H_3=y=\phi_1=0;  \nn \\
\underline{\text{horizon}, {\rm x}=0}:&& ~H_1=0:  \partial_{\rm x}=0 \text{~for the rest} \nn \\
\underline{\text{infinity}, {\rm x}=1}: &&~\phi_2=1; \text{~0 for the rest}. 
\ee
It turns out that these boundary conditions automatically ensure that at the symmetry axis one has 
\be          \label{bcc+}
\underline{\vartheta=0}: &&~H_2=H_4,~~~ \partial_\vartheta\Ko=0, 
\ee
as needed for the regularity. 

We solve the field equations with these boundary conditions to determine the components of the ``state vector''
\be       \label{state}
\Psi=[H_1,H_2,H_3,H_4,y,\phi_1,\phi_2,\Uo,\Ko,\So],
\ee
which are functions of $\rr,\vartheta$. We solve the equations with the 
FreeFem  numerical solver based on the finite element method  \cite{MR3043640}. 
This solver uses the weak form of  differential equations obtained by transforming them into integral equations,
and the functions to be determined are expanded over elementary 
basis functions obtained by triangulating  the integration domain. To handle 
the non-linearities, we use the Newton-Raphson algorithm.  
To estimate the numerical accuracy of the solutions, we use the 
virial relation that should hold on-shell. 

\subsection{Virial relations}

Consider the 1-form generating a spatial dilatation, 
\be            \label{vz}
\zeta_\mu dx^\mu= \rr\,e^{\Ko-\Uo}\, d\rr\,. 
\ee
The tensor  $\nabla_\mu\zeta_\nu+\nabla_\nu\zeta_\mu$ is diagonal and describes 
 the variation of the spacetime metric induced by the dilatation. 
The virial relation can be expressed in terms of $\zeta^\mu={\rm g}^{\mu\nu}\zeta_\nu$\,:
\be            \label{vir}
v_m \equiv \int_{\rr>\rr_{\rm H}} T^\mu_{~\nu}\nabla_\mu\zeta^\nu \,\sqrt{\rm -g} \,d^3x=0. 
\ee
Since $T^\mu_{~\nu}\nabla_\mu\zeta^\nu=\nabla_\mu (T^\mu_{~\nu}\zeta^\nu)-(\nabla_\mu T^\mu_{~\nu}) \zeta^\nu$
and the energy-momentum tensor is conserved, the volume integral can be transformed  to surface 
integrals over a two-sphere at infinity and over the horizon,
\be
v_m=\oint_{\rr\to\infty} T^\mu_{~\nu}\zeta^\nu d^2 S_\mu -\oint_{\rr=\rrh} T^\mu_{~\nu}\zeta^\nu d^2 S_\mu\,.
\ee
As $T^\mu_{~\nu}$ is bounded for $\rr\geq \rrh$ and approaches zero at infinity, 
the surface integral at infinity vanishes, while the integral over the horizon is zero  because $\zeta^\nu\sim \N$. 
As a result, the integral in \eqref{vir} vanishes,  although its integrand does not. 

It is instructive to see what happens in the spherically symmetric case when 
the metric reduces to \eqref{geom}, while   the energy-momentum tensor  is diagonal, so that we have 
$T^\mu_{~\nu}={\rm diag}\left[T^0_{~0}(r),T^r_{~r}(r),T^\vartheta_{~\vartheta}(r),T^\varphi_{~\varphi}(r)\right]$. This 
can correspond to any matter type, not only to the electroweak fields. 
The virial relation \eqref{vir}  reduces to:
\be             \label{vir2}
v_m=4\pi \int_{r_H}^\infty \sigma r^2\left( 
N\,T^k_{~k}
+\frac{r\,N^\prime }{2}\,T^r_{~r}
+\frac{r}{2\sigma^2}\,\left(\sigma^2 N \right)^\prime T^0_{~0}
\right)dr=0,
\ee
where $T^k_{~k}=T^r_{~r}+T^\vartheta_{~\vartheta}+T^\varphi_{~\varphi}$ is the sum of the principal pressures. 
Taking into account the conservation condition $\nabla_\mu T^\mu_{~\nu}=0$, it is not difficult to see that 
this is equivalent to:
\be
v_m=4\pi \int_{r_H}^\infty \left(r^3 \sigma N \, T^r_{~r} \right)^\prime dr=0,
\ee
which  indeed holds, 
since at the lower limit $N$ vanishes, while for $r\to\infty$, the components $T^\mu_{~\nu}$ 
decay faster that $1/r^3$ (otherwise  the energy would diverge).

In flat space, where $N=\sigma=1$, \eqref{vir2}  reduces to:
\be
v_{\rm m}=\int T^k_{~k}\,\sqrt{\rm -g}\,d^3 x=0,
\ee
which 
should be  fulfilled for any finite energy soliton configuration \cite{Deser:1976wq}. This  is equivalent 
to the classical scaling relation of Derrick \cite{Derrick:1964ww}. Therefore, 
\eqref{vir} provides the generalization of Derrick's argument  for black holes. 
This generalisation is  not unique, as $\zeta_\mu$ in \eqref{vz} can be multiplied by 
a scalar function without changing the value of the virial integral.

A similar virial relation  in the gravity sector 
can be obtained by replacing $T^\mu_{~\nu}$ with the Einstein tensor  $G^\mu_{~\nu}$:
\be            \label{virg}
v_g \equiv \int_{\rr>\rrh} G^\mu_{~\nu}\nabla_\mu\zeta^\nu \,\sqrt{\rm -g} \,d^3x=0. 
\ee

We adopt  \eqref{vir} and \eqref{virg} 
as virial relations to test the quality of our numerical solutions.  
These relations are satisfied for all our solutions, with a 
precision depending on the numbers of discretization points $N_{\rm x}$ 
and $N_\vartheta$ along the ${\rm x},\vartheta$ axes 
(these numbers determine the triangulation pattern for the FreeFem solver). 
Taking $N_{\rm x}=130$ and  $N_\vartheta=30$ typically yields: 
$v_m\sim 10^{-7}$, $v_g\sim 10^{-8}$.   

To the best of our knowledge, the above formulation of the virial relations 
has not been previously discussed in the literature, where alternative forms of 
virial relations can be found (see, e.g., \cite{Herdeiro:2022ids}).

 \begin{figure}
\hbox to \linewidth{ \hss
	\includegraphics[width=8 cm]{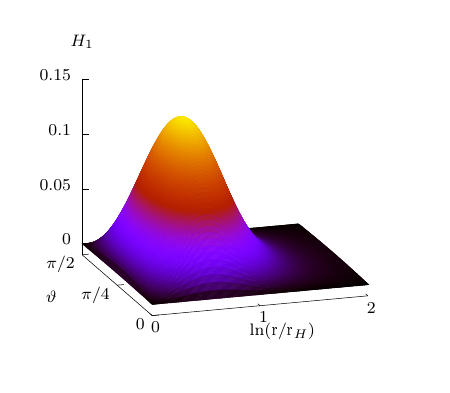}
	\includegraphics[width=8 cm]{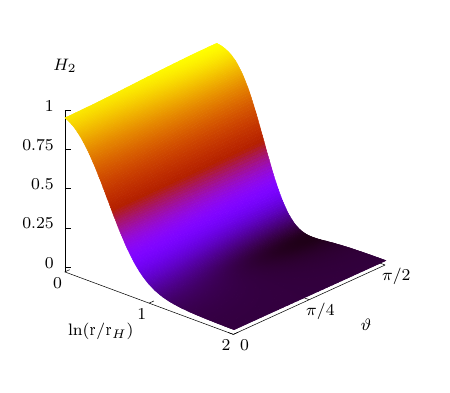}	

\hss}
\hbox to \linewidth{ \hss
	\includegraphics[width=8 cm]{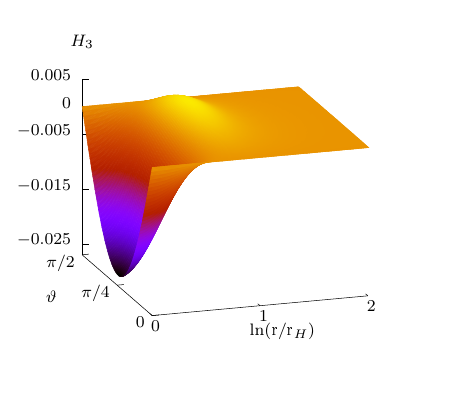}
	\includegraphics[width=8 cm]{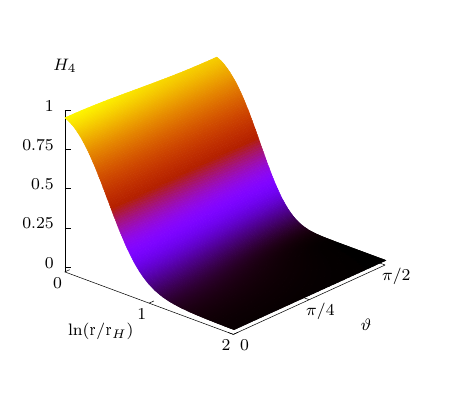}	
\hss}
\caption{The SU(2) amplitudes  $H_1,H_2,H_3,H_4$ for the hairy  
black hole solution with $\kappa=10^{-3}$, $\rrh=0.7$, and $n=4$ as functions of  $\ln(\rr/\rrh)$ and $\vartheta$.}
\label{Fig1a}
\end{figure}

\section{HAIRY SOLUTIONS  WITH $n=\pm 4$\label{Sechairy}}
 \setcounter{equation}{0}
 
 Before considering axially symmetric solutions with $|n|>2$, we have reproduced 
 the spherically symmetric solutions with $|n|=2$ within the axially symmetric setting. 
 This analysis, described in  \ref{AppB}, is needed to clarify  the choice of the radial gauge. 
 Here is  the summary of the essential points. 
 
  \begin{figure}
\hbox to \linewidth{ \hss
	\includegraphics[width=8 cm]{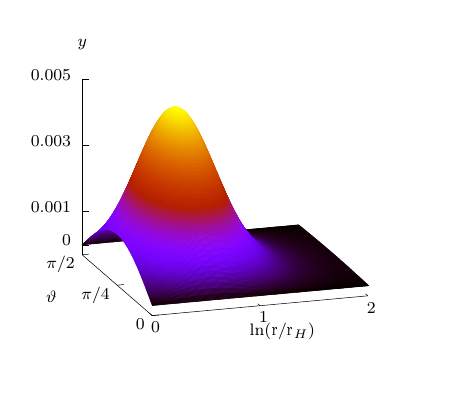}
	\includegraphics[width=8 cm]{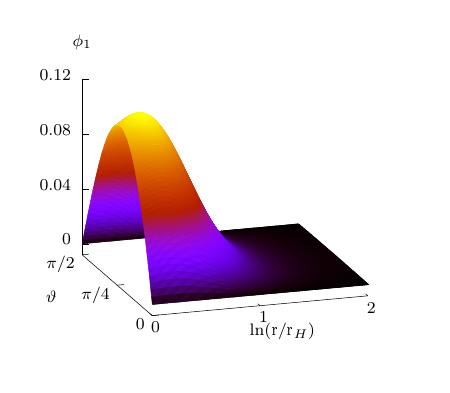}
\hss}
\hbox to \linewidth{ \hss
	\includegraphics[width=8 cm]{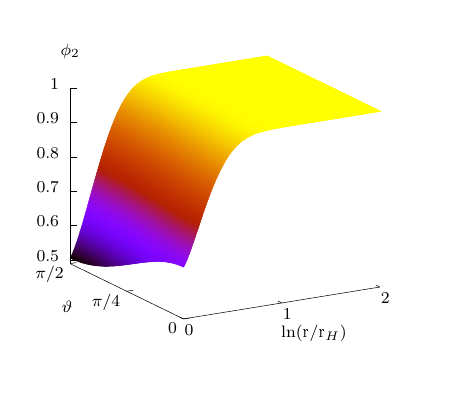}
	\includegraphics[width=8 cm]{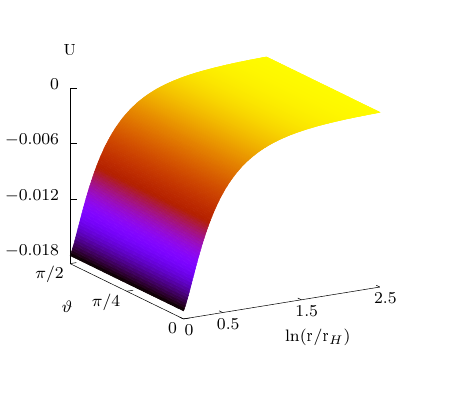}
\hss}
\caption{The  U(1) amplitude $y$, the Higgs amplitudes $\phi_1$ and $\phi_2$, and the metric amplitude $\Uo$
for the hairy  
black hole solution with $\kappa=10^{-3}$, $\rrh=0.7$, and $n=4$.}
\label{Fig2a}
\end{figure}

 \begin{figure}[th]
\hbox to \linewidth{ \hss
	\includegraphics[width=8 cm]{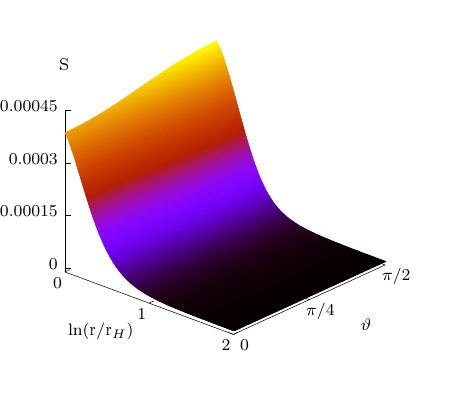}
	\includegraphics[width=8 cm]{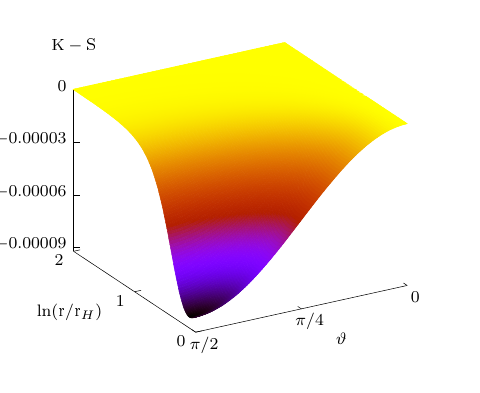}
\hss}
\caption{The  metric amplitude $\So$ and the difference $\Ko-\So$ 
for the hairy  
black hole solution with $\kappa=10^{-3}$, $\rrh=0.7$, $n=4$.}
\label{Fig3a}
\end{figure}

 \subsection{Choice of the radial gauge}
The axially symmetric fields  in Eqs.\eqref{metr11},\eqref{RR}, and \eqref{var} depend on the 
radial  coordinate $\rr$. This is not the same as the Schwarzschild coordinate  $r$, and it is 
defined up to reparametrizations. This gauge freedom is fixed  by specifying 
the auxiliary function $\N(\rr)$ in the line element \eqref{metr11}. 

For the non-extremal solutions, we choose 
\be              \label{Ngauge}
\N(\rr)=1-\frac{\rrh}{\rr},~~~~~~~~\rrh\in [0,\rrh^0],
\ee
where $\rr=\rrh$ corresponds to the position of the event horizon, and we refer to $\rrh$ as the  
``black hole size''. 
However, since $\rr$ is not 
the Schwarzschild coordinate, the parameter $\rrh$ does not have a direct physical meaning, although its value determines the 
 horizon area. 
 When $\rrh\to 0$, the solutions approach the extremal limit with a non-zero horizon area. 
When $\rrh\to \rrh^0$, the solutions merge with the RN solution, 
and we have  $\rrh^0=r^0_{+}-r_{-}^0$, where 
$r^0_{+}=r^0_h$ and $r^0_{-}$ are the two  RN horizons (see Eq.\eqref{rH_lm1}). 

The RN solution \eqref{QQ}, when transformed to the axially symmetric form \eqref{metr11},\eqref{Ngauge},
 is expressed by Eq.\eqref{RNax} in \ref{AppB}.

 The horizon values of the metric functions 
$\Uo,\Ko,\So$ grow without bounds in the extremal limit $\rrh\to 0$.
Therefore, to study the extremal limit, we use a different  gauge in which everything is finite at the horizon. 
This gauge is defined by:
\be              \label{kk_ex}
\N(\rr)={\rm k}^2(\rr)\times\left(1-\frac{\rrh}{\rr}\right)^2,~~~~~~
{\rm k}^2(\rr)=1-\left.\left.\frac{\Lambda}{3}\right[\rr^2+2\,\rr_{\rm H}\,\rr+3\,\rr_{\rm H}^2\right]\times 
\frac{1+\rr_{\rm H}^4}{1+\rr^4}\,,
\ee
with $\rrh=r_{\rm ex}$, which is the horizon radius value for the extremal RNdS solution in Eq.\eqref{horRNdS}.
The horizon is located at $\rr=\rrh$,  and it is degenerate, meaning the surface gravity vanishes. 
This gauge choice  is justified by the fact that, 
for some extremal hairy solutions (phase I), the horizon geometry coincides 
with that of the extremal RNdS geometry \eqref{RN},\eqref{Nex}.   
The radial coordinate $\rr$ then matches the Schwarzschild coordinate $r$ 
 at the horizon, where   the 
function $\N(\rr)$ coincides  with  $N(r)$ from Eq.\eqref{Nex}. 

The gauge choices \eqref{Ngauge} and \eqref{kk_ex} are justified in \ref{AppB} for spherically symmetric 
hairy solutions with $|n|=2$, but they also apply to axially symmetric solutions with  $|n|> 2$.

 \subsection{Structure of hairy solutions for $n=\pm 4$}

\begin{figure}[th]
\hbox to \linewidth{ \hss
	\includegraphics[width=8 cm]{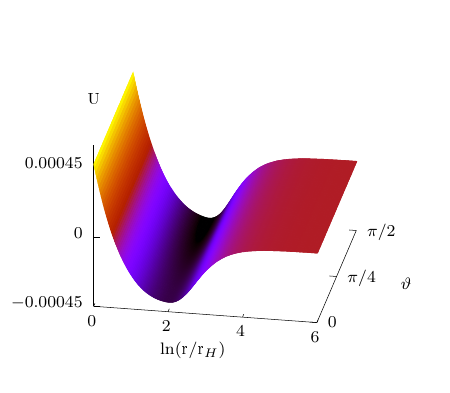}
	\includegraphics[width=8 cm]{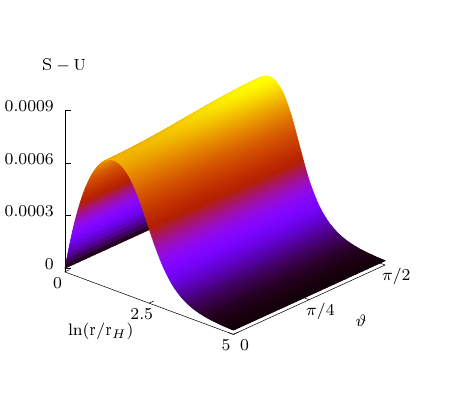}
\hss}
\hbox to \linewidth{ \hss
	\includegraphics[width=8 cm]{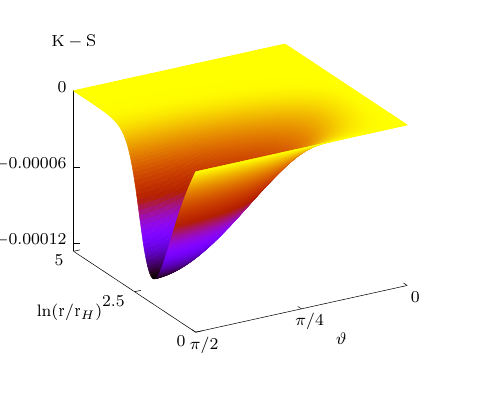}
	\includegraphics[width=8 cm]{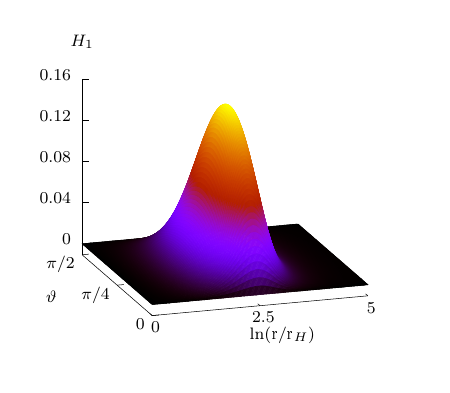}
\hss}
\caption{The  profiles of $\Uo$, $\So-\Uo$,  $\Ko-\So$, and $H_1$ 
for the extremal  hairy  black hole with $\kappa=10^{-3}$ and $n=4$.}
\label{Fig4b}
\end{figure}

The spherically-symmetric solutions with $n=2$  can be used as an   initial guess 
to construct axially symmetric solutions with $n>2$.  
Of course, $n$ should be integer to avoid  the 
line singularities in the gauge fields, but the latter are not seen by the equations, which can be solved for 
any real $n$. Therefore, one can change the value of $n$  by small steps, and 
we were able to proceed this way up to  $n=200$,
beyond which the virial relations deteriorate.

We choose $n=4$ to illustrate the solutions profiles, as those for higher values of $n$ are qualitatively the same. 
Solutions with $n=3$ are also qualitatively similar,
 but we choose $n=4$ because, as mentioned above in Section \ref{Secpert} 
and  reiterated below, the solutions for $n=4$ (but not for $n=3$) appear  to be stable.

The 10 amplitudes $H_k,y,\phi_1,\phi_2,\Uo,\Ko,\So$  are invariant under $n\to -n$, but the fields in the ansatz \eqref{RR},
along with  the electric  currents \eqref{cur} and the magnetic charge density \eqref{cur1a}, 
 are sensitive to the 
sign of $n$. 
We set 
$\kappa=10^{-3}$ instead of choosing the physical value $\kappa\sim 10^{-33}$, as otherwise, 
the backreaction on the spacetime geometry would be difficult to observe. However, later 
we extrapolate the results to the physical value of $\kappa$. 
The radial gauge is fixed according to \eqref{Ngauge}, 
where the black hole size parameter  
$\rrh$ ranges from zero 
up to the maximal value $\rrh^0\approx 1.4587$, which corresponds  to the bifurcation
with the RN solution. 

If $\rrh$ is close to the upper bound $\rrh^0$, then the black holes are only ``slightly hairy'', and the geometry 
is close to the RN with magnetic charge $P=2/e$. We therefore choose an intermediate  
value, 
$\rrh=0.7$,  and show in Figs.\ref{Fig1a}--\ref{Fig3a} 
the profiles of the 10 amplitudes  $H_k,y,\phi_1,\phi_2,\Uo,\Ko,\So$
as functions of  $\ln(\rr/\rrh)$ and $\vartheta$. 
The symmetry axis is at $\vartheta=0$, and 
the equator  is at $\vartheta=\pi/2$.

\begin{figure}[th]
\hbox to \linewidth{ \hss
	\includegraphics[width=8 cm]{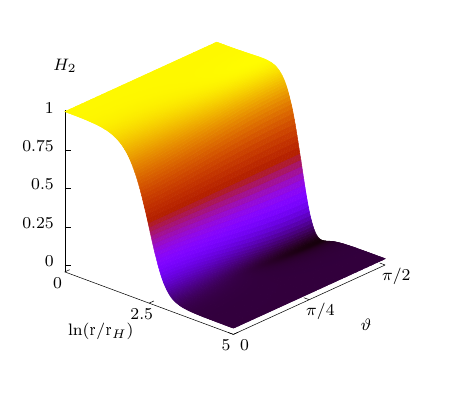}
	\includegraphics[width=8 cm]{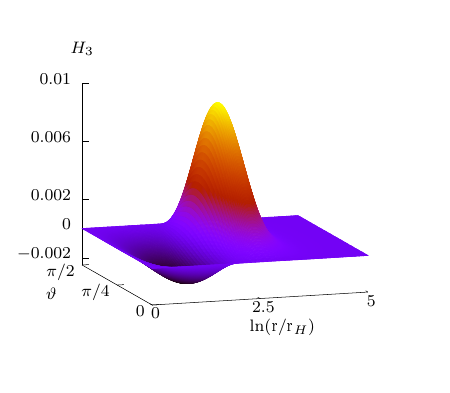}
\hss}
\hbox to \linewidth{ \hss
	\includegraphics[width=8 cm]{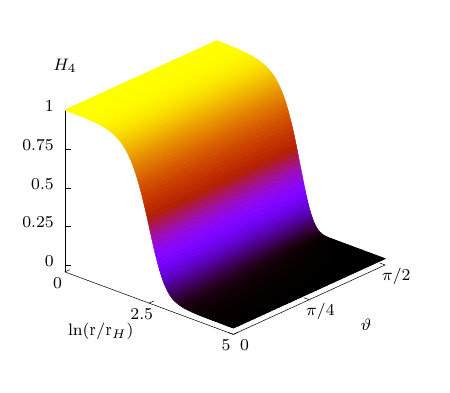}
	\includegraphics[width=8 cm]{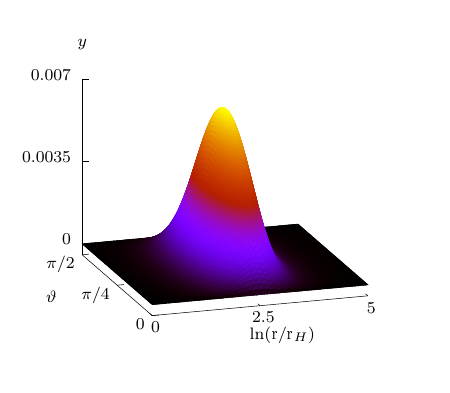}
\hss}
\caption{The  profiles of $H_2$,  $H_3$, $H_4$, and $y$ 
for the extremal  hairy  black hole  with $\kappa=10^{-3}$ and $n=4$.}
\label{Fig4d}
\end{figure}
\begin{figure}[th]
\hbox to \linewidth{ \hss
	\includegraphics[width=8 cm]{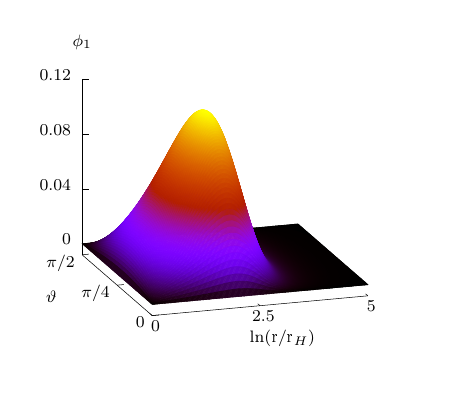}
	\includegraphics[width=8 cm]{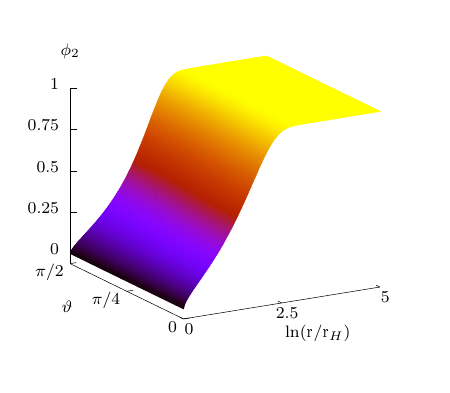}
\hss}
\caption{The  Higgs amplitudes $\phi_1$ and $\phi_2$ for the extremal  
hairy  black hole with $\kappa=10^{-3}$ and $n=4$.}
\label{Fig4f}
\end{figure}

The functions $H_2,H_4$,  and $\phi_2$, 
which do not vanish in the spherically symmetric case  $n=2$ (see Eq.\eqref{sps}), 
remain essentially the same and exhibit only  weak $\vartheta$-dependence for  $n=4$.  
On the other hand, the functions $H_1,H_3,y,\phi_1$, which vanish  for 
$n=2$, are no longer zero   for $n>2$. They exhibit strong 
$\vartheta$-dependence, with a pronounced maximum or minimum between the 
symmetry axis and the equatorial plane, and 
are antisymmetric under $\vartheta\to\pi-\vartheta$. 
Their  $\vartheta$-dependence remains significant even 
at the horizon, except for $H_1$, which vanishes there. 

\begin{figure}[b]
\hbox to \linewidth{ \hss
	\includegraphics[width=7 cm]{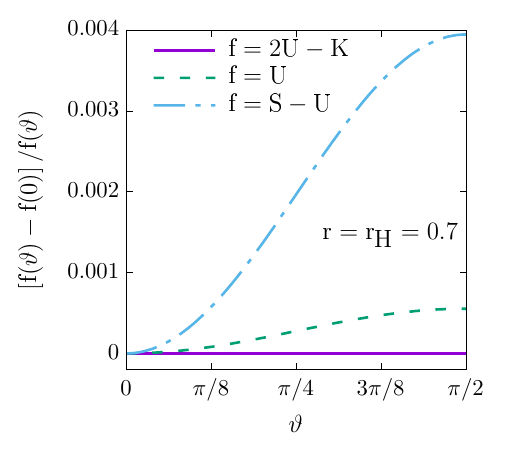}
	\includegraphics[width=7 cm]{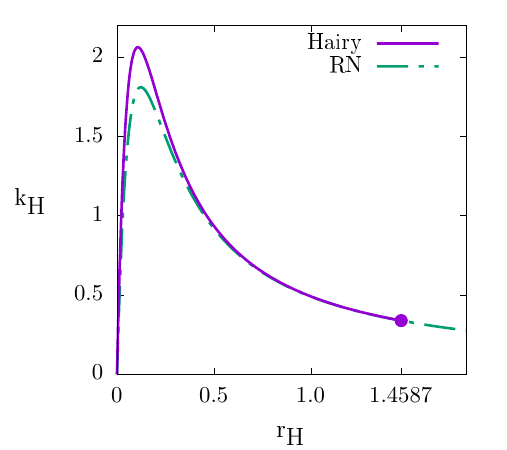}
	\hss}
\caption{Left: the  angular dependence of $2\Uo-\Ko$, $\Uo$, and $\So-\Uo$ at the horizon
for the hairy  
black hole solution with $\kappa=10^{-3}$, $\rrh=0.7$, and $n=4$. Right: 
the surface gravity as a function of  $\rrh$ for the hairy solutions with $\kappa=10^{-3}$, $n=4$, 
and for the RN solution. The maximal value of the surface gravity, and hence of the Hawking temperature ${\rm T}={\rm k}_{\rm H}/(2\pi)$, 
is higher for the hairy solutions than  for the RN solution. 
}
\label{Fig_hor}
\end{figure}

The metric function $\Uo,\Ko,\So$ almost do not depend on the angle $\vartheta$ and are close to 
the values defined by Eq.\eqref{RNax} in the RN case. In particular, one has $\Ko\approx \So$.
The difference $\Ko-\So$ is maximal at the equator and 
 can be seen as a measure of the deviation from spherical symmetry. 
For $\vartheta=0$, one has $\Ko=\So$  and also $\partial_\vartheta \Ko=\partial_\vartheta \So=0$. 
In view of Eq.\eqref{CNax}, this guarantees  that the two gravitational constraints 
${\cal C}_1$ and ${\cal C}_2$ vanish at the symmetry axis,
and Eq.\eqref{CN} then implies that they should vanish everywhere, which is indeed confirmed with a good precision. 
All three metric functions  $\Uo,\Ko,\So$  show a weak $\vartheta$-dependence  at the horizon, 
but as required  by Eq.\eqref{kAA},  the combination $2\Uo-\Ko$, which determines the surface gravity,
\be            \label{kH}
{\rm k}_{\rm H}=\left.\frac{\N^\prime}{2}e^{2\Uo-\Ko}\right|_{\rr=\rrh}=\left. \frac{1}{2 \rrh}\, e^{2\Uo-\Ko}\right|_{\rr=\rrh},
\ee
should be constant at the horizon. This is indeed confirmed by the numerics (see Fig.\ref{Fig_hor}) 
and shows that the constraints 
are fulfilled. 

For the solution with  $\rrh=0.7$ displayed  in Figs.\ref{Fig1a}--\ref{Fig3a}, one has
$(2\Uo-\Ko)|_{\rr=\rrh}=-0.0377$. Injecting this to \eqref{kH}  yields the value ${\rm k}_{\rm H}=0.6878$. 

When the horizon size decreases, the surface gravity first increases  but then reaches a maximum value and start decreasing,
approaching zero in the extremal  limit (see Fig.\ref{Fig_hor}). In  the extremal  limit, $\rrh\to 0$, the factor $1/\rrh$  in 
\eqref{kH} diverges, but its growth  is damped by the exponent  $e^{2\Uo-\Ko}$, which approaches zero 
 because the horizon value of $2\Uo-\Ko$ becomes large and negative. 
Therefore, as in the spherically symmetric case, the horizon values of $\Uo$ and $\Ko$
grow without bound as the extremal  limit is approached. The same conclusion holds 
for the $\So$ amplitude.

To analyze the extremal  limit, we choose $\N(\rr)$ according to \eqref{kk_ex}. 
Then, ${\rm N}^\prime (\rrh)=0$, and the surface gravity vanishes without requiring 
 large values of the  metric amplitudes at the horizon. 
The profiles of the extremal  solution with $n=4$ and $\kappa=10^{-3}$ are shown 
in Figs.~\ref{Fig4b}--\ref{Fig4f}. 
The most notable change compared to the non-extremal  solution in Figs.~\ref{Fig1a}--\ref{Fig3a} 
is in the profiles of $\Uo$ and $\So-\Uo$. They show a non-monotonic dependence on $\rr$ and 
 become very similar to those for the extremal spherically symmetric  solution  with $n=2$ shown 
in Fig.\ref{Fig1c} in \ref{AppB}
(for the non-extremal  solution, one has $\So-\Uo\approx -\Uo$). 

Another important change 
is that all amplitudes are now constant at the horizon: $\So-\Uo$, $\Ko-\So$, $H_1$, $H_3$, $y$, $\phi_1$, and 
$\phi_2$ all vanish,
$H_2$ and  $H_4$ equal to unity, whereas  $\Uo$ assumes a constant value. 
Therefore, the electroweak fields are in the false vacuum, $\WW^a_\mu=\Phi=0$ (when expressed in the gauge \eqref{gauge2},\eqref{gauge2a}),
and the horizon 
geometry is perfectly spherical, as will be shown in the next section. 
Thus, the horizon geometry coincides with that of
the extremal RNdS solution \eqref{RN},\eqref{Nex}. 
This motivates  the gauge choice \eqref{kk_ex} with 
$\rrh=r_{\rm ex}$ given by \eqref{horRNdS}.

We have described  above hairy solutions for $n=2$ and $n=4$. 
The $n=3$ solution is qualitatively the same as the one for $n=4$.  
However,  one may also wonder
what happens when $n=1$. The perturbation theory developed in Sections \ref{SecIV} and \ref{Secpert} is valid only for $|n|>1$, 
since for $|n|=1$, the perturbations are unbounded. This implies that there are no  hairy solutions with $|n|=1$ bifurcating 
from the RN family. However, there could be hairy solutions that do not intersect  with the RN family. To detect them, 
we start from the hairy solutions with $n=2$ and then iteratively decrease $n$  down to 
$n=1$. The numerical procedure converges, but the properties of the result, such as the virial 
relation, suggest  that this is not a true solution but rather 
a numerical artifact. Therefore, it seems that there are no  hairy solutions with $n=\pm 1$ within the axially symmetric ansatz.

\section{PROPERTIES OF HAIRY BLACK HOLES \label{Sprop}} 
 \setcounter{equation}{0}

Let us now provide a more general description of hairy solutions for higher values of the magnetic charge. 
Non-extremal hairy black hole  form a two-parameter family, characterized by an integer $n$ that determines  the 
magnetic charge and a real parameter $\rrh$ that sets  the horizon size. 
At large distances from the horizon, the fields asymptotically approach those of the RN solution with magnetic charge $P=n/(2e)$. 
To begin with, we assume  that the charge is not too large, such that
$|Q|=\sqrt{\kappa/2}\,|P|\ll Q_{\rm m}$, where $Q_{\rm m}$ is defined in \eqref{horRNdS1}. 

We emphasize once again that 
the size parameter $\rrh$ does not have a direct physical meaning but merely serves as a label for the solutions. 
Its value determines the event horizon area ${\rm A}_{\rm H}$ according to \eqref{kA}, and as  $\rrh$  
decreases, the horizon area shrinks. 

When the size parameter reaches its maximal value, $\rrh=\rrh^0(n)$, the hairy solution reduces to the RN black hole 
of charge $Q$ and horizon radius  $r_h=r^0_h(n)$. The values of  $ r^0_h(n)$ for  small $n$ are shown in Table \ref{TabII}
(for $\kappa=5.3\times 10^{-33}$), 
while for large $|n|$, we find that   $ r^0_h(n)\approx \sqrt{|n|}/g$. 

\begin{figure}
    \centering
       \includegraphics[scale=0.8]{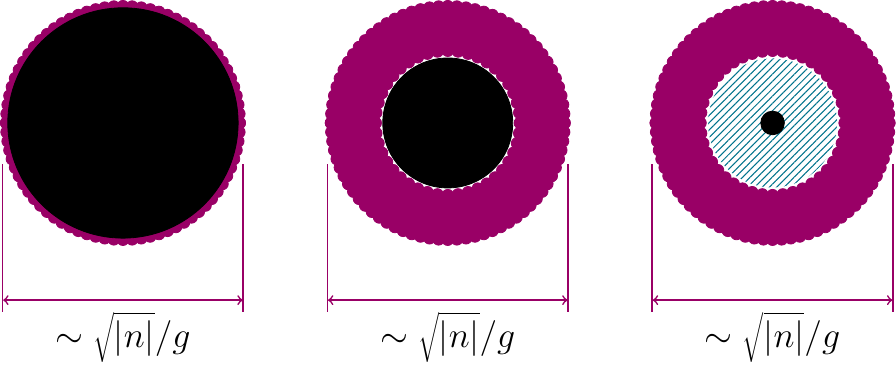}
 
       \caption{\small 
      As the horizon size decreases, 
 the hair first emerges (left), then extends further (center), and eventually, 
a bubble of symmetric phase forms around the horizon (right). 
        }
    \label{Fig22}
\end{figure}

The relation between the maximal value $\rrh^0(n)$ and the Schwarzschild radial coordinate  $r^0_h(n)$ of the 
RN horizon is given by 
\be                   \label{size}
\rrh^0(n)=r^0_h(n)-\frac{Q^2}{r^0_h(n)}= r^0_h(n)+\mathcal{O}(\kappa). 
\ee
For this value of $\rrh$, the strength of the hypermagnetic field at the horizon is just enough to trigger the electroweak 
condensation.  In this regime, the condensate can be described perturbatively by linearizing the W-equations 
 around the RN background, which leads to Eq.\eqref{Proca}. 
 The solution to this equation, given by Eqs.\eqref{Wpert}, \eqref{RNss}, approaches zero outside the horizon 
 more rapidly as $|n|$ increases  (see Fig.\ref{Figw}). 
  Hence, close to the bifurcation point and for large $|n|$, 
  the electroweak condensate is strongly localized  near the horizon, as illustrated 
 schematically in Fig.\ref{Fig22} (left panel). 
 
 When the size $\rrh$ becomes smaller than $\rrh^0(n)$,  the condensate begins to backreact,
 changing the geometry so that it is  no longer purely RN.  In this regime, the electroweak condensate 
 extends further,  spanning the interval $\rr\in[\rrh,\rrh^0]$, where the hypermagnetic field satisfies the condition 
 $|{\rm B}|>\mw^2$. The condensate terminates at approximately $\rr\approx \rrh^0(n)\approx r^0_h(n)$,
  where the hypermagnetic field reduces  to the threshold value  $|{\rm B}|\approx \mw^2$
 (Fig.\ref{Fig22}, central panel). 
 
When $\rrh$ decreases below a 
critical value $\rrh^{\rm b}$, where the hypermagnetic field reaches the second threshold $|{\rm B}(\rrh^{\rm b})|\approx \mh^2$,  
the massive condensate  ceases to expand and  remains  confined within the region 
$\rrh^{\rm b}<\rr<\rrh^0$, where the field strength satisfies $\mh^2>|{\rm B}|>\mw^2$. 
In the near-horizon region
$\rrh<\rr<\rrh^{\rm b}$, the hypermagnetic field becomes strong enough, $|{\rm B}|>\mh^2$, 
to suppress the  $\WW$ and Higgs fields. As a result, the full electroweak gauge symmetry is restored  in this region 
(see Fig.\ref{Fig22}, right panel). This bubble of symmetric phase 
expands as $\rrh$  decreases further, reaching its maximal extent in the $\rrh\to 0$ limit.
In this extremal limit, the horizon surface gravity vanishes, yet the horizon area  remains  finite. 

\subsection{The horizon oblateness} 

The hairy solutions are not spherically symmetric because the condensate is compressed toward the equatorial plane. Since 
the condensate backreacts  and influences the spacetime geometry, the event horizon itself is not perfectly spherical. 
 As a result, there is no uniquely defined event horizon radius.
 
However, the horizon area ${\rm A}_{\rm H}$  remains a well-defined quantity and can be computed using Eq.\eqref{kA}. 
This allows us to 
{\it define} an effective average event horizon radius  as 
\be          \label{rhA}
r_h=\sqrt{\frac{{\rm A}_{\rm H}}{4\pi}}.
\ee
As illustrated in Fig.\ref{Fig22a}, the average radius $r_h$ (and consequently the horizon area ${\rm A}_{\rm H}$)  
is a monotonic  function of 
$\rrh$.  In the extremal limit $\rrh\to 0$, the radius and area 
approach  finite values.  Fig.\ref{Fig22a} also displays the surface gravity 
 ${\rm k}_{\rm H}$, computed from Eq.\eqref{kA}, which tends to zero in the extremal limit, as expected.

Since the horizon is not spherically symmetric, one can define 
 the equatorial radius and polar radius as follows:
\be
r_{h}^{\rm eq}&&=\left.\sqrt{\rm g_{\varphi\varphi}}\right|_{\rr=\rrh,\vartheta=\pi/2}=
\rrh \left.e^{\So-\Uo}\right|_{\rr=\rrh,\vartheta=\pi/2},~~~~\nn \\
r_{h}^{\rm pl}&&=\frac{1}{\pi} \int_0^\pi \left. \sqrt{\rm g_{\vartheta\vartheta}}\right|_{\rr=\rrh}\,d\vartheta
=\frac{\rrh}{\pi} \int_0^\pi \left. e^{\Ko-\Uo}\right|_{\rr=\rrh}\,d\vartheta. 
\ee
This allows us to introduce the horizon oblateness, which is defined as: 
\be
\delta =\frac{r_{h}^{\rm eq}}{r_{h}^{\rm pl}}-1. 
\ee
As seen in Fig.\ref{Fig22a}, the oblateness is positive and vanishes in the RN limit as expected.  
Interestingly, it also vanishes in the extremal limit, a feature that will be further analyzed  below.

\subsection{The charge \label{SpropB}}

Within the Nambu description, 
the total magnetic charge of the black hole splits into two components: 
\be
P=P_{\rm H}+P_{\rm h}=\frac{n}{2e},
\ee
where $P_{\rm h}$ is  the charge contained in the condensate outside the horizon. It is computed 
in the same way as in the 
spherically symmetric case in Eq.\eqref{Pext}:
\be                \label{Pext1}
P_{\rm h}&=&\int_{\rr>\rrh}\tilde{J}^0 \sqrt{\rm -g}\, d^3x=
\frac{e}{8\pi}\int_{\rr>\rrh}\epsilon^{ijk}\partial_i\psi_{jk}\, d^3x 
=-\frac{e}{8\pi}\oint_{\rr=\rrh}\psi_{jk}\, dx^j\wedge dx^k \nn \\
&=&-\frac{e}{4\pi}\oint_{\rr=\rrh}\psi_{\vartheta\varphi}\, d\vartheta\, d\varphi
\equiv \gamma\, g^{\prime 2}P\equiv \lambda P.
\ee
Here, $\psi_{\mu\nu}$ is defined in \eqref{Hooft}. 
The remaining charge, 
$P_{\rm H}=P-P_{\rm h}$, 
is contained inside the black hole. 

The ratio $\lambda=P_{\rm h}/P$ vanishes in the RN limit $\rrh\to \rrh^0(n)$ 
and reaches its maximal value of $g^{\prime 2}\approx 0.22$ in the extremal limit $\rrh\to 0$ 
(provided that the total charge $P$ is not too large, as discussed in Section  \ref{Sextr} below). 
In this case we have $P_{\rm h}\approx 0.22\times P$
(see Fig.\ref{Fig22a}). We also introduce 
the parameter $\gamma=P_{\rm h}/(g^{\prime 2}P)=\lambda/g^{\prime 2}$, which  varies from zero to one. 
In the spherically symmetric case, this parameter corresponds to  $\gamma=f_h^2$, as seen in Eq.\eqref{Pext}.

In summary, hairy black holes exist for $\rrh\in [0,\rrh^0]$. They lose their hair in the RN limit $\rrh\to \rrh^0$,   
where all the charge is contained inside the black hole. In contrast, 
they become  maximally hairy in the extremal limit $\rrh\to 0$, with approximately
$22\%$ of the magnetic charge contained in   the hair outside the horizon.

\subsection{The ADM mass} 

The ADM mass $M$ can be determined either from the asymptotic  expansion of the metric coefficient, 
${\rm g}_{00}=-1+2M/\rr+{\cal O}(1/\rr^2)$,
or from the  integral expression derived using the Einstein equations in \eqref{M0},
\be          \label{mass} 
M=\frac{{\rm k}_{\rm H}{\rm A}_{\rm H}}{4\pi}+\frac{\kappa}{8\pi}\int_{\rr>\rrh}\left(-T^0_{~0}+T^k_{~k}\right)\,\sqrt{-\rm g}\, d^3x,
\ee
where the surface gravity ${\rm k}_{\rm H}$ and the horizon area ${\rm A}_{\rm H}$ are given  by Eq.\eqref{kA}. 
Both methods yield consistent results with the same level of precision  as for the virial relation. 

As in the spherically symmetric case,  
the total mass can be decomposed  as $M=M_{\rm H}+M_{\rm h}$, where $M_{\rm H}$ represents the mass of the RN 
black hole with the same area ${\rm A}_{\rm H}$ and charge $P_{\rm H}$. 
Some clarifications are necessary for computing $M_{\rm H}$.

 The horizon radius of the RN solution with the magnetic charge $P$ and mass $M$ is given by 
 $r_{h}=M+\sqrt{M^2-Q^2}$, 
 where $Q^2=(\kappa/2) P^2$. Expressing $M$ in terms of $r_h$, we obtain 
 \be            \label{MM}
 M=\frac{r_{h}}{2}+\frac{Q^2}{2 r_{h}}=\frac{r_{h}}{2}+\frac{\kappa P^2}{4 r_{h}}.
 \ee
 On the other hand, the total magnetic charge is carried by both the U(1) gauge field $\A_\mu$ 
 and the SU(2) field $\WW^a_\mu$, leading to a decomposition:
$
P=P_{\rm U(1)}+P_{\rm SU(2)}\,,
$
where  $P_{\rm U(1)}=g^2 P$ and $P_{\rm SU(2)}=g^{\prime 2} P$. 
Importantly, the square of the total charge $P^2$ in Eq.\eqref{MM}
can be represented in the form 
\be               \label{P22}
P^2=\frac{1}{g^2}\, P_{\rm U(1)}^2+\frac{1}{g^{\prime 2}}\, P_{\rm SU(2)}^2.
\ee
This formula is crucial: although $P$ is the sum of two charges, $P^2$ 
should not be computed as the square of the sum. Instead, one must use Eq.\eqref{P22} 
because the two charges are of different nature, 
and only the square of each charge contributes to the energy.

The magnetic charge of the  hairy black hole splits as 
$
P= P_{\rm H}+P_{\rm h}, 
$
where $P_{\rm H}$ is the charge inside the black hole, and 
$P_{\rm h}$ is the hair charge defined by Eq.\eqref{Pext1}.
The hair charge is always a part of the non-Abelian charge $P_{\rm SU(2)}$,
leading to the relations
\be
P_{\rm h}=\gamma P_{\rm SU(2)},~~~~P_{\rm H}= P_{\rm U(1)}+(1-\gamma)P_{\rm SU(2)},
\ee
where $\gamma\in [0,1]$ is the same parameter as that introduced in Eq.\eqref{Pext1}.
Comparing with Eq.\eqref{P22}, the square of the charge contained inside the 
black hole  is given by 
\be               \label{P222}
P_{\rm H}^2&=&\frac{1}{g^2}\, P_{\rm U(1)}^2+\frac{(1-\gamma)^2}{g^{\prime 2}}\, P_{\rm SU(2)}^2 
=
(g^2+(1-\gamma)^2g^{\prime 2})P^2.
\ee
Using the definition \eqref{rhA} of the average horizon radius $r_h$
and comparing with Eq.\eqref{MM}, yields the following formula for computing the horizon mass:
\be            \label{MMM}
 M_{\rm H}=\frac{r_{h}}{2}+\frac{\kappa P^2_{\rm H}}{4 r_{h}}=
 \frac12 \left( r_{h}+(g^2+(1-\gamma)^2g^{\prime 2})\frac{Q^2}{r_{h}}\right). 
 \ee
 This expression   matches the formula in Eq.\eqref{MHhor} 
 derived in the spherically symmetric case, upon replacing $\gamma\to f_h^2$. 
 In the RN limit, where $\gamma=0$,  the expression reduces to Eq.\eqref{MM}. In the extremal limit,  
 where $\gamma=1$, the average radius is $r_h=r_{\rm ex}\approx g|Q|$, yielding  
 \be             \label{MMMex}
 M_{\rm H}=
 \frac12 \left( r_{h}+g^2\,\frac{Q^2}{r_{h}}\right)\approx g|Q|.
 \ee

\begin{figure}
    \centering
           \includegraphics[scale=0.8]{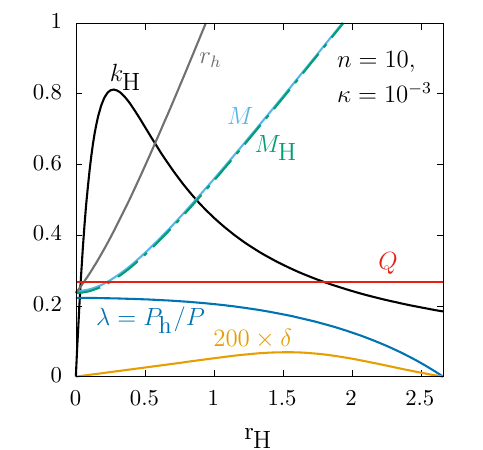}
    \includegraphics[scale=0.8]{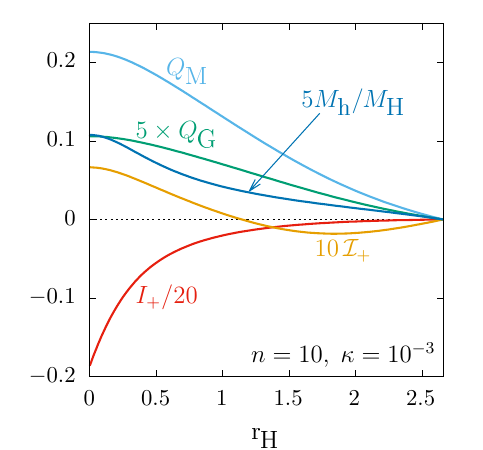}

       \caption{
       Parameters of the non-extremal solutions with $n=10$ and  $\kappa=10^{-3}$. 
       As $\rrh\to 0$, the solutions become extremal, while  for $\rrh\to 2.66$, they lose their hair 
      and merge with the RN solution. The mass $M$ of extremal and near extremal solutions is {\it less}
      than the charge $Q$. 
         }
    \label{Fig22a}
\end{figure}

As seen in Fig.\ref{Fig22a}, 
the mass $M$ increases with 
$\rrh$. For $\kappa=10^{-3}$, 
one finds that  $M\approx M_{\rm H}$ (when $|Q|\ll Q_{\rm m}$). However, 
the relative contribution of the hair mass,
$M_{\rm h}/M_{\rm H}$,  grows with $|n|$. 
This contribution reaches its maximum 
in the extremal limit, where  the black hole is maximally hairy, 
and vanishes in the RN limit.

\subsection{Quadrupole moments}

Far away from the horizon, all massive fields approach their vacuum values, and one has 
\be
B_\mu=\WW^3_\mu\equiv eA_\mu,~~~\WW^1_\mu=\WW^2_\mu=0, ~~~ \Phi=\begin{pmatrix}
0  \\
1
\end{pmatrix},~~~
\ee
thus  the theory reduces  to the electrovacuum:
\be                        \label{ev}
{\cal L}=\frac{1}{2\kappa}\, R-\frac{1}{4 }\, F_{\mu\nu}F^{\mu\nu},
\ee
where $F_{\mu\nu}=\partial_\mu A_\nu-\partial_\nu A_\mu$. 
Therefore, 
the hairy solutions approach an electrovacuum configuration at large distances,
which to leading order corresponds to a magnetic RN black hole of mass $M$ 
and charge $Q = \sqrt{\kappa/2}\,P$.
However, the asymptotic behavior also contains subleading terms 
that encode deviations from the pure RN solution and give rise to gravitational 
and magnetic multipole moments. This allows us to define the gravitational 
$Q_{\rm G}$ and magnetic $Q_{\rm M}$  quadrupole moments.

Determining these moments proved to be particularly challenging, as the existing literature on 
gravitational multipoles often presents conflicting results. The correct formalism was developed by 
Geroch \cite{Geroch:1970cd} and Hansen \cite{Hansen}, but its direct application is difficult in our 
setting. Fortunately, Fodor et al. \cite{Fodor,Fodor1} showed that the Geroch–Hansen procedure can
 be reformulated in terms of the asymptotic expansion of the Ernst potentials in Weyl coordinates. 
 Adopting this method, we were able to compute the quadrupole moments through the following steps.

Choosing the gauge where $\N(\rr)=1$, the line element in Eq.\eqref{metr11}  simplifies to:
\be              \label{ds4a}
ds^2&=& -e^{2\Uo} dt^2+e^{-2\Uo} dl^2,~~ 
\ee
where the 3-metric is:
\be              \label{ds4}
dl^2&\equiv &h_{ik}dx^i dx^k=e^{2\Ko}(d\rr^2+\rr^2 d\vartheta^2)+\rr^2 e^{2\So}\sin^2\vartheta\,d\varphi^2\,. 
\ee
The purely magnetic Maxwell field $F_{\mu\nu}$ can be expressed in terms of a magnetic potential $\Psi$ as: 
\be            \label{dual} 
\sqrt{\frac{\kappa}{2}}\,F_{ik}=e^{-2\Uo} \sqrt{h} \,\epsilon_{iks}\, \partial^s\Psi~\,.
\ee
In the far field region, the hairy solutions reduce to: 
\be                   \label{metr1}
e^{2\Uo}&=&e^{2\delta \Uo}\frac{\X^2-\mu^2}{(\X+M)^2},~~~~
e^{\Ko}=e^{\delta \Ko}\left(1-\frac{\mu^2}{4\xx^2}\right),\nn \\
e^{\So}&=&e^{\delta \So}\left(1-\frac{\mu^2}{4\xx^2}\right),~~~~~
\Psi=\frac{Q}{\X+M}+\delta\Psi,~
\ee
where $\delta\Uo,\delta\Ko,\delta\So,\delta\Psi$ are the subleading terms describing  deviations from the RN 
background. The variable $x$ and the radial coordinate $\rr$ are related by 
$x=\rr+\mu^2/(4\rr)$, where  $\mu^2=M^2-Q^2$. 
As seen  in Fig.\ref{Fig22a} (left panel), 
$\mu^2$ is not necessarily  positive. 

In the leading order,  the expressions in \eqref{metr1} correspond to the RN solution expressed 
in isotropic coordinates. By solving the equations of the electrovacuum theory \eqref{ev} in power series, 
we obtain the deviations:
\be     \label{damRN}
\delta \So&=&-\frac{s_2}{2\xx^2}+\ldots,   \nn \\
\delta \Ko&=&s_2\,\frac{2\sin^2\vartheta-1}{2\xx^2}+\ldots , \nn \\
\delta \Uo&=&\frac{Q_g\,(1-3\cos^2\vartheta)-Ms_2/3 }{2\xx^3}+\ldots , \nn \\
\delta \Psi&=&\frac{Q_m\,(1-3\cos^2\vartheta)+Qs_2/3 }{2\xx^3}+\ldots  ,
\ee
where $Q_g,Q_m$ and $s_2$ are integration constants, and the dots indicate higher-order  terms. 
These constants  can be determined  from our numerical solutions, keeping in mind that 
the numerical solutions were obtained in a different radial gauge where $\N(\rr)\neq 1$.

Introducing the Ernst potential \cite{Ernst,Ernst:1967by}, defined as
\be               \label{Ernst}
{\cal E}=e^{2\Uo}-\Psi^2,
\ee
the gravitational $Q_{\rm G}$ and magnetic $Q_{\rm M}$ quadrupole moments
are determined by the asymptotic behaviour of 
\be          
\xi_{\rm G}=\frac{1-{\cal E}}{1+{\cal E}},~~~~~\xi_{\rm M}=\frac{2\Psi}{1+{\cal E}}
\ee
near spatial infinity  \cite{Fodor,Fodor1}. To achieve this, one must 
switch  to the Weyl coordinates $(\rho,z)$, where the 3-metric in \eqref{ds4} becomes 
\be                  \label{Weyl} 
dl^2&=&\left.\left.e^{2\Ko}\right(d\rr^2+\rr^2d\vartheta^2\right)+e^{2 \So}\,\rr^2\sin^2\vartheta d\varphi^2
=e^{2K(\rho,z)}(d\rho^2+dz^2)+\rho^2 d\varphi^2,
\ee
with $K(\rho,z)$ expressed in terms of $\Ko,\So$. 
The next  step is to consider 
the asymptotic expansions at the symmetry axis ($\rho=0$,  $z\to\infty$):
\be            \label{GH}
\xi_{\rm G}=\sum_{k\geq 0} \frac{a_k}{z^{k+1}},~~~~~~
\xi_{\rm M}=\sum_{k\geq 0} \frac{b_k}{z^{k+1}}. 
\ee
The gravitational and magnetic quadrupole moments are then given by
$Q_{\rm G}=-a_2$ and  $Q_{\rm M}=-b_2$. 
Therefore, the task is to compute the coefficients $a_2$ and $b_2$ in the series \eqref{GH}. 

To pass to the Weyl coordinates, we notice that 
since  $T^r_r+T^\vartheta_\vartheta=0$ in the electrovacuum, Eq.\eqref{SSSS} with $\N(\rr)=1$ yields the exact 
solution for the amplitude $\So$ in \eqref{ds4}, 
 \be             \label{SSSS1}
 e^\So=1 +\frac{const}{\rr^2}=1-\frac{\mu^2+2s_2}{4\rr^2},
 \ee
 where we have also used  \eqref{metr1},\eqref{damRN}. 
 This implies that the relation between the Weyl coordinates $\rho,z$ in  \eqref{Weyl} and 
 the original coordinates 
 $\rr,\vartheta$  in \eqref{ds4} is provided by the complex Joukowski transformation:
 \be                    \label{Jouk}
 \rho+ iz=\varpi-\frac{\mu^2+2s_2}{4\varpi},~~~\text{with}~~~\varpi=i\rr e^{-i\vartheta}. 
 \ee
 Using this relation, the 3-metric in \eqref{ds4} can be transformed to the Weyl coordinates. 
 After computing the potentials $\xi_{\rm G}$ and $\xi_{\rm M}$, 
expanding them according to \eqref{GH}, and using \eqref{metr1},\eqref{damRN},
we obtain the expressions for the gravitational and magnetic quadrupole moments:
\be
Q_{\rm G}=-Q_g-\frac23\, M\, s_2,~~~Q_{\rm M}=Q_m-\frac23\, Q\, s_2.~~~
\ee
It is important to note that the radial coordinate $\rr$ used in \eqref{damRN} 
corresponds to the gauge where $\N(\rr)=1$, while  our numerical solutions were obtained in a different gauge 
where $\N(\rr)=1-\rrh/\rr$. Taking all this into account, we can determine  the quadrupole moments,
as   shown in Fig.\ref{Fig22a}. 

The quadrupole moments vanish for large $\rrh$ when the solution reduces to the RN.
They  grow as  $\rrh$ decreases and the solution becomes more and more hairy.
They reach  maximal values in the extremal  limit.

\subsection{Condensate, vector fields and currents} 

In the unitary gauge, 
the electroweak condensate around the black hole  is  described by a complex vector $w^\mu$ that satisfies 
Eq.\eqref{eqWSc}. The charged condensate  induces an electric current $J^\mu$,  
defined by Eqs.\eqref{psi1},\eqref{cur33}, which flows  along the horizon. 
The current, in turn,  generates the electromagnetic and Z fields governed by \eqref{eqWSa},\eqref{eqWSb}, 
forming a  corona  of vortices orthogonal to the horizon. 
 
 However, using the unitary gauge may lead to complications when the Higgs field approaches zero
 (see Appendix B in \cite{GVII}), which occurs in the extremal limit. To avoid this issue, 
we do not impose the unitary gauge and use the fields $B_{\mu\nu},\WW^a_{\mu\nu},\Phi$ obtained 
from our numerical solutions. Then we 
compute the unit vector 
\be
n^a=\frac{(\Phi^\dagger\tau^a\Phi)}{(\Phi^\dagger\Phi)},
\ee 
whose components remain finite 
when $\Phi$ approaches zero. 
The condensate density 
can be characterized by the  tensor 
\be
\psi_{\mu\nu}=-\frac{1}{g^2}\, \epsilon_{abc}\,n^a{\cal D}_\mu n^b{\cal D}_\nu n^c\,,
\ee
as defined in Eq.\eqref{Hooft}.
Using $n^a$, $\psi_{\mu\nu}$,  one can compute  the electromagnetic and Z  
fields. These fields are described by  ${\cal F}_{\mu\nu}$ and  ${\cal Z}_{\mu\nu}$, as defined   
in  Eq.\eqref{Nambu}  following Nambu,  or by $F_{\mu\nu}$ and  $Z_{\mu\nu}$,  
as defined in Eq.\eqref{FF} following t'Hooft. 

The divergences of ${\cal F}_{\mu\nu}$  and ${F}_{\mu\nu}$ 
yield, according to Eq.\eqref{cur}, the currents ${\cal J}^\mu$ and  ${J}^\mu$, respectively. 
The definitions of Nambu and 't\,Hooft   
lead to two distinct yet dual interpretations of the near-field  structure of solutions.  
However, global parameters such as the total charge and 
quadrupole moment are determined by the far-field behaviour. Since the difference 
between ${\cal F}_{\mu\nu}-{F}_{\mu\nu}=e\psi_{\mu\nu}$ decays exponentially at infinity, 
these global quantities remain the same in both descriptions. 

\subsubsection{'t\,Hooft   description}

Let us consider first the fields $F_{\mu\nu}$ and $Z_{\mu\nu}$, and the associated  current $J^\mu$. 
According to Eq.\eqref{eqWSb}, the neutral current is given by $J^\mu_Z=-(g/g^\prime) J^\mu$.  

As discussed in Section \ref{Secpert}
within the perturbative approach, the current ${J}^\mu$ flows along the horizon, generating vortices
orthogonal to it.  This behaviour persists at the non-perturbative level as well. 
In the axially symmetric case under consideration, the current is purely azimuthal. 

\begin{figure}
    \centering
      \includegraphics[scale=0.8]{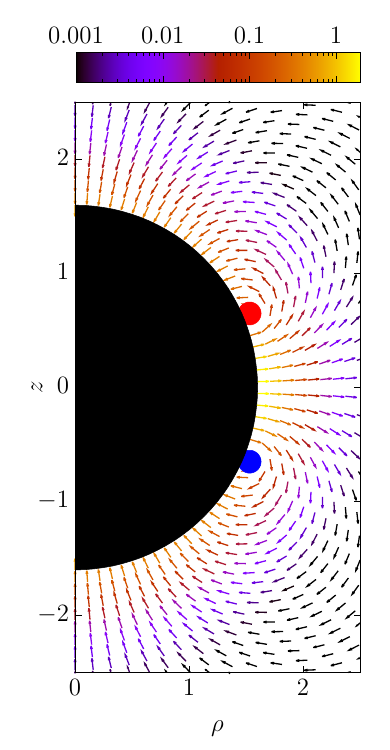}
           \includegraphics[scale=0.8]{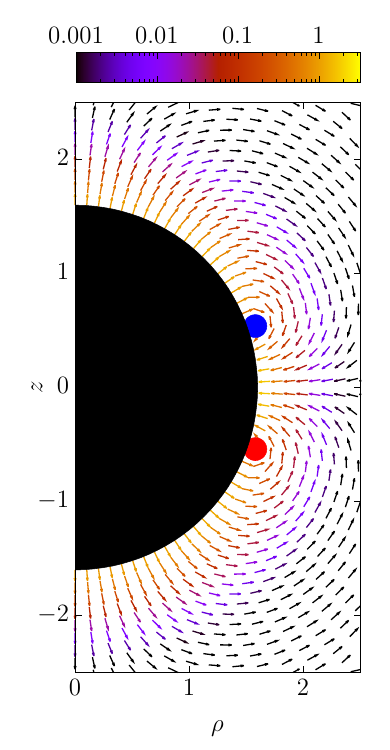}
      \includegraphics[scale=0.8]{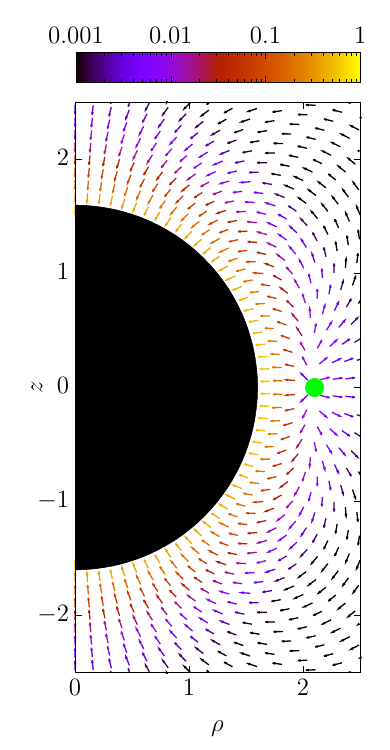}

       \caption{
   Profiles of the 't\,Hooft   magnetic field $\Delta{\cal B}$ (left), the 't\,Hooft   massive magnetic field ${\cal B}_Z$ (center),
   and the Nambu magnetic field $\Delta {\cal B}$ (right) for the 
   non-extremal hairy black hole with $n=10$, $\rrh=1.6$ and  $\kappa=10^{-3}$. 
   In the first two panels, the blue and red (online) spots indicate, respectively,  regions of 
 positive and negative azimuthal currents, $J^\varphi$ and  $J^\varphi_Z$. 
 In the third panel,  the green spot marks the location of the 
   maximal magnetic charge density $\tilde{\cal J}^0$. The field strength is represented by the arrow colour. 
        }
    \label{Fig23a}
\end{figure}

The electromagnetic tensor  has the structure 
\be                 \label{delF}
\frac12\, F_{\mu\nu}\,dx^\mu\wedge dx^\nu=P\sin\vartheta\, d\vartheta\wedge d\varphi+\Delta F_{\rr\varphi}\, d\rr\wedge d\varphi
+\Delta F_{\vartheta\varphi}\, d\vartheta\wedge d\varphi\,,
\ee
where the first term on the right describes the radial magnetic field surrounding  the black hole in the absence of  the condensate. 
The remaining two terms, involving $\Delta F_{\rr\varphi}$ and $\Delta F_{\vartheta\varphi}$, 
are generated by the condensate and are  obtained from the numerical solution. 
These terms  describe the magnetic field,
\be               \label{delB}
\Delta {\cal B}=\frac{1}{\sqrt{-\rm g}}\left(\Delta {F}_{\vartheta\varphi}\,\partial_\rr+\Delta{F}_{\varphi\rr}\,\partial_\vartheta\right). 
\ee
The azimuthal current density $J^\varphi$ is  given by 
\be                \label{delJ}
4\pi \sqrt{-\rm g}\,{J}^\varphi=\partial_\rr(\sqrt{\rm -g}\,F^{\varphi\rr})+\partial_\vartheta (\sqrt{-\rm g}\,F^{\varphi\vartheta})\,.
\ee
The current density changes sign across the equatorial plane and thus can be viewed as describing two distinct currents: 
one negative (for $n>0$) in the upper hemisphere and one positive in the lower hemisphere.
As a result, two loops of oppositely directed azimuthal currents emerge.
Together, these currents support the field $\Delta {\cal B}$, whose 
numerical  profile  is shown in Fig.\ref{Fig23a} (left panel). This field  
is purely solenoidal, winding around regions where the currents attain extremal values. 
The red spot in the upper half-space marks a negative extremum of the upper current, while the blue spot
in the lower half-space indicates a positive extremum of the lower current. 

The field $\Delta {\cal B}$ ressembles  that of 
 a finite segment of a vortex oriented upwards in the upper half-space and a similar vortex oriented downwards 
in the lower half-space.  
Each vortex originates at the horizon in the polar region, extends outwards,  
and eventually returns back to the horizon in the equatorial region. 

When superposed with the 
background magnetic field $P\sin\vartheta\,d\vartheta\wedge d\varphi$, the field  $\Delta {\cal B}$
exhibits an anti-screening effect: it enhances 
the background field in the equatorial region, where the condensate density $\sqrt{\psi_{\mu\nu}\psi^{\mu\nu}}$ is maximal, 
and screens the background field along the axis, where the condensate density vanishes.  
While the vortices do not change the 
total magnetic flux $4\pi P$, they redistribute it,  making it  inhomogeneous. 

Integrating the current density \eqref{delJ} over a 
slice $\rr\in [\rrh,\infty)$, $\vartheta\in [0,\pi/2]$, with $\varphi=const$ (compare with \eqref{cur1}) 
yields the total current in the upper half-space:
\be                 \label{9.30}
4\pi {I}_{+}&=&4\pi \int_{\rrh}^\infty d\rr \int_0^{\pi/2} d\vartheta\, \sqrt{\rm -g} \,{J}^\varphi \,
\nn \\
&=&\left.\int_0^{\pi/2}d\vartheta\,  \sqrt{-\rm g}\,{F}^{\varphi\rr}  \right|_{\rr=\rrh}^{\rr=\infty}+
\left.\int_{\rrh}^{\infty}d\rr\,  \sqrt{\rm -g}\, F^{\varphi\vartheta} \right|_{\vartheta=0}^{\vartheta=\pi/2}\, .
\ee
A similar computation in the lower half-space yields the current ${I}_{-}=-{I}_{+}$. 

The dependence of $I_{+}$ on 
$\rrh$ is shown in Fig.\ref{Fig22a}. It vanishes when the solutions lose hair and bifurcate with the RN configuration, 
and it reaches its maximal  amplitude 
in the extremal limit. 

The direction of the currents depend on the sign of $n$:
\be
{I}_{\pm}=\mp\frac{n}{|n|}\,|{I}_{+}|.
\ee
Thus,  for $P=n/(2e)>0$,  
the current 
$I_{+}$ is always negative (see Fig.\ref{Fig22a}), meaning it flows  opposite to the $\varphi$-direction, 
whereas the current $I_{-}$ is positive, flowing along the $\varphi$-direction. 
This result is consistent with the perturbative limit  (see Eq.\eqref{J} and \ref{AppC}).

\begin{figure}
    \centering
      \includegraphics[scale=0.8]{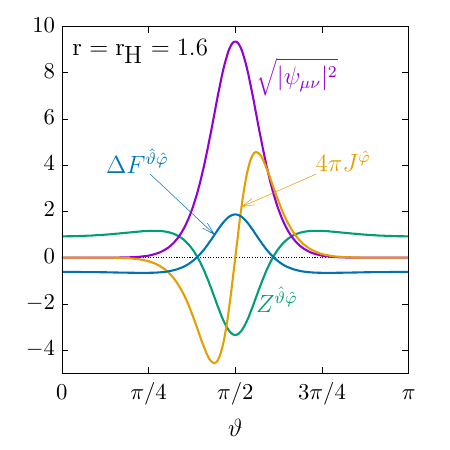}
       \includegraphics[scale=0.8]{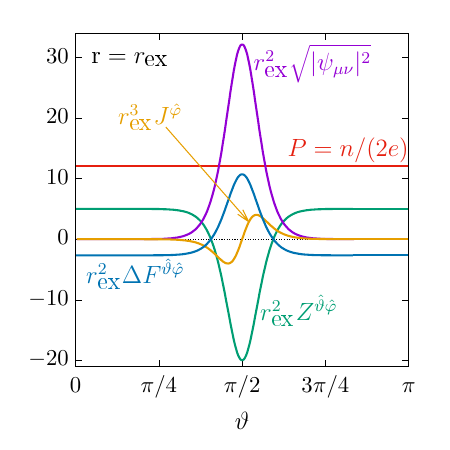}

       \caption{
       The horizon values of the 't\,Hooft fields, current density, and condensate 
      for the non-extremal hairy solution 
       with $n=10$, $\rrh=1.6$  and  $\kappa=10^{-3}$ (left) and for the extremal solution with $n=10$ (right). 
       The hatted indices denote projections on vectors of the orthonormal tetrad.
         }
    \label{Fig24}
\end{figure}

 The central panel of Fig.\ref{Fig23} shows the profile of the massive  magnetic field, 
 \be               \label{delBZ}
{\cal B}_Z=\frac{1}{\sqrt{-\rm g}}\left(Z_{\vartheta\varphi}\,\partial_\rr+Z_{\varphi\rr}\,\partial_\vartheta\right). 
\ee
This time, there is no need to subtract the value in the absence of the condensate because it is zero. 
 The profile  is similar to  that of $\Delta{\cal B}$, except that it flows in the opposite direction, 
 since the Z field is sourced by $J^\mu_Z=(-g/g^\prime)J^\mu$.  
 The vortex structure is now even more evident because the field ${\cal B}_Z$ is directed outwards 
 in the polar regions, resembling two vortices oriented upwards and downwards. 
 The vortices wind around the currents 
 and return back to the horizon in the equatorial region. 
 The field strength decreases rapidly some distance away from the horizon,  and 
  the arrows 
 in Fig.\ref{Fig23a} become black where  the field strength decreases by a factor of 
 $\sim 10^{-3}$. 
 Both $\Delta{\cal B}$ and ${\cal B}_Z$ 
 are purely radial at the horizon, and Fig\ref{Fig24} (left panel)  shows their horizon value, 
 along with those for the current and condensate densities. 

Although $J^\mu_Z=(-g/g^\prime)J^\mu$, meaning  the extrema of the two currents coincide, 
the fields $\Delta {\cal B}$ and ${\cal B}_Z$ in Fig.\ref{Fig23a} do not wind around exactly the same 
regions. This discrepancy arises because each field equation contains an additional term:
\be                \label{delJJ}
\partial_\rr(\sqrt{\rm -g}\,\Delta F^{\varphi\rr})+\partial_\vartheta (\sqrt{-\rm g}\,\Delta F^{\varphi\vartheta})&=& 
4\pi \sqrt{-\rm g}\,{J}^\varphi+\partial_\vartheta (\sqrt{-\rm g}\,P\sin\vartheta\, {\rm g}^{\vartheta\vartheta}g^{\varphi\varphi}), \nn \\
\partial_\rr(\sqrt{\rm -g}\,Z^{\varphi\rr})+\partial_\vartheta (\sqrt{-\rm g}\,Z^{\varphi\vartheta})&=& 
4\pi \sqrt{-\rm g}\,{J}_Z^\varphi-\sqrt{\rm -g}\,\frac{(\Phi^\dagger\Phi)}{2}\, Z^\varphi\,.
\ee

To summarize, within the 't\,Hooft   description, the electroweak condensate  generates azimuthal currents 
that flow along the horizon. These currents create a magnetic field, which forms two vortices orthogonal to the horizon,  
anchored in the polar and equatorial regions. 
These vortices cause  the magnetic field to become inhomogeneous, leading 
to the formations of  the black hole ``corona''.

\subsubsection{Nambu description}

In the Nambu case, the massless and massive magnetic fields are given by: 
\be
{\cal B}^i=\frac{1}{2\sqrt{-\rm g}}\,\epsilon^{ijk}{\cal F}_{jk},~~~~~
{\cal B}_{\cal Z}^i=\frac{1}{2\sqrt{-\rm g}}\,\epsilon^{ijk}{\cal Z}_{jk}.
\ee
These fields are not purely solenoidal,  as their divergences do not vanish (see Eq.\eqref{cur1a}):
\be
\nabla_k {\cal B}^k=\frac{e}{2\sqrt{-\rm g}}\,\epsilon^{ijk}\partial_i\psi_{jk},~~~~~
\nabla_k {\cal B}_{\cal Z}^k=-\frac{g^2}{2\sqrt{-\rm g}}\,\epsilon^{ijk}\partial_i\psi_{jk}. 
\ee
The Nambu field generated by the condensate, $\Delta {\cal B}$, is defined similarly to the previous case, with the substitution 
$F\to {\cal F}$ in Eqs.\eqref{delF},\eqref{delB}.

As seen in the right panel of Fig.\ref{Fig23a}, 
the field $\Delta {\cal B}$ emanates from the region where the magnetic charge density is maximal
(green spot in the figure) and then converges to the horizon. 
Thus, taking into account the rotational invariance, 
the magnetic charge $P_{\rm h}$ is distributed over a ring --  a toroidal region 
outside the event horizon. This ring is especially pronounced in the extremal limit, 
as represented by the green area in Fig.\ref{ex} below, where the level surfaces of $\sqrt{\rm -g}\,\tilde{J}^0$ are shown. 
The magnetic field created by the ring produces a {\it negative} flux through the horizon, 
ensuring  that the latter contains only a part 
of the total  magnetic charge.

The Nambu field contains also a solenoidal part associated with  the electric current, 
defined  by $\nabla_k  {\cal F}^{ik}=4\pi {\cal J}^i$.  
However, a similar relation for a ${\cal Z}$  current, analogous to that in \eqref{eqWS}, cannot be obtained 
since this would require a potential for ${\cal B}_{\cal Z}$, which does not exist since $\nabla_k {\cal B}_{\cal Z}^k\neq 0$. 
A potential for ${\cal B}$ does not exist either, 
but the current ${\cal J}$ is defined without the use of a potential. 

The current density forms   two rings of oppositely directed currents, which  
``sandwich''  the magnetically charged ring (see Fig.\ref{ex}). 
The effect of the Nambu currents is  subleading compared   to that of the magnetic charge density.
As seen in Fig.\ref{Fig22a}, the integrated  current ${\cal I}_{+}$ 
is much weaker than the 't\,Hooft   current ${I}_{+}$. 

One should note  that the curve  ${\cal I}_{+}(\rrh)$ in Fig.\ref{Fig22a} differs from that 
shown in Fig.3 of our previous report  \cite{Gervalle:2024yxj}. 
The difference arises because  ${\cal I}_{+}$  in \cite{Gervalle:2024yxj}  was evaluated using the surface 
integral,  the same one as in the first line in  \eqref{9.30} but with $J^\varphi\to{\cal J}^\varphi$,
which contains second derivatives, leading to large numerical errors. 
Using  instead the contour integrals as in the second line in  \eqref{9.30} 
increases the precision, yielding 
 ${\cal I}_{+}(\rrh)$ shown in  Fig.\ref{Fig22a}. 
 On the contrary, to compute 
 the 't\,Hooft   current ${I}_{+}(\rrh)$,  we always use   the 
contour integral.

Summarizing, within the 't\,Hooft   approach, all the magnetic charge is contained inside the black hole.
However,  there are strong azimuthal 
currents in the condensate that generate a corona made of vortices. In contrast, within the 
 Nambu description, a portion of the magnetic charge is contained in a ring outside the horizon, 
 which generates a  magnetic field which ``imitate'' the vortex structure, as can be seen by 
 comparing  the left and right panels in Fig.\ref{Fig23a}.

\section{EXTREMAL SOLUTIONS \label{Sextr}}  
\setcounter{equation}{0}

The extremal hairy black holes are particularly interesting because, as argued by Maldacena, the presence of the corona should enhance the evaporation rate, causing non-extremal black holes to quickly relax to the extremal state with zero Hawking temperature \cite{Maldacena:2020skw}. For this reason, we discuss the extremal solutions in more detail.

Hairy black hole configurations are squashed toward the equatorial plane. However, as 
shown in Fig.\ref{Fig22a}, the horizon oblateness $\delta$ 
vanishes in the extremal limit, indicating that the horizon becomes spherical --  provided the black hole 
charge is not too large. 

Specifically, the extremal solutions exhibit two distinct phases depending on the value of their charge,
$Q=\sqrt{\kappa/2}\times  n/(2e)\approx 0.85\sqrt{\kappa}\, n$, in comparison to  the critical value, 
\be                  \label{Qstar}
Q_\star\approx 0.62\, Q_{\rm m}=\frac{0.62}{2g\sqrt{\Lambda}}\approx \frac{0.72}{\sqrt{\kappa}}. 
\ee

If $|Q| < Q_\star$ (or equivalently, $|n| < 0.85/\kappa$), the horizon remains spherical, and its oblateness  
$\delta$ vanishes. In this case, the horizon geometry coincides with that of the 
extremal RNdS solution  \eqref{horRNdS}. We refer to this regime as phase I. 

For $|Q| > Q_\star$, the oblateness becomes nonzero, $\delta > 0$, and the horizon takes  on an oblate shape,
so the horizon geometry is no longer that of the RNdS solution. We refer to this regime as phase II.

Thus, the horizon oblateness $\delta$ serves as the order parameter for this phase transition. 
The parameter $Q_{\rm m}$ in \eqref{Qstar} is the same as in \eqref{horRNdS1}, 
but it does not represent the maximum charge for extremal hairy solutions. 
Instead, in \eqref{Qstar} and throughout the discussion below, $Q_{\rm m}$ is used as a natural scale for the charge.

The non-extremal solutions were obtained in the gauge \eqref{Ngauge}, where $\N(\rr) = 1 - \rrh/\rr$. 
However, the fields in this gauge become unbounded in the extremal limit. To study the extremal case, 
 we therefore switch to the radial gauge \eqref{kk_ex}, where $\N(\rr) \sim (1 - \rrh/\rr)^2$. 
In this gauge, the horizon becomes degenerate, and the surface gravity vanishes.
As explained in \ref{AppB}, in the extremal case the size parameter $\rrh$ no longer distinguishes between 
different solutions, but rather corresponds to different radial gauges of the same solution. 

In phase I, where the horizon geometry coincides with that of the 
extremal RNdS solution,  we use the gauge \eqref{kk_ex} with $\rrh = r_{\rm ex}$, 
where $r_{\rm ex}$ is the radius of the extremal RNdS solution, determined by \eqref{horRNdS}. 
With this choice, the radial coordinate $\rr$ coincides with the Schwarzschild coordinate $r$ at the horizon, 
and the function $\N(\rr)$ matches the near-horizon behavior of $N(r)$ for the extremal RNdS solution in \eqref{Nex}. 
The horizon area is then given by  ${\rm A}_{\rm H} = 4\pi r_{\rm ex}^2$.
Since the value of $\rrh$ is fixed, the only remaining essential physical parameter for extremal solutions is the charge $n$.

In phase II, the horizon geometry is no longer spherical. Moreover, the value of $r_{\rm ex}$, as defined by Eq.\eqref{horRNdS1}, 
becomes complex if the charge is large, for $|Q|>Q_{\rm m}$.  Therefore, we abandon the condition $\rrh = r_{\rm ex}$ 
and instead use the gauge \eqref{kk_ex} with  $\rrh = 0.2$. 

Below, we provide a detailed description of extremal hairy solutions in phases I and II.

\subsection{Phase I -- spherical horizon, $|Q|<Q_\star$ }
If the total magnetic charge of the extremal black hole is not too large such 
that   $|n|<0.85/\kappa$, then
the hypermagnetic field  is strong enough to create around the horizon 
a bubble of false vacuum where $\WW^a_{\mu\nu}\approx0$, $\Phi\approx 0$ and the electroweak symmetry is restored.

These field values are the same as those for the extremal RNdS solution \eqref{Nex}. Moreover, as shown in Fig.\ref{Fig22a},
the horizon oblateness vanishes in the extremal limit, 
indicating that the horizon geometry is spherically symmetric -- just as in the extremal RNdS case.
Therefore, the field configuration approaches the extremal RNdS solution at the horizon,
with radius $r_{\rm ex}$ determined by \eqref{horRNdS}, and we accordingly set $\rrh = r_{\rm ex}$ in Eq.\eqref{kk_ex}.

Since 
 $\rrh=r_h=r_{\rm ex}\approx g|Q|\approx 0.75 \sqrt{\kappa}\,|n|$, thus, 
according to \eqref{BRNdS}, the horizon value of the  hypermagnetic field is
\be             \label{Brh}
|{\rm B}(r_h)|=\frac{|n|}{2 r_h^2}\approx \frac{0.89}{\kappa |n|},
\ee
therefore, $|{\rm B}(r_h)|> \mh^2$ in phase I. 

The bubble of symmetric  phase extends up to $r\approx r_{\rm b}$, where $|{\rm B}(r_{\rm b})|\approx\mh^2$,
which occurs at $r_{\rm b}\approx \sqrt{|n|/\beta}\approx 0.73\sqrt{|n}|$.  This corresponds to 
the yellow  area in the left panel of Fig.\ref{Fig25} below. The Higgs field is very close to zero in this region and, 
as explained in \ref{AppE}, it can be approximated by 
\be             \label{smallphi}
\phi_1\approx \left(\frac{r-r_{\rm ex}}{r_{\rm ex}}\right)^{\sigma/\lambda} S_1(\vartheta),~~~~
\phi_2\approx \left(\frac{r-r_{\rm ex}}{r_{\rm ex}}\right)^{\sigma/\lambda} S_2(\vartheta),~
\ee
where $\lambda,\sigma,S_1,S_2$ are defined by \eqref{lam},\eqref{E13}--\eqref{E15}. 
Let us now analyze  the situation from two different perspectives: the approaches of 't\,Hooft and  of Nambu. 

\subsubsection{Structure of vortices forming the corona} 

\begin{figure}
    \centering
            \includegraphics[scale=1.05]{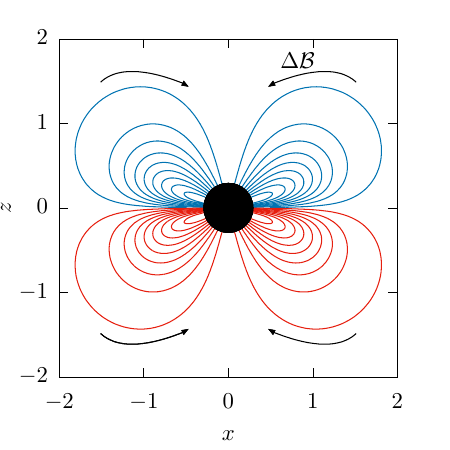}
       \includegraphics[scale=1.05]{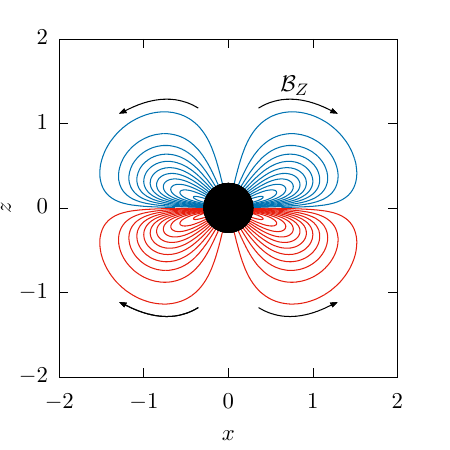}

       \caption{Electroweak corona around the black hole: 
       the level lines of $\Delta A_\varphi=A_\varphi+P\cos\vartheta$ (left) and of $Z_\varphi$ (right) in ``cartesian'' 
       coordinates $x=\pm \rr\sin\vartheta$, $z=\rr\cos\vartheta$ for the extremal hairy solution  with $n=10$  and  $\kappa=10^{-3}$. 
       They coincide, respectively, with the 
        field lines of  $\sqrt{\rm -g}\,\Delta {\cal B}^k$ and $\sqrt{\rm -g}\,{\cal B}^k_Z$, and with those  of     
        $\Delta {\cal B}^k$ and ${\cal B}^k_Z$.   The arrows show the direction of the fields.     }
    \label{Fig24a}
\end{figure}

Within the 't\,Hooft   definition, the electromagnetic and Z fields are defined by Eqs.\eqref{Hooft}--\eqref{2_29}.
They admit potentials \eqref{2_29} whose non-zero components are only azimuthal, hence $F_{j\varphi}=\partial_j A_\varphi$
and $Z_{j\varphi}=\partial_j Z_\varphi$. Subtracting from $F_{ik}$ the background contribution, that is
$P\sin\vartheta d\vartheta\wedge d\varphi=-d(P\cos\vartheta\, d\varphi)$,  and using the definition of the magnetic field in 
Eq.\eqref{magn} yields 
 \be
\sqrt{-\rm g}\, \Delta{\cal B}^i=\epsilon^{ij\varphi}\partial_j (A_\varphi+P\cos\vartheta)\,,~~
~~\sqrt{-\rm g}\, {\cal B}^i_Z=\epsilon^{ij\varphi}\partial_j Z_\varphi\,. 
 \ee
 It follows that field lines of  
 $\sqrt{\rm -g}\,\Delta {\cal B}^k$ and $\sqrt{\rm -g}\,{\cal B}^k_Z$ coincide with the 
level lines of  $\Delta A_\varphi=A_\varphi+P\cos\vartheta$ and $Z_\varphi$ shown in 
Fig.\ref{Fig24a}. They also coincide with lines of $\Delta {\cal B}^k$ and ${\cal B}^k_Z$. A  disadvantage of this representation
 is that the density of field lines reduces  in the polar regions where the factor 
$\sqrt{\rm -g}$ approaches zero, whereas the actual values of the fields  $\Delta {\cal B}^k$ and ${\cal B}^k_Z$ 
are not small at the poles, as seen in  Fig.\ref{Fig23a}. 

Nevertheless, Fig.\ref{Fig24a} clearly shows the structure of the 
  the black hole corona, consisting of two vortices: one in the upper hemisphere (coloured in blue) 
and the other in the lower hemisphere (coloured in red). 

Each vortex starts in the polar region, extends outwards, and then returns to the horizon  in the equatorial region. 
Therefore, each vortex forms loops attached to the horizon by both ends. This is reminiscent of 
the coronal  loops in the solar atmosphere  formed by the prominences.
However, unlike the latter, each vortex is not a localized jet but 
spreads out in all azimuthal directions,  like a fountain, before  returning  to the horizon.

The vortices  carry  fluxes, 
 but since they leave the horizon and then return, 
 the total flux through the horizon is zero  (excluding the background contribution $4\pi P$). 
 
Each vortex  carries  a magnetic dipole moment, 
and the vortices in the upper and lower hemispheres can be viewed 
as two magnetic dipoles  oriented against each other 
so that the total dipole moment vanishes (but not the quadrupole moment). 

The vortices  mutually repel,  which is why the field lines are squeezed when approaching 
the equatorial region. 

\subsubsection{Vortex fluxes and the similarity with Nambu strings}

The fluxes carried by vortices can be evaluated analytically for the extremal solutions. 
Specifically, the near-horizon behavior of the Higgs field expressed by 
Eq.\eqref{smallphi} implies that  the unit vector 
$n^a=(\Phi^\dagger\tau^a\Phi)/(\Phi^\dagger\Phi)$ depends on $\vartheta$ but is independent of $r$
(within the lowest perturbative order). 

Using this information in \eqref{Nambu}--\eqref{FF}, one can evaluate  the non-zero components of the 
condensate density and magnetic field at the horizon (see \ref{AppE}): 
\be             \label{horF2a}
-e\psi_{\vartheta\varphi}= g^{\prime 2}\,\frac{n}{2e}\,\times
\frac{|n|\, \left(\sin \frac{\vartheta}{2}  \cos\frac{\vartheta}{2}  \right)^{|n|-1}}
{\left[ \left(\cos\frac{\vartheta}{2} \right)^{|n|}+\left(\sin\frac{\vartheta}{2} \right)^{|n|}
\right]^2
 }, ~~~~~
 F_{\vartheta\varphi}=
g^2\,\frac{n}{2e}\,\sin\vartheta-e\psi_{\vartheta\varphi}. 
\ee
It is worth noting  that, introducing ${\rm t}=\tan(\vartheta/2)$, one has 
\be               \label{tpsi}
g^2\, \psi_{\vartheta\varphi}=n\,\frac{d}{d\vartheta}\left(\frac{1}{1+{\rm t}^{|n|} } \right)\,.
\ee
The derivative of $F_{\vartheta\varphi}$ determines the horizon value of the current density $J_\varphi$ (see Eq.\eqref{E23}). 
The flux of $F_{\vartheta\varphi}$ through the horizon is  
$4\pi P=4\pi\times n/(2e)$. 

One can split the electromagnetic field as in \eqref{delF}, 
\be             \label{10_7}
F_{\vartheta\varphi}=P\,\sin\vartheta+\Delta F_{\vartheta\varphi}\,,
\ee
where the first term on the right can be associated with the charge contained inside the black hole, 
while the second term can be interpreted  as the magnetic field created by the condensate. One has 
\be              \label{10_8}
\Delta F_{\vartheta\varphi}=-\frac{g^{\prime 2}}{e}\, Z_{\vartheta\varphi},~~~~~~~~
Z_{\vartheta\varphi}=\frac{n}{2}\,\sin\vartheta+g^2\psi_{\vartheta\varphi}.
\ee
The total  fluxes of $\Delta F_{\vartheta\varphi}$ and 
$Z_{\vartheta\varphi}$ through the horizon vanish, but locally they are not zero   because 
$\Delta F_{\vartheta\varphi}$ and $Z_{\vartheta\varphi}$ are sign definite in 
 the polar regions where $0<\vartheta<\vartheta_0$ and $\pi-\vartheta_0<\vartheta<\pi$  (see Fig.\ref{Fig24}, right panel). They 
 change sign at $\vartheta=\vartheta_0$, $\vartheta=\pi-\vartheta_0$, 
where
\be               \label{10_9}
Z_{\vartheta\varphi}(\vartheta_0)=0. 
\ee
To analyze this condition, we use \eqref{tpsi} and \eqref{10_8}, which yields 
\be
Z_{\vartheta\varphi} %
=n\,  \frac{d}{d\vartheta}\left(\frac{1}{1+{\rm t}^{|n|}}-\frac{1}{1+{\rm t}^{2}}  \right)
=n\,(1+{\rm t}^2) \left(\frac{\rm t}{(1+\rm{t}^2)^2}-\frac{|n|}{2}\,\frac{\rm t^{|n|-1}}{(1+\rm{t}^{|n|})^2}\right), 
\ee
hence $Z_{\vartheta\varphi}$ changes sign at the point where 
\be            \label{iter}
{\rm t}=\left(\sqrt{\frac{2}{|n|}}\,\frac{1+{\rm t}^{|n|}}{1+{\rm t}^2}\right)^{2/(|n|-2)}.
\ee
This equation can be solved by iterations, which gives  a value ${\rm t}={\rm t}_0(|n|)$. 
In the large $|n|$ limit, one has ${\rm t}_0\to 1$, whereas $({\rm t}_0)^{|n|}\to 0$. 
It follows then from \eqref{iter} that  $2|n|{\rm t}_0^{|n|}\to 1$, therefore 
\be             \label{theta0}
{\rm t}_0=1-\frac{\ln(2|n|)}{|n|}+\ldots~~~~\Rightarrow~~~~
\vartheta_0=\frac{\pi}{2}-\frac{\ln(2|n|)}{|n|}+\ldots,
\ee
the dots denoting subleading terms. As a result, the point $\vartheta_0$ where the Z field changes sign approaches equator for 
large $|n|$. 
The flux through the northern polar region is
\be
\Psi_Z=2\pi \int_0^{\vartheta_0}Z_{\vartheta\varphi} \,d\vartheta 
=2\pi\, n\,c (n),
\ee
where
\be
c(n)=\frac{{\rm t}_0^2-{\rm t}_0^{|n|}}{(1+{\rm t}_0^2)(1+{\rm t}_0^{|n|})}. 
\ee
This is a monotonous function of $|n|$, such that 
$c(2)=0$,  $c(4)\approx 0.15$, and $c(n)\to 1/2$ for $|n|\to\infty$. 
This gives the fluxes leaving  the horizon in the polar region:
\be                   \label{fluxes}
\Psi_Z=+4\pi\, c(n)\,\times \frac{n}{2},~~~~~\Psi_F=-4\pi\, g^{\prime 2}c(n)\,\times \frac{n}{2e}=-4\pi g^{\prime 2}\,c(n)P\,. 
\ee
Integrating over the equatorial region where $\vartheta_0<\vartheta<\pi/2$ yields the same values, up to the sign,
because fluxes leaving the horizon  are the opposite of those entering the horizon. 

Similarly, one obtains the fluxes in the southern hemisphere, which are the same due to the invariance of the system under the reflection
 $\vartheta\to \pi - \vartheta$.

 The vortex fluxes arise from  the electroweak condensation triggered  by the background 
 magnetic field.  Using the value of the  background flux  through the northern hemisphere, 
 $\Psi_F^0\equiv 2\pi P=\pi n/e$, one can {\it define} the 
 magnetic susceptibility as the ratio of the flux carried by the vortex to the background flux, 
\be              \label{sp} 
\Upsilon(n)=\Psi_F/\Psi_F^0=-2g^{\prime 2}c(n). 
\ee
The factor $c(n)$  has a  geometrical origin because  the vortices exist on top of the horizon 
which is not flat but spherical. The horizon radius increases with $|n|$, and therefore, for large $|n|$, 
the horizon locally approaches a plane, 
 whereas $c(n)\to 1/2$. Therefore, for large  $|n|$, one has $\Upsilon(n)\to -g^{\prime 2}$ (diamagnetism), whereas 
the fluxes carried by each vortex are 
\be                        \label{Nambuv}
\Psi_Z=\pi n,~~~~~\Psi_F=-\pi n \times \frac{g^{\prime 2}}{e}.
\ee
The value   $\Psi_F/\Psi_Z=-g^{\prime 2}/e$  is explained by the fact that the Z-field is massless close to the horizon, 
hence it is proportional to the F-field of the vortex, because the two fields are sourced by proportional currents $J^\mu$ and $J^\mu_Z$ whose ratio is 
$-g^\prime/g=-g^{\prime 2}/{e}$ (see Eqs.\eqref{eqWSa},\eqref{eqWSb}).

It is interesting to compare  the  fluxes \eqref{Nambuv} with those of other known electroweak vortices reviewed 
 in \ref{AppF}. By setting $n=-4m$, the 
  value of $\Psi_F$ matches that in Eq.\eqref{fHooft}, which corresponds 
to the chiral string -- a Z-string with a condensate in its core.
Therefore, horizon vortices are similar to chiral strings. 

As shown in   \ref{AppF}, the chiral string  closely resembles  the Nambu string, which
 terminates at the Nambu monopole \cite{Nambu:1977ag}. 
 The chiral string exhibits exactly the same Nambu fluxes \eqref{fNambu} as the Nambu string,
 and it carries the flux $\Psi_F$ proportional to  $g^{\prime 2}/e$, similarly to the flux produced by the Nambu monopole. 
 This suggests that the Nambu string can  be identified with the chiral string (rather than  with the Z-string as often
 assumed in the literature). 
 Therefore, being similar to chiral strings, the  horizon vortices are similar 
 to Nambu strings. 
 This analogy  is actually quite natural, since the charge $P_{\rm h}$ contained in the hair is a multiple 
 of the $g^{\prime 2}/e$. 
 
At the same time, the similarity cannot be complete, because
Nambu/chiral strings interpolate along the orthogonal direction between the symmetric and broken phases,
whereas for horizon vortices, the orthogonal direction lies along the horizon,
where the system remains entirely in the symmetric phase.
We now proceed to discuss this aspect in more detail.

\subsubsection{The origin of vortices within the symmetric phase \label{Sym_vort}}

The non-extremal hairy black holes carry the hypermagnetic field that creates 
an electroweak condensate, giving rise to vortices. 
The condensate is made of fields that are massive because the gauge symmetry is broken. 
The spherical symmetry is broken as well.

\begin{figure}
      \includegraphics[scale=0.8]{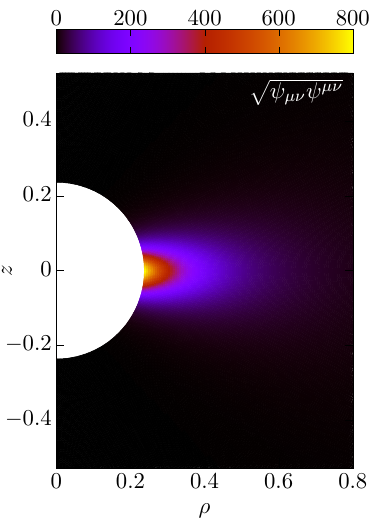}
         \includegraphics[scale=0.8]{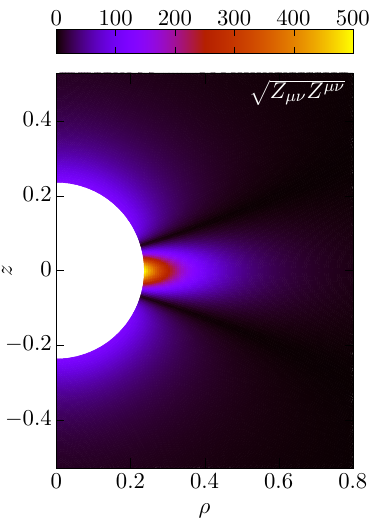}       
       \includegraphics[scale=0.8]{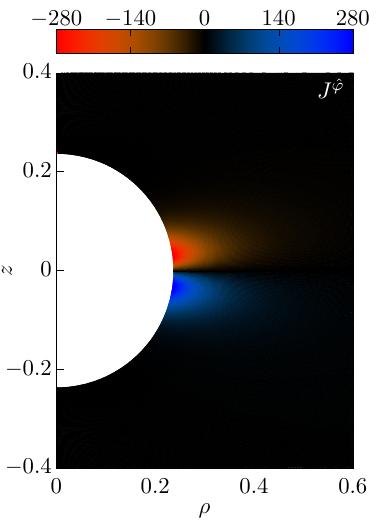}       
       \caption{
       The norms of the condensate $\sqrt{\psi_{\mu\nu}\psi^{\mu\nu}}$ (left), of the Z field 
       $\sqrt{Z_{\mu\nu}Z^{\mu\nu}}$ (center), and the azimuthal current density $J^{\hat{\varphi}}$ (right)
       for the extremal solutions with $n=10$ and  
       $\kappa=10^{-3}$. 
         }
    \label{Fignorm}
\end{figure}

For the extremal solutions, both the electroweak symmetry and spherical symmetry are restored
at the horizon. The geometry there is perfectly spherical, the SU(2) field strength 
and the Higgs field vanish, while the hypercharge field is spherically symmetric,
$B=(n/2)\sin\vartheta\,d\vartheta\wedge d\varphi$.
However, the $F$ and Z fields are non-spherical and form vortices,
whose horizon values are described by the above formulas.
According to our numerics, their amplitudes are maximal at the horizon, where the condensate density also reaches its 
maximal value (see Fig.\ref{Fignorm}). 
How can this be, and what plays the role of the condensate responsible for vortex formation?

In fact, a similar issue has already been analyzed in flat space in
Refs.\cite{Olesen:1991df,VanDoorsselaere:2012zb}.
It was found that, assuming the unitary gauge and approaching the symmetric phase, where $|{\rm B}|>\mh^2$ and
$\Phi=\WW^a_{\mu\nu}=0$,
the SU(2) field potential $\WW^a_\mu$ is not zero but corresponds to
a non-trivial “twisted” pure gauge. It is this pure gauge that plays the role of a {\it massless} condensate,
rendering the $F,Z$ fluxes inhomogeneous and creating vortices even in the symmetric phase.

This is precisely what happens in our case. At the horizon, where $\WW^a_{\mu\nu}=0$,
Eqs.\eqref{Nambu}, \eqref{FF} yield
\be \label{10_14}
F_{ik}=\frac{g^2}{e}\,B_{ik}-e\,\psi_{ik},~~~~Z_{ik}=B_{ik}+g^2\psi_{ik},
\ee
where the condensate field $\psi_{ik}$ is not spherically symmetric (see Eq.\eqref{horF2a}),
although the horizon geometry is. 

The tensor $\psi_{\mu\nu}$ is given by the gauge invariant expression \eqref{Hooft}. 
Let us first consider the 
gauge \eqref{gauge2} -- \eqref{gauge2a},  in which  $\WW^a_\mu$  vanishes  at the horizon. 
Then 
Eq.\eqref{Hooft} reduces at the horizon to 
\be                 \label{psi_ik}
g^2\psi_{ik}=-\epsilon_{abc}\,n^a\partial_i n^b \partial_k n^c\,, 
\ee
where  $n^a$ is the $\Phi\to 0$ limit of $(\Phi^\dagger\tau^a\Phi)/(\Phi^\dagger\Phi)$. 
Using $\Phi$ 
given by Eq.\eqref{gauge2a} 
with $\phi_1,\phi_2$ from \eqref{smallphi} yields 
\be                   \label{10.20}
n^a=\frac{(S_1)^2-(S_2)^2}{(S_1)^2+(S_2)^2}\, e_r^a+
\frac{2S_1S_2}{(S_1)^2+(S_2)^2}\, e_\vartheta^a\,,
\ee
with unit vectors $e^a_r,e^a_\vartheta$ defined in \eqref{3vec}  and $S_1(\vartheta),S_2(\vartheta)$ 
given by \eqref{E13},\eqref{E14}. Inserting  this into \eqref{psi_ik} and using the relation $\nu=-n/2$ 
yields exactly the same components of the condensate density 
$\psi_{\vartheta\varphi}=-\psi_{\varphi\vartheta}$  as in \eqref{horF2a}. 

One notices that $\psi_{ik}$ in \eqref{psi_ik} is proportional to the volume form on the two-sphere spanned by the unit vector $n^a$. 
When $\vartheta,\varphi$  run over the horizon, the unit vector $n^a$ spans the two-sphere $(-\nu)$ times, 
where the minus sign  arises  because  $n^a$ runs in the opposite direction, since one has 
$n^a=\mp \delta^a_3$ for $\vartheta=0,\pi$, respectively. 
This corresponds to the ``twist'' of the condensate. 

The horizon flux of $\psi_{ik}$ is therefore proportional to $(-\nu)\times 4\pi$, 
\be
e\oint \psi_{\vartheta\varphi}\, d\vartheta d\varphi
=-e\times\frac{(-\nu)\times 4\pi }{g^2}=
e\times \frac{4\pi \nu}{g^2}=-g^{\prime 2}\times 4\pi\, \frac{n}{2e}=-4\pi\,g^{\prime 2}P. 
\ee
This reproduces Eq.\eqref{E20} in \ref{AppE} and ensures that the flux of $F_{\vartheta\varphi}$, defined according to 
 \eqref{10_14}, is equal to $4\pi P$, while the flux of $Z_{\vartheta\varphi}$ is zero.  
 
 However, the gauge in which  $\WW^a_\mu$  vanishes  at the horizon
 is not unitary. The Higgs field is given in this gauge by Eq.\eqref{gauge2a}, where both the lower and upper components do not vanish,
 although they approach zero at the horizon. To pass to the unitary gauge where $\Phi=(0,\phi)^{\rm T}$, one can first apply the 
 transformation inverse to that in \eqref{Ureg}, which brings the fields back to the form \eqref{RR}, where $\Phi=(\phi_1,\phi_2)^{\rm T}$. 
 
 Next, one applies another  gauge 
 transformation generated by $\exp(2i\alpha\,\T_2)$, whose effect on the upper component of the Higgs field is 
 $\phi_1\to \tilde{\phi}_1=\phi_1\cos\alpha+\phi_2\sin\alpha$. Requiring that $\tilde{\phi}_1=0$ yields 
  \be
 \tan(\alpha)=-\frac{\phi_1}{\phi_2}=-\frac{S_1(\vartheta)}{S_2(\vartheta)}=\frac{2}{|\nu|+1}\,\frac{S_2^\prime(\vartheta)}{S_2(\vartheta)}=
 {\rm t}\, \frac{{\rm t}^{|\nu|-1} -1}{{\rm t}^{|\nu|+1} +1 },
 \ee
with ${\rm t}=\tan(\vartheta/2)$, and  we used the formula \eqref{smallphi}  for the Higgs field in the horizon vicinity, 
and also Eqs.\eqref{E13},\eqref{E14} from \ref{AppE}. 

As a result, the gauge transformation bringing  the fields  \eqref{gauge2} -- \eqref{gauge2a} to the unitary gauge 
is generated by 
\be
{\rm \U}=e^{i(\vartheta+2\alpha)\T_2 }e^{i\nu\varphi \T_3 }
\ee
(we have omitted the U(1) part).
The horizon value of the SU(2) field in this gauge is 
\be
\T_a \WW^a_\mu dx^\mu&=&i\,{\rm \U}\,\partial_\mu {\rm \U}^{-1}dx^\mu      \nonumber \\
&=&\T_2\, d(2\alpha+\vartheta)
+\nu\left\{\cos(2\alpha+\vartheta) \T_3 -\sin(2\alpha+\vartheta) \T_1\right\}d\varphi.       \label{10.24} 
\ee
Since in the unitary gauge we have $n^a=-\delta^a_3$, the formula \eqref{Hooft} reduces to \eqref{psi} and yields 
\be
g^2\psi_{\vartheta\varphi}=\WW^1_\vartheta \WW^2_\varphi-\WW^1_\varphi \WW^2_\vartheta=
-\nu \left(\cos(2\alpha+\vartheta)\right)^\prime=
n\,\frac{d}{d\vartheta}\left(\frac{1}{1+{\rm t}^{|n|} } \right)\,,~~~
\ee
which precisely agrees with \eqref{tpsi}. 
 Therefore,  the condensate density 
 $\psi_{ik}$ responsible for the vortex formation within the symmetric phase 
 can be viewed as  a non-trivial  pure gauge, in agreement  with 
\cite{Olesen:1991df,VanDoorsselaere:2012zb}. 

To recapitulate, in the gauge defined by Eqs.\eqref{gauge2},\eqref{gauge2a}, one has $\WW^a_\mu=0$ at the horizon, 
but the condensate is supported by the non-trivial angular dependence of the unit vector $n^a$ in Eq.\eqref{10.20}. 
Upon passing to the unitary gauge, the unit vector $n^a$ becomes trivial, while $\WW^a_\mu$ becomes the 
pure gauge configuration given by Eq.\eqref{10.24}, inheriting the non-trivial angular dependence.

\subsubsection{Agreement between the different descriptions \label{SecAdj}}

One more issue remains to be addressed.  The equatorial region, 
where the vortices converge  back to the horizon, $\vartheta_0\leq \vartheta\leq \pi-\vartheta_0$,  is notably  narrow: 
its width decreases    as $\ln(|n|)/|n|$  when   $|n|$ increases (see \eqref{theta0}). At the same time, the fluxes \eqref{Nambuv} 
scale as $|n|$, hence the vortex fields in this region grow  as $n^2/\ln(|n|)$. As illustrated  in Fig.\ref{Fig24a}, 
the density of field lines becomes very large in the equatorial region. This has interesting consequences.

As mentioned above, 
horizon vortices bear a resemblance to Nambu strings.  When the latter terminate, 
the magnetic flux $\Psi_F$  trapped inside the string spreads outward from the termination point, 
generating a radial magnetic field that mimics a magnetic monopole with charge proportional to $g^{\prime 2}/e$    \cite{Nambu:1977ag}. 
Based on this, Maldacena proposed that the black hole corona consists of vortices attached to the horizon at one end and terminating on monopoles some distance away, as depicted in Fig.4b of \cite{Maldacena:2020skw}. 
This interpretation is supported by the Nambu description, which suggests that 
a portion of the magnetic charge resides outside the horizon.

At the same time, we know that the corona rather consists of vortex segments attached to the horizon at both ends, forming loops. 
This might appear to contradict the interpretation of \cite{Maldacena:2020skw}. 
However, it turns out that the two interpretations are not mutually exclusive and can, in fact, be reconciled.

Specifically, let us compute the outgoing flux through the equatorial region of the total magnetic field, which includes both  
the background contribution and the vortex field, i.e., 
$F_{\vartheta\varphi}=P\sin\vartheta+\Delta F_{\vartheta\varphi}$ (see \eqref{10_7}).
Assuming that $|n|\gg 1$ and using \eqref{theta0}, \eqref{Nambuv}, we obtain 
\be
\Psi_{\rm equator}&=&2\pi\int_{\vartheta_0}^{\pi-\vartheta_0} 
(P\sin\vartheta+\Delta F_{\vartheta\varphi})\,d\vartheta= 4\pi P\cos(\vartheta_0)-2\Psi_F \nn \\
&=&4\pi P\left( g^{\prime 2}+\frac{\ln (2|n|)}{|n|}+\ldots\right) \approx 4\pi g^{\prime 2}P, 
\ee
where the dots denote subleading terms. Since the vortex field $\Delta F_{\vartheta\varphi}$ is very large, it provides 
the dominant contribution,  
while the effect of the background field $P\sin\vartheta$ is subleading. 

As a result, the total flux emitted by the narrow equatorial region corresponds to that 
of a magnetic monopole with charge $g^{\prime 2}P\propto g^{\prime 2}/e$,  
aligning with the interpretation  of  \cite{Maldacena:2020skw}. 
Due to their strong confinement in the equatorial region, the vortex fields  mimic  an ``effective'' 
magnetic charge $g^{\prime 2}P$  at the horizon, even though 
within  the 't\,Hooft framework, 
all of the charge remain confined inside the horizon. 

This effective charge is not localized at a single point but is instead distributed along 
a ring formed by the intersection of the horizon and the equatorial plane, 
as the flux $\Psi_{\rm equator}$ remains uniform along the azimuthal direction.
One can therefore interpret the equatorial region of the event horizon 
as being effectively charged. Remarkably, this aligns with the Nambu perspective, 
which posits that the $g^{\prime 2}P$ portion of the total magnetic charge $P$  resides in a ring outside the horizon (see below).
As a result, what the Nambu approach considers to be the actual charge 
$g^{\prime 2}P$ is, in the 't Hooft approach, effectively mimicked by the distribution of field lines, 
producing an equivalent charge of the same strength. The only  distinction lies in their locations: in the 't Hooft approach, 
the charge is located at the horizon, whereas in the Nambu approach, it resides some distance away -- 
a point that will be discussed  below. 

It is instructive to revisit  the perturbative picture shown in Fig.\ref{Fig6}. The right panel illustrates the 
axially symmetric solution, where  the yellow equatorial zone appears  charged within the Nambu approach. 
We now understand that this picture extends beyond the perturbative regime, 
showing that the equatorial zone corresponds to the region of the horizon 
that is effectively charged within the 't\,Hooft framework.

The left panel in  Fig.\ref{Fig6} depicts  a non symmetric minimum energy solution.
At present, we have not tried  to extend this solution at the non-perturbative level. 
However, if such a generalization exists, 
then the six yellow zones in the figure should likewise become effectively charged regions of the horizon, 
corresponding to monopoles.  Consequently, the non-perturbative extension of the non-symmetric solution 
in Fig.\ref{Fig6} should feature eight vortices, interpolating between to the dark zones, where the field lines 
leave the horizon, and the six monopoles, where the field lines  converge back.

These observations establish a connection between the 't\,Hooft perspective, 
according to which all the magnetic charge is confined within the black hole, 
and the Nambu perspective, which suggests that part of the charge resides outside. 
Furthermore, they provide a link between the interpretation of vortices proposed in \cite{Maldacena:2020skw} 
and our findings.

\subsubsection{Summary of the 't\,Hooft  description}

Let us recapitulate everything  said above. Within the 't\,Hooft   description,  
the electroweak condensate around the black hole supports 
currents flowing 
along the horizon. 
The condensate and the current attain their maximal values at the horizon \tc{(see Fig.\ref{Fignorm})}. 
These currents create 
vortices attached to the horizon at both ends, resembling the coronal loops 
in the Sun's atmosphere (see Fig.\ref{Fig24a}). 
The vortices originate in the polar regions of the horizon, extend outside,  spread out  like fountains, 
and eventually return   back to the 
horizon in the equatorial region. These vortices carry a magnetic flux proportional to $g^{\prime 2}/e$,
similar to Nambu strings. The narrow equatorial region, where the vortices converge to the horizon, 
can be viewed as carrying an effective charge $g^{\prime 2}P$, since the field strengths 
are significantly high there. 
The vortices collectively form the black hole corona, while 
the condensate and corona constitute  together the black hole hair.

\begin{figure}
      \includegraphics[scale=0.7]{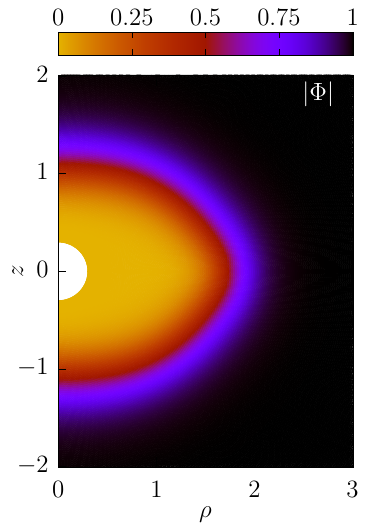}
         \includegraphics[scale=0.7]{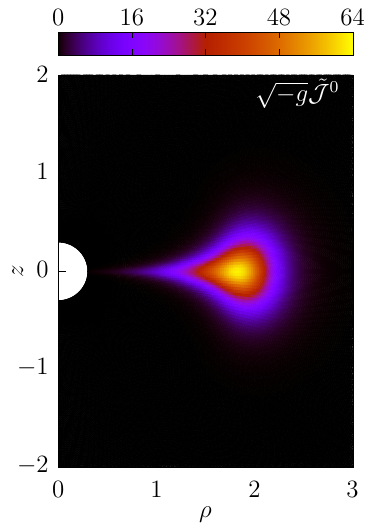}       
                     \includegraphics[scale=0.7]{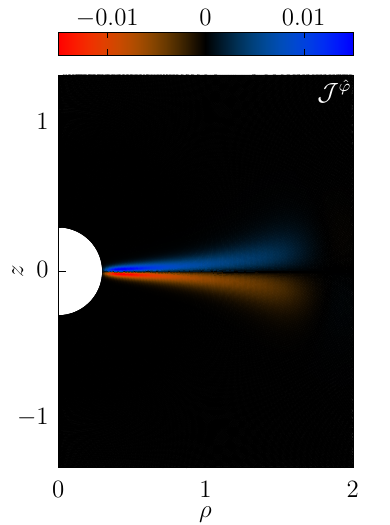}      
       \caption{
       The norm of the Higgs field $\sqrt{\Phi^\dagger\Phi}$ (left),
       the Nambu magnetic charge density $\sqrt{\rm -g}\,\tilde{{\cal J}}^0$ 
       (center), and the Nambu electric current density ${\cal J}^{\hat{\varphi}}$ 
       (right) for the extremal solutions with $n=40$ and  
       $\kappa=10^{-4}$. 
         }
    \label{Fig25}
\end{figure}

\subsubsection{The magnetic ring, ADM mass, and the weakness of gravity}

Within the Nambu description, the electroweak condensate around the black hole is magnetically charged. 
For extremal solutions in phase I, the condensate carries  the charge $g^{\prime 2}P$, which represents  $22\%$ 
of the total charge, the remaining  
$78\%$ being  contained inside the black hole (see Section \ref{SpropB} above).

Although  the 't\,Hooft   current  density is maximal  at the horizon, the Nambu current density 
and the magnetic charge density  both vanish there. 
However, as seen in Fig.\ref{Fig25}, the magnetic charge density $\sqrt{\rm -g}\,\tilde{{\cal J}}^0$ and the Nambu current density 
${\cal J}^{\hat{\varphi}}$ become non-zero and achieve maximal values some distance away from the  horizon,
where
the Higgs field 
deviates from zero.  Therefore, one may say that, in the 't\,Hooft description, the charge and current are located at the horizon,
whereas in the Nambu description, they are situated some distance away from it.

\begin{figure}
    \centering
		\includegraphics[width=9 cm,angle =0 ]{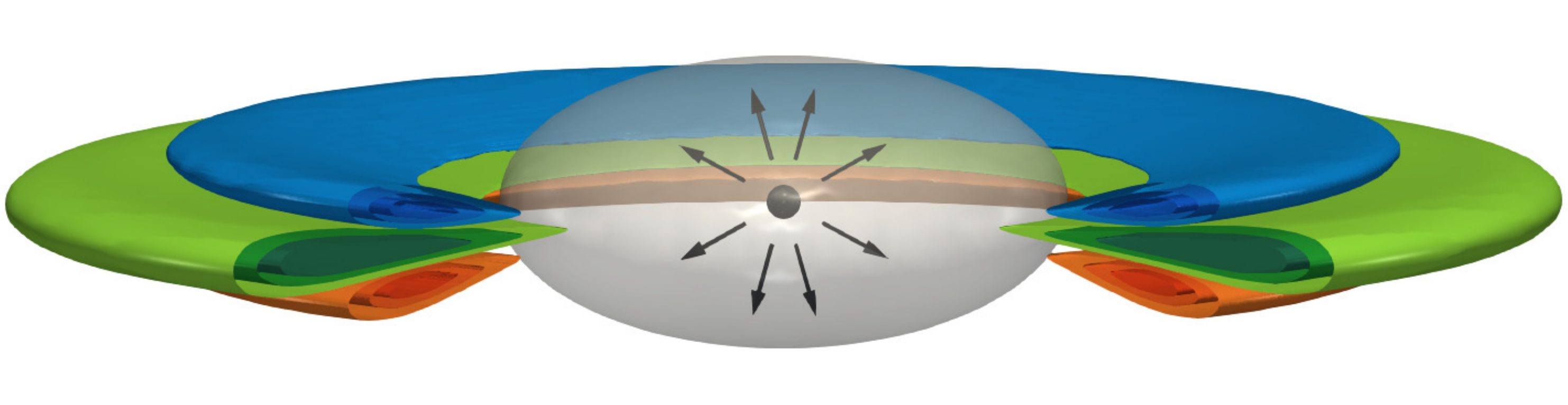}
						
       \caption{
The extremal  solutions in phase I contain  a small charged black hole inside a bubble of the symmetric phase (grey online), surrounded 
by a ring-shaped electroweak  condensate supporting $22\%$ of the total magnetic charge (green) and two opposite superconducting currents
(blue and red). 
Farther away the condensate disappears and  the magnetic field becomes coulombian. 
        }		
    \label{ex}
\end{figure}

The 3D profiles 
are shown in  Fig.\ref{ex}, where  the central ring (green online) corresponds to the level surfaces of the magnetic charge density.  
This ring is sandwiched between two others (blue and red online) 
containing  superconducting currents $\mathcal{I}_{+}$ and $\mathcal{I}_{-}=-\mathcal{I}_{+}$ 
represented in Fig.\ref{ex}  by  level surfaces of $\sqrt{\rm -g}\,{\cal J}^\varphi$. 
The  Nambu currents  are, however, relatively weak, 
the absolute values of $\mathcal{I}_{\pm}$ being  about two orders of magnitude smaller than 
those for  the 't\,Hooft   currents ${I}_{\pm}$ (see Fig.\ref{Fig23}).

The condensate 
terminates where $|{\rm B}|\approx\mw^2$,  at $r_{\rm c}\approx \sqrt{|n|}/g\approx 1.13\sqrt{|n|}$. 
Still farther away, 
the configuration approaches the RN solution \eqref{QQ},\eqref{RN} whose ADM  mass turns out to be 
{\it  smaller} than the charge,
\be                   \label{MQ}
M<|Q|. 
\ee
We have already observed this phenomenon in the spherically symmetric case (Section \ref{SspherB}). 
To repeat, the hair  carries $22\%$ of the total charge, hence its charge is 
 $Q_{\rm h}=0.22\times Q$, but its  mass $M_{\rm h}$ is small  due to the negative Zeeman energy. 
 This happens because 
 the condensate contains charged W bosons whose interaction 
 with the magnetic field of the black hole shifts their mass as $\mw^2\to \mw^2-|{\rm B}|$ \cite{Ambjorn:1989sz}. 
 As a result, one has $M_{\rm h}<|Q_{\rm h}|$, which explains  \eqref{MQ}.

If $|n|$ is not very large  such that  $|Q|\ll Q_\star$, then the 
mass-to-charge  ratio for the hair is very small, 
$M_{\rm h}/|Q_{\rm h}|\sim \sqrt{\kappa}\ll 1$. For example, for $n=2$, one has $M_{\rm h}/Q_{\rm h}\sim 10^{-16}$ 
(see the estimate below Eq.\eqref{MH}). 
The horizon contribution to the mass is then dominant, 
$M=M_{\rm H}+M_{\rm h}\approx M_{\rm H}$, where $M_{\rm H}$ is determined by 
\eqref{MMMex} with $r_h=r_{\rm ex}\approx g|Q|$. One obtains in this case  the same estimate  as in \eqref{MH},
\be             \label{MQo}
M\approx r_h\approx g|Q|=0.88\times |Q|. 
\ee
As $|n|$ increases, the ratio $M_{\rm h}/|Q_{\rm h}|$ also grows, 
and the estimate \eqref{MQo} ceases to be  valid, 
but the inequality $M < |Q|$ continues to hold for all extremal solutions.

The smallness of the ratio $M_{\rm h}/|Q_{\rm h}|$  can be interpreted as a manifestation 
of the weak gravity conjecture \cite{Arkani-Hamed:2006emk}: gravity is the weakest force because 
the magnetic repulsion between the condensate and the black hole is stronger than the gravitational attraction.
However,  the condensate cannot be pushed away to infinity due to the Yukawa 
potential barrier at large distances, as the condensate consists of massive fields. 

\subsubsection{Solutions with $|Q|\ll Q_\star$ versus  electroweak monopoles}

We  used in our numerics values of $\kappa$ ranging between 
$10^{-10}$  and $10^{-2}$ since otherwise, it is challenging to achieve numerical convergence. 
Extrapolating our  results to the physical value 
 $\kappa\sim 10^{-33}$ then yields the following picture of extremal solutions with  $|Q|\ll Q_\star$. 
 
 Their horizon size $r_{\rm ex}\approx 0.75\,\sqrt{\kappa}\, |n| $ 
 is {\it parametrically} small compared to the size of the region where the massive fields change, 
 $r_b\approx 0.73\,\sqrt{|n|} <r<r_c\approx 1.13\,\sqrt{|n|} $. 
 Therefore, the tiny black hole does not gravitationally affect the electroweak configuration 
 but only supports the strong hypermagnetic field, which is necessary to 
 create the electroweak condensate. It also regularizes the energy of this field to a 
 finite value, $M_{\rm H} \approx g|Q| \sim \sqrt{\kappa} ,|n|$. The massive condensate 
 resides far from the horizon, where the geometry is nearly flat. 
 The condensate carries mass $M_{\rm h} \sim \kappa ,|n|^{3/2}$ 
 (as explained below after Eq.\eqref{10_30}), meaning the ratio $M_{\rm h}/\kappa$ 
 remains finite as $\kappa \to 0$. This leads to the following consequence.

 The dimensionful ADM mass in units of the electroweak mass scale is  expressed by  \eqref{MM1},\eqref{M1}, 
 \be
 {\bf M}=\frac{e^2}{4\pi\alpha}\,{\cal M}\times {\bm{m}_0}\,,
 \ee 
where 
  \be
 {\cal M}\equiv \frac{8\pi}{\kappa}\times M=  \frac{8\pi}{\kappa}\times(M_{\rm H}+M_{\rm h})=\frac{4\pi g|n|}{\sqrt{2\kappa}\,e}+\frac{8\pi}{\kappa}\,M_{\rm h}\equiv {\cal M}_{\rm H}+{\cal M}_{\rm h}. 
 \ee
 Therefore, 
 \be
 {\bf M}={\bf M}_{\rm H}+{\bf M}_{\rm h},
 \ee
 where the horizon contribution is 
 \be
 {\bf M}_{\rm H}=\frac{e^2}{4\pi\alpha}\,{\cal M}_{\rm H}\times{\bm{m}_0}=\frac{e g|n|}{\sqrt{2\kappa}\,\alpha}\,{\bm{m}_0}=
 5.1\,|n|\,{\bf M}_{\rm Pl}\,.
 \ee
 Notice that this diverges  when  $\kappa\to 0$. At the same time, the hair contribution, 
 \be
 {\bf M}_{\rm h}=\frac{e^2}{4\pi\alpha}\,{\cal M}_{\rm h}\times{\bm{m}_0},
 \ee
 remains finite when  $\kappa\to 0$ because 
 ${\cal M}_{\rm h}$ reduces in this limit precisely to the regularized mass of the flat space electroweak monopole:
 the Cho-Maison monopole for $|n|=2$ \cite{Cho:1996qd}  and its  generalizations for $|n|>2$ \cite{GVII}. 
 We have already discussed in Section \ref{SspherC} the correspondence  between  the Cho-Maison monopole and the extremal hairy 
 black hole with $|n|=2$. It turns out that a similar  correspondence applies  to  higher charge solutions as well. 
 
 As a result, the  extremal hairy solution with $|Q|\ll Q_\star$ can be viewed
 as the flat space monopole with  the same charge $Q$ harboring in the center a tiny charged  black hole that  almost  does not 
 affect the electroweak fields. The total mass of  the flat space  monopole
 is infinite because of the Coulombian singularity of the hypermagnetic field ${\rm B}$ at the origin, 
 but gravity renders the mass  finite due to the cutoff at the horizon. Therefore, the extremal hairy black holes 
 provide  ``a classical UV completion'' for the  electroweak  monopoles. 
 
 \subsubsection{Increasing the charge} 
 
 The size of the black hole is small compared to the size of the surrounding condensate 
 only if  the charge is not too large: $|Q|\approx 0.85 \sqrt{\kappa}|n|\ll Q_\star\approx 0.72/\sqrt{\kappa}$. 
Increasing  $|n|$, the size of the vacuum bubble  
$r_{\rm b}\approx 0.73\,\sqrt{|n|}$ increases 
and the condensate size  $r_{\rm c}\approx 1.13\,\sqrt{|n|}$ grows as well,
but the horizon radius  $r_{\rm ex}\approx 0.75 \sqrt{\kappa}\,|n|$ 
 grows faster, and when $n$ becomes of the order of $1/\kappa$, 
  the black hole starts absorbing  the false vacuum region where the Higgs field is very small.  
  
The horizon mass and hair mass are 
\be           \label{10_30}
M_{\rm H}\approx r_{\rm ex}\approx 0.75 \sqrt{\kappa}\,|n|,~~~~~~
M_{\rm h}\approx 0.10\,\kappa\,|n|^{3/2},
\ee
where the latter follows from the estimate obtained in Minkowski space, 
because 
the rescaled hair mass ${\cal M}_{\rm h}=(8\pi/\kappa)M_{\rm h}$ 
essentially coincides with  the regularized energy $E_{\rm reg}$ of the flat space monopole 
given by Eq.(5.24) in \cite{GVII}. Taking gravity into account, our numerics confirm that $M_{\rm h}\propto |n|^{3/2}$
if $|Q|<Q_\star$. 

As a result, although for small $|n|$  the horizon mass $M_{\rm H}$ is totally 
dominant as compared to the hair mass $M_{\rm h}$, 
the latter increases  faster with growing $|n|$ and eventually overtakes the horizon mass. 
This means that 
the ratio $M_{\rm h}/|Q_{\rm h}|$ increases, but according to our numerics it always remains  less than unity. 
The estimate \eqref{MQo} is no longer valid when $Q$ approaches $Q_\star$, but the inequality $M<|Q|$ remains 
valid for ell extremal solutions. 

On the other hand, 
the horizon value of the hypermagnetic field, $|{\rm B}(r_h)|\sim 1/|n|$  (see \eqref{Brh}), 
{\it decreases} with growing $|n|$, and when $|Q|\to Q_\star$
then $|{\rm B}(r_h)|\to\mh^2$ and 
the Higgs field starts deviating  from zero at the horizon. 
The solution then enters the next phase in which the electroweak symmetry  is broken everywhere. 
Analyzing  the condition $|{\rm B}(r_h)|=|n|/(2r_h^2)=\mh^2$ together with Eq.\eqref{horRNdS1} 
determining  the value of $r_h=r_{\rm ex}$ 
yields 
\be
Q_\star=\frac{2\sqrt{2} g^\prime}{g(g^{\prime 2}+\beta)}\frac{1}{\sqrt{\kappa}}=\frac{0.7203}{\sqrt{\kappa}}=0.6157\,Q_{\rm m},
\ee
which reproduces \eqref{Qstar} and agrees with our numerics.

\subsection{ Phase II -- oblate  horizon, $|Q|>Q_\star$}

If  $|Q|<Q_\star$ then the horizon is spherical and one has $\delta=0$, 
 but for  $|Q|>Q_\star$ one has $\delta>0$, 
 the horizon geometry  is no longer spherical and is not described by the extremal RNdS solution \eqref{RNdSo},\eqref{Nex}. 
 Therefore, the horizon symmetry changes through the transition point $|Q|=Q_\star$.

The horizon remains degenerate, and we continue to use the radial gauge defined by Eq.\eqref{kk_ex}. 
However, since the horizon is no longer spherical, there is no longer any reason to enforce 
similarity with the RNdS case by imposing $\rrh = r_{\rm ex}$ in Eq.\eqref{kk_ex}. 
Moreover, the expression for $r_{\rm ex}$ in Eq.\eqref{horRNdS1} becomes 
complex-valued when $|Q| > Q_{\rm m}$. For these reasons, we fix $\rrh = 0.2$ 
for all solutions in phase II.
 
 Close to the transition  point for $|Q|>Q_\star$  one has  
\be
\delta\propto (|Q|-Q_\star)^s.
\ee
  This behavior resembles a second-order phase transition or possibly a crossover, given the rather large value 
 of the critical exponent: according to our numerics, one finds $s \approx 10.8$ for $\kappa = 10^{-2}$. 
 
 Additionally, the fraction of the magnetic charge contained in the hair, which was 
 constant in the previous phase, $\lambda = P_{\rm h}/P \approx 0.22$, starts to decrease for $|Q| > Q_\star$ 
 (see Fig.\ref{Fig23}, left panel). As the charge increases, the black hole gradually absorbs the 
 condensate and becomes less hairy; the ratio $M/|Q|$ increases, the geometry approaches that of the 
 extremal RN solution, and ultimately merges with it when the horizon size overtakes the condensate size. 
 This occurs at $r_h = |Q| = r_{\rm c} \equiv r_h^0(n) = \sqrt{|n|}/g$, corresponding to the maximal charge
 \be
  Q_{\rm max}=2g^\prime\sqrt{\beta}\,Q_{\rm m}=1.29\,Q_{\rm m}\,,
  \ee
 which in turn corresponds to the maximal value $|n|=8g^{\prime 2}/\kappa=1.78/\kappa$.
 This estimate of the maximal charge is supported by our numerics.
 No hairy solutions exist beyond this charge; black holes with $|Q| > Q_{\rm max}$ are thus RN.

 The estimate \eqref{MQo} should  be reformulated as an inequality valid  for  all extremal 
 hairy solutions, 
 \be             \label{MQoo}
 0.88< M/|Q|\leq 1. 
 \ee
 Here the lower bound corresponds to the minimal charge when $|n|=2$, whereas  the upper bound 
 is reached when the charge is maximal, $|Q|=Q_{\rm max}$. 
 
 \subsection{Maximally hairy black holes}
 
 The black hole is maximally hairy for $|Q|\approx Q_\star\approx 0.62\,Q_{\rm m}$ when 
 the  ratio $M_{\rm h}/M$ 
 is maximal (see Fig.\ref{Fig23}, the insertion in the right panel).   
 This corresponds to the following values of the black hole charge, size and mass: 
 \be
 |n|\approx \frac{0.85}{\kappa}\approx 1.6 \times 10^{32},~~~{\bf r}_h\approx 1.41~{\rm cm},~~~~{\bf M}\approx 2\times 10^{25}~{\rm kg},
 ~~~\frac{{\bf M}_{\rm h}}{\bf M}\approx 0.11.
 \ee 
 These size and mass are typical for the planetary mass black holes. The hair account for $\approx 11\%$ of the total mass,
 which means the black hole is surrounded by $\approx 2\times 10^{24}$ kg of  electroweak condensate  
 (the Earth mass is ${\bf M}_{\bigoplus}\approx 6\times 10^{24}$ kg). 
 These estimates for the mass and radius 
are obtained for the physical value of $\kappa\sim 10^{-33}$
by noting that for maximally hairy black holes one has 
$M\approx |Q|$ while $r_h=r_{\rm ex}$ is given by \eqref{horRNdS}. 
The value $\approx 11\%$  is obtained numerically for $10^{-3}<\kappa<10^{-2}$, but  it seems to be stable when  varying  $\kappa$, 
therefore we expect it to apply  for  the physical value of $\kappa$ as well. 

It is also interesting to estimate the condensate size. The condensate terminates where $|{\rm B}|\approx \mw^2$ at 
$r_{\rm c}\approx 1.13\sqrt{|n|}$, which corresponds to the dimensionful value 
\be
{\bf r}_{\rm c}\approx 2.20~{\rm cm}.
\ee 
One may wonder how massive fields can extend over macroscopic distances.
The key point is that they form a Bose-Einstein condensate. A familiar example is the condensate 
in a superconductor, which also can be macroscopically large despite involving massive fields. 
When the electroweak condensate terminates, the corona does not vanish abruptly but instead decays as $\sim Q_{\rm M}/r^4$,
 due to the presence of a magnetic quadrupole moment $Q_{\rm M}$ carried by the magnetic field.

One might think that the magnetic charge with $|n|=1.6\times 10^{32}$ is extremely large. However, 
as explained in \ref{AppD}, a charge of this magnitude placed at the center of the Sun 
would create a magnetic field of $\approx 10^3$~Gauss at the Sun surface. This  is  1000 times larger 
than the average proper magnetic field of the Sun, but  2-4 times smaller than the field in sunspots.  

\subsection{Summary}

Summarizing, the extremal hairy solutions show two phases depending on value 
of  their  charge $Q\approx 0.85\sqrt{\kappa}\,n$.  
If the charge is small  then the hypermagnetic field ${\rm B}$  is strong  enough  to restore 
the full electroweak symmetry at the horizon where $\WW^a_{\mu\nu}$ and $\Phi$ vanish.  
The horizon geometry is then perfectly spherical, but the magnetic fields ${\cal B}$ and ${\cal B}_Z$ keep a non-trivial 
dependence on the polar angle $\vartheta$ 
and form a corona of vortex loops  attached to the horizon. 
These vortices  do not change  the magnetic flux $4\pi P$ through the horizon 
but render it  inhomogeneous, highly increasing the field strength in the narrow equatorial region. 
Although all the charge $P$ is contained inside the black hole in the 't\,Hooft description,
the total flux can be interpreted as a superposition of two contributions: the flux produced by an effective charge
$g^{\prime 2} P$ distributed along a ring at the horizon in the equatorial plane, and the remaining flux
$g^{2} P$ distributed more or less homogeneously over the entire horizon.

When  the  charge increases, the horizon value of ${\rm B}$  decreases. 
If the charge exceeds the critical value $Q_\star\approx 0.62\, Q_{\rm m}$ where 
$Q_{\rm m}\approx 1.17/\sqrt{\kappa}$, then ${\rm B}$ becomes  insufficient to 
create  a region of symmetric phase around the horizon. As a result, 
the Higgs field starts deviating from zero at the horizon.   
The horizon geometry is then no longer spherical. However, the ${\rm B}$-field is still strong enough 
to create a condensate around the horizon. The condensate disappears and the black hole loses its hair 
when the charge approaches the maximal value $Q_{\rm max}\approx 1.29\, Q_{\rm m}$. 

The mass of all extremal hairy solutions is smaller than the charge, $M/|Q|\leq 1$,  which can be related 
to the Zeeman interaction between the W bosons in the condensate and the black hole magnetic field. 

\begin{figure}
    \centering
    
     \includegraphics[scale=0.8]{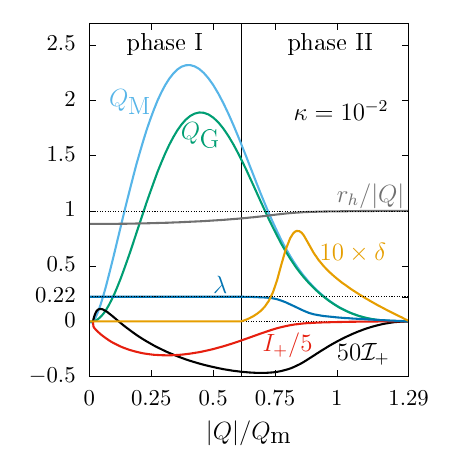}
     \includegraphics[scale=0.8]{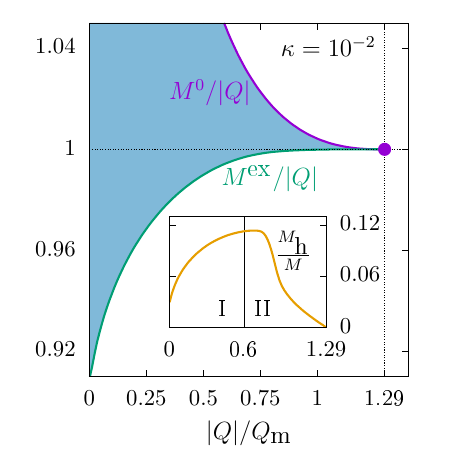}
               
       \caption{
 The parameters of  extremal  solutions (left) and the existence diagram for hairy solutions (right)
 for $\kappa=10^{-2}$. The upper violet curve on the right corresponds to solutions bifurcating with the RN while  the lower 
 green curve corresponds to extremal solutions. The blue shaded area between these two curves is the domain of existence 
 of non-extremal hairy solutions. 
 The scale for the charge is set by 
 $Q_{\rm m}\approx 1.17/\sqrt{\kappa}$ defined in \eqref{horRNdS1}.
        }
    \label{Fig23}
\end{figure}

Various parameters of the extremal solutions against the charge are shown in Fig.\ref{Fig23}. 
One can see that the quadrupole moments and 't\,Hooft currents attain their maximal values in phase I.
They  vanish for the minimal value of the charge $|n|=2$ when the configuration is
spherically symmetric, and for the maximal charge when the hair disappears. 
The horizon oblateness and the Nambu currents attain maximal values in phase II, while the fraction of the hair charge 
$\lambda =P_{\rm h}/P$ is constant in phase I and decreases down to zero in phase II. 
As seen in  Fig.\ref{Fig23}, one has $r_{h}/|Q|\approx M/|Q|\leq 1$.

The domain of existence for these  hairy solutions is shown in the right panel of 
Fig.\ref{Fig23},  which presents the ratio $M/|Q|$ against the  magnetic charge $|Q|=\sqrt{\kappa/2}\, |n|/(2e)$. 
The hairy solutions occupy  the shaded region  between two curves: 
  below the upper curve corresponding to
 the bifurcation with the RN branch, and above the lower curve 
 corresponding to the extremal hairy solutions. 
 Spherically symmetric solutions with $|n|=2$ correspond to the left edge of the region,
 they were previously known \cite{Bai:2020ezy}, whereas all  solutions with $|n|>2$ are new. 
 Descending from the upper curve to the lower one at fixed $Q$ 
 increases the amount of charge stored in the condensate and decreases the ratio $M/|Q|$. 
 As shown in the figure, the inequality $M/|Q|\leq 1$ holds not only for extremal solutions but also for  near-extremal ones. 
 For the the extremal solutions in Fig.\ref{Fig23}, 
one finds $0.91\leq M/|Q|\leq 1$,  
which slightly differs from the estimate 
 in Eq.\eqref{MQoo}  because the figure 
 is obtained not for the physical value of the gravitational coupling 
 $\kappa\sim 10^{-33}$,  but for $\kappa=10^{-2}$.

\section{CONCLUDING REMARKS \label{Ssum}}
\setcounter{equation}{0}

We have constructed solutions describing static and axially symmetric black holes that support a  magnetic field strong enough to change 
the structure of the electroweak vacuum. This results in  the formation  of  an electroweak condensate outside 
the black hole horizon. The condensate carries currents that flow along the horizon and generate a corona of fountain-like vortex loops 
around the horizon.

 According to Maldacena \cite{Maldacena:2020skw}, the presence of the corona should significantly enhance the evaporation rate.
As a result, non-extremal black holes are expected to rapidly relax toward the extremal state,
which correspond to the lower edge of the existence region in Fig.\ref{Fig23} where $M \leq |Q|$.
In this extremal state, the black holes cannot shed their hair to evolve into RN black holes, for which $M \geq |Q|$.
In principle, they could decay by emitting pairs of hairy black holes with $n = \pm 2$,
which minimize the ratio $M/|Q|$,
but such a process would require quantum tunnelling and is therefore expected to be strongly suppressed.

 However, axially symmetric black holes can reduce their energy by developing a fully spread corona 
of vortices, which may retain  at most only discrete symmetries. 
In the axially symmetric case, the corona consists of two oppositely directed multi-vortices.
Each multi-vortex is composed of many elementary vortices that repel each other, making the configuration inherently unstable
(except for $|n| = 4$). 
This instability is evident at the perturbative level,
as shown in Section \ref{Secpert}, where the axially symmetric configuration corresponds to a saddle point of the energy.
In contrast, the absolute energy minimum corresponds to a configuration of elementary 
vortices that maximize their mutual separation, as illustrated in Fig.\ref{Fig6} (left panel). 
Non-perturbative generalizations of this configuration are expected to describe stable hairy black holes 
surrounded by an electroweak corona.
However, the corresponding solutions are currently unknown,
except for the spherically symmetric case with $n = \pm 2$, which contains no vortices,
and the axially symmetric case with $n = \pm 4$, where the vortices are already elementary.

The transition from the  axially symmetric state to a fully spread corona should not only 
reduce the energy, but also increase the entropy. 
One may expect that, for very large $|n|$, the condensate and the associated vortices in the corona
form a gas-like or liquid-like phase \cite{Chernodub:2022ywg}.
The ground state of such a system is likely to be highly degenerate, with a high density of low-lying excitations above it.
As a result, there should be a nontrivial entropy associated with the black hole hair.

As we have seen above, the black hole mass and charge split into two parts, associated 
respectively with the black hole itself and with the surrounding hair.
It is therefore natural to assume that the entropy also splits as $S = S_{\rm H} + S_{\rm h}$, 
where the first term is determined by the horizon area.
As shown in Fig.~\ref{Fig23}, for extremal hairy black holes one finds $r_h < |Q|$,
whereas for extremal RN black holes $r_h = |Q|$,
so the horizon entropy satisfies $S_{\rm H}<S_{\rm RN}$.

On the other hand, as shown in Fig.\ref{Fig23}, the ratio $r_h/|Q|$ is close to unity.
It is therefore possible that including the hair entropy may reverse the inequality to $S_{\rm H} + S_{\rm h} > S_{\rm RN}$.
This would imply that extremal hairy black holes with a fully spread corona are not only energetically preferred over RN black holes,
but are also favoured entropically, which would enhance their formation rate.
However, for this to be true, the hair entropy $S_{\rm h}$ should be sufficiently large.
It is expected to be maximal for maximally hairy black holes, where $r_h \sim Q_\star \sim 1/\sqrt{\kappa} \sim \sqrt{|n|}$,
implying that $S_{\rm H} \sim S_{\rm RN} \sim r_h^2/\kappa\sim |n|/\kappa \sim n^2$.
Therefore, $S_{\rm h}$ should also scale as $n^2$.
Using, for instance, the ideal gas approximation for the hair entropy gives $S_{\rm h} \sim |n|\ln|n|$, which is not large enough. 
On the other hand, the vortices in the corona are not point particles but extended objects (fountains) that can become 
mutually entangled
in a complex way, as we observed in the perturbative description of Section \ref{Secpert}. This should increase their  entropy. 
Since we cannot currently give a reliable estimate for $S_{\rm h}$, we leave this issue for future analysis.

Our solutions provide the first example of hairy black holes described by well-tested  physical theories -- 
the Standard Model and General Relativity.  Therefore, these solutions may  be physically relevant 
(the $n = 2$ solutions of \cite{Bai:2020ezy} are too close to the Planck scale to be considered realistic).
Such black holes could plausibly originate from primordial black holes, 
which may acquire charge from the early electroweak plasma.

Suppose that the fluctuating hypermagnetic field in the ambient plasma becomes, 
at some moment, predominantly orthogonal to the black hole horizon.
This would induce a net flux through the horizon, and hence generate a charge.
Since the  total magnetic charge is conserved, the flux generated on one black hole must be 
compensated by an opposite flux on another black hole (or several others).
The resulting oppositely charged black holes would attract one another, 
but they might not necessarily annihilate, as cosmic expansion can drive them apart.
Moreover, oppositely charged black holes may potentially form bound systems
stabilized by scalar repulsion, as illustrated by the examples considered in \cite{Kleihaus:2000hx,Herdeiro:2023mpt}.

It is also possible that, during the electroweak phase transition, 
a network of electroweak Nambu strings formed in the early Universe, 
as suggested in \cite{Urrestilla:2001dd,Patel:2021iik}. 
Nambu strings terminate at Nambu monopoles, therefore, magnetic monopoles may have been dynamically created 
at early times. When these monopoles are absorbed by primordial black holes, the latter acquire a magnetic charge.

Mechanisms for magnetic charge generation remain to be verified, but if confirmed, it would have significant implications.
Magnetic monopoles are typically associated with Grand Unified Theories (GUT), 
which involve physics beyond the Standard Model --
the existence of which remains an open question.
However, if primordial black holes can indeed acquire magnetic charge through a process similar to the one described above,
this would suggest that no new physics beyond the Standard Model is required to account for monopoles.
In this scenario, the magnetic charge does not require a conventional carrier particle but is 
instead associated with the magnetic flux threading the black hole event horizon.
From this perspective, extremal hairy electroweak black holes naturally emerge as compelling 
candidates for magnetic monopoles -- objects that may, in principle, exist in Nature 
without invoking physics beyond the Standard Model.

These black holes can exhibit a variety of physical effects. For instance, 
they are expected to catalyze proton decay \cite{Maldacena:2020skw}.
They could potentially be detected if captured by a neutron star, 
leading to a sudden change in the star’s rotation period \cite{Estes:2022buj}.
Their phenomenology is quite rich and has been explored  in
Refs.~\cite{Bai:2020spd,Bai:2021ewf,Ghosh:2020tdu,Estes:2022buj,Diamond:2021scl}.

Finally, it is worth noting that there are numerous directions in which our results can be extended and generalized.
Most obviously, one should explicitly construct non-axially symmetric solutions with $|n| > 4$ 
and demonstrate that they indeed minimize the energy.
It would also be natural to explore their stationary generalizations.
Crucially, one should clarify the physical mechanism by which black holes can acquire magnetic charge.
In addition, it is important to account for the role of fermions, verify the enhancement of Hawking evaporation,
examine the Rubakov-Callan effect, further develop the phenomenology, etc.

{\bf Acknowledgements.} We thank Eugen Radu, Maxim Chernodub, Julien Garaud, Christos Charmousis,
and Valerie Domcke   for valuable discussions. 
M.S.V. is grateful to the Theory Division of CERN for its hospitality during the final stage of this work.

\renewcommand{\thesection}{Appendix A}
\section{Energy-momentum tensor \label{AppA}}
\renewcommand{\theequation}{A.\arabic{equation}}
\setcounter{equation}{0}

In this Appendix, we explicitly present  the components of the electroweak energy-momentum tensor in the axially symmetric case. 
Substituting   \eqref{metr11},\eqref{RR} 
into \eqref{TT} yields the following diagonal components  of the energy-momentum tensor: 
\begin{align}
    -\tensor{T}{^0_0}\sqrt{\rm -g}&=a_1+a_2+a_3+a_4+a_5, \nn \\
    -\tensor{T}{^{\text{r}}_{\text{r}}}\sqrt{\rm -g}&=-a_1+a_2-a_3+a_4+a_5,\nn \\
    -\tensor{T}{^\vartheta_\vartheta}\sqrt{\rm -g}&=a_1-a_2-a_3+a_4+a_5,\nn \\
    -\tensor{T}{^\varphi_\varphi}\sqrt{\rm -g}&=-a_1-a_2+a_3+a_4-a_5,
\end{align}
where $\sqrt{\rm -g}=e^{2\Ko+\So-2\Uo}\rr^2\sin\vartheta$, and 
\begin{align}              \label{a5}
    a_1&=\N e^{2\Uo-\So}\left[\frac{1}{2g'^2}(\partial_\text{r}Y)^2+\frac{1}{2g^2}\left((\mathcal{D}_\rr F_3)^2+(\mathcal{D}_\rr F_4)^2\right)\right]\frac{\nu^2}{\sin\vartheta},\nn \\
    a_2&=e^{2\Uo-\So}\left[\frac{1}{2g'^2}(\partial_\vartheta Y)^2+\frac{1}{2g^2}\left((\mathcal{D}_\vartheta F_3)^2+(\mathcal{D}_\vartheta F_4)^2\right)\right]\frac{\nu^2}{\text{r}^2\sin\vartheta},\nn \\
    a_3&=\N e^\So\left[\frac{1}{2g^2}e^{2\Uo-2\Ko}(\partial_\vartheta F_1-\partial_\text{r} F_2)^2+\text{r}^2\left(\mathcal{D}_\rr \phi_1\right)^2
    +\text{r}^2\left(\mathcal{D}_\rr \phi_2\right)^2\right]\sin\vartheta,\nn \\
    a_4&=e^\So\left[\left(\mathcal{D}_\vartheta \phi_1\right)^2+\left(\mathcal{D}_\vartheta \phi_2\right)^2+\frac{\beta\,\text{r}^2}{8}e^{2\Ko-2\Uo}\left(\phi_1^2+\phi_2^2-1\right)^2\right]\sin\vartheta,\nn \\
    a_5&=e^{2\Ko-\So}\bigg[\left((F_3+Y)\phi_1-F_4\phi_2\right)^2+\left((F_3-Y)\phi_2+F_4\phi_1\right)^2\bigg]\frac{\nu^2}{4\sin\vartheta}.
\end{align}
Here, the following abbreviations have been introduced: 
\be
\mathcal{D}_\rr F_3=\partial_\rr F_3+F_1 F_4,~
\mathcal{D}_\vartheta F_3=\partial_\vartheta F_3+F_2 F_4,~
\mathcal{D}_\rr F_4=\partial_\rr F_4-F_1 F_3,~
\mathcal{D}_\vartheta F_4=\partial_\vartheta F_4-F_2 F_3,~\nn \\
\mathcal{D}_\rr \phi_1=\partial_\rr \phi_1-\frac{F_1}{2} \phi_2,~
\mathcal{D}_\vartheta \phi_1=\partial_\vartheta \phi_1-\frac{F_2}{2} \phi_2,~
\mathcal{D}_\rr \phi_2=\partial_\rr \phi_2+\frac{F_1}{2} \phi_1,~
\mathcal{D}_\vartheta \phi_2=\partial_\vartheta \phi_2+\frac{F_2}{2} \phi_1\,. \nn
\ee
The only off-diagonal components are $T_{\rr\vartheta}=T_{\vartheta\rr}$ with 
\be
T^\rr_{~\vartheta}\sqrt{\rm -g}=\N e^{2\Uo-\So}\left[
\frac{1}{g^{\prime 2}}\,\partial_\rr Y\partial_\vartheta Y+\frac{1}{g^2}\left(\mathcal{D}_\rr F_3\,\mathcal{D}_\vartheta F_3 
+\mathcal{D}_\rr F_4\,\mathcal{D}_\vartheta F_4\right)
\right]\frac{\nu^2}{\sin\vartheta} \nn \\
+2\rr^2 \N e^{\So} \left(\mathcal{D}_\rr \phi_1\, \mathcal{D}_\vartheta \phi_1+\mathcal{D}_\rr \phi_2\,\mathcal{D}_\vartheta \phi_2
\right)\sin\vartheta. 
\ee
The energy-momentum tensor is gauge-invariant. Its diagonal components are invariant under the reflection 
$\vartheta\to\pi-\vartheta$, while the off-diagonal components then change sign. 

\renewcommand{\thesection}{Appendix B}
\section{Axially-symmetric form of the $n=\pm 2$ solutions \label{AppB}}
\renewcommand{\theequation}{B.\arabic{equation}}
\setcounter{equation}{0}

 Before considering axially symmetric solutions with $|n|>2$, it is instructive to first reproduce the spherically symmetric 
 solutions with $|n|=2$ within the axially symmetric formalism. 
 A spherically symmetric solution is described by  Eqs.\eqref{geom},\eqref{RRa}, 
 with functions $N(r),\sigma(r),f(r),\phi(r)$ depending on the Schwarzschild radial coordinate $r$. 
 
 The same solution, expressed  within   the axially-symmetric formalism,  is described by 
 the line element \eqref{metr11} and electroweak fields \eqref{RR},\eqref{var} parameterized 
 by functions $\N,\Uo,\Ko,\So,H_k,y,\phi_1,\phi_2$. These functions depend 
 on the radial coordinate $\rr$, distinct from $r$. 
 
Notice that $N(r)$ is obtained by solving the equations, whereas 
$\N(\rr)$ is the auxiliary function whose choice determines the radial gauge. For example, one can choose 
$\N(\rr)=1$.

 The relation between the coordinates $r$ and $\rr$  is determined by 
 \be            \label{r00}
 \frac{d\rr}{\rr \sqrt{\N(\rr)}}=\frac{dr}{r\sqrt{{N}(r)}},~~
 \ee
 integrating which yields $\rr=\rr(r)$. One has then 
\be                  \label{sps}
&&~~~
e^{2\Uo}=\frac{N(r)}{{\rm N}(\rr)}\,\sigma^2(r),~~~~~e^{\Ko-\Uo}=e^{\So-\Uo}=\frac{r}{\rr} \,, \nn \\
&&H_2=H_4=f(r),~~~~
H_1=H_3=y=\phi_1=0,~~~~\phi_2=\phi(r).
\ee

\subsection{Non-extremal solutions} 
For non-extremal solutions the metric coefficient ${\rm g}_{00}$ shows a simple zero at the horizon, which can be ensured 
by choosing  the auxiliary function $\N(\rr)$ to be 
$$
\N=1-\frac{\rrh}{\rr}. 
$$
For the non-extremal RN solution \eqref{QQ}, 
Eqs.\eqref{r00},\eqref{sps} then yield 
\be              \label{RNax}
\rr=r-r_{-},~~~\rr_{\rm H}=r_{+}-r_{-},~~~e^{\Uo}=\frac{\rr}{\rr+r_{-}},~~\nn \\
\Ko=\So=H_1=H_2=H_3=H_4=y=\phi_1=0,~~\phi_2=1,
\ee
where $r_\pm=M\pm \sqrt{M^2-Q^2}$ are the two RN horizons. One has 
\be
r_{+}r_{-}=Q^2\equiv \frac{\kappa}{2e^2}=r_{-}^2+\rrh\, r_{-}~~~\Rightarrow~~~
r_{-}=\frac12\left(\sqrt{\rrh^2+4Q^2}-\rrh\right).
\ee
In the extremal limit, when $M\to Q$, one has $\rrh\to 0$, while the horizon value 
$\Uo_{\rm H}=\ln(\rrh/r_{+})$ diverges. 

\begin{figure}[th]
\hbox to \linewidth{ \hss
	\includegraphics[width=8 cm]{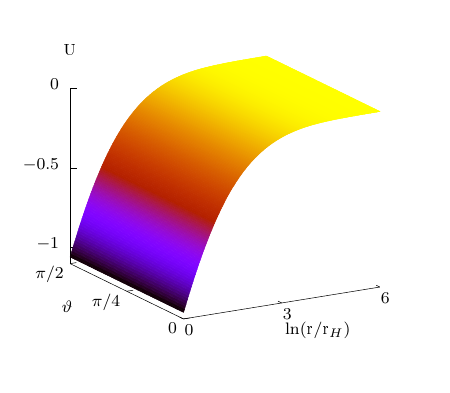}
	\includegraphics[width=8 cm]{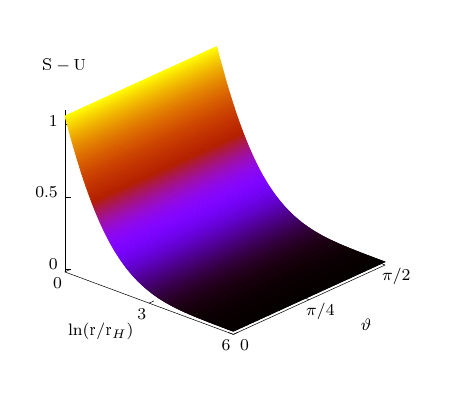}	

\hss}
\caption{The metric amplitude  $\Uo$ and the difference $\So-\Uo$ for the non-extremal spherically symmetric hairy 
black hole with $n=2$, $\kappa=10^{-3}$, and  $\rr_{\rm H}=0.02$ as functions of  $\ln(\rr/\rr_{\rm H})$ and $\vartheta$.}
\label{Fig1b}
\end{figure}

For the non-extremal  hairy solutions one can represent $N(r)$ in \eqref{geom} as
\be
N(r)=\frac{1}{\alpha^2(r)}\left( 1-\frac{r_h}{r}\right),
\ee
where $\alpha(r_h)\equiv \alpha_h$ and $\alpha(\infty)=1$. 
 Integrating then Eq.\eqref{r00} 
 yields the relation between $\rr$ and $r$, expressed by 
\be
X(\rr,\rr_{\rm H})=(X(r,r_h))^{\alpha(r)}\times{\cal F}(r)~~\text{with}~~
{\cal F}(r)=\exp\left(\int_r^\infty \alpha^\prime(r)\ln\left( X(r,r_h)\right)dr\right),
\ee
where the following definition has been used: 
\be
X(x,x_1)\equiv \frac{\sqrt{x}+\sqrt{x-x_1}}{2}.
\ee 
This  implies that at infinity, where $r,\rr\to\infty$, one has  $\rr/r\to 1$ and $\Uo,\So,\Ko\to 0$. 
At the horizon, where $\rr=\rr_{\rm H}$ and $r=r_h$, one has 
\be
\sqrt{\rr_{\rm H}}=2\left(\frac{\sqrt{r_h}}{2}\right)^{\alpha_h}\times {\cal F}(r_h),
\ee
hence $e^{\So_{\rm H}-\Uo_{\rm H}}=r_h/\rr_{\rm H}\neq 1$. 
With some algebra, one finds 
\be
\lim_{r\to r_h}\frac{dN(r)}{dr}=\frac{1}{\alpha_h^2 r_h},~~~~~~
\lim_{r\to r_h}\frac{d\N(\rr(r))}{dr}=\frac{\alpha_h^2}{r_h},~~~~
~\Rightarrow~~~
e^{\Uo_{\rm H}}=\frac{\sigma_h}{\alpha_h^2}. 
\ee
Applying the surface gravity formula \eqref{kA} separately within the spherically symmetric  and  axially symmetric 
formalisms  yields 
\be
{\rm k}_{\rm H}=\frac{\sigma_{h}}{2\alpha_{h}^2 r_h}
=\frac{1}{2r_h}\, e^{\Uo_{\rm H}}
=\frac{1}{2\rr_{\rm H}}\, e^{2\Uo_{\rm H}-\Ko_{\rm H}}\,.
\ee
In the extremal  limit,  $r_h\to r_{\rm ex}>0$ when the horizon becomes 
degenerate and ${\rm k}_{\rm H}\to 0$, 
the value $\sigma_h$ approaches 
a non-zero limit, so $\alpha_h\to \infty$.  It follows that 
\be           \label{toex}
r_h\to r_{\rm ex},~~~~~~\rr_{\rm H}\to 0,~~~~e^{\Uo_{\rm H}}\to 0,~~~~
e^{\Ko_{\rm H}-\Uo_{\rm H}}=e^{\So_{\rm H}-\Uo_{\rm H}}=\frac{r_h}{\rr_{\rm H}}\to\infty. 
\ee

Having computed $\Uo,\So,\ldots $ from \eqref{sps}, the same values 
are obtained by solving the axially symmetric equations numerically.
Fig.\ref{Fig1b} shows the profiles  of $\Uo$ and $\So-\Uo$ for $\kappa=10^{-3}$ and $\rr_{\rm H}=0.02$. 
One finds that  $r_h=\rr_{\rm H}\, e^{\So_{\rm H}-\Uo_{\rm H}}=0.058>r_{\rm ex}=0.047$,
while $\rrh<r_{\rm ex}$. 
As $\rrh$ decreases further and approaches zero, $\Uo_{\rm H}$ shifts 
more and more to negative values, while 
$\So_{\rm H}-\Uo_{\rm H}$ grows, as suggested by \eqref{toex}.

\subsection{Extremal solutions}

Since $\Uo_{\rm H}$ becomes unbounded when approaching the extremal  limit, 
this case should be considered separately.  In this limit, the horizon becomes degenerate, and 
the function $N(r)$ takes   structure shown in \eqref{Nextr}. 
Equations \eqref{r00} and \eqref{sps} then imply that 
the amplitudes $\Uo,\Ko,\So$ will be finite at the horizon
and approach zero at infinity, provided that 
${\rm  N}(\rr)$ 
has the same structure as $N(r)$: 
\be               \label{NNs}
{\rm N}(\rr)={\rm k}^2(\rr)\,\left(1-\frac{\rr_{\rm H}}{\rr}\right)^2\,,~~~~~~
{N}(r)=k^2(r)\,\left(1-\frac{r_{\rm ex}}{r}\right)^2\,,~~~~~
\ee 
where  $r_{\rm ex}$ is defined in \eqref{horRNdS} with 
${\rm k}(\rr_{\rm H})=k(r_{\rm ex})\equiv k$,  while  ${\rm k}(\infty)=k(\infty)=1$. 

It is convenient  to set 
\be           \label{rrex}
\rr_{\rm H}=r_{\rm ex}\,,
\ee
so that the radial coordinate $\rr$ coincides  with the Schwarzschild coordinate $r$ at the horizon. 
Integrating 
Eq. \eqref{r00} then  yields 
\be               \label{logs}
\frac{\ln(r-\rr_{\rm H})}{k(r)}-\frac{\ln(\rr-\rr_{\rm H})}{{\rm k}(\rr)}
=\int_r^\infty \ln(r-\rr_{\rm H})\,\frac{k^\prime(r)}{k^2(r)}\, dr-
\int_\rr^\infty \ln(\rr-\rr_{\rm H})\,\frac{{\rm k}^\prime(\rr)}{{\rm k}^2(\rr)}\, d\rr. 
\ee
The right-hand side of  this relation approaches zero when $r,\rr\to\infty$,
so that $\rr/r\to 1$. According to \eqref{sps}, this 
implies that  
$\Uo,\Ko,\So\to 0$. When $r,\rr\to \rr_{\rm H}$, 
the right-hand side of \eqref{logs} approaches a finite value, which we denote by $C$. 
Hence, close to the horizon, we have  
\be
\frac{r-\rr_{\rm H}}{\rr -\rr_{\rm H}}=\frac{\sqrt{N}}{\sqrt{\N}}=e^{kC}\,. 
\ee
Using Eq.\eqref{sps}, this implies 
\be           \label{U_ex}
e^{\Uo_{\rm H}}=e^{\So_{\rm H}}=e^{\Ko_{\rm H}}=\sigma_h\, e^{kC},
\ee
so that the metric amplitudes are now finite at the horizon. 

Applying the above formulas to the extremal  RN solution yields a very simple result:
\be
\N=\left(1-\frac{\rrh}{\rr}\right)^2,~~\Uo=\Ko=\So=H_1=H_2=H_3=H_4=y=\phi_1=0,~~\phi_2=1.~~~~
\ee
For the extremal  hairy black hole, the function $k(r)$ takes  the 
form  \eqref{Nex} at the horizon. Hence, by introducing  an additional cutoff factor, we set 
\be              \label{k_ex}
{\rm k}^2(\rr)=1-\left.\left.\frac{\Lambda}{3}\right[\rr^2+2\,\rr_{\rm H}\,\rr+3\,\rr_{\rm H}^2\right]\times 
\frac{1+\rr_{\rm H}^4}{1+\rr^4}\,,
\ee
so that  ${\rm k}^2(\rr_{\rm H})=k^2(r_{\rm ex})$ and ${\rm k}^2(\infty)=1$. 
\begin{figure}[th]
\hbox to \linewidth{ \hss
	\includegraphics[width=8 cm]{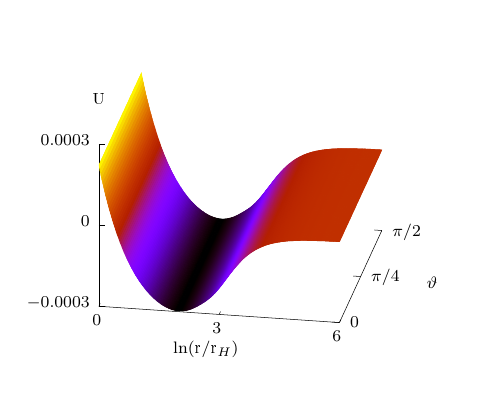}
	\includegraphics[width=8 cm]{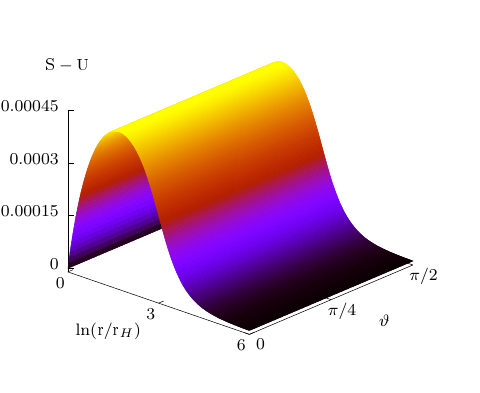}	

\hss}
\caption{The metric amplitude  $\Uo$ and the difference $\So-\Uo$ for the extremal  spherically symmetric hairy  
black hole with $n=2$ and  $\kappa=10^{-3}$, in which case $\rrh=r_{\rm ex}=0.047$.
}
\label{Fig1c}
\end{figure}
Integrating the axially symmetric equations numerically 
then yields profiles of $\Uo$ and $\So-\Uo$,  
shown in Fig.\ref{Fig1c}. These profiles are quite  different from those 
for the non-extremal  solution in Fig.\ref{Fig1b}, but they agree with the result in \eqref{U_ex}. 
In particular, $\So-\Uo=\ln(r/\rr)$ now vanishes both at the 
horizon and at infinity. 

\subsection{Fixing the radial gauge -- summary}

The upshot  is as follows. The spherically symmetric hairy
solutions can be labelled by the Schwarzschild radius of their event horizon, $r_h\in[r_{\rm ex},r_h^0]$. 
The same solutions within the axially symmetric description are labeled by $\rr_{\rm H}$, which we call 
event horizon size, even though it does not have a direct physical meaning. The parameter $\rrh$ 
and the auxiliary function $\N(\rr)$ in  \eqref{metr11}
take the following values:
\be                  \label{rH_lm}
\text{Non-extremal  case:}&&~~~
0<\rr_{\rm H}\leq \rrh^0=r_{+}^0-r^0_{-},~~~~~~\N(\rr)=1-\frac{\rr_{\rm H}}{\rr}\,; \nn \\
\text{Extremal  case:}&&~~~
\rr_{\rm H}=r_{\rm ex},~~~~~~~~~~~\N(\rr)={\rm k}^2(\rr)\left(1-\frac{\rr_{\rm H}}{\rr}\right)^2\,;
\ee
with ${\rm k}(\rr)$ given by \eqref{k_ex} and $r_{\rm ex}$ defined in \eqref{horRNdS}. 
Here, $r_{+}^0\equiv r^0_{h}$  and $r^0_{-}$ are the two horizons of the RN solution, to which the hairy 
black holes reduce at the bifurcation point $r_h=r^0_h$ (note that for the RN solution, 
$\rrh=r_{+}-r_-$). Since $r_{+}r_{-}=Q^2$ with $Q$ from $\eqref{QQ}$, 
one has 
\be     \label{rH_lm1}
\rrh^0=r_{+}^0-r^0_{-}=r_h^0-Q^2/r_h^0.
\ee
We use the same gauge conditions \eqref{rH_lm}, \eqref{rH_lm1} also for $|n|>2$. 

\subsection{Choice of $\rrh$ and diffeomorphisms}
The value of $\rrh$ can be changed by a diffeomorphism  in the extremal case, while for the non-extremal 
case this is not possible. Suppose that $\rr\to \rr(\tilde{\rr})$ and consider Eq.\eqref{change},
\be               \label{B.21}
\frac{d\rr}{\rr\sqrt{\N(\rr)}}=\frac{d\tilde{\rr}}{\tilde{\rr}\sqrt{\tilde{\N}(\tilde{\rr})}}. 
\ee
Integrating this equation yields a relation between $\rr$ and $\tilde{\rr}$. 
In the non-extremal case, one chooses 
$\N=1-\rrh/\rr$ and $\tilde{\N}=1-\tilde{\rr}_{\rm H}/\tilde{\rr}$. Performing the integration then shows that the only possibility 
is $\rr=\tilde{\rr}$ and $\rrh=\tilde{\rr}_{\rm H}$. Therefore,  the value of $\rrh$ cannot by changed by a diffeomorphism. 

In the extremal case,  $\N$ is chosen as in Eq.\eqref{NNs}, and similarly for  $\tilde{\N}$, with 
 the replacement $\rr\to \tilde{\rr}$, $\rrh\to \tilde{\rr}_{\rm H}$. 
In this case  the integration of \eqref{B.21} yields a relation similar to that in \eqref{logs}:
\be
\frac{\ln(\rr-\rrh)}{{\rm k}(\rr)}+\int_{\rrh}^{\rr} \ln(\rr-\rrh)\,\frac{{\rm k}^\prime(\rr)}{{\rm k}^2(\rr)}\,d\rr=
\frac{\ln(\trr-\trr_{\rm H})}{\tilde{{\rm k}}(\trr)}+\int_{\trr_{\rm H}}^{\trr} \ln(\trr-\trr_{\rm H})\,\frac{\tilde{{\rm k}}^\prime(\trr)}{\tilde{{\rm k}}^2(\trr)}\,d\trr.
\ee
This establishes a non-trivial correspondance between $\rr$ and $\tilde{\rr}$ for any given $\rrh$ and $\trr_{\rm H}$. In the vicinity of the horizon, 
it simplifies to 
\be
\trr-\trr_{\rm H}=(\rr-\rrh)^{\tilde{\rm k}/{\rm k}}(1+\ldots),
\ee
where ${\rm k}\equiv {\rm k}(\rrh)$, $\tilde{\rm k}\equiv \tilde{{\rm k}}(\trr_{\rm H})$, and the dots denote subleading term. 

Assuming that $\rrh=r_{\rm ex}$, which is the radius of the RNdS solution defined in Eq.\eqref{horRNdS1}, 
the metric amplitudes $\Uo,\Ko,\So$ remain everywhere regular.  However, 
since ${\rm k}\neq \tilde{\rm k}$, it follows from the relation $e^{\Uo}\sqrt{\N}=e^{\tilde{\Uo}}\sqrt{\tilde{\N}}$ (see Eq.\eqref{change})
that the amplitude $\tilde{\Uo}$ becomes unbounded at the horizon. 
This implies that, while the value of $\rrh$ can indeed be changed  by a diffeomorphism in the extremal case, 
the resulting gauge choice leads to a singularity at the horizon.

At the same time, the singularity  is very mild and located in a narrow region near the horizon. 
We have verified that choosing in our numerical code $\rrh=r_{\rm ex}$, in which case everything is regular, 
or selecting a different value such as $\rrh=1$, in which case there is a singularity, 
returns the same values of the  physical parameters such as the total mass, horizon area, etc. 

Therefore, we adopt the following strategy. As described in Section \ref{Sextr}, extremal hairy solutions exhibit two distinct phases 
depending on their charge.  For extremal solutions in phase I, where  the horizon remains spherical, as for the RNdS solution, 
we always choose 
$\rrh=r_{\rm ex}$ to ensure regularity at the horizon. 

In contrast,  for extremal solutions in phase II,  the horizon is still degenerate but no longer spherical. In this regime, 
 there is no compelling reason to preserve similarity with the RNdS case. 
Furthermore, hairy solutions in phase II exist for $|Q|>Q_{\rm m}$, in which case 
the value of $r_{\rm ex}$ defined by Eq.\eqref{horRNdS1} becomes complex. 
Therefore, we abandon the condition $\rrh=r_{\rm ex}$ and 
instead set $\rrh=0.2$ for all solutions in phase II. Numerical solutions obtained in this gauge exhibit no irregularities, 
and the virial relation is satisfied. We therefore conclude that this gauge choice is acceptable.
Moreover, we observe a smooth transition between phase I and phase II.

\renewcommand{\thesection}{Appendix C}
\section{Anti-screening  property of the condensate \label{AppC} }
\renewcommand{\theequation}{C.\arabic{equation}}
\setcounter{equation}{0}

It is instructive to examine simple models that illustrate the anti-screening property of the electroweak condensate.
We originally developed these models to enhance our own understanding of the phenomenon.
However, they can also be considered independently, as they are of interest in their own right, regardless of their connection to the main discussion.

The anti-screening property of the 
electroweak condensate property is the opposite of the well-known 
screening  within the conventional Abelian model of superconductivity defined by Eq.\eqref{F1} in  \ref{AppF}. 
This model admits vortices consisting of a 
magnetic field and a complex scalar  (see \ref{AppF}). 
Their magnetic field is  maximal at the vortex center, where scalar describing the superconducting condensate vanishes.
However, the magnetic field decays to zero away from the center,
 where the condensate approaches a constant value. 
Therefore, the field is screened by the condensate: it is zero where the condensate is maximal. 

As we shall now see, the structure of vortices appearing via the electroweak condensation is just the opposite: 
their magnetic field is maximal where the condensate is maximal, so that the field is enhanced by the condensate
\cite{Ambjorn:1988tm,Ambjorn:1989sz}.

In the linear approximation, the electroweak condensate is described by a complex vector $w_\mu$ subject to 
Eq.\eqref{Proca}, 
\be              \label{Proca1}
{\cal D}^\mu w_{\mu\nu}+ie F_{\nu\sigma}w^\sigma =\mw^2\,w_\nu\,,
\ee
with $w_{\mu\nu}={\cal D}_\mu w_\nu-{\cal D}_\nu w_\mu$ and
${\cal D}_\mu=\nabla_\mu +ieA_\mu$, where $F_{\mu\nu} = \partial_\mu A_\nu - \partial_\nu A_\mu$ is the background 
electromagnetic field. 
In some special cases,
this equation can be solved by 
\be              \label{Pr2}
w_{\mu\nu}=0,~~~~~ie F_{\nu\sigma}w^\sigma =\mw^2\,w_\nu\,. 
\ee
The condensate generates the current $J_\nu$ and produces 
a second order correction $f_{\mu\nu}$ to the 
background electromagnetic field $F_{\mu\nu}$. One has (see \eqref{psi1},\eqref{cur33},\eqref{eqWS2})
\be              \label{Pr3}
4\pi J_\nu=\frac{ie}{2}\nabla^\sigma(w_\sigma\bar{w}_\nu-w_\nu\bar{w}_\sigma),~~~~~~~~
\nabla^\mu f_{\nu\mu}=4\pi J_\nu\,.
\ee

\subsubsection{Magnetic field along $z$-axis.} Let us consider a constant magnetic field inside a cylinder 
of radius $R$ in flat space,
\be
ds^2=-dt^2+d\rho^2+\rho^2d\varphi^2+dz^2,~~~~
A_\varphi=\frac{B}{2}\, \rho^2,~~~{\cal B}^z=\frac{1}{\rho}\, \epsilon^{z\rho\varphi}F_{\rho\varphi}\equiv B>0, 
\ee
where $\rho\in [0,R]$. 
The equations \eqref{Pr2} are solved by 
\be
B=\frac{\mw^2}{e},~~~~w_0=w_z=0,~~~~
w_\rho=\frac{i}{\rho}\,w_\varphi=
\exp(-i\varphi)\exp\left(-\frac{\mw^2}{4}\rho^2\right), 
\ee
hence, the magnetic flux in the cylinder is equal to $\pi \mw^2 R^2/e $ and condensate density is 
\be
{\rm w}(\rho)\equiv 
w^\mu\bar{w}_\mu=2|w_\rho|^2= 2\exp\left(-\frac{\mw^2}{2}\rho^2\right).
\ee
This gives rise to the current with the only non-vanishing component in the azimuthal direction, 
\be              \label{C6}
4\pi J_\varphi=-e\rho\, \left(|w_\rho|^2\right)^\prime_\rho=
e\mw^2 \rho^2 |w_\rho|^2,
\ee
whose sign is opposite to that of the Larmor precession.  Therefore, the field 
$\delta{\cal B}$ created by the current rather 
enhances the background field instead of screening it. 

The Maxwell equation \eqref{Pr3} reduces to 
\be              \label{C7}
\frac{1}{\rho}\left(\rho f^{\varphi\rho} \right)^\prime_\rho=4\pi J^\varphi=-\frac{e}{\rho}\left(|w_\rho|^2 \right)^\prime_\rho~~~~
\rightarrow~~~~
\delta{\cal B}^z=\frac{1}{\rho}\, f_{\rho\varphi}=e|w_\rho|^2+const.,
\ee
where the  constant is fixed by the requirement  that the field $\delta{\cal B}_z$ should not change the 
magnetic flux, 
hence 
\be
\int_0^R \rho \,\delta{\cal B}^z \, d\rho=0~~~\Rightarrow~~~
\delta{\cal B}^z=e\left(\frac{1}{2}\,{\rm w}(\rho)-\frac{2-{\rm w}(R)}{\mw^2 R^2}\right).
\ee
This field is maximal at $\rho=0$ where the condensate density ${\rm w}(\rho)$ is maximal and where  $\delta{\cal B}^z>0$, therefore 
the background field is enhanced by the condensate (anti-screening). At the cylinder surface, for $\rho=R$,  the condensate is minimal 
and one has $\delta{\cal B}^z<0$.

\subsubsection{Magnetic field on a sphere.} Let us consider a radial magnetic field in a spacetime 
with the direct product   geometry,
\be             \label{C8}
ds^2=-dt^2+d r^2+R^2(d\vartheta^2+\sin^2\vartheta d\varphi^2),~~
A_\varphi=-P\cos\vartheta,~~{\cal B}^r=\frac{F_{\vartheta\varphi}}{R^2\sin\vartheta}=\frac{P}{R^2},~~~~
\ee
where $r\in(-\infty,\infty)$ while $R=const.$    The magnetic flux through the 2-sphere is $4\pi P$, and we assume for 
notational simplicity  that $P>0$. 

The equations \eqref{Pr2} are solved by 
\be             \label{C9}
R^2=\frac{eP}{\mw^2},~~w_0=w_r=0,~~~w_\vartheta=\frac{i}{\sin\vartheta}\,w_\varphi=\left(\sin\vartheta\right)^j,~~~~j=eP-1. 
\ee
The 
condensate density is ${\rm w}(\vartheta)=w^\mu\bar{w}_\mu=(2/R^2)|w_\vartheta|^2$. 

\begin{figure}[th]
\hbox to \linewidth{ \hss
	\includegraphics[width=7 cm]{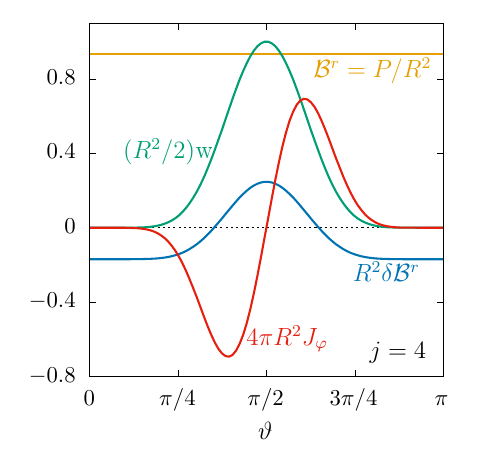}	

\hss}
\caption{The magnetic field $\delta {\cal B}^r$ in \eqref{C_14}, created by the condensate \eqref{C9} with $j=4$, 
the corresponding azimuthal 
current $4\pi J_\varphi$ in \eqref{C10}, and the condensate density $(R^2/2)\,{\rm w}$ are shown 
against $\vartheta$. Also shown the background magnetic  $P/R^2=\mw^2/e$, which corresponds  to the threshold 
of the electroweak condensation. 
}
\label{Fig22b}
\end{figure}

The current  has only one component in the azimuthal direction, 
\be           \label{C10}
4\pi J_\varphi=-\frac{e}{R^2}\,\sin\vartheta \left(|w_\vartheta|^2\right)^\prime_\vartheta=-\frac{2ej}{R^2}\cos\vartheta |w_\vartheta|^2\,.
\ee
The sign in \eqref{C10} agrees with that in  \eqref{J}, but in the upper hemisphere, for  $0\leq \vartheta\leq \pi/2$,
 it is opposite to that in \eqref{C6} because $|w_\vartheta|^2$ increases with $\vartheta$, 
whereas $|w_\rho|^2$ in \eqref{C6} decreases with $\rho$. This produces 
the second order correction for  the magnetic field, 
\be           \label{C11}
\delta{\cal B}^r= \frac{f_{\vartheta\varphi}}{R^2\sin\vartheta} =\frac{e}{R^2}|w_\vartheta|^2+const.,
\ee
where the integration constant is fixed by the requirement that the total magnetic flux should not change, $\oint \delta{\cal B}_r d^2\Omega=0$ 
(with $d^2\Omega=\sin\vartheta\, d\vartheta\, d\varphi$), 
hence 
\be               \label{C_14}
\frac{R^2}{e}\,\delta{\cal B}^r=\left(\sin\vartheta\right)^{2j}-\frac{\sqrt{\pi}\,\Gamma(j+1)}{2\,\Gamma(j+3/2)}.
\ee
Therefore, the total field is 
\be           \label{C15}
\frac{P}{R^2}+\delta {\cal B}^r=\frac{\mw^2}{e}
+\frac{e}{R^2}\left((\sin\vartheta)^{2j}-\frac{\sqrt{\pi}\,\Gamma(j+1)}{2\,\Gamma(j+3/2)} \right). 
\ee
For $j=0$, this reduces to the background field $P/R^2=\mw^2/e$. For $j>0$, 
the total  field  is maximal and positive at the equator $\vartheta=\pi/2$, where the condensate is maximal, and it is 
minimal and negative at the  poles $\vartheta=0,\pi$, where 
the condensate vanishes (see Fig.\ref{Fig22b}). Therefore, 
 the background magnetic field is again enhanced by the condensate  and the total field is maximal where the condensate 
is maximal. 

The current flow lines are parallel to $\vartheta=const.$ on the sphere, encircling  two vortices centered at the poles 
$\vartheta=0,\pi$. The condensate vanishes at the vortex centres, where the total magnetic field $P/R^2+\delta{\cal B}^r$ 
is minimal, while  the condensate is maximal at the equator, where the total magnetic field is maximal. 
Therefore, these vortices resemble ``hollow tubes''. This agrees with the analysis in Sec.\ref{Secpert}.

The solution \eqref{C9}--\eqref{C11} is still valid if we set $R=Q=\sqrt{\kappa/2}\,P$ 
and replace the geometry in \eqref{C8} with the 
$AdS_2\times S^2$ geometry \eqref{BR} of Bertotti-Robinson. 
The latter describes the near-horizon limit of the extremal RN geometry. 
Therefore, Eqs.\eqref{C9}--\eqref{C11} describe the condensate on the extremal  RN black hole, 
hence the bifurcation 
of the hairy black holes with the extremal RN black holes. 
This bifurcation corresponds to the rightmost point 
on the existence diagram in Fig.\ref{Fig23} (right panel), which relates 
to the maximal value of the charge 
$Q_{\rm max}=1.29 \,Q_{\rm m}=2g^\prime\sqrt{\beta}\,Q_{\rm m}=\sqrt{8/\kappa}\,g^\prime /g$. 
The same value is obtained by setting $R=Q$ in \eqref{C9}, indeed, 
\be
Q^2=\frac{\kappa}{2}\,P^2=\frac{eP}{\mw^2}~~\Rightarrow~~~P=\frac{2e}{\kappa \mw^2}~~~
\Rightarrow~~~Q=\sqrt{\frac{\kappa}{2}}\,P=\sqrt{\frac{8}{\kappa}}\frac{g^\prime}{g}. 
\ee

\renewcommand{\thesection}{Appendix D}
\section{Estimates of the magnetic field strength \label{AppD}}

\renewcommand{\theequation}{D.\arabic{equation}}
\setcounter{equation}{0}
It is instructive to evaluate 
the dimensionful  magnetic field expressed in Teslas (T),
\be
\left[\frac{\boldsymbol{\mathcal{B}}}{\pmb{c}}\right]=\frac{1\, \text{kg}}{1\, \text{Coulomb}\times 1\,\text{sec}}=1\,\text{T}=10^{4}\, \text{Gauss};
\ee
($[\pmb{e}\,\pmb{\cal B}]$ has the dimension of force). 
Consider the magnetic field corresponding to the onset of the W-condensation, 
$\mathcal{B}=\mw^2/e$. Its dimensionful version is 
\be           \label{Bwww} 
\frac{\pmb{\mathcal{B}}}{\pmb{c}}=\frac{1}{\pmb{c}}\,\frac{(\pmb{\mw}\pmb{c}^2)^2}{\pmb\hbar\,\pmb{c}\,\pmb{e}}\,=
1.097\times 10^{20}\,\frac{1\, \text{kg}}{1\, \text{Coulomb}\times 1\,\text{sec}}\approx 1.1\times 10^{20}\,\text{T}\,, 
\ee
where we used the numerical values $\pmb{e}=1.6\times 10^{-19}$\, Coulombs, $\pmb{\mw}\pmb{c}^2=80.4$\, GeV, 
$1\, \text{GeV}=1.78\times 10^{-27}\,\text{kg}\times\pmb{c}^2$, $\pmb{c}=3\times 10^8\,\text{m}/\text{sec}$ and 
$\pmb{\hbar}=1.05\times 10^{-34}\,\text{kg}\,\text{m}^2/\text{sec}$.

Consider now the magnetic field of the Dirac monopole, ${\cal B}=n/(2er^2)$. Its dimensionful version is 
$\pmb{\mathcal{B}}=(\pmb{\Phi}_0/\pmb{l}_0)\,\mathcal{B}$ whereas $r=\pmb{r}/\pmb{l}_0$ while  
$\pmb{\Phi}_0\pmb{l}_0=1/\pmb{g}_0=e\pmb{e}/(4\pi\alpha)=e\,\pmb{\hbar}\,\pmb{c}/\pmb{e}$ (see Eqs.\eqref{2.6},\eqref{2.9}), hence 
\be 
\frac{\pmb{\mathcal{B}}}{\pmb{c}}=n\times \frac{\pmb{\hbar}}{2\pmb e}\times \frac{1}{\pmb{r}^2}
=
  \frac{n}{\pmb{r}^2}\times  3.28\times 10^{-16}\, \text{T}\times \text{m}^2 \,.
\ee
Setting here $\pmb{r}=1.13\times \sqrt{|n|}\,\pmb{l}_0$ according to \eqref{Bww} with $\pmb{l}_0=1.53\times 10^{-18}\,\text{m}$
yields again the same value as in \eqref{Bwww}. 
Setting $\pmb{r}$ to be the Sun's radius, $\pmb{r}=\pmb{R}_\odot\approx 6.96\times 10^8$~m, and choosing 
$n=1.6\times 10^{32}$, which corresponds to the charge of the  maximally hairy black hole, yields 
\be 
\frac{\pmb{\mathcal{B}}}{\pmb{c}}=n\times \frac{\pmb{\hbar}}{2\pmb e}\times \frac{1}{\pmb{R}_\odot^2}
=
 n\times  6.77\times 10^{-34}\,\text{T}\approx 0.1\,\text{T}\approx 10^3\,\text{Gauss}.
\ee
Therefore, a magnetic charge with $n=1.6\times 10^{32}$ placed at the center of the Sun 
would create at the surface a magnetic field of $\approx 10^3$~Gauss. 
For comparison, the average magnetic field at the Sun's surface is $\sim 1$~Gauss, but that 
in the sunspots can reach  $2 - 4\times 10^3$~Gauss.

\renewcommand{\thesection}{Appendix E}
\section{Horizon limit of extremal black holes \label{AppE}}

\renewcommand{\theequation}{E.\arabic{equation}}
\setcounter{equation}{0}
\setcounter{subsection}{0}

The extremal hairy solutions in phase I reduce, in the vicinity of the horizon, to the extremal RNdS 
solution  \eqref{geom},\eqref{RRa},\eqref{RNdSo},\eqref{horRNdS}--\eqref{Nex}, up to small corrections that vanish at the horizon. 
To determine these corrections, it is convenient to transform the solution to coordinates 
similar to those used in \eqref{extrBR} for the extremal RN case. 

The geometry of the extremal hairy solutions is described by the line element  \eqref{metr11}, with $\N(\rr)$ given by \eqref{kk_ex}. 
Let us perform  the coordinate transformation \eqref{change}, assuming that in the new gauge, we have $\tilde{\N}=1/\tilde{\rr}^2$, 
and denote the new radial coordinate as $\tilde{\rr}=\eta$. This amounts to setting 
\be                          \label{E1}
d\eta=\frac{d\rr}{\rr\sqrt{\N}}=\frac{d\rr}{{\rm k}(\rr)\,(\rr-\rrh) }~~\Rightarrow~~\eta-\eta_0=\frac{1}{{\rm k}(\rr)}\ln\frac{\rr-\rrh}{\rrh}
+\int_{\rrh}^\rr \frac{\rm k^\prime }{{\rm k^2}(\rr)}\ln\frac{\rr-\rrh}{\rrh}\,d\rr.~~
\ee
As $\rr\to\rrh$, we have $\eta\to -\infty$. 

Re-expressing  the metric amplitudes $\Uo,\Ko,\So$ in   \eqref{metr11}  in terms of 
three other functions, $\delta\Uo,\delta\Ko,\delta\So$, as 
\be
e^{2\Uo}=\rrh^2\,\eta^2\, e^{2\lambda \eta+2\delta \Uo},~~~~~
\eta^2 e^{2\Ko-2\Uo}=\rrh^2\, e^{2\delta \Ko},~~~~~
\eta^2 e^{2\So-2\Uo}=\rrh^2\, e^{2\delta \So}, 
\ee
with a constant $\lambda$, 
the line element takes the form 
\be                \label{extrBR1}
ds^2&=&\rrh^2\left\{-e^{2\lambda \,\eta}e^{2\delta\Uo} dt^2 +e^{2\delta \Ko}(d\eta^2+d\vartheta^2)+e^{2\delta \So}\sin^2\vartheta d\varphi^2 \right\}.
\ee
We also re-express  the electroweak  amplitudes in \eqref{gauge2},\eqref{gauge2a} as
\be            \label{extrBR2}
H_1&=&\delta H_1,~~H_2=1+\delta H_2,~~~H_3=\delta H_3,\
H_4=1+\delta H_4,~~~\nn \\
y&=&\delta y,~~~\phi_1=\delta\phi_1,~~~\phi_1=\delta\phi_2. 
\ee
In these formulas, the deviations  $\delta \Uo,\ldots,\delta\phi_2$ are not necessarily small, but they approach zero at the 
horizon, where  $\eta\to-\infty$. Setting them to zero, 
$\delta\Uo=\delta\So=\ldots =\delta\phi_2=0$,  all field equations are satisfied provided that the constants $\lambda$ and $\rrh$ 
obey the relations 
\be             \label{lam}
\rrh=r_{\rm ex},~~~~~
\lambda^2=1-\frac{\kappa\beta}{4}\, r_{\rm ex}^2= 
\sqrt{1-\frac{\beta\kappa^2 n^2 }{16 g^{\prime 2}}}.
\ee
These are precisely the same relations as for the extremal RNdS solution in \eqref{horRNdS1}. 
The argument of the square root is positive for $|Q|< Q_{\rm m}$,
which applies to all extremal hairy solutions in phase I, where    $|Q|\leq 0.6\,Q_{\rm m}$. 
This yields 
\be              \label{BR22}
ds^2&=&r_{\rm ex}^2\left\{-e^{2\lambda \eta}dt^2 +d\eta^2+d\vartheta^2+\sin^2\vartheta d\varphi^2 \right\},~~
B=-\frac{n}{2}\,\cos\vartheta d\varphi,~\WW=\Phi=0,
\ee
which is an exact solution 
for any $\eta\in(-\infty,\infty)$. This is the same as  the horizon limit 
of the extremal RNdS solution in \eqref{BRa}. 

In the vicinity of the horizon, the deviations $\delta \Uo,\ldots,\delta\phi_2$ are expected 
to be small, rather than zero. 
Therefore, one can expand the field equations around the 
limiting solution  \eqref{BR22}, assuming $\eta$ to be large and negative. This yields, 
in the first perturbative order, linear equations for the deviations. 
These equations  split into three 
independent groups. 

\subsection{The first group of perturbations} 
This group contains   the three metric amplitudes 
as well as $\delta y$, 
whose dependence on $\eta$ and $\vartheta$ can be separated as follows:
\be
\delta\Uo=S_{\rm U}(\vartheta)e^{\lambda \eta},~~~
\delta\Ko=S_{\rm K}(\vartheta)e^{\lambda \eta},~~~
\delta\So=S_{\rm S}(\vartheta)e^{\lambda \eta},~~~
\delta y=S_{y}(\vartheta)e^{\lambda \eta}. 
\ee
The functions $S_{\rm U}(\vartheta),S_{\rm K}(\vartheta),S_{\rm S}(\vartheta),S_y(\vartheta)$ satisfy equations,
whose  simplest solution is 
$\vartheta$-independent,
\be
S_{\rm K}=S_{\rm S}=const.,~~~S_{\rm U}=-\frac{\lambda^2+2 }{3\lambda^2}\, S_{\rm S},~~~~~S_y=0. 
\ee
The condition $S_{\rm K}=S_{\rm S}$ implies that the geometry remains spherically symmetric in the first perturbative order. 
One can set $S_{\rm S}=1$ by adjusting the integration constant $\eta_0$ in \eqref{E1} 
and then absorbing the constant $e^{\lambda\eta_0}$
by redefining the time coordinate. As a result, 
 the Schwarzschild radius of the 2-sphere, $r=\sqrt{{\rm g}_{\vartheta\vartheta}}$, 
 is no longer constant:
\be
r=r_{\rm ex}\,e^{\delta \So},~~~\text{where}~~~~\delta \So=e^{\lambda\eta}=\ln\frac{r}{r_{\rm ex}}\approx \frac{r}{r_{\rm ex}}-1\,.
\ee
The same results can be obtained by directly applying \eqref{E1} to the 
line element of the extremal RNdS solution \eqref{Nex}.

\subsection{The second group of perturbations} 
This group includes the two Higgs amplitudes, which can be expressed as: 
\be              \label{E10}
\delta\phi_1=S_1(\vartheta)\,e^{\sigma\eta},~~~\delta\phi_2=S_2(\vartheta)\,e^{\sigma\eta},~
\ee
where $\sigma$ is a constant. The equations governing $S_1$ and $S_2$ are given by:
\be
&&\left(
l(l+1)+\frac{d^2}{d\vartheta^2}+\cot\vartheta\,\frac{d}{d\vartheta}-\frac{\nu^2}{\sin^2\vartheta}
+\frac{3\nu^2-1}{4}
\right)S_1
-\left(\frac{d}{d\vartheta}+\frac{1-\nu^2}{2}\,\cot\vartheta
\right)S_2=0, \nn \\
&&\left(
l(l+1)+\frac{d^2}{d\vartheta^2}+\cot\vartheta\,\frac{d}{d\vartheta}
-\frac{\nu^2+1}{4}
\right)S_2
+\left(\frac{d}{d\vartheta}+\frac{1+\nu^2}{2}\,\cot\vartheta
\right)S_1=0,
\ee
where 
\be
\nu=-\frac{n}{2}\,,~~~~~~l(l+1)=\sigma^2+\lambda\,\sigma+\frac{\beta r_{\rm ex}^2}{4}\,.
\ee
This defines the eigenvalue problem to determine $l$, whose solution is 
\be            \label{E13}
l=\frac{\sqrt{1+2|\nu|}-1 }{2},~~~~~S_1(\vartheta)=-\frac{2}{|\nu|+1}\,\frac{d}{d\vartheta}\,S_2(\vartheta)\,,
\ee
where 
\be           \label{E14}
\left(
\frac{d^2}{d\vartheta^2}-|\nu| \cot\vartheta\,\frac{d}{d\vartheta}
+\frac{1-\nu^2}{4}
\right)S_2=0~~~\Rightarrow~~
S_2=\left(\sin\frac{\vartheta}{2}\right)^{|\nu|+1}+\left(\cos\frac{\vartheta}{2}\right)^{|\nu|+1}.~~
\ee
As a result, the exponent $\sigma$ in \eqref{E10} is given by: 
\be           \label{E15}
\sigma=\frac{-\lambda+\sqrt{\lambda^2+|n|-\beta r_{\rm ex}^2 } }{2}, 
\ee
and
\be              \label{E16}
\phi_1\propto\phi_2\propto e^{\sigma\eta}=
 \left(\ln \frac{r}{r_{\rm ex}}
 \right)^{\sigma/\lambda}
\approx  \left(\frac{r-r_{\rm ex}}{r_{\rm ex}} \right)^{\sigma/\lambda}.
\ee
When the magnetic charge $|n|$ increases, $\sigma$ grows as $\sqrt{|n|}$, while $\lambda$ decreases. Therefore, the  Higgs field 
approaches zero faster and faster in the vicinity of the horizon,  and the size of the false vacuum region around the horizon increases. 

If $|n|\ll 1/\sqrt{\kappa}$, then the horizon size $r_{\rm ex}\ll 1$, and the spacetime geometry at the electroweak scale, where  $r\sim 1$,
 is essentially flat. In that case, 
the size of the vacuum region  can be estimated to be $\sim \sqrt{|n|}$, as expressed  by Eq.(5.24) in \cite{GVII}. 

The relation \eqref{E16} implies that the components of the unit vector 
$n^a=(\Phi^\dagger\tau^a\Phi)/(\Phi^\dagger\Phi)$ used in \eqref{Nambu} to define 
the electromagnetic field do not depend on the radial coordinate:
\be                    \label{E17}
n^1=\frac{2S_1S_2}{(S_1)^2+(S_2)^2},~~~~n^2=0,~~~~~n^3=\frac{(S_1)^2-(S_2)^2}{(S_1)^2+(S_2)^2},
\ee
where we used $\Phi$ in the gauge \eqref{RR}. 
These relations are exact at the horizon, while in the vicinity of the horizon, they may receive corrections only in 
higher perturbative orders. 

This allows us to determine the horizon value of the electromagnetic field. 
Since the perturbations $\delta H_1,\delta H_2,\delta H_3,\delta H_4$ approach zero at the horizon, 
 it follows from \eqref{RR},\eqref{var} that the SU(2) gauge field strength 
vanishes at the horizon, $\WW^a_{\mu\nu}=0$. 
The Nambu field ${\cal F}_{\mu\nu}$, defined in \eqref{Nambu},
then  reduces to the U(1) hypercharge field, up to a constant factor. 
 
 Therefore, the only non-zero  components of the Nambu field at the horizon are 
 ${\cal F}_{\vartheta\varphi}=-{\cal F}_{\varphi\vartheta}$, where 
 \be             \label{horF1}
 {\cal F}_{\vartheta\varphi}= \frac{g}{g^\prime}\,B_{\vartheta\varphi}= 
 \frac{g}{g^\prime}\,\frac{n}{2}\,\sin\vartheta=g^2\,\frac{n}{2e}\,\sin\vartheta=g^2 P\sin\vartheta. 
 \ee
 This corresponds to the purely radial magnetic field of total  flux $4\pi g^2 P$, created by the magnetic charge $g^2 P$ 
 inside the black hole. The remaining part of the charge $g^{\prime 2}P$ is contained in the black hole hair.

The 't\,Hooft   field, $F_{\mu\nu}={\cal F}_{\mu\nu}-e\psi_{\mu\nu}$, defined in \eqref{FF}, includes the condensate tensor $\psi_{\mu\nu}$, 
which is constructed from the unit vector $n^a$ according to \eqref{Hooft}. Using $n^a$ in \eqref{E17} and the SU(2) field $\WW^a_\mu$ 
in the gauge \eqref{RR}, we find that the only components of the condensate which do not vanish 
at the horizon are $\psi_{\vartheta\varphi}=-\psi_{\varphi\vartheta}$, where 
\be             \label{horF2}
-e\psi_{\vartheta\varphi}= g^{\prime 2}\frac{n}{2e}\,\times
\frac{|n|\, \left(\sin \frac{\vartheta}{2}  \cos\frac{\vartheta}{2}  \right)^{|n|-1}}
{\left[ \left(\cos\frac{\vartheta}{2} \right)^{|n|}+\left(\sin\frac{\vartheta}{2} \right)^{|n|}
\right]^2
 }.
\ee
It is instructive to compare this with the formula $\psi_{\vartheta\varphi}\sim (\sin\vartheta)^{|n|-1}$, which follows from 
\eqref{PSI},\eqref{PSI22} in the axially symmetric case (when $c_{\rm k}=\delta_{0\rm k}$ in \eqref{PSI}). 
This formula 
was obtained using perturbation theory at the bifurcation with the RN solution. 
In contrast,  the relation in   \eqref{horF2} 
is non-perturbative and corresponds to the fully non-linear solution restricted to the event horizon. According to 
\eqref{PSI},\eqref{PSI22}, the value of $|\psi_{\vartheta\varphi}|$ 
(or more precisely, $|\psi_{\vartheta\varphi}|/\sin\vartheta$) can be interpreted as a measure of the condensate strength. 
Remarkably, both the perturbative and non-perturbative expressions for 
 the condensate exhibit the same qualitative behavior:
they reach their maximum at the equator and vanish at the poles.

The flux of $\psi_{\mu\nu}$ is given by 
\be                 \label{E20}
-e \oint \psi_{\vartheta\varphi}\, d\vartheta d\varphi=4\pi g^{\prime 2}P. 
\ee
As a result, 
 the horizon value of the 't\,Hooft   field is the sum of \eqref{horF1} and \eqref{horF2}, 
\be                  \label{E21} 
F_{\vartheta\varphi}= \frac{g}{g^\prime}\,B_{\vartheta\varphi}-e\,\psi_{\vartheta\varphi}=
g^2\,\frac{n}{2e}\,\sin\vartheta+g^{\prime 2}\frac{n}{2e}\,\times
\frac{|n|\, \left(\sin \frac{\vartheta}{2}  \cos\frac{\vartheta}{2}  \right)^{|n|-1}}
{\left[ \left(\cos\frac{\vartheta}{2} \right)^{|n|}+\left(\sin\frac{\vartheta}{2} \right)^{|n|}
\right]^2
 },
\ee
whose flux is $4\pi P$.  
Since the 2-form $F_{\mu\nu}$ is closed, its flux is independent of the integration surface. As a result, the flux though 
a 2-sphere at spatial infinity is the same as that though the horizon, implying that 
the entire magnetic charge is contained within the black hole.
In contrast, for the Nambu field \eqref{horF1}, the flux through the horizon is 
$4\pi g^2 P$, while the flux at infinity remains $4\pi P$. The difference is $4\pi g^{\prime 2 } P$, corresponds to the 
magnetic charge $g^{\prime 2 } P$ distributed outside the black hole.

The $Z$ field, defined by \eqref{FF}, has non-zero components 
\be                \label{E22}
Z_{\vartheta\varphi}=-Z_{\varphi\vartheta}=B_{\vartheta\varphi}+g^2 \psi_{\vartheta\varphi}. 
\ee
Its horizon flux 
vanishes, as expected,  since the 2-form $Z_{\mu\nu}$ is also closed, and its flux at infinity is zero because this field is massive.  

Let us now consider the current densities. Taking the divergence of the Nambu field \eqref{horF1} gives zero, meaning that the Nambu 
current  vanishes at the horizon. The magnetic charge density in \eqref{cur1a} is proportional to the radial derivative of $\psi_{\vartheta\varphi}$, 
which is also zero at the horizon. Therefore, both the Nambu current density ${\cal J}^\mu$ 
and the magnetic charge density $\tilde{{\cal J}}^0$ vanish at the horizon, although  they 
deviate form zero at some distance away, as indicated by our numerics. 

On the other hand, taking the divergence of the 't\,Hooft   field \eqref{E21} gives a non-zero value of the 
azimuthal current density at the horizon:
\be               \label{E23}
4\pi r_{\rm ex}^2 J_\varphi=-\sin\vartheta \left(\frac{F_{\vartheta\varphi}}{\sin\vartheta} \right)^\prime_\vartheta 
=-\frac{g^\prime\,n|n|}{g\,2^{|n|}}\,\sin\vartheta  \left(\frac{\left(\sin\vartheta\right)^{|n|-2}}{\left[ \left(\cos\frac{\vartheta}{2} \right)^{|n|}+\left(\sin\frac{\vartheta}{2} \right)^{|n|}
\right]^2 } \right)^\prime_\vartheta\,.
\ee
According to our numerics (see Fig.\ref{Fignorm}, right panel)
the  current density is maximal at the horizon and starts to decrease  as we move away from it. 
Notice that the formulas \eqref{horF1}--\eqref{E23} do not contain $\kappa$.

\subsection{The third group of perturbations } 
This group contains four SU(2) amplitudes that we re-express as: 
\be
\delta H_1=e^{q\eta}\,h_1(\vartheta),~~~\delta H_2=e^{q\eta}\,h_2(\vartheta),~~~
\delta H_3=e^{q\eta}\,h_3(\vartheta),~~~
\delta H_4=e^{q\eta}\,h_4(\vartheta),~~~
\ee
where $q$ is a constant. The corresponding equations for the perturbations are: 
\be          \label{4eqs}
&&\left(\frac{d^2}{d\vartheta^2} +\cot\vartheta\, \frac{d}{d\vartheta}-\frac{\nu^2}{\sin^2\vartheta}+q^2
\right)h_1+q\cot\vartheta\,h_2+q\nu^2\,h_3-q\nu^2\cot\vartheta\, h_4=0, ~~~\nn \\
&&\left(\frac{d^2}{d\vartheta^2} -\frac{\nu^2}{\sin^2\vartheta}+q(q+\lambda)
\right)h_2+\lambda \frac{d}{d\vartheta}\,h_1-\nu^2 (\frac{d}{d\vartheta}+2\cot\vartheta)h_3 \nn \\
&&\hspace{9 cm}+
\nu^2(\cot\vartheta\,\frac{d}{d\vartheta}+\cot^2\vartheta-1)h_4=0,~~~~~~  \\
&&\left(\frac{d^2}{d\vartheta^2} +\cot\vartheta\, \frac{d}{d\vartheta}-\frac{1}{\sin^2\vartheta}+q(q+\lambda)-1
\right)h_3-\lambda h_1-\cot\vartheta\,h_2+(2\frac{d}{d\vartheta}+\cot\vartheta) h_4=0, \nn \\
&&\left(\frac{d^2}{d\vartheta^2} +\cot\vartheta\, \frac{d}{d\vartheta}-\frac{1}{\sin^2\vartheta}+q(q+\lambda)-1
\right)h_4+\lambda \cot\vartheta\,h_1+\cot^2\vartheta\,h_2-(2\frac{d}{d\vartheta}+\cot\vartheta) h_3=0. \nn 
\ee 
By using the transformations:
\be                    \label{h34}
h_4(\vartheta)=\cos\vartheta\, \tilde{h}_3(\vartheta)+\sin\vartheta\, \tilde{h}_4(\vartheta),~~~
h_3(\vartheta)=-\sin\vartheta\, \tilde{h}_3(\vartheta)+\cos\vartheta\, \tilde{h}_4(\vartheta),~~~
\ee
the system of equations \eqref{4eqs} splits into three coupled equations involving $h_1,h_2,\tilde{h}_3$, 
plus one independent equation for $\tilde{h}_4$: 
\be
\left(\frac{d^2}{d\vartheta^2} +\cot\vartheta\, \frac{d}{d\vartheta}-\frac{1}{\sin^2\vartheta}+q(q+\lambda)
\right)\tilde{h}_4=0. 
\ee
The solution to this equation is given by:
\be
\tilde{h}_4= C_4\, P^1_l(\cos\vartheta),~~~~q(q+\lambda)=l(l+1),~~~~l=1,2,\ldots ,
\ee
where $P^1_l(\cos\vartheta)$ are the Legendre polynomials and $C_4$ is an integration constant. 

By setting $h_1=h_2=\tilde{h}_3=0$, and using the equation \eqref{h34}, we obtain 
\be
h_4=C_4\, \sin\vartheta\, P_l^1(\cos\vartheta),~~~
h_3=C_4\, \cos\vartheta\, P_l^1(\cos\vartheta),~~~            \label{E.29}
\ee
and the value of $q$ is
\be
q=\frac{-\lambda+\sqrt{\lambda^2+4l(l+1)}}{2}\,,~~~~
\ee
where $C_4$ is an integration constant. 

The analysis of  the the remaining equations for  $h_1,h_2,\tilde{h}_3$ involves applying  the gauge fixing condition \eqref{fix}, 
which now simplifies  to: 
\be
qh_1(\vartheta)=h_2^\prime(\vartheta).
\ee
This allows us to eliminate $h_1$ since its 
equation becomes a differential consequence of  the two other equations. The remaining equations reduce to 
\be
&&\left(\frac{d^2}{d\vartheta^2} +\cot\vartheta\, \frac{d}{d\vartheta}-\frac{1}{\sin^2\vartheta}+q(q+\lambda)
\right)\tilde{h}_3+\frac{1}{\sin\theta}\left(\frac{\lambda}{q} \frac{d}{d\vartheta}+\cot\vartheta\right)h_2=0, \nn \\
&&\left((q+\lambda)\,\frac{d^2}{d\vartheta^2} +q^2(q+\lambda)-\frac{q\,\nu^2}{\sin^2\vartheta}
\right)h_2+\frac{q\nu^2}{\sin\vartheta}\left(\frac{d}{d\vartheta}+\cot\vartheta \right)\tilde{h}_3=0.           \label{E.32}
\ee
Unfortunately, we were unable to find an exact solution of these equations, although it is clear that 
the solution should be expressible  in terms of 
Legendre polynomials. We leave this problem for future work. 

\renewcommand{\thesection}{Appendix F}
\section{Vortices \label{AppF}}
\renewcommand{\theequation}{F.\arabic{equation}}
\setcounter{equation}{0}
\setcounter{subsection}{0}

Vortices are frequently mentioned in the main text, and for completeness, 
we provide in this Appendix a brief overview of the vortex solutions that may be relevant to our analysis. 
The results presented in subsections \ref{subvort} and \ref{subEW} are known and can be found in the literature, 
while subsection \ref{subchir} contains observations that, to the best of our knowledge, are new.

\subsection{Vortices in the Abelian Higgs model \label{subvort}}

 Vortices were originally described  by Abrikosov  \cite{Abrikosov:1956sx} 
 within the phenomenological theory of superconductivity developed by  Ginzburg and Landau  \cite{Ginzburg:1950sr}.
 Some time later, exactly the same solutions were rediscovered by Nielsen and Olesen in the context of  relativistic field theory
 \cite{Nielsen:1973cs}. These vortices  arise in a model that  includes 
 an Abelian vector field ${\cal C}_\mu$ and a complex 
 scalar $\phi$, with the Lagrangian 
 \be              \label{F1}
 {\cal L}=-\frac14\,{\cal C}_{\mu\nu}{\cal C}^{\mu\nu}-\overline{{\cal D}_\mu\phi }\,{\cal D}^\mu\phi
 -\frac{\beta}{8}(|\phi|^2-1)^2. 
 \ee
 Here ${\cal C}_{\mu\nu}=\partial_\mu {\cal C}_\nu-\partial_\nu {\cal C}_\mu$ and ${\cal D}_\mu\phi=\left(\nabla_\mu-iq\,{\cal C}_\mu\right)\phi$
 with $q$ being the electric charge. The theory is invariant  under U(1) gauge transformations, 
 \be             \label{Ug}
 {\cal C}_\mu\to {\cal C}_\mu+\frac{1}{q}\,\partial_\mu\alpha,~~~~\phi\to e^{i\alpha}\phi.
 \ee
 The equations of the theory, 
 \be                    \label{ANO}
 \nabla^\mu {\cal C}_{\mu\nu}=iq\left(\overline{\phi}{\cal D}_\nu\phi-\phi\overline{{\cal D}_\nu\phi}\right),~~~~~~
 {\cal D}_\mu {\cal D}^\mu\phi=\frac{\beta}{4}(|\phi|^2-1)\phi,
 \ee
 admit solutions describing  a localized magnetic flux along the $z$-direction.  
 Passing to cylindrical coordinates, where the metric is 
 $ds^2=-dt^2+d\rho^2+\rho^2 d\varphi^2+dz^2$, and setting 
  \be             \label{C2}
{\cal C}_{\mu} dx^\mu =-{\rm v}(\rho)\,d\varphi,~~~~~~\phi=f(\rho),
 \ee
 Eqs.\eqref{ANO} reduce to 
 \be              \label{eqANO}
 \rho\left(\frac{{\rm v}^\prime}{\rho} \right)^\prime =2q^2\,  f^2 {\rm v} ,~~~~~~~~
 \frac{1}{\rho}\left(\rho f^\prime\right)^\prime =q^2\, \frac{{\rm v}^2}{\rho^2}\, f+\frac{\beta}{4}(f^2-1)f.
 \ee
 These equations admit globally regular solutions with the boundary conditions 
 \be              \label{bc}
 {\rm v}(0)\leftarrow {\rm v}(\rho)\to 0,~~~~~0\leftarrow f(\rho)\to 1~~~~\text{as}~~~~0\leftarrow\rho\to \infty. 
 \ee
 The value of ${\rm v}(0)$ can be arbitrary, but for generic values, the vector field in \eqref{C2} is singular 
 at the symmetry axis, since the azimuthal component ${\cal C}_\varphi$ does not vanish as $\rho\to 0$. 
 However, if
${\rm v}(0)=k/q$ with $k\in\mathbb{Z}$, then performing  the  gauge transformation \eqref{Ug} with $\alpha=k\varphi$ 
brings the fields to the form 
 \be              \label{C1}
 {\cal C}_{\mu} dx^\mu =\left(\frac{k}{q}-{\rm v}(\rho)\right) d\varphi,~~~~\phi=f(\rho) e^{ik\varphi}~~~\text{with}~~~k\in \mathbb{Z}. 
 \ee
 The fields now vanish at $\rho=0$, so the singularity is gauged away. A similar ``desingularization'' procedure 
 will be used below for other vortex solutions. 
 
The magnetic flux is determined by the asymptotic value of ${\cal C}_\varphi$ in \eqref{C1},
 \be               \label{fluxANO}
 \Psi_{\cal C}=\oint_{\partial{\mathbb R}^2} {\cal C}_k dx^k =\frac{2\pi k}{q}. 
 \ee
The vortices have a finite energy per unit length,
\be
E[q,k]
=2\pi \int _0^\infty \left(\frac{{\rm v}^{\prime 2}}{2\rho^2} +f^{\prime 2}
+q^2\, \frac{{\rm v}^2 f^2}{\rho^2}+\frac{\beta}{8}(f^2-1)^2\right)\rho\, d\rho. 
\ee

\subsection{Electroweak vortices \label{subEW}}
Returning to the parametrization \eqref{RR} of the electroweak fields and setting 
\be          \label{anz}
F_1=F_2=0,~\nu F_3=-v_3(\rho),~\nu F_4=v_1(\rho),~\nu Y=-v(\rho),~\phi_1=f_1(\rho),~\phi_2=f_2(\rho),~~
\ee
the electroweak equation \eqref{P2} reduce to 
\be               \label{EWcyl}
\rho\left(\frac{v^\prime}{\rho}\right)^\prime&=&\frac{g^{\prime 2}}{2}\left\{
(v+v_3)f_1^2+2v_1 f_1 f_2+(v-v_3)f_2^2
\right\}\,,   \nn \\
\rho\left(\frac{v_1^\prime}{\rho}\right)^\prime&=&\frac{g^{2}}{2}\left\{
v_1 (f_1^2+f_2^2)+2v f_1 f_2
\right\}\,,   \nn \\
\rho\left(\frac{v_3^\prime}{\rho}\right)^\prime&=&\frac{g^{2}}{2}\left\{
(v_3+v) f_1^2+(v_3-v)f_2^2
\right\}\,,   \nn \\
\frac{1}{\rho}\left(\rho f_1^\prime\right)^\prime&=&
\left\{
\frac{1}{4\rho^2}\,\left[(v+v_3)^2+v_1^2 \right]+\frac{\beta}{4}(f_1^2+f_2^2-1)
\right\}f_1+\frac{vv_1}{2\rho^2}\,f_2\, \nn \\
\frac{1}{\rho}\left(\rho f_2^\prime\right)^\prime&=&
\left\{
\frac{1}{4\rho^2}\,\left[(v-v_3)^2+v_1^2 \right]+\frac{\beta}{4}(f_1^2+f_2^2-1)
\right\}f_2+\frac{vv_1}{2\rho^2}\,f_1\, ,
\ee
plus a first order constraint, that is compatible with the other equations:
\be               \label{EWcons}
\frac{1}{\rho^2}\left(v_1v_3^\prime-v_3v_1^\prime \right)+g^2(f_2 f_1^\prime-f_1 f_2^\prime)=0.
\ee
It is sufficient to impose this constraint  only at $\rho=0$. 
These equations (actually a more general system including also a non-zero $F_2$) were originally 
obtained  in Ref.\cite{Garaud:2009uy}, where one can find the analysis of the constraint \eqref{EWcons}.

\subsubsection{Z-strings } 
Setting in \eqref{EWcyl} 
\be            \label{RRZ0}
v=g^{\prime 2}{\rm v}(\rho),~~v_1=0,~~~v_3=-g^{2}{\rm v}(\rho),~~~f_1=0,~~~f_2=f(\rho), 
\ee
the  equations reduce  to \eqref{eqANO} with $q=1/2$. 
Their solution 
determines the electroweak fields 
\be               \label{RRZ}
\WW&=&g^2 \T_3\, {\rm v}(\rho)d\varphi\,, ~~~
B=-g^{\prime 2}{\rm  v}(\rho)d\varphi\,,~~~
\Phi=\begin{pmatrix}
0 \\
f(\rho)
\end{pmatrix}\,.
\ee
This fields can be desingularized 
by choosing ${\rm v}(0)=2k$ and performing 
the gauge transformation \eqref{gauge},\eqref{U} with 
${\rm \U}=\exp\{i(g^{\prime 2}-g^{2}\tau_3)\,k\varphi\}$. This  converts the fields to 
\be               \label{RRZ1}
\WW&=&g^2 \T_3\, \left( {\rm v}(\rho)-2k\right) d\varphi\,, ~~~
B=g^{\prime 2} \left( 2k-{\rm v}(\rho)\right) d\varphi\,,~~~
\Phi=\begin{pmatrix}
0 \\
f(\rho)e^{ik\varphi}
\end{pmatrix}\,,
\ee
in which form  they  vanish at $\rho=0$. 

Using \eqref{Hooft} yields $\psi_{\mu\nu}=0$, implying that 
the definitions of Nambu and 't\,Hooft give the same result for the physical 
fields,  ${\cal F}_{\mu\nu}=F_{\mu\nu}=0$, with the only non-zero component of ${\cal Z}_{\mu\nu}=Z_{\mu\nu}$ being 
$
Z_{\rho\varphi}=-Z_{\varphi\rho}=-{\rm v}^\prime(\rho). 
$
This determines the fluxes, 
\be              \label{PsiZ}
\Psi_F=0,~~~~~\Psi_Z=4\pi k. 
\ee
The solution was discovered in \cite{Vachaspati:1992fi}. Since it 
contains  only the massive $Z$-field  and no usual 
magnetic field, it is called Z-string. 

\subsubsection{W-strings} 
Setting in \eqref{EWcyl} 
\be
v=v_3=f_1=0,~~v_1=g{\rm v}(\rho),~~~f_2=f(\rho), 
\ee
the equations again reduces to \eqref{eqANO}, but this time  with $q=g/2$. 
Their solution determines the electroweak fields, 
\be               \label{RRZ3}
B=0,~~~
\WW&=&-\T_1\, g\,{\rm v}(\rho)d\varphi\,, ~~~
\Phi=\begin{pmatrix}
0 \\
f(\rho)
\end{pmatrix}\,.
\ee
The singularity at the axis can be removed by setting ${\rm v}(0)=2k/g$ and performing 
the gauge transformation \eqref{gauge},\eqref{U} generated by 
${\rm \U}=\exp\{i\tau_1\,k\varphi\}$, which 
 yields 
\be               \label{RRZ4}
\WW&=&\T_1 \WW^1_\mu dx^\mu=\T_1\, (2k-g\,{\rm v}(\rho))d\varphi\,, ~~~
\Phi=f(\rho)\begin{pmatrix}
i\sin (k\varphi) \\
\cos(k\varphi)
\end{pmatrix}\,,
\ee
in which form the fields vanish at $\rho=0$. Getting back to the gauge
\eqref{RRZ3} and using \eqref{Hooft} shows that ${\cal W}_\mu=0$, therefore, both the magnetic and Z fields vanish:
\be              \label{PsiZ}
\Psi_F=\Psi_Z=0
\ee
(this corrects the values quoted in \cite{Garaud:2009uy}).
The solution can be characterized by a non-zero W-flux,
\be
2\pi \int_0^\infty \WW_{\rho\varphi}\,d\rho=4\pi k\,\T_1,
\ee
which is, however, not gauge-invariant. This solution was obtained in  \cite{Vachaspati:1992pi}.

\subsubsection{Dressed Z-strings} 

The Z-strings and W-strings are {\it Abelian}  and  correspond to solutions of the Abelian Higgs model \eqref{F1} 
embedded 
into the electroweak theory. The equations \eqref{EWcyl} admit also more general solutions, which 
can be viewed as Z-strings  containing  a condensate at their  core. 
Such ``dressed'' Z-strings were discussed qualitatively long ago  \cite{Perkins:1993qz,Olesen:1993ra,Achucarro:1993bu},
and later they were explicitly obtained in Ref.\cite{Garaud:2009uy} (see Section 9.3 of that article). 

These solutions are characterized by the boundary conditions at $0\leftarrow\rho\rightarrow\infty$ 
(our notation is related to that of \cite{Garaud:2009uy} by the replacement $v_3\to -v_3$, 
$f_1\leftrightarrow f_2$):
\be            \label{bdc}
2k-m\leftarrow v\to c&&,~~~~~0\leftarrow v_1\to -c\,\sin\gamma,~~~~~~-m\leftarrow v_3\to c\,\cos\gamma,~~~\nn \\
q\rho^{|k-m|}&&\leftarrow f_1\to \sin\frac{\gamma}{2},~~~~a\rho^{k}\leftarrow f_2\to \cos\frac{\gamma}{2}.
\ee
Here $k$ is an integer that we assume positive for simplicity, one has  $m=1,2,\ldots 2k$, while  $c,\gamma,a,q$ are real. 

The parameter $\gamma$ can be called vacuum angle. 
If $\gamma=0$, then $f_1=0$ and the 
the solution corresponds to the Z-string configuration  \eqref{RRZ0}, but with shifted values of $v,v_3$:
\be
v=g^{\prime 2}{\rm v}(\rho)+c,~~v_1=0,~~v_3=-g^2{\rm v}(\rho)+c,~~f_1=0,~~f_2=f(\rho). 
\ee
Here  ${\rm v}(\rho),f(\rho)$ are the same as for the Z-string, so that 
${\rm v}(0)=2k$, while  
$c=2kg^2-m$. 
This is the Z-string \eqref{RRZ}, gauge-transformed by  \eqref{gauge},\eqref{U} with 
${\rm \U}=\exp\{-i c\, (1+\tau_3)\,\varphi/2\}$. Therefore, for the Z-string the second integer $m$ 
is generated by the gauge transformation. 

\begin{figure}
   \includegraphics[scale=0.8]{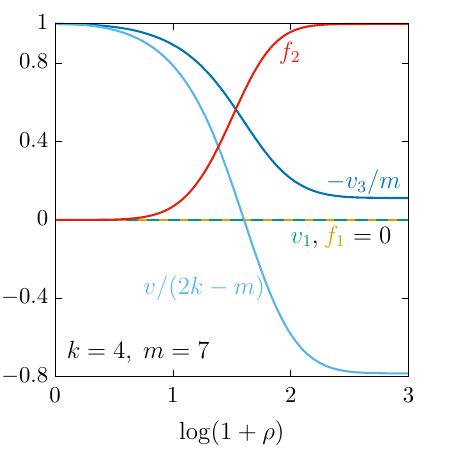}  
      \includegraphics[scale=0.8]{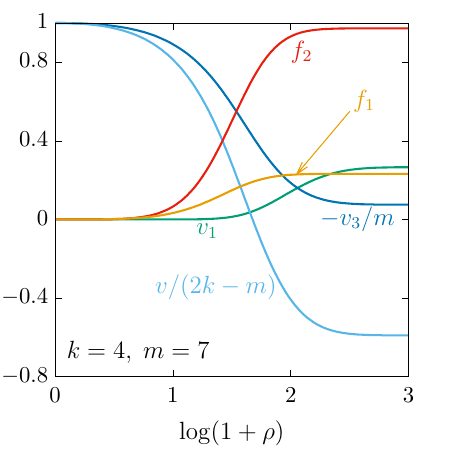}

       \caption{The Z-string (left) and the dressed Z-string (right) with  $k=4$, $m=7$ for the physical values 
       $\sin^2\thetaw=0.223$, $\beta=1.88$. One has $\Psi_F=0$ and $\Psi_Z=4\pi k$ for the Z-string, 
       whereas for the dressed solution 
      $\Psi_F\approx 0.9\times 4\pi m\times g^{\prime 2}/e$ and  $\Psi_Z=4\pi (k-m)$.
         }
    \label{FigZ}
\end{figure}

The dressed Z-strings correspond to
more general solutions with  $v_1\neq 0$ and $f_1\neq 0$. They are  labeled by pairs $(k,m)$
and they  are 
physically different for different $m$. 
For some choices of  $(k,m)$ these solutions exist only for non-physical values of the theory parameters 
$\thetaw,\beta$ \cite{Garaud:2009uy}. Choosing the physical values $\sin^2\thetaw=0.223$ and $\beta=1.88$, 
we were able to reproduce such solutions for $k=4,5,6,\ldots$ with $m=2k-1$.However,  they probably exist also 
for other values of $(k,m)$. 
Their profiles for $k=4,m=7$ are shown in Fig.\ref{FigZ}. 

These solutions can be desingularized with the gauge transformation \eqref{gauge},\eqref{U} generated by 
${\rm \U}=\exp(ik\varphi)\exp\{-i (1+\tau_3)\,m \varphi/2\}$. This   brings the fields to the form 
\be               \label{RRZdr}
\WW&=&\left\{-v_1\tilde{\T}_1-\T_3 (v_3+m)\right\} d\varphi\,, ~
B=\left( 2k-m-v\right) d\varphi\,,~
\Phi=\begin{pmatrix}
f_1\, e^{i(k-m)\varphi} \\
f_2\,e^{ik\varphi}
\end{pmatrix}\,,
\ee
with $\tilde{\T}_1=\T_1\cos(m\varphi)+\T_2\sin(m\varphi)$. In view of the boundary 
conditions \eqref{bdc}, these  fields vanish at $\rho=0$. 

The dressed solutions carry non-zero magnetic and Z fluxes. To compute  them, we consider  the unit vector $n^a$ defined in \eqref{Nambu1}. 
It reduces to $n^a=-\delta^a_3$ for Z-strings, while for  the dressed solutions one has 
for $0\leftarrow \rho \to\infty$:
\be
\delta ^a_3 \leftarrow n^a=(n^1,n^2,n^3)=\left(\frac{2f_1 f_2}{f_1^2+f_2^2},0,\frac{f_1^2- f_2^2}{f_1^2+f_2^2}\right)\to 
(\sin\gamma,0,-\cos\gamma), 
\ee
hence, $n^a$ rotates by an angle $\pi-\gamma$ close to the origin, where  $f_1$ approaches zero slower than 
$f_2$. Using Eqs.\eqref{Hooft} yields 
\be
n^a \WW^a_{\rho\varphi}&=&-n^1 (v_1)^\prime -n^3 (v_3)^\prime, \nn \\
g^2\psi_{\rho\varphi}&=&(n^1(n^3)^\prime-n^3 (n^1)^\prime)(v_1 n^3-v_3 n^1),\nn \\
{\cal W}_\varphi&=&-n^1 v_1-n^3 v^3\,,
\ee
hence, 
\be
{\cal W}_{\rho\varphi}\equiv 
n^a W^a_{\rho\varphi}+g^2\psi_{\rho\varphi}=\left({\cal W}_\varphi\right)^\prime\,,
\ee
which confirms the identity in \eqref{id}. 
This allows us to compute the fluxes. 
Using the boundary conditions \eqref{bdc}, the flux of the U(1) hyperfield is
\be
\Psi_B=2\pi \int_0^\infty B_{\rho\varphi}\, d\rho=-2\pi \int_0^\infty v^\prime (\rho) d\rho=2\pi\,(2k-m-c).
\ee
Defining 
\be
\Psi_{\cal W}&=&2\pi \int_0^\infty {\cal W}_{\rho\varphi}\, d\rho=-2\pi \int_0^\infty (n^1 v_1 +n^3 v_3)^\prime\, d\rho
=2\pi\,(c-m), \nn \\
\Psi_W&=&2\pi \int_0^\infty n^a\WW^a_{\rho\varphi}\, d\rho=-2\pi \int_0^\infty (n^1 v_1^\prime +n^3 v_3^\prime)\, d\rho, 
\ee
the 't\,Hooft fluxes are totally determined by the boundary conditions for the fields: 
\be            \label{FT}
\Psi_F&=&\frac{g}{g^\prime}\,\Psi_B-\frac{g^\prime}{g}\,\Psi_{\cal W}=\frac{2\pi}{e}\,\left(m-c+2(k-m)g^2\right),~~~\nn \\
\Psi_Z&=&\Psi_B+\Psi_{\cal W}=4\pi(k-m). 
\ee
The Nambu fluxes are 
\be          \label{FN}
\Psi_{\cal F}=\frac{g}{g^\prime}\,\Psi_B-\frac{g^\prime}{g}\,\Psi_W,~~~~
\Psi_{\cal Z}=\Psi_B+\Psi_W. 
\ee

\subsection{Chiral strings versus Nambu strings \label{subchir}} 

The results mentioned above are not  new and can be found in  \cite{Garaud:2009uy}, but 
now  we can add something  else. 
Let us consider chiral solutions -- the dressed strings  in the limit where they approach Z-strings.
This  means that $\gamma$ and $v_1,f_1$ are small,
while $v,v_3,f_2$ are close to the Z-string values, so that one has almost everywhere $n^3\approx -1$ and  $n^1\approx 0$. 
As $\gamma$ decreases, the configurations approach Z-strings pointwise. However, 
at $\rho=0$ one always has $n^3=+1$, hence,  the vector $n^a$ always  flips sign 
close to the origin. Therefore, 
the topology is not the same as for the Z-strings. 
In the limit where $\gamma\to 0$, one has 
\be 
c\to 2kg^2-m,~~~~~~\Psi_W\to 2\pi \int_0^\infty v_3^\prime\, d\rho=2\pi (c+m)=4\pi k g^2.
\ee
Therefore, the condensate in the vortex core does not vanish in the limit, 
\be
2\pi g^2\int_0^\infty \psi_{\rho\varphi}\,d\rho=\Psi_{\cal W}-\Psi_W=-4\pi m,
\ee
the t'\,Hooft fluxes \eqref{FT} reduce to 
\be            \label{fHooft}
\Psi_F=4\pi\, \frac{g^{\prime 2}}{e}\times m,~~~\Psi_Z=4\pi(k-m),
\ee
while the Nambu fluxes \eqref{FN} become 
\be            \label{fNambu}
\Psi_{\cal F}=0,~~~\Psi_{\cal Z}=4\pi k. 
\ee

 
 Notice that the magnetic flux $\Psi_F$ is proportional to $4\pi g^{\prime 2}/e$,
which matches the flux produced by the Nambu monopole \cite{Nambu:1977ag}.
This monopole is attached to a string and corresponds to the point where the string terminates,
allowing the fields that were confined within the string 
to spread outward in all directions
(see \cite{Patel:2021iik,Patel:2023sfm,Domcke:2025itc}).

Assuming that a chiral string terminates at some point, the magnetic field previously 
confined within the string 
will escape to the exterior, carrying with it the same total flux $\Psi_F$. 
This follows from the fact that the 't\,Hooft fluxes $\Psi_{F}$ and $\Psi_{Z}$ are conserved, 
and therefore must be identical both inside and outside the string. 
Consequently, the endpoint of a chiral string should manifest as a Nambu monopole.

On the other hand, the Nambu fluxes $\Psi_{\cal F}$ and $\Psi_{\cal Z}$ are not conserved, 
and the fluxes inside the string differ from those produced by the monopole outside the string. 
Remarkably, the Nambu fluxes in \eqref{fNambu} match exactly (for $k=1$) 
those obtained in Nambu's original analysis \cite{Nambu:1977ag} for a string terminating on a monopole.

This suggests that the Nambu string -- the electroweak string attached to the Nambu monopole --
can be interpreted as a chiral string, since the Nambu fluxes coincide in both cases and the chiral string
carries the same magnetic flux $\Psi_F$ as the Nambu monopole.
To the best of our knowledge, the Nambu string has never been explicitly described in the literature,
and it is typically assumed to be identical to the Z-string \cite{Patel:2023sfm},
as the Nambu fluxes also match in that case.
However, because the Nambu string must carry the same magnetic flux $\Psi_F$ as the monopole,
it cannot be a Z-string, although it may closely resemble one in the sense described above.

Nevertheless, one should note  that, unlike Z-strings,
 chiral solutions   exist only for special parameter values 
corresponding to the chiral curves  shown in Fig.4 in \cite{Garaud:2009uy}.
For the physical values of $\thetaw,\beta$,  chiral solutions do not exist. 
However,  there are dressed solutions that also carry a magnetic flux, although 
its value is not exactly a multiple of $4\pi g^{\prime 2}/e$.
For example, choosing the physical parameters  $\sin^2\thetaw=0.223$, $\beta=1.88$, 
the solution with $k=4$, $m=7$ shown in Fig.\ref{FigZ} 
has  $\gamma=0.177$ and the flux $\Psi_F\approx 0.9\times 4\pi m\times g^{\prime 2}/e$.
By contrast, a
 solution with the same $(k,m)$ and with $\gamma=0.01$, 
 which is quite close to the chiral limit, is found only for 
$\sin^2\thetaw=0.199$. This illustrates that  chiral solutions are not generic.

It is therefore logically possible that Nambu monopoles arising in realistic scenarios,
such as those produced via the Kibble mechanism \cite{Urrestilla:2001dd,Patel:2021iik}, 
are connected to dressed strings, in which case their magnetic flux is not exactly equal to 
$4\pi g^{\prime 2}/e$, but only approximately so.
At the same time, the magnetic flux associated with chiral strings -- 
being an exact multiple of $4\pi g^{\prime 2}/e$ -- appears to emerge generically in various contexts with physical parameter values.
For instance, such fluxes are present in the horizon vortices described by \eqref{Nambuv}, 
as well as in the vortex connecting a monopole-antimonopole pair with charges 
$\pm g^{\prime 2}/e$ in the flat-space sphaleron configuration \cite{GVII}.
These examples involve vortices of finite length that are not necessarily straight.
This suggests that while chiral strings are exceptional and non-generic in the form of straight, 
infinite configurations, finite segments of such strings may well exist under realistic
physical conditions.

In summary, the electroweak string described by Nambu,
often identified with the Z-string,
appears to be more accurately interpreted as a chiral string -- 
a configuration that closely resembles a Z-string but includes a condensate in its core.
Nevertheless, a more detailed analysis of this identification remains necessary.

\newpage


\providecommand{\href}[2]{#2}\begingroup\raggedright\endgroup


\end{document}